\documentclass[twoside,11pt,titlepage,a4paper,english]{scrbook}

\usepackage{verbatim}				%
\usepackage[toc,page]{appendix}			%
\usepackage[withpage]{acronym}			%
\usepackage{amsthm}				%

\usepackage{xcolor}					%
\definecolor{DarkRed}{cmyk}{0.20,1,1,0}

\definecolor{FillRed}{cmyk}{0,0.14,0.15,0.05}
\definecolor{FillPurple}{cmyk}{0.03,0.07,0,0.10}
\definecolor{FillGreen}{cmyk}{0.06,0,0.07,0.09}

\definecolor{FillBlue}{cmyk}{0.10,0.05,0,0.18}
\definecolor{FillPetrol}{cmyk}{0.19,0.02,0,0.10}
\definecolor{FillDarkBlue}{cmyk}{0.10,0.05,0,0.18}

\definecolor{LineRed}{cmyk}{0,0.48,0.51,0.33}
\definecolor{LinePurple}{cmyk}{0.11,0.29,0,0.36}
\definecolor{LineGreen}{cmyk}{0.21,0,0.38,0.30}
\definecolor{LineBlue}{cmyk}{0.39,0.25,0,0.27}
\definecolor{LinePetrol}{cmyk}{0.57,0.06,0,0.47}
\definecolor{LineDarkBlue}{cmyk}{0.53,0.26,0,0.64}

\usepackage{geometry}
\usepackage[onehalfspacing]{setspace}

\newcommand{\setwidesite}				%
{
	\fancyhfoffset[LE,RO]{0cm}
	\fancyheadoffset[LO,RE]{0cm}
	\fancyfootoffset[RE]{2cm}
	\newgeometry{inner=3.3cm, outer=3.3cm, bottom=5.4cm, top=3.3cm}
}

\usepackage[hyphens]{url}				%
\usepackage{hyphenat}				%

\addtokomafont{disposition}{\rmfamily}

\usepackage[T1]{fontenc}
\usepackage{cochineal}

\usepackage{textcomp} 				%
\makeatletter						%

\usepackage{tabularx}
\usepackage{booktabs}
\usepackage{multirow}
\usepackage{longtable}				%

\usepackage{amsmath}				%
\usepackage{amssymb}				%
\usepackage{setspace}				%

\usepackage[english]{babel}				%

\usepackage[T1]{fontenc}
\usepackage[utf8]{inputenc}				%

\usepackage[babel,english=american]{csquotes}	%

\usepackage{units}					%

\usepackage{appendix}

\usepackage{listings}

\definecolor{violet}{cmyk}{0.45,0.97,0.27,0.21}
\definecolor{lstblue}{cmyk}{1,0.80,0,0}
\definecolor{lstgreen}{cmyk}{0.71,0.21,0.65,0.22}
\definecolor{bluegrey}{cmyk}{0.56,0.24,0.11,0.05}
\definecolor{javadoc}{cmyk}{0.88,0.59,0,0}
\definecolor{lstgrey}{cmyk}{0.55,0.44,0.42,0.32}

\lstdefinelanguage{SQL}{
     keywords={},
     keywordstyle=\color{bluegrey}\bfseries,
     morekeywords=[2]{CREATE,TABLE,IF,NOT,EXISTS,NULL,SET,DEFAULT,PRIMARY,KEY,COLLATE,CHARACTER,AUTO_INCREMENT,ENGINE,CHARSET},
     keywordstyle={[2]\color{violet}\bfseries},
     otherkeywords={int,varchar,double,text,tinyint},
     sensitive=false,
     morecomment=[l][\color{lstgreen}]{//},
     morecomment=[s][\color{lstgreen}]{/*}{*/},
     morecomment=[s][\color{javadoc}]{/**}{*/},
     morestring=[b]',
     morestring=[b]"
  }
\lstdefinelanguage{PHP}{
     keywords={},
     keywordstyle=\color{bluegrey}\bfseries,
     morekeywords=[2]{static,function,if,return,pow,sin,cos,asin,min,sqrt,int},
     keywordstyle={[2]\color{violet}\bfseries},
     otherkeywords={@param, @returns, @author, @type, @link, @see},
     sensitive=false,
     morecomment=[l][\color{lstgreen}]{//},
     morecomment=[s][\color{lstgreen}]{/*}{*/},
     morecomment=[s][\color{javadoc}]{/**}{*/},
     morestring=[b]',
     morestring=[b]"
  }
\lstdefinelanguage{JavaScript}{
     keywords={},
     keywordstyle=\color{bluegrey}\bfseries,
     morekeywords=[2]{attributes, class, classend, do, empty, endif, endwhile, fail, function, functionend, if, implements, in, inherit, inout, not, of, operations, out, return, set, then, types, while, use},
     keywordstyle={[2]\color{violet}\bfseries},
     otherkeywords={@param, @returns, @author, @type, @link, @see},
     sensitive=false,
     morecomment=[l][\color{lstgreen}]{//},
     morecomment=[s][\color{lstgreen}]{/*}{*/},
     morecomment=[s][\color{javadoc}]{/**}{*/},
     morestring=[b]',
     morestring=[b]"
  }
\lstdefinelanguage{Java}{
     keywords={},
     keywordstyle=\color{bluegrey}\bfseries,
     morekeywords=[2]{abstract,boolean,break,byte,case,catch,char,class,
      const,continue,default,do,double,else,extends,false,final,
      finally,float,for,goto,if,implements,import,instanceof,int,
      interface,label,long,native,new,null,package,private,protected,
      public,return,short,static,super,switch,synchronized,this,throw,
      throws,transient,true,try,void,volatile,while},
     keywordstyle={[2]\color{violet}\bfseries},
     morekeywords=[3]{@SuppressWarnings, @Capability, @Override},
     keywordstyle={[3]\color{lstgrey}},
     otherkeywords={@param, @return, @returns, @author, @link, @see},
     sensitive,
     morecomment=[l]//,
     morecomment=[s]{/*}{*/},
     morecomment=[s][\color{javadoc}]{/**}{*/},
     morestring=[b]",
     morestring=[b]',
  }[keywords,comments,strings]

\lstdefinelanguage{HTML5} {morekeywords={a, abbr, address, area, article, aside, audio, b, base, bb, bdo, blockquote,  body, br, button, canvas, caption, cite, code, col, colgroup, command, datagrid, datalist, dd, del, details, dialog, dfn, div, dl, dt, em, embed, eventsource, fieldset, figure, footer,  form,  h1, h2,  h3,  h4, h5,  h6,  head,  header,  hr, html,  i, iframe,  img,  input,  ins, kbd,  label,  legend,  li,  link,  mark,  map,  menu,  meta,  meter,  nav,  noscript,  object,  ol,  optgroup,  option,  output,  p,  param,  pre,  progress,  q,  ruby,  rp,  rt,  samp,  script,  section,  select,  small,  source,  span,  strong,  style,  sub,  sup,  table,  tbody,  td,  textarea,  tfoot,  th,  thead,  time,  title,  tr,  ul,  var,  video},
sensitive=false, morecomment=[s]{<!--}{-->}, morestring=[b]", morestring=[d]'}

\lstdefinelanguage{CSS} {morekeywords={azimuth,  background-attachment,  background-color,  background-image,  background-position,  background-repeat,  background,  border-collapse,  border-color,  border-spacing,  border-style,  border-top, border-right, border-bottom, border-left,  border-top-color, border-right-color, border-bottom-color, border-left-color,  border-top-style, border-right-style, border-bottom-style, border-left-style,  border-top-width, border-right-width, border-bottom-width, border-left-width,  border-width,  border,  bottom,  caption-side,  clear,  clip,  color,  content,  counter-increment,  counter-reset,  cue-after,  cue-before,  cue,  cursor,  direction,  display,  elevation,  empty-cells,  float,  font-family,  font-size,  font-style,  font-variant,  font-weight,  font,  height,  left,  letter-spacing,  line-height,  list-style-image,  list-style-position,  list-style-type,  list-style,  margin-right, margin-left,  margin-top, margin-bottom,  margin,  max-height,  max-width,  min-height,  min-width,  orphans,  outline-color,  outline-style,  outline-width,  outline,  overflow,  padding-top, padding-right, padding-bottom, padding-left,  padding,  page-break-after,  page-break-before,  page-break-inside,  pause-after,  pause-before,  pause,  pitch-range,  pitch,  play-during,  position,  quotes,  richness,  right,  speak-header,  speak-numeral,  speak-punctuation,  speak,  speech-rate,  stress,  table-layout,  text-align,  text-decoration,  text-indent,  text-transform,  top,  unicode-bidi,  vertical-align,  visibility,  voice-family,  volume,  white-space,  widows,  width,  word-spacing,  z-index},
sensitive=false, morecomment=[s]{/*}{*/}, morestring=[b]", morestring=[d]'}

\lstdefinelanguage{JavaFX} {morekeywords={abstract, after, and, as, assert, at, attribute, before, bind, bound, break, catch, class, continue, def, delete, else, exclusive, extends, false, finally, first, for, from, function, if, import, indexof, in, init, insert, instanceof, into, inverse, last, lazy, mixin, mod, new, not, null, on, or, override, package, postinit, private, protected, public-init, public, public-read, replace, return, reverse, sizeof, static, step, super, then, this, throw, trigger, true, try, tween, typeof, var, where, while, with },
sensitive=false, morecomment=[l]{//}, morecomment=[s]{/*}{*/}, morestring=[b]", morestring=[d]'}

\lstdefinelanguage{MXML} {morekeywords={mx:Accordion, mx:Box, mx:Canvas, mx:ControlBar, mx:DividedBox, mx:Form, mx:FormHeading, mx:FormItem, mx:Grid, mx:GridItem, mx:GridRow, mx:HBox, mx:HDividedBox, mx:LinkBar, mx:Panel, mx:TabBar, mx:TabNavigator, mx:Tile, mx:TitleWindow, mx:VBox, mx:VDividedBox, mx:ViewStack, mx:Button, mx:CheckBox, mx:ComboBase, mx:ComboBox, mx:DataGrid, mx:DateChooser, mx:DateField, mx:HRule, mx:Image, mx:Label, mx:Link, mx:List, mx:Loader, mx:MediaController, mx:MediaDisplay, mx:MediaPlayback, mx:MenuBar, mx:NumericStepper, mx:ProgressBar, mx:RadioButton, mx:RadioButtonGroup, mx:Spacer, mx:Text, mx:TextArea, mx:TextInput, mx:Tree, mx:VRule, mx:VScrollBar, mx:Application, mx:Repeater, mx:UIComponent, mx:UIObject, mx:View, mx:FlexExtension, mx:UIComponentExtension, mx:UIObjectExtension, mx:Fade, mx:Move, mx:Parallel, mx:Pause, mx:Resize, mx:Sequence, mx:WipeDown, mx:WipeLeft, mx:WipeRight, mx:WipeUp, mx:Zoom, mx:EventDispatcher, mx:LowLevelEvents, mx:UIEventDispatcher, mx:CurrencyFormatter, mx:DateFormatter, mx:NumberFormatter, mx:PhoneFormatter, mx:ZipCodeFormatter, mx:CursorManager, mx:DepthManager, mx:DragManager, mx:FocusManager, mx:HistoryManager, mx:LayoutManager, mx:OverlappedWindows, mx:PopUpManager, mx:SystemManager, mx:TooltipManager, mx:CreditCardValidator, mx:DateValidator, mx:EmailValidator, mx:NumberValidator, mx:PhoneNumberValidator, mx:SocialSecurityValidator, mx:StringValidator, mx:ZipCodeValidator, mx:DownloadProgressBar, mx:ArrayUtil, mx:ClassUtil, mx:Delegate, mx:ObjectCopy, mx:URLUtil, mx:XMLUtil, mx:CSSSetStyle, mx:CSSStyleDeclaration, mx:CSSTextStyles, mx:StyleManager, mx:HTTPService, mx:RemoteObject, mx:Service},
sensitive=false, morecomment=[s]{<!--}{-->}, morestring=[b]", morestring=[d]'}

\lstdefinelanguage{LZX} {morekeywords={a, alert, animator, animatorgroup , attribute, audio , axis, axisstyle , b, barchart, basebutton , basebuttonrepeater , basecombobox , basecomponent , basedatacombobox , basedatepicker , basedatepickerday , basedatepickerweek , basefloatinglist , basefocusview , baseform , baseformitem , basegrid , basegridcolumn , baselist , baselistitem , basescrollarrow , basescrollbar , basescrollthumb , basescrolltrack , baseslider , basestyle , basetab , basetabelement , basetabpane , basetabs , basetabsbar , basetabscontent , basetabslider , basetrackgroup , basetree , basevaluecomponent , basewindow , br , button , canvas , chart , chartbgstyle , chartstyle , checkbox , class , columnchart , combobox , command , connection , connectiondatasource , constantboundslayout , constantlayout , datacolumn , datacombobox , datalabel , datamarker , datapath , datapointer , dataselectionmanager , dataseries , dataset , datasource , datastyle , datastylelist , datatip , datepicker , debug , dragstate , drawview , edittext , event , face , floatinglist , font , font , form , frame , grid , gridcolumn , gridtext , handler , hbox , horizontalaxis , hscrollbar , i , image , img , import , include , inputtext , javarpc , label , labelstyle , layout , legend , library , linechart , linestyle , list , listitem , LzTextFormat , menu , menubar , menuitem , menuseparator , method , modaldialog , multistatebutton , node , p , param , piechart , piechartplotarea , plainfloatinglist , plotstyle , pointstyle , pre , radiobutton , radiogroup , rectangularchart , regionstyle , remotecall , resizelayout , resizestate , resource , reverselayout , richinputtext , rpc , script , scrollbar , security , selectionmanager , sessionrpc , simpleboundslayout , simpleinputtext , simplelayout , slider , soap , splash , stableborderlayout , state , statictext , style , submit , swatchview , SyncTester , tab , tabelement , tabpane , tabs , tabsbar , tabscontent , tabslider , Test , TestCase , TestResult , TestSuite , text , textlistitem , tickstyle , tree , u , valueline , valuelinestyle , valuepoints , valuepointstyle , valueregion , valueregionstyle , vbox , verticalaxis , view , view , vscrollbar , webapprpc , window , windowpanel , wrappinglayout , XMLHttpRequest , xmlrpc , zoomarea},
sensitive=false, morecomment=[s]{<!--}{-->}, morestring=[b]", morestring=[d]'}

\usepackage{enumitem}
\setdescription{font=\color{DarkRed}\rmfamily\itshape}

\deffootnote[1em]{1em}{1em}{\textsuperscript{\thefootnotemark}\,}

\usepackage{graphicx}
\graphicspath{{images/}}				%
\usepackage[format=hang,font={small,it}]{subcaption}
\usepackage{eso-pic}					%
\usepackage{chngpage}				%
\usepackage{tikz}					%
\usepgflibrary{arrows}

\usetikzlibrary{arrows.meta}
\usepackage{pgfplots}
\pgfplotsset{compat=newest}
\pgfplotsset{every axis plot/.append style={very thick}}
\pgfplotsset{every axis/.append style={
					very thick,
                    }}

\usepackage{float}
\usepackage{placeins}

\usepackage{calc}
\usepackage{fancyhdr}
\fancyhfoffset[RO,LE]{0.1cm} %
\fancyhfoffset[RE,LO]{0.1cm}

\usepackage[T1]{fontenc}
\usepackage{cabin}
\newcommand{\allparagraphformat}{\color{DarkRed}\rmfamily\normalfont}
\newcommand{\allsectionformat}{\rmfamily\cabin}

\addtokomafont{part}{\Huge\allparagraphformat}
\addtokomafont{chapter}{\Huge\allsectionformat}
\addtokomafont{section}{\allsectionformat}
\addtokomafont{subsection}{\allsectionformat}
\addtokomafont{subsubsection}{\allsectionformat}
\addtokomafont{paragraph}{\allparagraphformat}
\addtokomafont{subparagraph}{\allparagraphformat}

\RedeclareSectionCommand[
  beforeskip=-.75\baselineskip,
  afterskip=.5\baselineskip]{section}

\RedeclareSectionCommand[
  beforeskip=-5\baselineskip,
  afterskip=.5\baselineskip]{chapter}

\addtokomafont{section}{\LARGE}
\addtokomafont{subsection}{\large}

\usepackage[doi=false, url=false]{biblatex}
\AtEveryBibitem{\clearfield{pages}} 

\usepackage{url}
\usepackage{acronym}
\usepackage{makecell}
\usepackage{tablefootnote}
\usepackage[a-2b]{pdfx}
\usepackage{amssymb}%
\usepackage{pifont}%
\newcommand{\cmark}{\ding{51}}%
\newcommand{\xmark}{\ding{55}}%
\newcommand{\omark}{\ding{108}}%
\usepackage{color, colortbl}
\definecolor{Gray}{gray}{0.9}
\usepackage{adjustbox}

\def\thickhrulefill{\leavevmode  
\leaders \hrule height 1pt \hfill \kern \z@}
\def\@makechapterhead#1{%
  {\parindent \z@ \centering \cabin%
       \color{black}
       \Huge \bfseries #1\par\nobreak
       \vspace*{10\p@}%
       \normalfont \small
       \color{DarkRed} \thickhrulefill\quad
       \color{DarkRed} \scshape \@chapapp{} \thechapter
       \color{DarkRed} \quad \thickhrulefill
       \par\nobreak
       \interlinepenalty\@M
  }
}
\def\@makeschapterhead#1{%
  {\parindent \z@ \centering  \cabin \color{DarkRed} %
        \hrule
        \vspace*{10\p@}%
        \Huge \bfseries #1\par\nobreak
        \par
        \vspace*{10\p@}%
        \hrule
    \vskip 10\p@
  }}
\usepackage[tight]{minitoc}

\nomtcrule
\mtcsetfeature{minitoc}{after}{\color{DarkRed}\hrule height 1pt \vspace*{15\p@}}
\usepackage{microtype}
\usepackage{hyperxmp}
\hypersetup{
  pdftitle={Real-Time Aware IP-Networking for Resource-Constrained Embedded Devices},
	pdfauthor={Ilja Behnke},
	pdfkeywords={real-time, cyber-physical systems, networking, embedded systems, industrial internet of things, distributed scheduling},
	pdfsubject={dissertation on real-time network packet processing},
  pdflang={en},
  pdfcopyright={CC BY 4.0},
  pdflicenseurl={https://creativecommons.org/licenses/by/4.0/},
  colorlinks = {true},
  linkcolor = {black}, %
  urlcolor  = {black}, %
  citecolor = {black}, %
  pdftex,
  pdfpagelabels,
  bookmarks,
  hyperindex,
  hyperfigures,
  bookmarksnumbered
}

\makeatletter
\def\@textbottom{\vskip \z@ \@plus 1pt}
\let\@texttop\relax

\def\@listI{\leftmargin\leftmargini
            \parsep 4\p@ \@plus2\p@ \@minus\p@
            \topsep\z@
            \itemsep4\p@ \@plus2\p@ \@minus\p@}
\makeatother

\usepackage{paralist}
\usepackage[nottoc,numbib]{tocbibind}
\usepackage{todonotes}

\newtheorem{mydef}{Definition}
\begin{document}

\frontmatter

\begin{titlepage}
    \setwidesite{
    \begin{center}
    \large
    
    \hfill
    \vfill
    
    \Huge
		\begin{spacing}{.9}
			\cabin{Real-Time Aware IP-Networking\\ for Resource-Constrained\\ Embedded Devices}\\
		\end{spacing}
    \large
    \bigskip
    \bigskip
    
    submitted by \\
    \bigskip
    Ilja Behnke, M.Sc. \\
    \bigskip
    \bigskip
    to the Department for Computer Science and Electrical Engineering \\
    Technische Universität Berlin \\
    for the attainment of the academic degree \\
    \bigskip
    	
    - Dr.-Ing. - \\
    \bigskip

    accepted dissertation

    \end{center}
    \bigskip
    \bigskip
    \large
    \noindent Doctoral committee: \\
    \newline
    Chair: Prof. Dr. Marianne Maertens \\
    Examinor: Prof. Dr. Odej Kao \\
    Examinor: Prof. Dr. Geir Mathisen \\
    Examinor: Prof. Dr. Carsten Griwodz
    \newline\\
    Day of the scientific defense: June 3, 2024 \\[2\baselineskip]

    \begin{center}
    \large Berlin, 2024
    \end{center}
    }
    \end{titlepage}

\chapter*{Acknowledgments}
\label{cha:acknowledgments}

A great number of people have accompanied me during my research for this thesis. I was fortunate to be surrounded by many bright people who inspired me during discussions, supported me during difficult times, and helped me realize ideas.

I would like to thank Odej Kao for supervising me and always keeping me on track and focused on what is relevant. Without the conversations and discussions with my colleagues, I would have lost the motivation to do this research. Sören Becker was a great office neighbor and friend. Philipp Wiesner pushed me to do better, created beautiful plots for our publications, and brightened my days outside and inside the office. Lauritz Thamsen mentored me in the early years of my research, taught me how to write scientific papers, and provided structure during the COVID pandemic. Special thanks to Jana Bechstein, who helped me with everything I brought to her attention and saw things through until they were solved. 
My colleagues set a high standard for my future workplaces and were the reason I enjoyed coming to the office every day (except Fridays). Thank you Dominik Scheinert, Jonathan Bader, Jonathan Will, Thorsten Wittkopp, Jasmin Bogatinovski, Kevin Styp-Rekowski and Alexander Acker for creating this atmosphere. Thanks also to whoever keeps my office plants alive.

Meaningful research is not the product of one person's work alone. I had the support of my colleagues and students, who helped me a lot with both the design of my publications and the implementation of the simulators. I am especially grateful to Leonard Weikopf and Kalin Iliev, who helped me implement prototypes and evaluate them. I would also like to thank my co-authors Henrik Austad, Christoph Blumschein, Paul Voelker, and Robert Danicki for their brilliant work. 

Finally, staying sane during the sometimes stressful times of completing a doctorate is not possible without a strong social backbone. I would like to thank my parents, Anja and Malte, and my sister Lea for always being there. Thanks also to all my friends who make my life extremely enjoyable and adventurous.

\chapter*{Abstract}
\label{cha:abstract}

This dissertation explores the area of real-time IP networking for embedded devices, especially those with limited computational resources. With the increasing convergence of information and operational technologies in various industries, and the growing complexity of communication requirements in (semi-)autonomous machines, there is a need for more advanced and reliable networking solutions. This research focuses on the challenge of integrating real-time embedded devices into packet-switched networks.

Through a comprehensive review of current real-time communication technologies, standards, and practices in the context of Industry 4.0, a notable gap is identified: the lack of a robust real-time communication standard tailored for wireless mobile machines, and insufficient research on real-time embedded devices in highly networked environments. The study includes detailed experimentation with commercially available off-the-shelf networked microcontrollers, revealing a priority inversion problem where network packet processing interrupts real-time tasks, potentially causing real-time violations.

To address this challenge, this thesis proposes mitigation methods and system designs that include software and hardware implementations. These include a new embedded network subsystem that prioritizes packet processing based on task priority, and a real-time-aware network interface controller that moderates interrupt requests. In addition, a hybrid hardware-software co-design approach is developed to ensure predictable and reliable real-time task execution despite network congestion.

Furthermore, the research extends to task offloading in wireless Industrial Internet of Things environments, presenting a system architecture and scheduler capable of maintaining real-time constraints even under heavy loads and network uncertainties.
\chapter*{Zusammenfassung}

Diese Dissertation untersucht den Bereich der Echtzeit-IP-Vernetzung für eingebettete Geräte, insbesondere solche mit begrenzten Rechenressourcen. Angesichts der zunehmenden Konvergenz von Informations- und Betriebstechnologien in verschiedenen Branchen und der wachsenden Komplexität der Kommunikationsanforderungen in (halb-)autonomen Maschinen besteht ein Bedarf an fortschrittlicheren und zuverlässigeren Netzwerklösungen. Diese Forschungsarbeit konzentriert sich auf die Herausforderung der Integration von echtzeitfähigen eingebetteten Geräten in paketvermittelte Netzwerke.

Durch eine umfassende Auswertung aktueller Echtzeit-Kommunikationstechnologien, Standards und Praktiken im Kontext von Industrie 4.0 wird eine wesentliche Lücke identifiziert: das Fehlen eines robusten Echtzeit-Kommunikationsstandards, der auf drahtlose mobile Maschinen zugeschnitten ist, und unzureichende Forschung zu eingebetteten Echtzeit-Geräten in stark vernetzten Umgebungen. Die Studie umfasst detaillierte Experimente mit handelsüblichen vernetzten Mikrocontrollern, die ein Problem der Prioritätsumkehrung aufzeigen, bei dem die Verarbeitung von Netzwerkpaketen Echtzeitaufgaben unterbricht, was zu dem Verpassen von Deadlines führen kann.

Um dieses Problem zu lösen, werden in dieser Arbeit Methoden und Systemdesigns vorgestellt, die sowohl Software- als auch Hardware-Implementierungen umfassen. Dazu gehören ein neues eingebettetes Netzwerk-Subsystem, das die Paketverarbeitung auf der Grundlage der Aufgabenpriorität priorisiert, und ein echtzeitfähiger Netzwerkschnittstellen-Controller, der Interrupt-Anfragen mäßigt. Darüber hinaus wird ein hybrider Hardware-Software-Co-Design-Ansatz entwickelt, der eine vorhersehbare und zuverlässige Echtzeit-Task-Ausführung trotz Netzwerküberlastung gewährleistet.

Weitergehend befasst sich die Forschung mit der Auslagerung von Aufgaben in drahtlosen Umgebungen des industriellen Internets der Dinge und stellt eine Systemarchitektur und einen Scheduler vor, die in der Lage sind, Echtzeitbedingungen auch bei hoher Belastung und Netzwerkunsicherheiten einzuhalten.
\dominitoc[n]
\begin{singlespace}
  \tableofcontents{}
\end{singlespace}

\mainmatter
\chapter{Introduction}
\label{cha:1_1_introduction}
\minitoc

Embedded devices that control machines in the physical world have been part of industrial processes as well as home and automotive appliances for decades~\cite{microcontrollers, robots, indusTempMonit}. In such scenarios, real-time systems are used to enable the digital control of physical movements. In contrast to general-purpose computing, these systems need to fulfill timing constraints in a predictable manner. 
Real-time systems are computing systems designed to process and respond to data with precise timing constraints, which is crucial for industrial automation and control applications. These systems ensure that critical tasks are executed within specified time intervals, guaranteeing reliability and accuracy. Examples include \acp{plc} in manufacturing lines, Distributed Control Systems in chemical plants, and \acp{mcu} in industrial robots. In short, anywhere, where timely and accurate responses to external triggers are essential for operational success~\cite{wang2017real}.
Machines such as robot arms, logistic robots, vehicles, and conveyor belts not only have timing requirements for propper functionality but also need to reliably guarantee control task deadlines for safety purposes. Depending on the exact use-case, reliable reaction times in the milli and nano second ranges are required to prevent catastrophic consequences~\cite{schweitzer2016millisecond, hamdan2022using, paventhan2022latency}.

The advent of the Fourth Industrial Revolution (Industry 4.0), the \ac{iiot}, and autonomous vehicles has ushered in a transformative era where interconnected embedded real-time systems play a pivotal role in shaping the landscape of modern industries. Many novel applications involve deploying a large number of devices, making cost-effectiveness a vital requirement. For devices deployed in mobile machines or remote locations their power consumption also plays a significant role, leading to longer CPU cycles and slower memory access. Another factor leading to this is the oftentimes small form factor required by embedded devices. 
Yet, the simplicity of embedded real-time devices is also due to the safety-criticality of their applications. Traditionally, a device is not used for more than one specific control or monitoring task. This specificity leads to simpler architectures and better verifiability~\cite{yen1998performance}. Hence, even when cost or energy-efficiency are of no concern, using more complex devices with advanced parallelization or heterogeneous architectures might not be a feasible option. Therefore, the embedded devices used in real-time systems are oftentimes resource contrained and computationally weak.

The availability of cost-effective microcontrollers as well as advances in communication and artificial intelligence technologies enables scenarios requiring flexible connectivity and high data rates in real-time contexts leading to a convergence between \ac{it} and \ac{ot}~\cite{ehie2020understanding}. This convergence is driven by the increasing need for seamless communication and data exchange between traditionally isolated IT systems (handling enterprise-level data) and OT systems (managing industrial processes and control). The goal is to create a unified ecosystem that enhances operational efficiency, facilitates data-driven decision-making, and enables a more responsive and agile industrial infrastructure~\cite{aakerberg2021future}. Factors such as Industry 4.0 initiatives, the rise of smart manufacturing, and the demand for real-time insights have catalyzed the convergence of IT and OT, breaking down silos and fostering a more connected and intelligent industrial landscape.

Examples for IT/OT convergence are applicable across diverse industries, such as:

\begin{description}
    \item[Smart Manufacturing] where IT/OT convergence enables the seamless integration of enterprise resource planning systems with production processes. This integration allows for real-time monitoring of manufacturing operations, predictive maintenance, and adaptive production planning~\cite{ehie2020understanding}.
    \item[Energy Systems] where IT solutions for data analytics can be integrated with OT systems controlling energy production and distribution. This convergence facilitates better monitoring of energy consumption, predictive maintenance of equipment, and optimization of energy efficiency~\cite{ahmad2021using}.
    \item[Public Infrastructures] where IT/OT convergence can be applied to enhance the management of water treatment plants, transportation systems, and public utilities~\cite{martinez2020use}. 
    \item[Healthcare Systems] where IT/OT convergence integrates electronic health records and operational processes within hospitals. This integration ensures secure and seamless communication between IT systems managing patient data and OT systems controlling medical equipment~\cite{kodali2015implementation}.
    \item[Automated Agriculture] where IT/OT convergence involves integrating information systems with farming equipment and processes for real-time monitoring of crop conditions, automated irrigation systems, and data-driven decision-making for optimal resource utilization~\cite{sahoo2023integration}.
\end{description}

The given examples show how new advancements in many industries lead to or are dependent on a flexible connection between real-time embedded systems and management networks. Since all bigger networks utilize the \ac{ip} and are connected to the internet, this means that for most new scenarios the real-time systems of the OT-layer are also connected via IP~\cite{xu2018survey}. Hence, packet-switched IP networking is or might be used for communciation between sensors, actuators, and controling components in safety-relevant real-time systems like \textbf{Power Plants} for stability control, \textbf{Medical Devices} for life support systems and robotic surgery, and \textbf{Industrial Control Systems} for remote control of moving machinery.

As a packet-switched communication technology, IP is not designed to provide timing predictability. The larger and more complex a network becomes, the higher the variance of latencies gets. While new technologies enable a new scenarios, a number of challenges has to be overcome before many of the envisioned scenarios in Industry 4.0 and autonomous machinery can be achieved.
Additionally, the paradigm of IT/OT convergence might introduce new safety and security risks to real-time systems. 

\section{Problem Statement}
Integrating IP networking into resource-constrained real-time embedded systems introduces complexity and unpredictable communication overhead. Furthermore, the introduction of \acp{nic} and IP subsystems to embedded real-time devices creates computational workload and a new input sources to secure and handle. 
This thesis focuses on scientific findings and approaches in the area of IP communication in resource-constrained real-time systems. The objective of this thesis is to

\begin{center}
    \textit{"Mitigate the real-time violation of IP networking in resource-constrained embedded systems and provide methods for real-time aware networking for them."}
\end{center}

This is addressed by decomposing the objective into the following four components.

\begin{compactenum}
    \item \textbf{Analysis and testing.} Incoming network packets are received and processed by computationally weak embedded devices. Real-time developers and researchers must be able to test their systems and quantify packet processing workloads under arbitrary network loads. Tipping points for real-time violations must be easily identifiable.
    \item \textbf{Run-time detection and mitigation.} Because network packets arrive in arbitrary intervals, real-time operating systems must be able to detect and respond to dangerous packet loads during run-time. This requires the identification of appropriate detection metrics and methods to reduce the amount of packet processing without disabling network tasks required for system functionality.
    \item \textbf{Real-time aware networking.} Existing \acp{nic} and network stack implementations for embedded systems are designed to be both simple and resource efficient. Although they are typically combined with \acp{rtos}, they lack real-time coordination. The resulting preemption of real-time tasks can lead to real-time violations.
    Therefore, network subsystems on embedded real-time devices need to be made real-time aware and protected against network induced overloading.
    \item \textbf{Distribution.} To overcome the limitations of computationally weak embedded devices, it is common practice to offload tasks to more powerful local computers. However, meeting real-time requirements in distributed systems is challenging, especially given the unpredictability of latencies in wireless networks with mobile devices. Managing real-time processes in distributed systems that include embedded devices, edge servers, and cloud computing requires a holistically designed architecture.
\end{compactenum}

\section{Contribution}
This thesis proposes methods and architectures to protect resource-constrained real-time systems from IP networking-caused timing violations. Concretely, the contributions are:

\paragraph*{1. Methods for the analysis of real-time performance of networked embedded systems.}
The effect of processing loads on real-time performance caused by packet reception is difficult to quantify due to the arbitrary nature of packet reception and largely unstudied. We present a method for research and development of networked embedded systems to analyze the timing impact of packet processing. The method is applied to a computationally weak embedded system running a real-time operating system in various experimental setups to present a quantitative correlation between packet rate, schedulable processing time, and unschedulable \ac{isr} executions.

\paragraph*{2. Algorithms for the detection and mitigation of network-caused real-time violations.}
This thesis introduces a network overload detection method and four reactive algorithms as an initial measure towards establishing a safer connection of resource constrained real-time devices to IP networks. These algorithms adaptively limit packet processing rates to protect critical real-time tasks from missing their deadlines.

\paragraph*{3. Approaches for real-time aware packet reception and processing.}
Many embedded system deployment scenarios require low latency and timely response to sensor or control signals. Indiscriminately limiting packet processing can contradict this requirement. We develop an embedded real-time aware network subsystem that protects task deadlines while ensuring timely processing of time-critical packets. It uses early packet demultiplexing and classification based on real-time task priorities. By inheriting the priority of the packet-associated process, the network driver is prevented from preempting higher-priority tasks. Additional filtering and discriminatory rate limiting protects the system from network processing overload. 
A multiqueue real-time aware \ac{nic} is implemented that demultiplexes and classifies received packets before \acp{irq} are invoked. Packets are placed into priority-specific ingress queues with queue-specific interrupt moderation parameters. Interrupt moderation reduces the number of \acp{irq} that are generated while prioritizing the timeliness of critical packets. Additional filtering of unwanted packets further reduces unnecessary \acp{irq} and \ac{isr} executions.

\paragraph*{4. Approaches for distributed real-time task offloading and scheduling.} 
Embedded real-time systems are used in mobile, semi-autonomous machines connected over wireless networks. Due to resource constraints, real-time tasks may need to be offloaded to more powerful local machines. We present a distributed system architecture that includes a partitioned \ac{edf}-based real-time scheduler. Network latencies and their uncertainties, as well as current capacity utilization, are considered in scheduling to ensure timely result delivery and task rejection, respectively.

\paragraph*{}

The research presented in this thesis has partly been published in the following peer-reviewed publications.
                                                                 
\sloppy{                                                                                                                                
    \begin{refsection}                                                                                                                      
    \nocite{goebel2018application, behnke2019hector, lorenz2020fingerprinting, beilharz2021towards, lorenz2020scalable, behnke2020interrupting, bender2021pieres, wiesner2021let, beilharz2021continuously, toll2022iotreeplay, behnke2023towards, behnke2023offloading, behnke2023realtime, behnke2022priority, danicki2021detecting, blumschein2022differentiating}  
    \printbibliography[heading=none] %
    \end{refsection}                                                               
}
\newpage
\section{Outline of this Thesis}
This doctoral thesis makes contributions towards a safer deployment of IP networking within real-time systems. It is structured as follows.
\begin{description}
    \item[Chapter 2] provides the theoretical and technical background required to follow this thesis. %
    \item[Chapter 3] performs a survey of industrial use-case scenarios enabled by recent advances in communication and computation technologies. The scenarios are classified and their real-time networking requirements are derived. A review of standardized communication technologies for industrial networks is performed and their feasibility for the scenario classes is examined. Following, recent research in the field of real-time networking are surveyed and put into perspective of the derived real-time networking requirements. %
    \item[Chapter 4] presents a method for the quantitative analysis of network packet reception impact on embedded real-time devices. In multiple experiment setups the real-time behavior of computationally weak embedded devices is evaulated. %
    \item[Chapter 5] presents software-based methods for the mitigation of the previously quantified networking time impact. It addresses the detection of deadline-violating packet loads and their mitigation. %
    Further, an IP subsystem for real-time aware packet reception and processing is designed and evaluated.
    \item[Chapter 6] presents a hardware extension that protects the system from real-time violating interrupt requests. A multiqueue \ac{nic} design is provided that differentiates incoming packets by mapping IP flows to real-time processes running on the system. A \ac{nic} simulator is implemented and used to test the viability of the design. 
    \item[Chapter 7] presents a hardware/software co-design unifying the approaches of the previous two chapters. The individual designs are modified to interlock. The co-design is further implemented on an FPGA-SoC. 
    \item[Chapter 8] presents a system architecture for offloading real-time tasks from embedded wireless machines to local edge servers. A scheduler based on partitioned EDF is proposed that takes network latencies and uncertainties into account and distributes the offloaded tasks to worker machines so that their deadlines can be met. The architecture is distributed to remove the single point of failure of a centralized scheduler.
    \item[Chapter 9] gives an extended overview of the related work to this thesis. The focus lies on interrupt management in real-tie systems, real-time aware packet processing, and distributed real-time scheduling as well as networking in industrial settings.
    \item[Chapter 10] concludes this thesis and proposes directions for future research. 
\end{description}

\chapter{Background}
\label{cha:1_2_background}
\minitoc
The underlying infrastructure supporting modern embedded systems, especially in the field of industrial applications, is based on the integration of \acp{rtos} and sophisticated embedded networking subsystems. These components form a technological foundation of the \ac{iiot}, enabling seamless communication and coordination among interconnected devices. This chapter explores the interplay between \acp{rtos}, embedded networking subsystems, and IIoT considerations. 
Section~\ref{sec:rtos} provides the technical background of real-time operating systems before delving into the scheduling of real-time and aperiodic tasks and interrupts. 
Section~\ref{sec:rx_path} explains the receive path of embedded networking subsystems, starting at the network interface and ending with embedded IP stack implementations. 
Section~\ref{sec:1_2_iiot} covers some of the \ac{iiot} considerations relevant to this thesis.

\section{Real-Time Operating Systems}
\label{sec:rtos}
Embedded systems often have special timing requirements. The correctness of these so-called real-time systems depends not only on the computation results, but also on the time at which the computation is completed. Tasks running on such systems have deadlines that must be met to avoid undesirable or even catastrophic behavior. For example, it is undesirable for a system to fail to respond to user input in a timely manner, but if the airbag controller of a car misses the deadline to deploy the airbag, the consequences can be severe~\cite{distributedRTOS}.
Given these special requirements, an \ac{rtos} can be distinguished from general-purpose \acp{os} by a number of unique characteristics~\cite{RTOScharacteristics}:

\begin{compactitem}
    \item Determinism
    \item Responsiveness
    \item Reliability
    \item User Control
    \item Fail-soft Operation 
\end{compactitem}

The system must be \textit{deterministic}, meaning that the execution time of operations is known and independent of external factors. \textit{Responsiveness} is the ability to respond quickly to an event, such as an I/O interrupt. In real-time systems, interrupt latency, i.e. the time between the generation of the interrupt signal and the execution of the interrupt service routine, is therefore a key feature. Furthermore, the need for high \textit{reliability} in \acp{rtos} requires measures to avoid system failures and reboots. In general-purpose \acp{os}, \textit{user-control} is very limited by the kernel, but in \acp{rtos} fine-grained control over task priorities and memory management is required. Finally, \acp{rtos} strive for so-called \textit{Fail-soft operation}. While general-purpose \acp{os} typically respond to critical failures by terminating system execution, \acp{rtos} should isolate the problem and minimize its effects if possible~\cite{RTOScharacteristics}.
Ultimately, these characteristics cannot be fully satisfied, but much effort has been made to develop \acp{rtos} that come close. 

\paragraph*{Memory Management}
General purpose \acp{os} usually support virtual memory, including memory protection. Embedded and \acp{rtos} do not always support these features. 
Memory protection is the concept of keeping address spaces separate between different contexts, such as the user space and the kernel space, or between different users. Without memory protection, \acp{rtos} often do not differentiate between user and kernel space because embedded applications must be highly trustworthy and efficient. However, some \acp{rtos} support memory protection (e.g. LynxOS) or make memory protection an option when configuring the RTOS (e.g. VxWorks)~\cite{LiuJane}.
When using virtual memory, the physical memory is divided into frames of fixed length, each of which is mapped to a virtual page. The sum of all virtual pages is the virtual memory. Applications can only access virtual addresses, which are then translated into physical addresses by specialized hardware. This introduces an abstraction between the actual layout of memory space and the application's view of it. While this allows for more efficient use of available memory, it introduces computational overhead and requires a hardware \ac{mmu}. It is therefore often not implemented in \acp{rtos}~\cite{distributedRTOS}.
In addition, caches introduce uncertainty in memory access times and are therefore either not used in hard real-time systems or the worst-case memory access time must be known~\cite{rtcache}.

\subsubsection*{FreeRTOS}
One of the prevalent open-source \acp{rtos} available is FreeRTOS\footnote{FreeRTOS Kernel Developer Docs and Kernel Secondary Docs, \url{https://www.freertos.org/}}. Initially developed by Richard Barry in 2003, the system is designed for low-power embedded devices with emphasis placed on a small size as well as easy maintainability and portability.

Its scheduler supports preemptive or co-operative multitasking, with determinism achieved through user-assigned task priorities. The scheduler runs on a fixed internal tick rate, relying on a tick interrupt for performing task switches, which are based on the current task states and their assigned priorities. Strictly abiding by these priorities, a running task will only be preempted by a higher-priority task in ready state. Furthermore, a task will transition to the blocked state when required to wait for a shared resource or when a blocking API function is called. These include API calls around events, semaphores, queues and notifications. In case multiple tasks with the same priority are ready to transition into the running state, the scheduler will employ a round-robin scheme to schedule each one for a certain time period. Conversely, when no task is ready, FreeRTOS will switch the idle task into running mode, which is the the default task with the lowest-priority (priority 0). Furthermore, FreeRTOS offers high flexibility in assignment and modification of task priorities, enabling changes to the priorities of tasks to be made during runtime.

\subsection{Real-Time Scheduling}
As is typical in scheduling theory, we classify tasks by their timing requirements into~\cite{LiuJane}:

\begin{compactitem}
    \item \emph{hard real-time} tasks, whose utility depends entirely on their results being delivered on time. They are critical to the operation of the system, and missing deadlines can have catastrophic effects.
    \item \emph{firm real-time} tasks, whose utility also drops to zero if their completion is delayed, even though they are not critical to the overall system.
    \item \emph{soft real-time} tasks, which have timing requirements where the value of a delayed result degrades more continuously.
    \item regular ("best-effort") tasks. Timing is still be somewhat desirable, but only as an end in itself.
\end{compactitem}
In the literature, we find varying definitions of firm real-time tasks, sometimes lumped into a broader notion of soft real-time tasks due to their low criticality, and sometimes referred to as non-critical hard real-time tasks.
In addition, we use the following terms to describe different periodicities: 

\begin{compactitem}
    \item \emph{periodic tasks} are executed at regular intervals, their period $P_i$. In addition, they may have a deadline before the end of their period.
    \item \emph{aperiodic tasks} as the opposite, appear irregularly at arbitrary points in time.
    \item \emph{sporadic tasks} are somewhere in between the first two types. As aperiodic tasks, they do not arrive at regular times. However, there is a lower bound on their inter-arrival time, the minimum interval between two occurrences.
\end{compactitem}

Over the last decades there evolved a multitude of scheduling strategies. Popular scheduling algorithms are e.g. \ac{rms} and \ac{edf}. In \ac{rms} tasks with a short period are assigned a high priority, while in \ac{edf} the task with the earliest deadline are assigned the highest priority~\cite{rtsched}.

\subsubsection*{Fixed-Priority Scheduling}
\label{sec:fixed}
In a fixed-priority scheduling strategy, each task is assigned a static or at least not frequently changing priority in a totally ordered space $P$, partitioning the set of tasks $T$ into a prioritized task set $(T_p)_{p \in P}$.
At each time, the scheduler then simply selects a ready task with the highest priority to run on the \ac{cpu}.
For the usually sufficient priority spaces that do not exceed the bit width of a machine word, the necessary priority queue data structure can be implemented in a way that allows deterministic constant-time access using only a few machine code instructions.
As a result, fixed-priority scheduling is widely used in practice, particularly in the domain of constrained embedded devices.

When a feasible schedule for a set of critical tasks is desired, \ac{rms}~\cite{liu1973scheduling} provides an optimal strategy using fixed priorities:
If there is a feasible schedule using fixed priorities, simply assigning priorities inversely to the task durations yields one.
Note that these assigned priorities do not necessarily correspond to the criticalities of the tasks, since every task in the set is certain to meet its deadline as long as the periods and \ac{wcet} assumptions are obeyed.
There are many theories about feasibility analysis.
For example, the famous sufficient condition of Liu and Leyland~\cite{liu1973scheduling} gives a lower bound of $\ln(2)$ on the compound utilization at which every possible set of tasks is schedulable.
However, an exact analysis can be tedious, especially when there are many inhomogeneous periods.
For this reason, and due to the fact that the \ac{wcet} analysis itself gives loose upper bounds, in practice we are usually left with a lot of unused idle time.

It is important to note that in a set of prioritized tasks, static feasibility analysis can be useful even though the task set as a whole is not schedulable.
By mathematical construction, meeting the deadline requirements of a task depends only on the set of tasks with higher or equal priority.
Thus, in a prioritized task set $(T_p)_{p \in P}$ there exists a bound $p_\text{feasible}$ such that the set of tasks with higher priority $\bigcup_{p \geq p_\text{feasible}} T_p$ has a particular schedule.

A secondary prioritization approach suitable for the tasks below this boundary is to naively assign priorities according to task criticality.
Without feasibility guarantees, this is still a legitimate approach for less critical soft real-time tasks.
The higher the priority of such a task, the lower the risk that the \ac{cpu} is already overbooked. 

\subsection{Aperiodic Scheduling}
\label{sec:aperiodic}
When dealing with sporadic tasks, a simple approach is to treat them as if they were regular periodic tasks, assigning their minimum inter-arrival time as their period, thus preparing for the worst case.
As lower inter-arrival times converge to the characteristics of an aperiodic task, this worst-case budget increases inversely, making such treatment infeasible.
While aperiodic tasks are also given deadlines in the literature \parencite{sha2004real}, these are impossible to meet in the general case due to the unpredictable nature of aperiodicity.

One approach to integrating aperiodic tasks into fixed-priority scheduling is to use so-called server tasks.
To the scheduler, these behave like normal prioritized tasks.
Unlike other tasks, they do not have an individual goal.
Instead, they use their budget to support the execution of aperiodic jobs.
Because of their limited budget in each period, they can be easily included in scheduling considerations.

\paragraph*{Deferrable Servers}
A very simple but effective aperiodic server is the deferrable server~\cite{strosnider1995deferrable}.
It has a limited \ac{cpu} time budget to serve aperiodic events.
When the budget is exhausted, it pauses execution until the next renewal.
At the end of each period its budget is restored to the initial amount.
A big advantage is the simplicity of the mechanism and therefore of an implementation for this server scheme.
Yet, the deferral of budget consumption incurs on a higher worst-case processing demand than one budget per period, due to possible back-to-back execution patterns.

Let $p$ be the server period and $e$ its execution budget for a period.
There may arrive jobs just before the end of a period consuming the whole capacity $e$ of the server for this period.
With the start of the next period and the consequent budget replenishment, another duration $e$ may be serviced to jobs.
In the worst case, we need to expect one extra execution budget. Thus, the highest possible server demand $d(\Delta)$ inside an arbitrary interval can be indicated as
\[ d(\Delta) = e \cdot \left(\left\lceil \frac{\Delta}{p} \right\rceil + 1\right) \]
Note that at least for a small period $p$, the CPU bandwidth $\frac{d(\Delta)}{\Delta}$ still approaches the theoretical server optimum $\frac{e}{p}$.

\paragraph*{Sporadic Servers}
The sporadic server tries to improve the worst-case demand-bound situation of the deferrable server~\cite{sprunt1989aperiodic}.
It does this by replenishing in chunks instead of restoring the full capacity at the beginning of a period.
There are different definitions in the literature that differ in the exact conditions under which replenishments are scheduled.
We mainly consider the \ac{posix} specification \texttt{SCHED\_SPORADIC}.

By replenishing chunks of the budget one period after their consumption, we can easily ensure that jobs are not served more than $e$ times per period $p$, while allowing job arrival bursts of size below $e$ to be processed immediately. The largest possible demand within an interval is then
\[ d(\Delta) = e \cdot \left\lceil \frac{\Delta}{p} \right\rceil \]
Usually, as in the \ac{posix} variant of the sporadic server, the replenishment for work chunks is scheduled relative to the server activation time.
This is the time at which the incoming job request makes the server ready, which it could then possibly serve if there were no higher-priority ready tasks.
Since a regular periodic task is allowed to run for more than $e$ times in any period $p$ after being preempted by higher priority tasks, we can mimic its behavior.
This improves the productivity of a frequently preempted server while still relying on a worst-case demand equivalent to a periodic task with the same properties.

\subsection{Interrupt Scheduling}
\label{sec:irq}

Hardware interrupts are used by peripheral components for signaling the need for action of the CPU. These interrupts are not subject to scheduling and depend on external hardware, which oftentimes requires a low response time to interrupts. In addition, the required time to handle an interrupt can vary. Therefore interrupts introduce uncertainty, threatening the real-time property of the system. As a counter measure interrupt handling is separated into a short \ac{isr} and a kernel task. The \ac{isr} executes only the absolutely necessary actions to acknowledge an interrupt, like saving the processor state to the stack and afterwards calling the kernel task. During execution of the \ac{isr}, further interrupts are disabled. The subsequent service task then executes the necessary steps depending on the interrupt source and type to fully service the interrupt. This task becomes subject to scheduling and is usually preemptable~\cite{LiuJane}.

For real-time scheduling theory, \acp{irq} pose several challenges.
First, they can take over \ac{cpu} resources at any time.
However, if they are systematically tamed to known minimum inter-arrival times and \acp{wcet}, an integration into the considerations of a schedulable task set, as discussed in Section~\ref{sec:fixed}, is possible again.
More problematic is that in all common \ac{os} implementations, they trigger the execution of an \ac{isr} in an elevated \ac{irq} context.
There, the execution can either be completely uninterruptible itself, or only by another higher priority \ac{irq} source (\enquote{interrupt nesting}).

Modern interrupt controllers commonly used in microcontroller designs\footnote{e.g., the Nested Vectored Interrupt Controller included in ARM Cortex-M core designs} allow \ac{irq} priorities to be assigned at run-time and a priority mask to be set so that only higher-priority \ac{irq} are scheduled.
By introducing an interrupt mask register write operation to the task scheduler, it is possible to partially integrate the launching of \acp{irq} into the scheduling of tasks~\cite{task_aware}. However, once an interrupt has passed through this mask, it cannot be preempted by a recently unblocked regular task, and thus exhibits atomic execution behavior.
Thus, \ac{irq} and task priorities form two different priority spaces.

To minimize the worst-case latency caused by priority inversion situations between interrupts and high-priority tasks, a common programming best practice is to minimize the work done in an \ac{isr} by only unblocking a deferred \ac{ist}, which then does the actual processing.
This compromises interrupt handling performance, especially for relatively short interrupt routines, for better scheduler control.
Therefore, in reality, the driver developer often has to weigh how much additional latency is acceptable until an \ac{isr}/\ac{ist} split is introduced.

\subsection{Execution Time} \label{sec:wcet}
A critical part to effectively schedule tasks with deadlines is that the task execution times must be known.
Especially the longest possible time a program needs until completed is of interest here, commonly refereed to as \textit{\ac{wcet}}~\cite{wcet_survey}. 

\subsubsection*{Worst-Case Execution Time}
In general, two approaches to finding the worst-case execution time of a program exist, which can also be combined:
\emph{Static analysis} and \emph{execution time measuring}.

Simplified, a static analysis aims to analyze an executable to find the longest control flow path of the program.
For this path the number of CPU instructions are counted, and with information about the clock speed and cycles per instruction of the executing processor the maximum execution time can be calculated.
The advantage of static analysis is that the actual worst-case execution time can be found.
However, for complex programs it is not always feasible.

The execution times can also be measured from the actual runtime of the program.
For this, the program must run on the real hardware, and in a realistic environment which can trigger all possible control flows of the program.
This might be more practical for complex programs, but does not guarantee that the actual worst-case execution was really found.
Both approaches can be combined, by finding control flow blocks with static analysis and then measuring the execution times of these blocks.
In summary, worst-case execution time analysis is a hard, but well studied problem~\cite{wcet_survey}.

\section{Embedded Networking Subsystem}
\label{sec:rx_path}
The \ac{nic} acts as an I/O device for any networked computer. While the sending of packets can be influenced by both the sending process and the networking subsystem of the operating system, the receiving of packets, like any other external input, is not predictable. Moreover, unlike most other external input sources, the frequency of reception is subject to an open complex network that makes prediction impossible at any time.
As a result, IP packet reception occurs arbitrarily and without definitive knowledge of packet rates and the corresponding triggered workload. In order to find feasible real-time protection against unscheduled packet loads, the entire receive path must be understood. This section explains that path in detail.

\begin{figure}[h]
    \centering
	\includegraphics[width=.9\columnwidth]{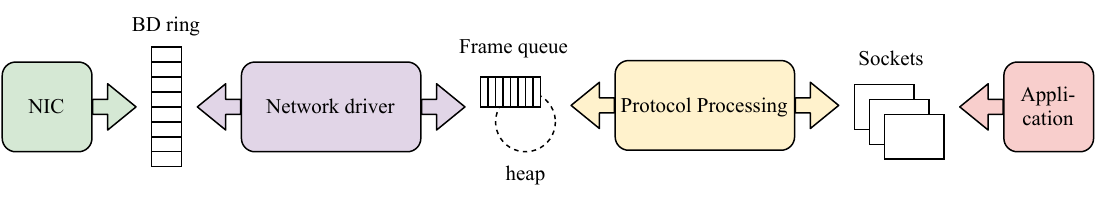}
	\caption{\textbf{\ac{rx}-Path:} Packets are handled by OS data structures, the network driver, and the networking stack implementation before they can be accessed by the receiving process.}
	\label{fig:rx_path_overview}
\end{figure}

At a high level, the \ac{rx} path is organized into subsequently executed stages as seen in Figure~\ref{fig:rx_path_overview}.
Upon packet reception, the \ac{nic} transfers the packet content to a previously prepared memory location via \ac{dma}, marks the corresponding \ac{bd} entry and triggers an interrupt.
The network driver, handling the interrupt, acknowledges the \ac{dma}-operation and exchanges the received frame buffer with a newly allocated one.
From here, protocol processing can commence disregarding the already finished \ac{mac}-layer operations.

The different network stack implementations vary in their set of features. The \textsc{lw}IP network stack is widely spread among embedded applications and sets its focus on memory efficiency~\cite{lwip}. It covers the majority of commonly used and necessary protocols from Layer 2 up to Layer 4 like \ac{arp}, \ac{ip}, \ac{tcp}, \ac{udp}, \ac{dns}, and \ac{dhcp}. It offers different \acp{api} for efficient access, multi-threading, and an implementation of the Berkley Socket \acp{api}. Internally, \textsc{lw}IP does not represent a complete data frame but stores a subset of data in \texttt{pbuf} structures. Other stack implementations like FreeRTOS+TCP\footnote{Documentation, \url{https://www.freertos.org/FreeRTOS-Plus/FreeRTOS_Plus_TCP/}} offer the complete frame and stick to the Berkeley sockets API while being thread-safe. Using sockets, application tasks can register to receive and transmit through their desired ports.

\subsection{Network Interface Conterollers}
\label{sec:nic}
The \ac{nic} facilitates communication in \acp{lan} by implementing the first two layers of the OSI model. The physical layer and data link layer. A \ac{nic} may support one or multiple transmission media, for example twisted pair copper cable, fiber optics or wireless communication as defined by the \ac{lan} standard it implements. Hence, the technologies implemented in \acp{nic} supporting distinct technologies will differ fundamentally. This is not only due to the difference in medium on the physical layer, but also due to the requirements reliable data delivery imposes on the data link layer. For example, Wireless LAN (IEEE 802.11) has a core requirement of collision avoidance due to contention on a shared medium, which in comparison would not be required when using point-to-point communication over cable on a Token Ring (IEEE 802.5). Consolidation and abstraction of these differences is the design objective of the NIC. 

The most prevalent LAN technology in use is Ethernet (IEEE 802.3), with Gigabit Ethernet (1000 Mb/s baseband networks) being a prominent exponent. The standard is designed in layers, with the physical layer components (PHY) and data link layer components (\ac{mac}) connected via a Media Independent Interface (MII). On the link layer, data is transmitted as packets (called \ac{mac} frames) with a size between 64 and 1518 octets. The integrity of the frames is guaranteed using a 4 octet Frame Check Sequence (FCS) containing a cyclic redundancy check (CRC) value. The CRC is used by the \ac{mac} to detect bit errors in the transmission, subsequently dropping any corrupted frames. The NIC's output are Ethernet Frames which contain the network-layer packets as payload.

\paragraph*{Checksum Offloading}
Checksums of (parts of) the frame typically occur at Layer 2 and Layer 4 (for \acs{tcp}/\acs{udp}). For \ac{ip}v4 even the header itself is checksummed again at Layer 3.
Modern \acp{nic} can fill in blank checksum fields with the respectively calculated values and test incoming packet checksums for correctness, relieving the software and therefore the \ac{cpu} from that task.

\paragraph*{TCP Offloading}
A more sophisticated feature is to have the \ac{nic} do \ac{tcp} processing, either by managing the segmentation of larger buffers into packets ("partial \ac{tcp} offloading") or also handling the management of \ac{tcp} connections ("full offloading") \parencite[107]{tcpoffloadengines}. These are however only used in high bandwith server systems.
For our scope of embedded devices, such features are not expected to be available.

\subsection{Interrupt Moderation}
\label{sec:interrupt_moderation}
To decrease the performance impact of incoming packets, high performance NICs employ interrupt moderation techniques. Instead of sending an interrupt for each received data frame, the NIC delays the delivery of an interrupt in order to receive and coalesce additional packets~\cite{makineni2006receive}. Different strategies to realize the delay exist. 

A simple approach is to use a \emph{packet counter} that triggers an interrupt and resets once a certain number of packets have arrived. This leads to a constant and homogeneous reduction of interrupts but also introduces the possibility of starving packets and very unpredictable packet delays. To have control over the time packets reside in memory unnoticed, different types of delay timers are applied: 

The \emph{absolute timer} begins a countdown once a packet has been received and only triggers an interrupt once reaching zero. All packets received in this time frame are announced by this interrupt and do not reset the timer. The obvious disadvantage of this approach is the high latency the first packet of each countdown experiences. In low traffic scenarios, this is highly inefficient. 

To this end, \emph{packet timers} can be introduced. Instead of having a relatively long countdown timer to trigger an interrupt for multiple packets, the counter is a lot smaller and resets with each incoming packet. In low traffic scenarios this leads to smaller delays while interrupts can be entirely impeded under high traffic. The mostly applied solution is therefore a combination of a longer absolute timer and a shorter packet timer. The tuning of the specific parameters is highly dependent on the expected load and subject to research in high performance computing~\cite{interruptmoderation_hpc}.

\subsection{Direct Memory Access}
\label{sec:dma}
In order to relieve the CPU from actively pulling data from or pushing data to the NIC memory, \ac{dma} has been established as a common feature in peripheral hardware. With it, the NIC can asynchronously place received packet data into main memory at previously assigned locations and afterwards only needs to notify the CPU about the arrival of new packets via interrupt. 

\begin{figure}
	\centering
	\includegraphics[scale=0.6]{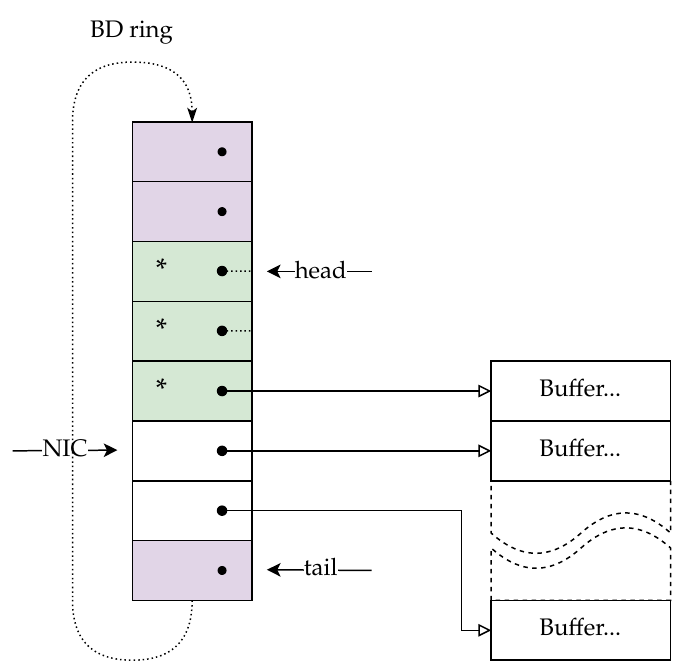}
	\caption{\textbf{BD Ring:} The data structure allows simultaneous access from the NIC adding entries and the CPU checking off processed descriptors.}
	\label{fig:BD_ring}
\end{figure}

The memory assignment usually happens in so-called \ac{bd} rings, as outlined in Figure~\ref{fig:BD_ring}. In memory such a ring comprises an array of Buffer Descriptors, interpreted as a ring buffer. Each \ac{bd} contains a memory pointer to the respective buffer and some metadata for cooperation. The latter typically includes an ownership bit, indicating whether the CPU or NIC is obligated to go on with processing, and a length field indicating how far into the buffer data shall be sent or has been received respectively. This way, both actors can track their current working position(s) individually.

\subsubsection*{Scatter/Gather DMA}
A more sophisticated mode is Scatter/Gather \ac{dma}, through which successive transfers into non-contiguous blocks of memory are realized. This technique uses memory-resident structures called block descriptors, each holding the memory address of a write destination buffer as well as a status field. Every block descriptor also stores the address of another block descriptor, consequently forming a single-linked list of descriptors called a descriptor chain \footnote{Xilinx, \ac{axi} \ac{dma} LogiCORE IP Product Guide (PG021), April 2022.}. When active, the \ac{dma} engine traverses the chain sequentially from the first descriptor (head) to the last descriptor (tail), processing each block descriptor:

\begin{compactenum}
	\item Fetch the current block descriptor from memory
	\item Transfer a received block of data to the memory address pointed to by the block descriptor
	\item Write a completion bit into the status word of the descriptor
	\item Replace the address of the current descriptor with the next descriptor
	
\end{compactenum}

Once the \ac{dma} controller reaches the tail pointer, it halts and signals an \ac{irq}. This allows the software to access the transferred data and update the descriptor chain.

\subsubsection*{Cache Coherency}
Since \ac{dma} modifies a processing system's memory independently of software, cache coherency needs to be established. When using \ac{dma} to receive data, this is required to avoid reading stale cache data in place of the transferred data in memory. This can be achieved by clearing the cache lines for the address range of the received data\footnote{ARM, ARM Cortex-A Series Programmer's Guide Version: 4.0, January 2014.}. Similarly, when transferring data to the peripheral via \ac{dma}, the cache lines in memory regions holding data need to be flushed beforehand. This ensures all data intended for transfer has been flushed from cache to memory.

\subsection{Network Driver}
The \acs{rx}-path begins with the triggering of an \ac{irq} by the \ac{nic}. The \ac{isr} is executed and either directly or after scheduling a deferred \ac{ist}, the driver does all the work necessary to have the packet data in an owned buffer, either by
\begin{compactitem}
	\item actively pulling the data from the \ac{nic} via \ac{mmio} and writing it into a custom buffer, or by
	\item acknowledging the \ac{dma}-operation and exchanging the respective buffer for a freshly allocated one.
\end{compactitem}

\subsection{IP Stack}

\begin{figure}[h]
    \centering
    \includegraphics[width=0.95\textwidth]{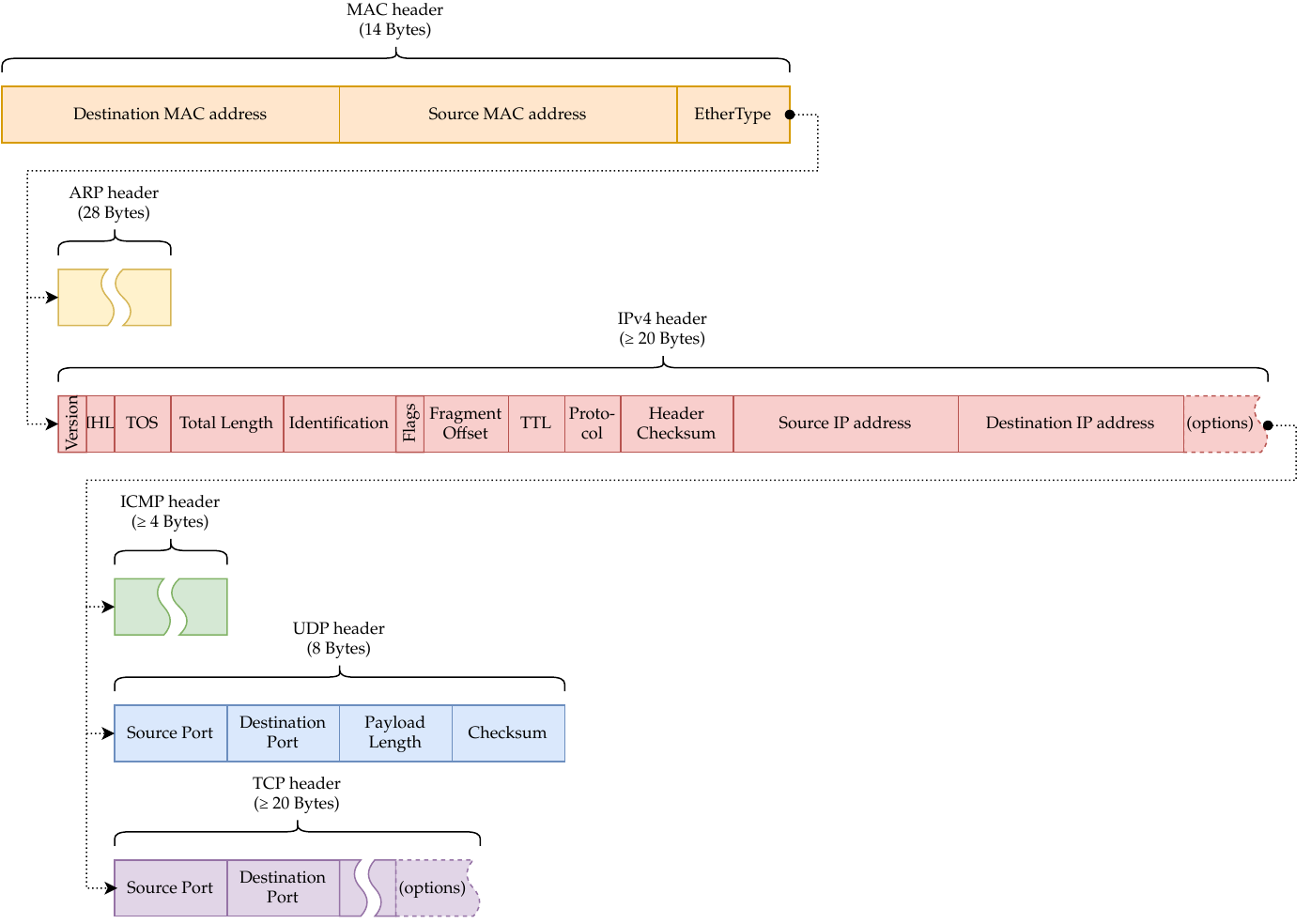}
    \caption{\textbf{IPv4 Headers:} Distribution of protocol headers relevant to this thesis.} 
    \label{fig:relevant_protocol_headers}
\end{figure}

As described in Section~\ref{sec:nic} the \ac{mac} layer does not need to be processed again.
The fact that the \ac{nic} received this packet already indicates its relevance.
Only the \emph{EtherType}-field (cf. Figure~\ref{fig:relevant_protocol_headers}) is important to guide the further processing.
In usual \ac{ip}-stacks, there are three relevant types: \acs{arp}, \acs{ip}v4 and \acs{ip}v6.

\subsubsection*{ARP}

The \ac{arp} protocol is used to resolve network addresses to hardware (\ac{mac}) addresses in local \acs{ip}v4 subnets.
Its packets are almost impossible to attribute to a single network application. On a well-behaved local network, they appear sporadically in small quantities.
A packet received may be a request for our \ac{ip} address, which we must respond to.
Alternatively, it could be a response containing another node's address pair that needs to be cached.

\subsubsection*{IPv4}

\paragraph*{Routing/Forwarding}
For the \acs{ip} layer, it must be checked whether the packet is destined for this node, based on the \emph{destination \acs{ip} address}.
If not, it can act as a router and forward it through another \ac{nic}.
In embedded devices that are leaves in the network topology, this step can be omitted by simply dropping misaddressed packets.

\paragraph*{IP Fragmentation}

The length of the payload of an \acs{ip} packet is limited only by the 16 bits of the \emph{Total Length} field.
Therefore, a packet can be as long as 65,535 bytes.
When transmitting over a local link, the length restrictions are often much finer.
For example, regular Ethernet has a maximum payload size of 1,500 bytes.
Packets that exceed this limit must be broken into multiple fragments.
As a stateful mechanism in an otherwise stateless \acl{ip}, this introduces robustness, reliability, and even security issues \parencite{kent1987fragmentation,gilad2011fragmentation}.
It is therefore rarely used. Reliable transport manages its own segmentation, and even runs into transmission quality issues if it inadvertently causes \ac{ip} fragmentation.
Untrustworthy datagram applications also like to handle efficient encoding of application logic data into single datagrams themselves, to increase performance and reliability.
In \acs{ip}v6, there is a guaranteed lower bound of 1,280 bytes for the MTU of any link, while \acs{ip}v4 only requires a minimum of 68 bytes.
When the stack supports \acs{ip} fragmentation, fragmented packets must be cached by the key tuple (\emph{source \acs{ip} address}, \emph{identification}) or reassembled with other fragments in that cache.
Given the general discouragement of using \acs{ip} fragmentation, and in order to avoid the reassembly overhead, embedded networking stacks tend to ignore fragmented packets, even though this makes them non-standard.
Then all you have to do to prove that a packet is complete is look up the \emph{Fragmentation Offset} and \emph{Flags}.

\paragraph*{IP Options}

The header of \acs{ip}v4 packets is actually not fixed in size.
In addition to the standard 20 bytes, there can be up to 40 bytes of additional options in the header.
These are rarely used outside the Internet.
In fact, packets with IP options are discriminated against by Autonomous System providers, having a higher drop rate and experiencing measurably worse latency and jitter \parencite{fonseca2005ipoptions}.
If these options are respected, the stack must process them.
Otherwise, they may simply be ignored, or packets using them may be dropped altogether, similar to how some stacks handle \acs{ip} fragmentation.

\paragraph*{Header Correctness}

If not already done by the \acs{nic}, check the correctness of \emph{Header Checksum}, \emph{Version}, and \emph{Total Length}.

\paragraph*{Next Protocol}

The \emph{Protocol} field directs further processing, again with three common types: \acs{icmp}, \acs{udp} and \acs{tcp}.

\subsubsection*{\acs{icmp}}

Control messages are mostly used for diagnostic purposes. For example, ping packets can be used to check the availability of a particular host. Also, packets destined for unreachable hosts or ports may trigger a corresponding \ac{icmp} response message.

Incoming \ac{icmp} packets may target a specific socket that's trying to reach its destination. Otherwise, similar to \ac{arp} packets, these packets are directed at the network stack itself, and thus have little accountability to a receiving process.

\subsubsection*{\acs{udp}}

\ac{udp} essentially allows demultiplexing of packets destined for the same host but different processes.
Incoming packets must therefore be assigned to the appropriate bound socket, or possibly responded to with a \ac{icmp} message.
If not already done by the \ac{nic}, the correctness of the \emph{Payload Length} and \emph{Checksum} fields must be checked again.

\subsubsection*{\acs{tcp}}

Besides allowing demultiplexing similar to \ac{udp}, \ac{tcp} also provides a reliable communication stream over unreliable segments.
This requires a sophisticated state machine to manage the exchange of segments.
This processing is usually done in the \enquote{host} network stack, which is provided by the \ac{os}, which usually resides in kernel space to facilitate efficient protocol processing.
For many years there has been the idea of custom user-level stacks that can be provided and configured by each application independently. \parencite{honda2014userlevelstacks}.
However, competitors to \ac{tcp}, such as QUIC, are usually implemented on top of \ac{udp}, which itself introduces little overhead over simple demultiplexing, since it is supported by virtually every \acs{tcp}/\acs{ip} stack and respected by all the Internet middleboxes that otherwise love to destroy end-to-end protocol agnosticity.

\subsubsection*{\acs{ip}v6}
\acs{ip}v6 is similar to \acs{ip}v4 and provides the same mechanisms as \parencite{hagen2006ipv6}.
Its header is more streamlined, omitting legacy complications such as variable header length for additional options and \ac{ip} fragmentation.
These, in turn, can be included by a special extension header that sits between the main \acs{ip}v6 header and higher protocols.
The second major difference is the address space, which uses 128 bits instead of 32.
Because of the scarcity of \acs{ip}v4 addresses alone, the importance of \acs{ip}v6 to the \ac{iot} cannot be overstated.
Nevertheless, \acs{ip}v4 is still much more widely used, and \acs{ip}v6 is generally derided as a future protocol, since it has been around for 20 years and will probably take another 20 years to really replace \acs{ip}v4.

\subsection{Implementations}

\subsubsection*{LwIP}
Lightweight IP\footnote{\url{https://www.nongnu.org/lwip/2_1_x/index.html}} (LwIP) is an open-source implementation of the Internet Protocol Suite for embedded devices. It was originally developed by Adam Dunkel at the Swedish Institute of Computer Science in 2003. Having been designed especially for low-power embedded devices, its implementation places an emphasis on reduced memory usage and code size~\cite{lwip}.

The IP stack takes advantage of the typically soft division between kernel and application space on embedded devices, which allows sharing of data structures and directly passing buffer between memory spaces without copy. Furthermore, LwIP does not adhere to the strict protocol layering in the TCP/IP suite. Instead, it is implemented in a flexible event-based architecture, combining data from different layers in shared data structures where advantageous and relying on callbacks for protocol processing between layers~\cite{lwip}.

Multiple APIs exist for implementing access to the network stack. The low-level raw API is designed for applications in bare-metal environments, offering an event-driven workflow while requiring no separate network thread. This API provides native access to the protocol processing functionality, since it is also used internally for protocol processing between layers of the network stack.

LwIP also implements more traditional higher-level sequential APIs to be used in environments with \acp{os}, such as the Netconn API or Berkeley Sockets. These are implemented with an internal user-space core process, which encapsulates all network stack data as well as the invocation of the native functions for protocol processing. Communication with the core process is implemented through a messaging bus, where the external API calls pass messages containing the internal functions as callbacks to the core process for invocation. This avoids the critical section problems inherent with multi-tasking \acp{os} as well as multi-threading, making the high-level APIs thread-safe. The messaging bus requires some platform-dependent means of inter-process communication.

Furthermore, LwIP is also supported in concurrent environment without the necessity for the core process. This is realized by core locking, where a platform-defined synchronization primitive is used for mutual exclusion. The mutex guards against concurrent access to the internal LwIP core data structures and functionality. It also accommodates priority inversion required when using LwIP with a preemptive multitasking operating system.

\subsubsection*{FreeRTOS+TCP}
The FreeRTOS+TCP stack is developed as an open source library extension to FreeRTOS.
As the naming suggests, it is designed to work with the mechanisms of FreeRTOS and is developed in a similar clean code style.
The internal architecture features a monolithic \acs{ip} task that handles all kinds of events, from packet reception to socket binding, and thus acts as a synchronization unit.
It is slightly lighter than \acs{lwip}.
Support for \acs{ip}v6 is still in preview.
Also, \acs{ip}v4 options are not evaluated. Depending on the preprocessor configuration, the options may be ignored or the whole packet discarded.
Outgoing packets can be fragmented, but reassembly is not supported.

\section{The Industrial Internet of Things}
\label{sec:1_2_iiot}
The \ac{iiot} builds on the same core ideas as the IoT and applies them to the industrial domain.
Whereas in \ac{iot}, devices are equipped with sensors, computing power and, most importantly, communication capabilities, in \ac{iiot} this is applied to all kinds of industrial assets.
This enables the collection and analysis of data generated by machines and their control systems, for example to optimize process flows or detect anomalies in order to prevent malfunctions.
In addition, the comprehensive networking of processing machines makes it possible to respond in real-time to external influences such as changing demand.

In contrast to IoT, where the networked devices and the network itself are generally considered unreliable and volatile, the industrial domain tends to assume static networks with low mobility and high reliability.
These are well structured and tailored to the expected load, since control applications, for example, require latencies in the millisecond range~\cite{iiot2}.

\subsection{Edge Computing}
\label{sec:edge}
Although a central component of both IoT and IIoT is the connection of individual devices to the Internet and the use of cloud technology, this is not the optimal approach in every scenario.
As just described, control applications in particular have very high communication latency and reliability requirements.
Particularly with the large amounts of data generated in the IIoT context, the point is quickly reached where cloud-based data processing no longer meets the requirements and, for example, real-time critical processes miss their deadlines.

For these reasons, the edge computing approach is becoming increasingly popular in the IIoT space.
Edge computing describes a paradigm in which computing capacity in physical proximity to the data source is used instead of cloud services for data processing of specific, primarily latency-critical applications~\cite{qiu2020edge}.
In the context of Chapter~\ref{cha:3_3_offloading}, we assume that such an edge computing cluster exists in the immediate vicinity of the real-time actors.

\subsection{Network Delay}
Since the later chapters of this work consider network constraints when scheduling and processing real-time tasks in a distributed fashion, the exact network latency conditions are of central importance.
In this thesis, a simple approach to dynamically determine the network characteristics is presented and implemented, although it is highly unlikely that this approach can be applied universally.
However, the topic of network delay estimation is well researched and many approaches have been published, such as in~\cite{delay_estimation_1},~\cite{delay_estimation_2} and~\cite{delay_estimation_3}.

We therefore assume that even in scenarios where the approach to network delay estimation used in this work cannot be used, one of the published approaches is applicable and that, in principle, well-founded statements about the performance of a network, especially in the area of delay, can be profiled.

\subsection{Clock Synchronization}
Another important factor in distributed real-time systems is time synchronization.
Meaningful scheduling of real-time tasks can only work if the actors generating the tasks and the scheduler have the same understanding of time.
Since control applications require latencies in the millisecond range, as described earlier, and wireless networks are typically in this range, time synchronization must be in the sub-millisecond range.

The IEEE 1588 standard \enquote{for a Precision Clock Synchronization Protocol for Networked Measurement and Control Systems} defines a protocol that enables sub-microsecond clock synchronization in local area networks~\cite{ptp_ieee}.

Another approach with a focus on distributed real-time systems, dating back to 1987, is presented in~\cite{clock_sync_1}.
Since these approaches clearly exceed the precision requirements, we assume in the further course of the work that the clocks of all actors are sufficiently synchronized and the system is not influenced by this aspect.
		
\chapter{Real-Time Performance of Industrial Communication Technologies}
\label{cha:1_3_survey}
\minitoc
Following the technical fundamentals for this thesis, we now take a deeper dive into communication technologies specifically designed to meet real-time requirements. This chapter performs a review of these technologies and presents recent academic works. 
With the growing need for automation and the ongoing merge of \ac{ot} and \ac{it}, industrial networks have to transport a high amount of heterogeneous data with mixed criticality such as control traffic, sensor data, and configuration messages.
Current advances in IT technologies furthermore enable a new set of automation scenarios under the roof of Industry 4.0 and IIoT where industrial networks now have to meet new requirements in flexibility and reliability. 
The necessary real-time guarantees place significant demands on the networks.
We present and discuss existing standards and technologies, as well as recent works related to real-time IIoT networking solutions. In doing so, we identify current IIoT scenarios addressed in research, derive their real-time networking requirements, and review established real-time networking technologies. This chapter presents the most notable recent research, and discusses the open challenges.

The remainder of this chapter is structured as follows.
Section \ref{sec:1_3_introduction} introduces IT/OT convergence and motivates this review.
Section \ref{sec:1_3_scenarios} identifies \ac{iiot} scenario classes and derives real-time networking requirements.
Section \ref{sec:1_3_background} reviews the path of networking in industrial systems and presents existing standards designed for industrial networks.
Section \ref{sec:1_3_recent} presents ongoing academic works on real-time networking.
Section \ref{sec:1_3_discussion} discusses the feasibility of existing and ongoing solutions.
Section \ref{sec:1_3_conclusion} concludes this chapter and indicates the research gaps the remainder of this thesis tries to partially bridge.

\section{IT/OT Convergence in Real-Time Systems}
\label{sec:1_3_introduction}

One of the main goals of Industry 4.0 is operational optimization.
In a recent study~\cite{rwsn_industrial_AI_smart_factory_2021} regarding the usage of pervasive \ac{iiot} sensor data, a large majority of the respondents expect Industry 4.0 to provide agility to scale the production to match demands, improved flexibility to customize products, and a reduced time to market for new products.
The same participants reported that the current usage of IoT sensors in their systems was primarily used for remote monitoring of equipment, asset and material tracking, and predictive maintenance.
Hence, to fully realize the potential of Industry 4.0, pervasive sensor coverage and interconnectedness are essential.
To achieve this, great demands will be placed on the networks used in industrial systems.
What used to work for \ac{ot} systems will require a much higher degree of flexibility and scalability than what has been available in the past.

Science and technology are bridging the mechanized industry with the digital world to automate as many processes as possible by connecting operational technologies and business networks.
Such ubiquitous sensor integration combined with a high degree of network connectivity, both amongst \acp{ot} and between \acp{ot} and back-office systems \ac{it}, is often referred to as \emph{IT/OT Convergence} and is the driving force behind Industry 4.0.
Examples of envisioned scenarios can be found in research on autonomous drones~\cite{lohan2018benefits}, remote control~\cite{liu2020latency}, real-time monitoring~\cite{zhan2022industrial}, and predictive maintenance~\cite{nordal2021modeling}.

Furthermore, advancements in Cyber-Physical Systems (CPSs) increase the real-time requirements for industrial networks.
CPSs combine embedded systems with cybernetic control systems to control the physical world from the networked space.
The physical processes involved often place strict real-time requirements on CPSs.
Command, control, and safety considerations of different criticality levels demand constant latency and predictability, often implemented on small and restricted embedded systems.
By attaching these embedded systems to larger networks and incorporating them into complex distributed systems, the same real-time requirements are extended to the network and its participants.
Use cases with time-critical control tasks are spread over a network, requiring predictable and reliable packet routing through it. 

However, networking concepts like \ac{ip} and packet routing are not created to consider real-time demands. as \ac{ip} networks work on a best-effort basis.
Packets are indiscriminately routed over a variable number of networking devices, with very limited predictability concerning latency and arrival.
The jitter introduced by wireless communication complicates a number of envisioned scenarios containing mobile devices.

To meet the network real-time demands introduced by envisioned \ac{iiot} scenarios, established specialized protocols and standards exist. At the same time, current works in research attend to IIoT-specific requirements such as scalability~\cite{mirani2022key}, integration of business and factory networks~\cite{hicking2021collaboration}, and mobile computing~\cite{ma2021tcda}.

\section{Regarded IIoT Scenarios}
\label{sec:1_3_scenarios}

In this section, we perform a preliminary literature survey to identify IIoT scenarios containing real-time and network requirements. To this end, we classify the scenarios into groups as depicted in Table~\ref{tab:scenarios}.
Network and real-time requirements are derived from the scenario groups to prepare the following survey on enabling technology standards and recent works.
A total of 31 research papers with IIoT scenarios of interest have been reviewed. The papers and articles chosen were peer-reviewed and published in the past 10 years.
The extracted use cases were either the main focus of a paper or example implementations for evaluation purposes.

\begin{table}[ht]
    \centering
    \caption{IIoT scenario classes}
    \label{tab:scenarios}
    \begin{adjustbox}{max width=\textwidth}
    \begin{tabular}{cl|c|c|c|c|c}
    &\thead{\textbf{Scenario Class}} & \thead{\textbf{References}} & \thead{\textbf{Predictability}\\ \textbf{Requirement}} &  \thead{\textbf{Latency} \\ \textbf{Requirement}} & \thead{\textbf{Bandwidth} \\ \textbf{Requirement}} & \thead{\textbf{Wireless} \\ \textbf{Communication}}\\ \hline
    \textbf{\textrm{I}} & \makecell[l]{Autonomous Vehicles \\ \& Mobile Robots} & \cite{lohan2018benefits, miljkovic_new_2013, singholi_review_2021, chavhan_smart_2021, lo_bello_perspective_2019,  barzegaran_fogification_2020} & high & medium & variable & essential\\ \hline
    \textbf{\textrm{II}}& Remote Control of Machines & \cite{zhohov_real-time_2018, kayan_cybersecurity_2022, kajan_control_2013, huang_sense_2019, bock2021performance} & high & high & low & not necessary\\ \hline
    \textbf{\textrm{III}}& \makecell[l]{Sensor Data Analysis\\ \& Monitoring} & \cite{kanawaday_machine_2017, blanes_80211n_2015, iqbal_cooperative_2017,  hicking2021collaboration, salhaoui2019smart} & low & low & high & not relevant\\ \hline
        \textbf{\textrm{IV}}& \makecell[l]{Worker Safety \\ \& Hazard Protection}  & \cite{kivela2018towards, zhou2017industrial, mcninch2019leveraging, reyes2013intelligent} & high & high & low & not necessary\\ \hline
        \textbf{\textrm{V}}& Task Offloading & \cite{hong_multi-hop_2019, hossain_edge_2020, deng_intelligent_2021, xu_joint_2022, ma2022reliability} & medium & medium & high & not relevant\\ \hline
        \textbf{\textrm{VI}}& \makecell[l]{Industrial Wearable Systems \\ \& Augmented Reality} & \cite{kong_industrial_2019, svertoka_wearables_2021, lorenz_industrial_2018, rosales_iiot_2021, batalla_analyzing_2020, li2021helically} & low & medium & high & essential\\% 
        \hline
    \end{tabular}
    \end{adjustbox}
\end{table}

Using the identified scenario classes, we hope to get a better understanding of network requirements concerning predictability, latency, and bandwidth.
These requirements will then be used in the following sections to assess proposed IIoT real-time networking solutions. Furthermore, we identify relevant network components for each scenario class.

\subsection{Scenario Classes}

\paragraph*{Autonomous Vehicles \& Automation}
\label{para:sc:autonomous}
Warehouses and manufacturing sites are getting under pressure to fulfill an ever-increasing demand. To this end, conventional rigid automation is being exchanged for more flexible autonomous solutions controlled via IoT-connected sensors~\cite{chavhan_smart_2021}. Autonomously moving machines such as transport robots, forklifts, and warehouse robots pose challenges yet to be solved, especially when employed at a large scale or in close proximity to human workers. Autonomous vehicles have to navigate through dynamic environments making decisions for real-time controlled electric motors~\cite{barzegaran_fogification_2020} while being dependent on communication networks to prevent collisions and improve efficiency~\cite{singholi_review_2021, lo_bello_perspective_2019}. 

Converging these real-time and networking requirements still requires research, especially considering the necessity for wireless communication. Since moving, potentially dangerous machines are involved, predictability and latency requirements are high.
Bandwidth requirements are implementation-specific, with the main factor being whether high-volume sensor data (e.g. video, LIDAR) is transmitted for analysis. 

\paragraph*{Remote Control}
Industrial Control Systems (ICS) can be changed to make use of the IIoT paradigm by physically shifting controlling entities away from the real-time device, possibly outside of the plant floor~\cite{bock2021performance}. To this end, the originally existing air gap between ICS and external networks has to be bridged~\cite{kayan_cybersecurity_2022}. Besides the resulting security implications, this also means opening up real-time control systems to conventional IP networks and protocols. Use-cases furthermore include the remote control of vehicles in inaccessible places~\cite{zhohov_real-time_2018}. 

Again, real-time requirements are to be considered as high, since we are regarding moving machines.
Some use cases furthermore need wireless communication technologies to work. Bandwidth requirements are low as only control commands need to be transmitted.

\paragraph*{Data Analysis \& Monitoring} 
With the help of sensors on factory floors, a high amount of real-time data can be used to improve business automation, monitoring, and decision-making. To this end, potentially broad data streams have to be transmitted sharing a medium with real-time control traffic~\cite{blanes_80211n_2015}. Use cases include data aggregation by wireless sensor networks~\cite{iqbal_cooperative_2017} and predictive maintenance on IIoT sensor data~\cite{kanawaday_machine_2017}. IT/OT integration also plays a role in monitoring and data analysis. Real-time information from IIoT sensors can help make business decisions and enable automating certain use-cases such as active energy management~\cite{hicking2021collaboration}.

This scenario class, while somewhat dependent on real-time data streams does not contain hard real-time tasks. Hence requirements in this regard are negligible. However, due to the large amount of collective data produced by the high number of sensors, bandwidth requirements are usually high. With data-intensive applications benefitting from offloading, bandwidth requirements can be considered high for many use cases.

\paragraph*{Worker Safety \& Hazard Protection}
Most considered works regard the introduction of IIoT to safety critical systems as a relevant risk factor. Safety in this case means the physical protection of workers and equipment.
Due to the attack vector granted by the new connectivity, safety is more conjoined with security than ever~\cite{kivela2018towards}. Yet, some works promote the use of interconnected devices for better safety in high-risk environments such as in the mining industry. IIoT technologies can be used to monitor access to hazardous areas around operating machinery, improve documentation/monitoring of maintenance that requires shutdown of the machinery, and prevent unexpected startup or movement during machine maintenance activities~\cite{mcninch2019leveraging, zhou2017industrial, reyes2013intelligent}. 

Safety solutions distinguish between informational and automatizing setups. Purely informational scenarios react to real-time sensor data by issuing alarms or notifying workers and human operators. Smart monitoring and data analysis in the cloud make these scenarios and their real-time requirements similar to the \textit{Data Analysis} class. Systems that automatically react to hazards such as fire, toxic gases, or dangerous proximity require the highest predictability and latency. However, reactive systems should work independently, making distributed solutions unsuitable.

\paragraph*{Task Offloading} 
IIoT devices, like traditional embedded systems, only have limited computing capabilities due to power and space constraints.
To this end, current research, especially in the field of mobile edge computing, considers the offloading of delay-sensitive tasks to local servers.
The main challenges are the guarantee of maximum delays over shared networking channels and compute resources~\cite{deng_intelligent_2021}.
Subject to offloading is mostly soft real-time tasks for better Quality of Service (QoS) where a balance between latency and reliability has to be found~\cite{ma2022reliability}.

Due to the nature of offloading, hard real-time tasks with high requirements in predictability are not considered.
At the same time, while offloading to more powerful machines can reduce computation time, latency requirements that are too tight to accommodate network delays cannot be accommodated.

\paragraph*{Wearable Systems \& Augmented Reality}
Industrial wearable systems are the means of human-machine interaction (HMI) for most of the other listed scenario classes. With manufacturing, maintenance, and control still requiring human interaction, wearables present the necessary interface to embedding human workers in smart factories. Wearable systems enable remote control~\cite{li2021helically}, real-time monitoring via augmented reality~\cite{batalla_analyzing_2020}, sensor data aggregation~\cite{rosales_iiot_2021}, and safety systems~\cite{svertoka_wearables_2021} while adding the requirements of energy efficiency, size (weight), and wireless networking. Examples roughly follow two streams: human interaction devices and data collection devices~\cite{kong_industrial_2019}. For this scenario class, only the network requirements specific to wearables and augmented reality are considered. If the devices are used for i.e. remote control, Scenario Class II covers the requirements.

\subsection{Deriving Network Requirements}
To derive network requirements for the scenario classes, we need to somewhat generalize the regarded use cases. In the end, it comes down to the specific requirements of the underlying real-time tasks and how messages (i.e. network packets) are integrated into the control flow.
The requirements broadly quantified in Table~\ref{tab:scenarios} therefore only give a rough overview of the focal points and limiting factors related to the classes.
The spectrum of feasible IIoT scenarios increases with the strictness of real-time guarantees networks can adhere to while not limiting their flexibility. 
Furthermore, wireless communication technologies that can fulfill these guarantees are the necessary enablers for many of the identified scenarios.
Interfacing general IP networking into real-time systems is necessary for use cases requiring external access or cloud computing.
However, for most real-time relevant scenarios IT/OT integration is not the enabler. Real-time scenarios become more complex the further away participants are, limiting real-time guarantees and criticality levels. 

\section{Industrial Networking}
\label{sec:1_3_background}

The introduction of the first PLC, Modicon model 084 in 1969, changed how production systems were designed, implemented, operated, and maintained. 
Improving mechanical systems and machinery with relay-based control loops was but a small step compared to what PLCs brought to the industry.
Whole cabinets of relays could be replaced with a relatively small number of compact devices saving both space and power.
Where the first PLC only had 16 inputs, 16 outputs, and a meager 1 kB of memory, contemporary PLCs have I/O in the order of hundreds, abundant memory, and much higher speeds.

The next sections will cover different industrial networks and common requirements.
Table \ref{tbl:network_tech} presents the connection between networks and use cases.

\subsection{Fieldbuses}
With improved industrial automation, communication became an obvious hurdle to tackle.
In 1979, Modicon developed and published the serial communication protocol Modbus, an application-layer protocol.
Shortly thereafter the work to standardize Profibus started in Germany and was finally published in 1989.
These first communication protocols would be the inaugural \ac{ot} networks.

As the first fieldbus protocols grew in popularity, so did issues concerning interoperability.
The different protocols were seldom compatible, and few vendors could support multiple protocols.
After a somewhat turbulent era often referred to as "the Fieldbus War"~\cite{felser_fieldbus_2002}, an agreement was made in the late 1990s to create a common standard.
When the first fieldbus standard (IEC 61158) was published, the final product was essentially a collection of \textit{all} the competing standards at the time.
The latest revision of IEC61158 \cite{iec61158} lists 26 different fieldbus protocols grouped into services and protocols.
Since then, it has become apparent that the industry is increasingly looking towards open standards \cite{kayan_cybersecurity_2022} to avoid vendor lock-in and a more transparent way to handle security vulnerabilities to name a few.

In this same decade, the Controller Area Network (CAN) was also standardized by the automotive industry.
CAN is a serial communication protocol where the message identifier also serves as the message priority.
The protocol is optimized for short messages and has the priority encoded in the message identifier.
This provides a non-destructive address arbitration which enables messages to pass without any delay induced by lower priority messages.

\subsection{Packet Switched Networks}
At the same time, the IT sector underwent a networking revolution of its own where Ethernet would ultimately end up as the de-facto standard.
After its initial version was published in 1985, it has since moved the throttle from 10 Mbps to 400 Gbps~\cite{ieee:802.3-2022} with its sight firmly set on 800 Gbps.
Data is split into discrete packets, each identified using a 48-bit Media Access Control (MAC) address giving ample room for growth.
To connect hosts, systems can be chained together to form a ring topology or by using ethernet bridges (where store-and-forward switches are the most common).
Each frame is individually routed, and the network can quickly adapt to changes in topology and assign new routes to frames while in transit.
 
In 1974, Cerf et al. published TCP/IP \cite{rfc675} which was to become the backbone of the "network of networks" lovingly called "the Internet".
TCP is a connection-oriented protocol that when traffic moves in one direction, control traffic will move both ways allowing TCP to resend lost data and adjust the rate to the slowest link along the path.
UDP is the connectionless sibling of TCP without any control traffic and thus offers no indication that traffic arrives at the destination.

\paragraph*{Industrial Ethernet}
Industrial Ethernet encompasses the usage of Ethernet in industrial settings.
The target applications typically have both latency and reliability requirements, which drives the design of protocols away from traditional Ethernet approaches for collision detection and avoidance.
The most common real-time Ethernet protocols are EtherNet/IP\footnote{IP is for
  \textbf{Industrial} Protocol.}, Profinet, and EtherCAT~\cite{watteyne_industrial_2016}.
PROFInet is a translation of PROFIbus to run over an Ethernet network.
EtherCAT is often used in industrial control and automation due to its speed and determinism.
TTEthernet is a congestion-free network based on Ethernet that provides Time Triggered service for critical traffic, Rate Constrained for event-triggered traffic as well as a best-effort service.
EtherNet/IP is an adaption of DeviceNet to Ethernet, it uses the Common Industrial Protocol (CIP) over TCP and UDP.

\paragraph*{Time Sensitive Networking} 
TSN is a series of IEEE Standards for switched Ethernet~\cite{ieee:802.1ba_2021, ieee:802.1Q-2018} where Commercial Off-the-Shelf (COTS) networks can be configured to give bounded latency and extremely low packet loss for critical traffic on the data link layer.
TSN removes some of the initial robustness of Ethernet by requiring static routes but gains lower jitter and less out-of-order delivery.
In addition, TSN defines strict and fine-grained Quality of Service (QoS) mechanisms which are covered in more detail in Section \ref{sec:1_3_psn_qos}.

\subsection{Wireless networks}
Wireless sensors are cheaper, easier to install and maintain than their wired counterparts and bear the promise of an infrastructure that is vastly more scalable.
In some scenarios, the cost-savings can be as high as 60-90\%~\cite{power_reduce_2009,8558500} compared to wired solutions.
It can even be the only viable option (e.g., the Tire Pressure Monitoring System (TPMS), tool deflection measurements~\cite{ostling_cutting_2018}).

The downside of wireless networks is the shared medium; the integrity of the network is fully dependent upon cooperative participants that only transmit during allocated transmit slots.
Frequency bands needed for wireless protocols are a tightly managed resource, and only a few bands are available for free use, most notably in the 2.4GHz and 5 GHz range.
Where a wired network can be fairly resilient to signal interference, have high and predictable bandwidth, and require an adversary to be close to the wired infrastructure to eavesdrop, wireless systems have no such luxury.

Over the years, many wireless protocols have been defined with different characteristics such as high bandwidth, low latency, long-range, a high number of addressable hosts, and robust resistance to interference.
Oftentimes, these attributes will be at odds; i.e. with longer ranges come lower bandwidth and higher latency, and high bandwidth can make the traffic more susceptible to interference due to the denser encoding.

Both WirelessHART and ISA 100.11a are common industrial protocols that are based on IEEE 802.15.4 and operate in the unlicensed 2.4GHz frequency band.
ZigBee, while also being based on 802.15.4, can be found in some industrial settings but is primarily intended for home automation and low-criticality systems.
Bluetooth is mentioned due to its pervasiveness in personal handheld devices.
With its low power, it is only capable of transmitting data over a few meters and is primarily intended for short-range personal area networks.
The upside is rather low power consumption ample access to complete modules, and comparably high available bandwidth.

In recent years, Wireless LAN (WLAN) has become a common technology in most households, but the technology has not been reliable or deterministic enough for industrial settings.
With the recent WLAN6(E), this has improved markedly, and with the availability of Multiple-Input/Multiple-Output (MIMO), access is more reliable and less prone to interference from other senders.
The latest version, WLAN6E has also markedly improved the clock accuracy for protocols such as PTP, which makes sensor aggregation more accurate.
In addition to the rather crowded 2.4GHz band, WLAN can also operate in the 5GHz and 6GHz range allowing for higher bandwidth at the cost of higher signal attenuation from obstacles along the path.

The fifth generation of mobile telephony (5G) from the 3GPP aims to cover all areas of wireless communications from cellular networks, IoT devices, and industrial networks.
5G is designed to operate in licensed bands where a site license specifying both assigned frequency range and spatial location is required.
Interference from other networks should therefore be at a minimum.
Three broad use cases have been defined for 5G: cellular data (\emph{enhanced Mobile Broadband, eMMB}), IoT (\emph{massive Machine Type Communications, mMTC}) and low-latency, critical traffic (\emph{Ultra-Reliable, Low Latency, URLLC}). 
It is important to note that these use cases are not introduced in any single release, but support for each is added as increments with different capabilities in each release.
With the first 5G release from 3GPP, Release 15, support for New Radio, rudimentary slicing, and the IoT profile from 4G were among the many parts included.
The next release saw the introduction of improved slicing ("VLAN for wireless networks") and redundant transmission for high-reliability communications.
Finally, with the latest approved release (Rel17, March 2022), improvements in backhaul networks, RAN slicing, public safety, and non-public networks (NPN) have been ratified.
NPN is often called ``private 5G`` and makes it possible to run a 5G network as a standalone network as a local service rather than a public mobile operator.
Running a 5G NPN is highly relevant in industrial automation as slicing, URLLC and access to licensed radio bands can greatly aid in security, reliability, and latency.
Whereas eMMB is the use case most commonly supported by public providers, URLLC will be most relevant over NPN.
Although the specification seems impressive, not much hard data is currently available to evaluate the performance of URLLC.

\subsection{Quality of Service}
\label{sec:1_3_psn_qos}
The key strength of Ethernet is its ability to seamlessly accept new systems, adapt to topological changes, and handle variable amounts of traffic.
For industrial networks where rigid deadlines are the norm rather than the exception, this dynamism can jeopardize transmission delays and be a major problem.
First, packets can take one of multiple \emph{routes} in such a network leading to different arrival orders, variance in delay, or being completely lost.
Secondly, once a packet has started transmitting, no other packet, regardless of importance can be transmitted.
Since most network bridges are \textit{store and forward}, the entire packet must be fully received before it can be forwarded to the next stop.
This requires buffer capacity before newly arrived packets can be forwarded to the next stop.
During times of high traffic, these buffers can be exhausted and the bridge will have no other choice than to drop packets.

\paragraph*{Standard PSN QoS measures}
QoS describes the treatment of frames belonging to a particular class (or stream) compared to other frames.
In this context, a stream, or a flow, is a set of packets that logically belong together and should be treated similarly by the network.
Several measures are available, yet whereas these are highly relevant to IT networks and the Internet, fewer apply to the needs of industrial networks.

Differential Services (\emph{DiffServ}) is a highly scalable class-based service where each stream is assigned to a particular service level and it is up to the network administrator to assign resources to each level.
DiffServ is intended to be used across large networks and networks of networks where there can be more than one operator.
However, the actual service provided to a class may differ between operators.
What is more, there is no way to differentiate between streams within a class, meaning as the network grows, the interference from other streams increases.

Integrated Services (\emph{IntServ}) maintains a 1:1 mapping between streams and granted service levels.
This ensures consistent QoS throughout the network.
Since IntServ is stream-based, its scalability is tied directly to the number of streams.
For large networks or networks of networks, the number of active streams will quickly outgrow the available resources in each bridge.

For industrial control applications, the uncertainties of DiffServ may not suffice, and for large-scale sensor networks as foreseen in Industry 4.0, IntServ will have scalability problems.
An effective solution demonstrated by Harju and Kivimaki~\cite{harju_co_operation_2000} is to use IntServ on the edges to shape and limit all incoming streams, and DiffServ in the core network to handle the bulk of traffic.
This hybrid approach can deliver adequate services, especially for large networks, which makes this one of the possible approaches presented by the Deterministic Networking (DetNet) group of IETF.

\paragraph*{Time Sensitive Networking}
TSN works by reserving buffer capacity for incoming streams and using different traffic queues on the egress port.
An outgoing queue can have a shaper attached to it, which changes the traffic pattern to conform to the desired behavior.
The Credit Based Shaper (CBS, ~\cite{ieee:802.1Qat}) is a \emph{class based} shaper that will group streams for a given traffic class into a single queue.
Traffic belonging to this queue is then shaped so that it does not exceed the reserved bandwidth over a short service interval.
Since \emph{every} bridge in the domain is required to support these features, the edges will effectively limit the inflow, and the core bridges will work to reduce bursts.
CBS is specifically designed to eliminate transient network overloads from bursty traffic and this has been proven to work using Network Calculus~\cite{azua_avb_modelling_2014, nc_intro_tsn_maile_2020}.

With the Time Aware Shaper (TAS~\cite{ieee:802.1Qbv-2015}), TSN provides Time Division Multiplexing (TDM) by aligning transmission windows along the path of the traffic.
For scheduled traffic (ST) to work, all bridges must be part of a tightly synchronized time domain.
When configured properly, TAS can yield extremely low delay variations and give an upper bound of 100$\mu$s transmission delay.

However, even with minor time errors, carefully adjusted transmit windows can become desynchronized and effectively delay scheduled traffic for a complete TDM cycle.
As the network complexity grows, this problem only worsens.
This makes TAS a complex and difficult scheduler to operate and thus suitable for only the most critical streams.
It also requires the network to be centrally managed.
To allow more sporadic, yet critical traffic and reduce the operational complexity at the same time, a new urgency-based scheduler has been adopted by the working group.
The Asynchronous Traffic Shaper~\cite{specht2016urgency} has per-class queues and per-stream shaping and uses its internal clock and allows for mixing traffic types such that it can handle sporadic, critical events whilst enforcing rate limits to reduce the impact of bursts.
It has been shown that ATS will not increase the worst-case delay~\cite{nasrallah_ultra-low_2019} of traffic through the network.

CBS is appropriate for traffic that should be regular but occasionally bursty, whereas TAS is best suited for critical traffic that is regular and must be expedited through the network.
For important, yet sporadic or aperiodic traffic (e.g. monitor alarms, events), ATS is appropriate.

It is worth noting that the reservation of streams may fail if one or more bridges are unable to accommodate the requirements and that can only reliably be detected at run-time.
It is possible to discover this \emph{a priori} to a certain extent (Maile et al.~\cite{nc_intro_tsn_maile_2020}), but as small changes to the network can result in new routes for traffic, previous scenarios may no longer be possible.
This makes it difficult to realize the full potential of a network using a fully distributed model where each node announces or reserves resources for a stream individually.
Instead. a centralized controller can handle end-station interaction and bridge configurations.
The \emph{Centralized User Configuration} (CUC) and \emph{Centralized Network Configuration} (CNC) provide single points of contact for end-stations.
New streams are evaluated and by using the total overview, an optimal path can be created and forwarded to the CNC, which in turn configures all the required bridges before the end stations are notified and traffic can begin.
For TAS, such a centralized controller is a requirement.

\subsection{Common Network Requirements for Industrial Automation}
\label{sec:1_3_rt_net_req}
Devan et. al \cite{devan_survey_2021} list 4 criteria as fundamental for a wireless industrial network: 1) Secure against malicious intruders and misconfigured devices, 2) easy and dependable access to sensor data, 3) Interoperability between vendors and protocols, and 4) active research community to adapt and grow to future needs.
Missing in this list is the need for deterministic transport.
From the discussion in Section \ref{sec:1_3_psn_qos}, we can extend this list to also include wired industrial networks and include bounded latency and time synchronization as relevant metrics.

An updated requirement is thus:
\begin{compactitem}
\item \textbf{Open Standards} Available standards and royalty-free technology should be the preferred approach as this will allow for better interoperability between vendors and a more open community that encourages academic research.
\item \textbf{Security} Not only should the content be shielded from external eyes, but the network should be able to detect if traffic has been altered, delayed, or injected into the network by a third party.
\item \textbf{Reliability and Availability} Network delivery must be reliable in that once sent, the sender should be confident that it will arrive at the destination within a known time frame. At the same time, a sender should expect the network to be available to accept new data within the pre-defined constraints (not overstepping BW bounds, etc.). The network should also be robust against jamming and other malicious attempts to disrupt the service~\cite{pirayesh_jamming_2022}
\item \textbf{Latency} Especially closed control loops are sensitive to delay variations, but for any streams, expecting delivery within a certain time frame is a requirement.
\item \textbf{Time synchronization} As networks are used to build distributed systems, a shared understanding of time (what is often called a shared time domain) is needed. The accuracy of the domain is dependent upon how well the network can forward time synchronization messages.
\end{compactitem}

\begin{table}
    \centering
    \caption{Technologies along Security, Determinism, and Bandwidth. Legend: \cmark: well suited, \xmark: not well suited}
    \label{tbl:network_tech}
    \begin{adjustbox}{max width=\textwidth}
    \begin{tabular}{r | l l r r | c c c c c c}
     & & & & & \multicolumn{6}{c}{\textbf{\textit{Relevant Scenarios}}}\\
        & \textbf{Security}
        & \textbf{Determinism}
        & \textbf{Bandwidth} &
        & \textbf{\textrm{I}} & \textbf{\textrm{II}} & \textbf{\textrm{III}} & \textbf{\textrm{IV}} & \textbf{\textrm{V}} & \textbf{\textrm{VI}}\\
    \hline
    \textit{Fieldbuses} &&&&&\\
    CAN         & E & 10 ms &  1 Mbps & \cite{godoyDesignCANbasedDistributed2010, rubiobenitoPerformanceEvaluationFour1999} & %
    \xmark & \cmark & \xmark &\cmark & \xmark & \xmark \\
    CAN-FD      & E & 100 $\mu$s &  5 Mbps & \cite{austermannConceptsBitrateEnhancement2020} & %
    \xmark & \cmark & \xmark & \cmark & \xmark & \xmark\\
    Modbus      & E & 10 ms & 115 kbps & \cite{rubiobenitoPerformanceEvaluationFour1999} & %
    \xmark & \cmark & \xmark & \cmark & \xmark & \xmark\\
    PROFIBUS    & E & 10 ms & 12 Mbps & \cite{junAnalysisPROFIBUSDPNetwork2005, rubiobenitoPerformanceEvaluationFour1999} & %
    \xmark & \cmark & (\cmark) & \cmark & (\cmark) & \xmark\\
    \midrule
    \textit{Wired network}&&&&&\\
    Modbus TCP  & E & 45$\mu$s (ideal conditions) & 100 Mbps & \cite{modbus_smartgrid} & %
    \xmark & (\cmark) & \cmark & \cmark & \cmark & \xmark\\
    PROFINET    & E     & 10-100 ms &  1 Gbps & \cite{profinet_chans} & %
    \xmark & (\cmark) & \cmark & (\cmark) & \cmark & \xmark\\
    EtherNet/IP & E   & 1-2 ms &  1 Gbps & \cite{alessandriaPERFORMANCEANALYSISETHERNET2007} & %
    \xmark & \cmark & \cmark & \cmark & \cmark & \xmark %
    \\
    TSN CBS     & E,A   & 2 / 50 ms & 40 Gbps & \cite{zhao2017timing} & %
    \xmark & (\cmark) & \cmark & \cmark & \cmark & \xmark\\
    EtherCAT    & E   & 34 $\mu$s & 10 Gbps & \cite{nguyenEtherCATNetworkLatency2016} & %
    \xmark & \cmark & \xmark & \cmark & (\cmark) & \xmark
    \\
    TTEthernet  & E,A    &  low, $\mu$s (offline schedule) & 1 Gbps & \cite{zhao2017timing} & %
    \xmark & \cmark& \cmark & \cmark & (\cmark) & \xmark\\
    TSN ST      & E,A   & 100 $\mu$s & 40 Gbps &\cite{zhao2017timing} & %
    \xmark & \cmark & (\cmark) & \cmark & (\cmark) & \xmark \\
    \midrule
    \textit{Wireless network}&&&&&\\
    WirelessHART & I,C,A & 10 ms  & 250 kbps & \cite{godoyEvaluatingSerialZigBee2012} &%
    \cmark & (\cmark) & \xmark & (\cmark) & \cmark & \xmark\\
    ISA 100.11a  & I,C,A & 10 ms  & 250 kbps & \cite{godoyEvaluatingSerialZigBee2012} &%
    \cmark & (\cmark) & \xmark & (\cmark) & \cmark & \xmark\\
    Bluetooth 5.0& I,C,A & 10-100 ms & 48 Mbps & \cite{rondonEvaluatingBluetoothLow2017} & %
    \xmark & \xmark & \cmark & \xmark & (\xmark) & (\cmark)\\
    WLAN 6       & I,C,A & 1 ms (ideal conditions) & 9.6 Gbps & \cite{bankovEnablingRealtimeApplications2019, oughton_revisiting_2021} &%
    (\cmark) & \xmark & \cmark & \xmark & (\cmark) & \cmark\\
    5G eMMB     & I,C,A & 10-100 ms  & 10 Gbps & \cite{oughton_revisiting_2021} & %
    \xmark & \xmark & \cmark & \xmark & (\cmark) & (\cmark)\\
    \hline
    \end{tabular}
    \end{adjustbox}
    \end{table}

In Table \ref{tbl:network_tech}, the most common solutions found in industrial networks are listed.
Each is briefly evaluated along 3 common axes:
\begin{compactenum}
\item Security,
\item Determinism (i.e. jitter), and
\item Bandwidth (upper limit).
\end{compactenum}
For brevity, Reliability, Availability, and Latency have been combined into ``Determinism`` as all the aforementioned requirements will directly affect the determinism of the traffic.
Security can be further divided into
\begin{compactenum}
    \item \textbf{E}xternal: the protocol relies on external security measures,
    \item \textbf{I}ntegrity: messages are accompanied by robust checksums to detect modification,
    \item \textbf{C}onfidentiality: each message is encrypted using a secret known only to authorized parties,
    \item Secure \textbf{A}uthentication and Authorization: each node on the system must be securely identified before being granted access to the network.\footnote{Both wired and wireless Ethernet can achieve better security by including standards such as Port-based authentication (802.1X) and Secure Device Identity (802.1AR).}
\end{compactenum}

In addition, each row is weighted against the list of use cases in Table \ref{tab:scenarios} with \cmark indicating a good fit, \xmark ~a poor.
A (\cmark) indicates that the protocol can support it under ideal conditions, but it is not well suited.
Likewise, (\xmark) indicates that the protocol \emph{could} be made to work but should not.

\section{Ongoing Efforts towards Real-Time Networks}
\label{sec:1_3_recent}
When it comes to the inherent predictability requirements of real-time applications, packet-switched networking alone is not adequate.
This is because the load on individual network hardware in larger networks is unpredictable, and packet transmission times (as well as routes) are variable.
In addition, the packet loss rates of wireless transmissions are always difficult to account for.
As described in the previous section, there are industry standards that address these issues, either by sticking to packet-less bus communication or by adding specific real-time capabilities to Ethernet.
However, with the convergence of IT and OT networks, as well as the highly mobile nature of the IIoT scenarios mentioned above, further real-time networking solutions need to be found.

In this section, we present recent work on real-time IIoT networking. We consider peer-reviewed articles and papers from the last 5 years. The works can be either novel techniques, adaptations, or evaluations of existing standards optimized for real-time IIoT scenarios. The reviewed works are divided into subsections of equal levels of abstraction and underlying technologies.
Table~\ref{tab:comparison_recent} compares the works qualitatively concerning their scope and applicability to our scenario classes. Entries marked with \cmark or \xmark ~determine whether the work is feasible for the scenario class.
Entries marked with \omark ~are works that do not directly facilitate a scenario class but also do not prohibit it.
It also indicates the type of source institution (academic or industry) and whether the work supports wireless communication.

\begin{table}
    \centering
    \caption{Recent works}
    \label{tab:comparison_recent}
    \begin{adjustbox}{max width=\textwidth}
    \begin{tabular}{c|c|c|c|c|c|c|c|c|c|c} 
       \textbf{Work} & \textbf{Scope} & \textbf{Issue(s) addressed} & \textbf{Source} &\textbf{\textrm{I}}&\textbf{\textrm{II}}&\textbf{\textrm{III}}&\textbf{\textrm{IV}}&\textbf{\textrm{V}} & \textbf{\textrm{VI}} & \textbf{Wireless} \\ \hline
         \cite{ishtaique_ul_huque_system_2019} & SDN architecture & IT/OT; flexibility & academia &\omark&\xmark&\cmark&\xmark&\cmark&\omark & \cmark\\ %
         \cite{kiangala_effective_2021}  & IP architecture & reliability \& determinism & academia & \xmark&\cmark&\omark&\cmark & \cmark & \xmark & \xmark \\ %
        \cite{foschini_sdn-enabled_2021} & SDN architecture & IT/OT; security & academia &\xmark&\xmark&\cmark&\xmark&\cmark&\xmark & \xmark\\ %
        \cite{li_practical_2020} & OPC UA & TSN integration & academia & \xmark & \cmark & \cmark & \omark & \omark & \xmark  & \xmark\\ %
        \cite{morato_assessment_2021} & OPC UA & TSN assessment & academia & \xmark & \cmark & \cmark & \omark & \omark & \xmark & \cmark \\ %
        
        \rowcolor{Gray}
        \cite{wang_adaptive_2018} & SDN for IIoT & offloading & academia &\xmark&\omark&\cmark&\xmark&\cmark&\xmark & \cmark \\%yes
        \rowcolor{Gray}
        \cite{lin_dte-sdn_2018} & SDN for traffic engineering & delay sensitivity & academia &\xmark&\cmark&\xmark&\cmark&\omark&\xmark & \xmark\\ %
        \rowcolor{Gray}
        \cite{zeng2019time} & SDN industrial Ethernet & flexibility; determinism; scale & academia &\xmark&\cmark&\cmark&\omark&\cmark&\xmark  & \xmark\\ %
        
        \cite{schriegel_migration_2021} & Time-aware forwarding & Profinet/TSN migration & academia &\xmark&\cmark&\cmark&\cmark&\omark&\xmark  & \xmark\\ %
        \cite{bujosa2022hermes} & TSN scheduler & scalability & academia &\xmark&\omark&\cmark&\omark&\cmark&\xmark  & \xmark\\  %
        \cite{chaine2022egress} & TSN traffic shaping & scalability; complexity & aviation &\xmark&\omark&\cmark&\xmark&\cmark&\xmark & \xmark\\ %
        \cite{bruckner_introduction_2019} & OPC UA with TSN & network infrastructure&automation&\xmark&\cmark&\cmark&\cmark&\cmark&\xmark  & \xmark\\ %
        \cite{8823854}& TSN using DDS & heterogeneous TSN; flexibility & academia & \xmark&\omark&\cmark&\cmark&\xmark&\xmark  & \cmark\\ %

        \rowcolor{Gray}
        \cite{cavalcanti2019extending} & Wireless TSN & determinism; reliability & chip manufacturer &\cmark&\cmark&\cmark&\omark&\xmark&\cmark  & \cmark\\ %
        \rowcolor{Gray}
        \cite{sudhakaran_enabling_2021} & Link layer/application mapping& wireless TSN applicability & chip manufacturer &\cmark&\cmark&\cmark&\omark&\xmark&\cmark  & \cmark\\ %
        \rowcolor{Gray}
        \cite{yun_rt-wifi_2022} & RT-WiFi over SDR & determinism; flexibility & academia&\cmark&\cmark&\cmark&\omark&\xmark&\cmark  & \cmark\\ %
        
        \cite{walia_5g_2019} & 5G slicing management & determinism; isolation & telecom &\omark&\omark&\omark&\cmark&\cmark&\cmark  & \cmark\\ %
        \cite{aijaz_private_2020} & 5G private networks & assessment & telecom &\omark&\omark&\omark&\omark&\omark&\omark  & \cmark\\ %
        \cite{wen_private_2022} & 5G private networks & assessment & telecom/car &\omark&\omark&\omark&\omark&\omark&\omark  & \cmark\\ %
        
        \rowcolor{Gray}
        \cite{behnke2023towards} & RT-aware packet reception & determinism & academia &\xmark&\cmark&\cmark&\omark&\omark&\xmark  & \cmark\\ %
        \rowcolor{Gray}
        \cite{behnke2022priority} & RT-aware NIC & determinism &  academia&\xmark&\cmark&\cmark&\omark&\omark&\xmark  & \cmark\\ %
        \rowcolor{Gray}
        \cite{johansson2022priority} & RT-aware MAC filtering & determinism & academia &\xmark&\cmark&\cmark&\omark&\omark&\xmark  & \cmark\\ %
        \rowcolor{Gray}
        \cite{austad_ftc}&Timed C extension & TSN-aware applications & academia & \xmark & \cmark &\cmark&\cmark&\omark&\xmark  & \xmark\\ %
        \hline
    \end{tabular}
    \end{adjustbox}
\end{table}

\subsection{Network Architectures}
The field of research with the broadest scope regards network architectures for real-time IIoT environments.
Next to architectures realized by SDN, this segment also includes research on IT/OT integration and OPC UA.

In~\cite{ishtaique_ul_huque_system_2019} Ishtaique ul Huque et al. present a system architecture for time-sensitive heterogeneous wireless distributed software-defined networks. They derive this system architecture from enabling state-of-the-art technologies and their requirements.
Besides the manageability of parameters in heterogeneous networks, the authors hope to facilitate TSN integration with legacy networks. 

Kiangala and Wang combine zero-loss redundancy protocols, TSN, and edge computing concepts to realize an intra-domain network architecture attending to the reliability and predictability needs of time-critical IIoT applications~\cite{kiangala_effective_2021}.
Their solution is entirely based on IP networking.
Instead of manageability via SDN they focus on technological implementations for high reliability and determinism. 

Network architectures have been a focus in recent works primarily due to the IT/OT convergence requirements of Industry 4.0 and IIoT applications. Foschini et al. present an SDN-enabled architecture for this and analyze its behavior during DDoS attacks~\cite{foschini_sdn-enabled_2021}.
With a multi-layered network architecture, they aim to provide layer-specific security and real-time properties while making data from IIoT devices usable inside the business network of an industrial plant with autonomous vehicles.
Within their DDoS attack, they showed that mere changes to the network architecture cannot secure the highly critical lower levels of the network (i.e. physical machines).

The Open Platform Communication Unified Architecture (OPC UA) was introduced in 2008 as an interoperability standard for reliable information exchange and has gained momentum with discussions on IIoT networks. In recent years Li et al. presented an OPC UA architecture with TSN capabilities for the IIoT~\cite{li_practical_2020} and Morato et al. published a general assessment of OPC UA implementations under current Industry 4.0 requirements~\cite{morato_assessment_2021}. Both works showed that OPC UA can be used to integrate real-time communication requirements into Ethernet-based networks in the manufacturing industry.

\subsection{SDN}
When industrial networks become more complex and contain mixed computing systems and real-time requirements Software-Defined Networking can help with managing the resulting heterogeneity.
Wang and Li present an SDN-based solution for task offloading in IIoT environments \cite{wang_adaptive_2018}. A computing mode selector is implemented in the SDN controller and offloaded tasks are given priority based on real-time parameters. The network transmits offloaded tasks to fog computing resources in order of this priority. 

Another solution that addresses mixed timing criticality in large networks is proposed by Lin et al. \cite{lin_dte-sdn_2018}. SDN is used to implement a traffic engineering approach to schedule the transfer of delay-sensitive traffic. This works using real-time analysis of the network and monitoring the QoS metrics of participating links. Based on these metrics a dynamic scheduler with multi-path routing capabilities manages the flows.

Zeng et al. address flaws of Industrial Ethernet, namely poor scalability, insufficient self-configuration capabilities, and increased costs due to the use of proprietary hardware and present a time-slotted software defined Industrial Ethernet \cite{zeng2019time}. The approach contains a time synchronization mechanism based on PTP and a system architecture for time slot-based industrial switches. Using SDN, the implementation becomes scalable, and reconfigurable, and is not dependent on frequent infrastructure changes.

\subsection{Time-Sensitive Networking}
In recent years, researchers have addressed various shortcomings of TSN by extending, improving, and integrating it with other systems and frameworks.
Schriegel and Jasperneite worked on a bridging mode called Time-Aware Forwarding (TAF) \cite{schriegel_migration_2021} to increase the flexibility of TSN and thereby accelerate the migration from Profinet to TSN in industrial communication.
With the approach, already existing networking hardware can be made compatible with TSN.  

Tackling the scheduling of time-triggered traffic in TSN, Bujosa et al. present Hermes, a heuristic multi-queue scheduler \cite{bujosa2022hermes}.
With this improvement to the synthesis algorithm of gate control lists in TSN networks, they increase network scalability by reducing its scheduling complexity. 

To control network jitter, incoming packets are buffered along the routing path in a TSN network. Chaine et al. propose a solution \emph{Egress TT} that performs this buffering only at the final network node \cite{chaine2022egress} and optimizes it for network scalability.
The solution allows to use of non-TSN networking on the routing path and reduces computational complexity while somewhat increasing latencies. 

In~\cite{bruckner_introduction_2019}, Bruckner et al. showed how the ubiquitous OPC-UA protocol benefits from using TSN.
Similarly, in~\cite{8823854} Agarwal et al. demonstrated improved reliability and reduced delay for Data Distribution Services (DDS) when TSN was used to protect and expedite the traffic through the network.

\subsection{Real-Time Aware Wireless Technologies}
The need for reliable and deterministic wireless communication can be derived from most scenarios presented in Section \ref{sec:1_3_scenarios}. The following works from the past years attend to real-time wireless communication. 

A seemingly natural step is to transfer TSN methods to work with wireless technologies.
The concomitant challenges and research objectives have been thoroughly investigated by Cavalcanti et al.~\cite{cavalcanti2019extending}.
Next to a comprehensive overview of state-of-the-art wireless technologies and IEEE 802.1 TSN standards, they discuss new approaches on top of next-generation wireless technologies (e.g. WiFi 6, WigGig, and 5G) to overcome the radio-inherent challenges concerning packet loss and jitter. 

Sudhakran et al. build on top of this work and present a methodology to map application layer timing requirements in a collaborative robot application to the link layer, based on wireless TSN over WiFi~\cite{sudhakaran_enabling_2021}.
Rather than going further into the TSN standards, they address how to classify traffic, extract time-critical flow parameters, and define an efficient schedule for the end-to-end QoS approach.

Yun et al. recognize the older real-time aware WiFi (RT-WiFi~\cite{wei2013rt}) technology which is implemented on common off-the-shelf hardware and hence hard to update and maintain.
In their work, they propose a software-defined radio approach of RT-WiFi implemented on an FPGA called SRT-WiFi ~\cite{yun_rt-wifi_2022}. They provide a fully open system architecture and perform extensive evaluations on a multi-cluster SRT-WiFi testbed.

\subsection{Fifth Generation Mobile Networks} 
The fifth generation of mobile networking (5G) promises several improvements and enabling technologies for Industry 4.0 and smart factory applications. Among them are network slicing, private management, and ultra-low latencies.

Network slicing allows for logically separated virtual networks over the same physical 5G network. Both, the integration of different business layers as well as connection-specific latency control could be realized with this.
Walia et al. studied the usage of network slicing in a smart factory and developed a management model for 5G slicing in the domain~\cite{walia_5g_2019}. 
As the effective usage of 5G for Industry 4.0 environments requires self-managed or "private" 5G networks, some recent works have focused on their implementation and feasibility.
Aijaz et al. present an overview of the motivation behind and functions of private 5G networks~\cite{aijaz_private_2020}.
They present a large number of potential use cases and benefits for the IIoT and portray private 5G networking as the future of industrial networks.
Wen et al. come to a similar conclusion while pointing out the early stage at which private 5G networking currently is~\cite{wen_private_2022}.
Actual implementation and management are currently only feasible for large companies with a networking background.
In their paper covering critical mMTC, Pokhrel et al.~\cite{pokhrel_towards_2020}, argue that although URLLC is capable of providing excellent real-time behavior, the cost of doing so means relatively few devices can be connected in such a fashion.
Similarly, where mMTC is capable of connecting a plethora of devices, the individual QoS is not compatible with sensor networks.
To be able to meet the need for a large, wireless sensor network, they propose ``critical mMTC``.

\subsection{Device-Layer Technologies}
Only a few works have considered making improvements to the lowest layer in IIoT networks. The following research papers are concerned with securing embedded devices that are (somewhat newly) connected to large networks and responsible to execute real-time tasks.

Johansson et al. use MAC layer filtering to mitigate the effects of best-effort traffic reception on real-time tasks \cite{johansson2022priority}. They extend the network driver of VxWorks to support multiple receive queues and an interface that supports the configuration of the Ethernet Controller's MAC filter.
This way, packets are filtered and enqueued based on their priority in hardware, making it possible to treat incoming traffic based on real-time considerations.

In~\cite{austad_ftc}, TSN was used to extend Timed C, an extension to the C programming language, and include network channels as a primitive in the language.
When faced with large network loads, the critical traffic was reliably forwarded with minimal delay and jitter demonstrating the usefulness of TSN in distributed real-time systems.

\section{Discussion}
\label{sec:1_3_discussion}
Traditionally, industrial networks have been isolated, mostly homogeneous, and designed with dedicated protocols and hardware to meet stringent reliability and real-time requirements.
With the coming transformation of Industry 4.0 and IT/OT convergence, industrial networks need to become more configurable and flexible, enabling a wider range of applications than before.
This, of course, requires tighter connectivity between networked systems, as well as a higher degree of dynamism for nodes joining and leaving the network.

\subsection{Bridging the Gap between IIoT Scenarios and Network Requirements}
Looking at Table \ref{tbl:network_tech}, the wireless networking requirements of Scenario Classes I and VI appear to be the most difficult to meet.
While several wireless communication standards exist, they either cannot meet the strict latency predictability requirements or cannot meet the bandwidth requirements of multimedia data.
Once high reliability and latency are taken into account, current wireless networks do not scale sufficiently, i.e., a research gap exists where critical real-time requirements meet a need for wireless communication.
Current wireless real-time protocols are generally unable to scale to the required size and the aggregated bandwidth expected in the coming years.
Similarly, WLAN, and especially WLAN6E, can be seen as an attractive alternative with high bandwidth, low latency, improved timing accuracy, and readily available COTS hardware.

Comparing this to Table \ref{tab:comparison_recent}, it can be seen that this is where wireless TSN technologies will play a significant role in enablement.
The main gap-filler distilled from this work is wireless technologies that can provide a measure of predictability.
However, the overall feasibility of this must be viewed with caution. While latency guarantees can be improved by real-time aware wireless packet switching, the inherent unpredictability of radio must not be overlooked.

While privately managed 5G networks seem to be a part of the solution, the high complexity and cost make it impractical for many companies and use cases.

The presented works regarding wired networks show that ``real-time networks`` are more static and carefully planned than traditional IT networks tend to be.
Even though technologies such as TSN help bring determinism to COTS Ethernet, achieving high flexibility and reconfigurability is a largely unsolved challenge.
Combining TSN with its centralized controllers (CNC, CUC) and SDN is one promising avenue that would extend to both WLAN and 5G solutions.
The downside of such a solution is the vast complexity of all systems; needless to say, the factory network of the future will be almost unrecognizable from today's, and be a major asset of its own.

Standardization and holistic large-scale implementations also are open challenges to the remaining subfields.
Even where research papers are feasible for specific scenario class requirements these only focus on individual aspects. 
The feasibility of the presented approaches still has to be validated in complex engineered systems.
Industry 4.0 is still missing a fixed set of standards and technologies which will be important for trust in safety and reliability. 
The approaches from the works presented have to be merged and implemented in holistic system architectures including all actors of the network, from embedded real-time devices, network devices, and compute nodes such as edge servers and cloud systems.

\section{Conclusion \& Thesis Integration}
\label{sec:1_3_conclusion}
Real-time demands on networks in industrial environments have risen with the advent of the IoT paradigm and Industry 4.0. In this chapter, we listed common and anticipated use cases for industrial networks and identified the criteria needed to meet the requirements of each. 
We reviewed established industrial networking technologies and protocols and presented current academic works regarding the topic. 
By comparing different networking technologies to these criteria and use cases, we give an overview of challenges and solutions concerning IIoT scenarios from a real-time networking perspective.

The remainder of this thesis provides contributions to real-time communication research, not only for industrial use-cases, but real-time embedded systems in general. In particular, embedded system networking subsystems are analyzed and extended to meet the demands of modern use case scenarios from classes I, II, \& IV.

In Chapter \ref{cha:2_3_software} we present a real-time aware IP networking stack for IoT devices \cite{blumschein2022differentiating}. By classifying packets as early as possible and inheriting real-time task priorities to the networking task, they prevent priority inversions between a high-priority task and the processing of lower-priority packets. Additionally, incoming packet rates are limited to prevent a system overload from incoming packets.

In Chapter \ref{cha:2_4_hardware} a hardware-based approach with similar goals is presented. Using a multi-queue NIC design, packets are reordered and filtered based on the priority of the receiving real-time task. Furthermore, real-time aware interrupt moderation is proposed that reduces the number of interrupts generated by incoming traffic while guaranteeing low latencies for high-priority packets.
A SW/HW co-design for real-time aware packet reception \cite{behnke2023towards} is presented in Chapter \ref{cha:2_5_codesign}. 

The challenge of task offloading in modern industrial systems with wireless mobile machines as identified in Scenarios I \& V is considered in Chapter \ref{cha:3_3_offloading}. We present a system architecture and distributed scheduler for the offloading of real-time tasks to local edge servers. 
				
\chapter{Connecting Real-Time Devices to IP Networks}
\label{cha:2_2_problem}
\minitoc
This chapter provides a method for the analysis of network load impact on critical real-time tasks running on state-of-the-art \acp{mcu} with modern \ac{rtos} used in \ac{iot}. To this end, we first introduce the issue of \acp{irq} in real-time systems and further specify the role of network generated interrupts. In Section~\ref{sec:2_2_methodology} we present the evaluation methodology of our analysis. We then evaluate the timing measurements of critical tasks on microcontrollers running vendor-supplied \ac{rtos}, network drivers, and network stack tasks under various network-triggered \ac{irq} loads. 
To uncover any existing mitigations in the hardware and closed-source drivers, a pseudo-network driver is designed to serve as a second \ac{irq} source. Thus, we performed experiments on two setups: One that treats NICs and drivers as black boxes, and one that simulates network interrupts on the machines.
Measurements are taken and compared between real and pseudo network packet processing. Section~\ref{sec:2_2_evaluation} presents the empirical results and Section~\ref{sec:2_2_discussion} discusses them. In the final section of this chapter, we present a testbed that facilitates on-device research on the impact of packet reception on real-time performance and helps developers test their applications accordingly. Parts of this chapter have been peer-reviewed and pre-published in two conference papers~\cite{behnke2020interrupting,bender2021pieres}.

\section{Interrupting Embedded Real-Time Devices}
\label{sec:2_2_intro}
As motivated in Chapter~\ref{cha:1_1_introduction}, embedded devices that control machines in the physical world have been part of industrial processes as well as home and automotive appliances for decades. Unlike general-purpose computers, these devices must meet timing constraints.
To this end, \acp{rtos} are used, which are lightweight and provide guarantees towards the timing predictability of tasks~\cite{rtos}.
Typically, a preemptive task scheduler allows the configuration of different priorities for concurrent tasks, so that the most time-critical tasks always precede less critical ones. While implementation flaws in \acp{rtos} and their tasks still occasionally occur, decades of research and development in this area have mitigated many reliability and security issues.

\acp{irq} are generated by the hardware and are essential to the functioning of systems.
At the same time, they introduce a degree of unpredictability into the process.
Since the corresponding \acp{isr} are handled by the processor, the scheduler of an \ac{os} has no control over their execution.

\begin{figure}[h]
	\centering
	\includegraphics[width=0.8\textwidth]{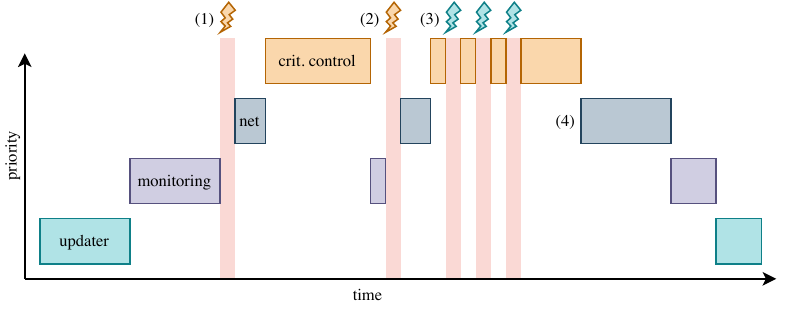}
	\caption{\textbf{Priority inversion:} Illustration of priority inversions caused by network packet \acp{irq} and processing.}
	\label{fig:prio_inv}
\end{figure}

A common example of an unwanted effect of this is priority inversion, where the execution of a high-priority task is preempted by an \acp{isr} belonging to a lower-priority task.
However, by keeping \acp{isr} execution times to a minimum and considering worst-case scenarios during design, most traditional embedded systems can handle \acp{irq} without missing deadlines.
However, in the past, the environment controlled by embedded systems tended to be self-contained, the number of environment-triggered \acp{irq} was typically small, and their impact therefore predictable.
With the advent of \ac{iot} in industrial applications, this premise has changed.
\ac{iot} microcontrollers come with built-in networking chips and are increasingly being connected to networks for remote control, monitoring, and maintenance~\cite{iiot_mcu}.
\ac{iot} networks are open by design and thus less controlled~\cite{zhang15communication}.
This is especially dangerous for critical real-time tasks on networked embedded microcontrollers:
Embedded systems have to handle the additional resource consumption of non-critical networking tasks, and the required \ac{nic} introduces a new source of unpredictability, as incoming packets trigger \acp{irq} that disrupt the flow of scheduled tasks~\cite{hip}. This can lead to a critical load of interrupts and triggered network tasks in the \ac{rtos}, invalidating real-time guarantees and thereby reducing system reliability.

Figure~\ref{fig:prio_inv} illustrates the resulting priority inversion. In this example, an embedded system controlling a machine is connected to a network for control and updating purposes. A low-priority update task sends a request to a remote server before being preempted by a monitoring task. While this task is running, a network packet is received that triggers an \ac{irq} (1). Since the network processing task has a higher priority, it preempts the monitoring task and processes the packet, which turns out to be a machine control instruction. The payload is forwarded to the critical control task, which executes the command. Monitoring is then preempted again for the same reason, with the same result (2). However, now the packets requested by the low-priority updater task begin to arrive, each triggering an \ac{irq} that preempts the critical control task. Since it has a higher priority than the network driver, the processing of these packets is postponed until the control task has finished (4). Before the control task can resume, the update packets are processed and forwarded to the updater, which, as the lowest priority process, finishes last.

In this example, we observe two instances of priority inversion. First, at (3), when the low-priority packet reception takes precedence over the critical control task, and second, at (4), when the processing of the low-priority packets takes precedence over the monitoring task. In the first case, the resulting time shift of the high-priority task per packet is equal to the execution time of the \ac{isr} and context switches (in red). The upper half of the protocol processing in the network driver can be expected to be longer, but somewhat manageable, since this task is subject to OS scheduling. In the following, this chapter analyzes the timing requirements of both situations.

\section{Challenges and Assumptions}
\label{sec:2_2_problem}
As motivated, the inclusion of network controllers in embedded real-time systems introduces an unpredictable source of interrupts. The goal of this chapter is to provide methods for the analysis of the interrupt impact in IoT environments. This includes the execution of ISRs, network drivers, and network stack tasks, as well as the robustness of the overall system under high network loads. This section outlines the challenges and assumptions for the analysis.

\subsection{IoT Environments}
Available IoT devices are microcontrollers with the means to wirelessly connect to networks and handle a variety of different network protocols, comparable to network stack implementations in general computing. At the same time, these systems are designed to be deployed in untrusted environments, more or less directly connected to the Internet~\cite{iotnetstack}. Packet floods caused by failures and security breaches also expose potentially critical systems deployed in remote networks. IPv6, whose large address space is seen as an enabler for the deployment of the large number of IoT devices~\cite{ipv6}, may also lead to the abandonment of the obfuscating properties of Network Address Translation (NAT) required in IPv4. Due to these circumstances, we consider networked embedded systems of increased vulnerability while controlling critical systems with real-time constraints.

The popularity of the IoT has led to the mass production of IP-enabled (mostly WiFi) devices that are cheaply available. To make the somewhat complex programming of embedded systems more accessible in the IoT context, devices come with programming frameworks that hide low-level software such as networking stacks and drivers. Examples include Espressif's IoT Development Framework, Arduino libraries, and Particle's DeviceOS. While it is technically possible to program the examined modules without their respective frameworks, the immense workload involved in implementing and integrating the network drivers and stacks makes this infeasible in most cases. An integration of the embedded WiFi chips into fully manageable and transparent real-time systems is hindered by the unavailability of the driver source code.

\subsection{Embedded Real-Time Development}
When designing and developing real-time systems, the developer is responsible for ensuring that all guarantees (deadlines) are met in the worst-case scenario. In the case of externally generated interrupts, it is therefore necessary for the developer to consider the maximum frequency of interrupts and their effects in the design of the system. Network generated interrupts, on the other hand, leave the designer with few options, such as disabling interrupts or reserving a physical core for network tasks~\cite{hermes}. While both actions may solve the problem of interrupt flooding over the network, the loss of a core (if even available) and remote access to the embedded system may not be feasible.

In addition, limited access to low-level functions results in a loss of control over the real-time system, which should generally be designed holistically. Changing network task priorities, choosing which parts of the network stack to run, and which protocols to support are all taken away from the developer in these frameworks. 

\subsection{Assumptions}
Concluding, we make the following assumptions for our analysis. 
\begin{compactitem}
\item \textit{A1 - Unreliable networks}: IoT devices are commonly deployed in large IP networks and/or might be subject to network faults.
\item \textit{A2 - Wireless}: While the issue of unpredictable interrupts is the same across network interfaces, driver availability and specific processing costs differ. WiFi connections belong to the most commonly used interfaces, since no modifications to available hardware has to be made and the technology is mature.
\item \textit{A3 - Interdependence}: Networking tasks, drivers, and real-time application tasks have interdependencies and might run on the same \ac{mcu} core.
\item \textit{A4 - Device-specific frameworks}: To implement real-time systems with WiFi capabilities on IoT devices, device-specific frameworks are used.
\end{compactitem}

The design of our experiment setup is derived from these assumptions. 

\section{Methodology}
\label{sec:2_2_methodology}
A certain impact of networking tasks and interrupts on \ac{mcu} utilization and timing predictability in real-time embedded systems seems to be inherent. With the experimental methodology and setup, we aim to analyze this impact on two current \acp{mcu} used in state-of-the-art \ac{cots} IoT devices.

\subsection{Design}

For the quantitative analysis, we have designed two sets of experiments that observe the timing metrics of a periodic task. The experiments are run under different network loads with two considered interrupt setups (see Section~\ref{sec:intrsetups}). The experiment permutations were additionally adjusted to account for differences in priorities of pre-existing network tasks.

\subsubsection*{Observed Metrics}
We defined two sets of experiments observing different metrics under changing network loads.

\begin{figure}[ht]
\centering
\includegraphics[scale=1]{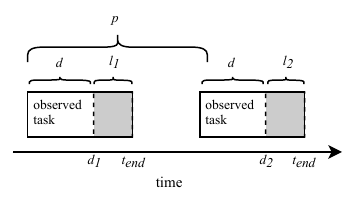}
\caption{\textbf{Lateness:} Measurement of processing delays.}
\label{fig:lateness}
\end{figure}

\paragraph*{Lateness}
In preemptive real-time systems, tasks are brought into the active state by interrupts, timers, or synchronization calls from other tasks. 
The critical task we observe is periodically called by a timer with period \textit{p} and deadline \textit{d}. We define lateness \textit{l} as the time the task takes longer to finish than its deadline allows, hence $ l = t_{end} - d $, where $t_{end}$ is the time at which one process cycle of the task has finished, as shown in the Figure~\ref{fig:lateness}. To see the relationship between the network load and the delay caused, the task does not miss its deadline without any external influence. 
Once we reach a load where the task misses its deadline, the lateness starts to accumulate over iterations. We plot the accumulated lateness per second.

\paragraph*{Relative MCU Utilization}
In the second set of experiments, the observed task runs in a closed loop for a duration of $\Delta t$ in an interval \textit{i}. By measuring the number of cycles the critical task completes in a $\Delta t$ under different network loads, we can determine the resource utilization of network operations as a function of the number of packets received.

\subsection{Test Environment}
The experiments were performed on two widely used IoT development boards in a similar performance and price range. One is equipped with a dual-core ESP32 chip running at 160~MHz, the other with a single-core Particle Photon (P0) ARM Cortex M3 running at 120~MHz. While both devices have ports of the FreeRTOS operating system, the vendor-provided programming frameworks differ significantly in terms of programmability and port-specific implementations. Both devices are programmed using their respective frameworks\footnote{Particle DeviceOS, \url{https://www.particle.io/device-os/}}\footnote{ESP-IDF, \url{https://docs.espressif.com/projects/esp-idf/en/latest/esp32/}}. While the frameworks themselves and the operating systems are open source, some low-level software, such as WiFi drivers, is only available as binary objects.
While the differences between the frameworks and the processing chips used limit the degree of direct comparability, they do allow us to evaluate the possibilities of development within the constraints of the frameworks. In this way, all results are realistic for the systems tested within their established workflows.

\paragraph*{Task Priorities}
\label{sec:priorities}
To keep connections alive, driver and network tasks must be kept running on the devices. These are called by NIC-triggered ISRs, but can be preempted by a higher priority task on the same core.
To evaluate the different shares of delay caused by ISRs and networking tasks, experiments are repeated with observed task priorities chosen above, equal to, and below the driver. The default network task and driver priorities are set by the frameworks as shown in Table~\ref{tab:priorities}. The WiFi driver priorities cannot be changed.

\begin{table}[h]
\caption{Fixed networking task priorities of MCU firmwares.}
\label{tab:priorities}
\centering
\begin{tabular}{r|c||c|l}
\hline
\textbf{OS Priority} & \textbf{ESP-IDF} & \textbf{DeviceOS} & \textbf{OS Priority}  \\
\hline
\textit{24} & - & WiFi driver & \textit{9} \\
\textit{23} &WiFi driver & network (high) & \textit{8} \\
\textit{22-19} & - & network (low) & \textit{7} \\
\textit{18} & network & - & \textit{6-1} \\
\hline
\end{tabular}
\end{table}

\subsection{Interrupt Workload}
\label{sec:intrsetups}
To analyze the impact on latencies and compute resource utilization, the devices under test are subjected to different network loads.

\paragraph*{Generation}
One of the main difficulties in studying the process flow triggered by a received packet is the unavailability of large portions of the driver source code. A quantitative analysis of the number of times an ISR is actually called is therefore difficult to perform. To compare the relative impact of ISRs and low-level packet processing, two interrupt setups are run on the devices.

\begin{figure}
\centering
\begin{subfigure}{.45\columnwidth}
	\centering
	\includegraphics[scale=1]{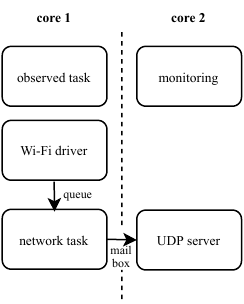}
	\caption{WiFi driver}
	\label{fig:setup_net}
\end{subfigure}
\hfill
\begin{subfigure}{.45\columnwidth}
	\centering
	\includegraphics[scale=1]{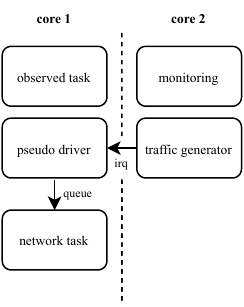}
	\caption{network simulator}
	\label{fig:setup_sim}
\end{subfigure}
\caption{\textbf{Analysis Method:} Experiment setups on ESP32 with core division and messaging channels.}
\label{fig:setups}
\end{figure}

\paragraph*{Real Network Packets}
In the first setup, network packets are sent to the devices over a WiFi connection and handled by the driver and networking tasks provided by the framework (Figure~\ref{fig:setup_net}). We use the second core of the ESP32 to run a minimally configured UDP server that binds a port. While the UDP server task does not directly affect our observed task, the impact of packets received on used ports differs from the impact generated by unused ports. Since P0 does not have a dual-core processor, no UDP server was running and interrupts were generated externally.

\paragraph*{Simulated Packets}
Next, we implemented an analogous task flow to perform WiFi driver-independent experiments (Figure~\ref{fig:setup_sim}). A network task simulation performs much of the same actions as a packet received over a network: Upon registering the interrupt, which in this case is triggered by an input pin, a short ISR is called that preempts the currently running process to copy a packet descriptor to a FreeRTOS queue. A pseudo driver task under the control of the OS scheduler waits for packets in this queue and unblocks it if it is not empty to process the entries.

\paragraph*{Traffic Load}
Changes in network load directly affect the number of ISR, driver, and network stack calls in the operating system. Depending on the system and interrupt type, network loads were increased in steps of 10, 100, or 1,000 packets per second and held for measurement. In addition, experiments were conducted with traffic bursts of 120,000 packets per second for one second.
Given the limited resources of the target devices, this can be considered a fault injection experiment.
The priority of the observed critical task is modified with respect to the priority of the WiFi driver between different runs as described in Section~\ref{sec:priorities}.

\pgfplotsset{every axis/.append style={very thick}, legend style={font=\small}, label style={font=\small}, tick label style={font=\small}}

\section{Experimental Results}
\label{sec:2_2_evaluation}
This section presents our preliminary empirical results of how interrupt loads impact the lateness of critical tasks and utilization of CPUs.

\subsection{Latency Experiments}
\begin{figure}[h]
	\centering
	\begin{subfigure}{.3\textwidth}
		\begin{tikzpicture}
			\begin{axis}[width=\textwidth,
				xmin=0,
				xmax=125000,
				x tick label style={/pgf/number format/1000 sep=},
				xlabel={packets / s},
				y tick label style={/pgf/number format/1000 sep=},
				ylabel={lateness [ms]},
				legend pos=north east,
				legend style={font=\tiny},
				grid=major,
				grid style={dashed},
				no markers
				]
	
				\addplot+ [dotted, color=teal] table[x=sent_low, y=late_low, col sep=comma]{data/plot1.csv};
				 \addlegendentry{p-}
				 \addplot+ [dashed, color=purple] table[x=sent, y=late_high, col sep=comma]{data/plot1.csv};
				 \addlegendentry{p+}
			\end{axis}
		\end{tikzpicture}
		\caption{ESP32: sim. network traffic}
		\label{fig:plot1}
	\end{subfigure}
	\hfill
	\begin{subfigure}{.3\textwidth}
		\begin{tikzpicture}
			\begin{axis}[width=\textwidth,
				xmin=0,
				xmax=4200,
				ymax=1200,
				x tick label style={/pgf/number format/1000 sep=},
				xtick={0,2000,4000},
				xlabel={packets / s},
				y tick label style={/pgf/number format/1000 sep=},
				ylabel style = {align=center},
				ylabel={lateness [ms]},
				legend style={cells={align=left}},
				legend style={font=\tiny},
				legend pos=north west,
				grid=major,
				grid style={dashed},
				no markers
				]
	
				\addplot+ [dotted, color=teal] table[x=sent, y=late_low, col sep=comma]{data/plot5.csv};
				 \addlegendentry{p- bound}
				 \addplot+ [dashed, color=purple] table[x=sent, y=late_high, col sep=comma]{data/plot5.csv};
				 \addlegendentry{p+}
				 \addplot+ [loosely dashdotted, color=gray] table[x=sent, y=late_low_noudp, col sep=comma]{data/plot5.csv};
				 \addlegendentry{p-}
			\end{axis}
	
		\end{tikzpicture}
		\caption{ESP32: real IP packets}
		\label{fig:plot5}
	\end{subfigure}
	\hfill
	\begin{subfigure}{.3\textwidth}
		\begin{tikzpicture}
			\begin{axis}[width=\textwidth,
				xmin=0,
				xmax=1020,
				x tick label style={/pgf/number format/1000 sep=},
				xtick={0,400,800},
				xlabel={packets / s},
				y tick label style={/pgf/number format/1000 sep=},
				ylabel={lateness [ms]},
				legend pos=north west,
				legend style={font=\tiny},
				grid=major,
				grid style={dashed},
				no markers
				]
	
				\addplot+ [dotted, color=teal] table[x=sent, y=late_low, col sep=comma]{data/plot6.csv};
				 \addlegendentry{p-}
				 \addplot+ [dashed, color=purple] table[x=sent, y=late_high, col sep=comma]{data/plot6.csv};
				 \addlegendentry{p+}
			\end{axis}
		\end{tikzpicture}
		\caption{P0: real IP packets}
		\label{fig:plot6}
	\end{subfigure}
\caption{\textbf{Lateness experiment results:} Observed task priorities under networking tasks (p-) and critical (p+).}
\label{fig:lateplots}
\end{figure}
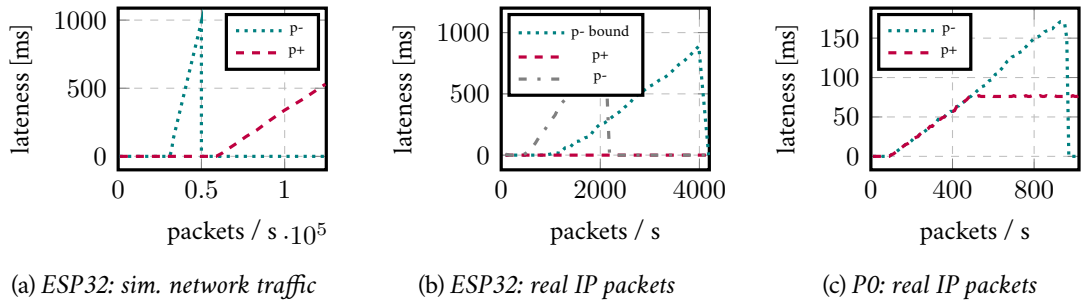

\paragraph*{ESP32}
The first set of experiments was conducted to measure the lateness of the observed task under increasing packet loads. Figures~\ref{fig:plot1} and~\ref{fig:plot5} show the lateness results under simulated and real network traffic on the ESP32. Both show a linear increase in lateness with increasing packet load once lateness occurs for priorities chosen below the driver priority, with real IP packet impact reaching 50\% lateness increase per packet per second. While the absolute values depend on the priority, the relative increase is universal to the system.
The differences between the priorities correspond to the priority of the fixed network tasks when the \acp{mcu} are fully utilized. If the monitored task has a lower priority than the network tasks, it will starve as soon as too many packets arrive. If the monitored task has a critical priority, it will starve the network tasks. As can be seen, this results in no impact, suggesting that no ISRs are triggered by the NIC in this case.

Figure~\ref{fig:plot5} also shows that the impact of incoming packets is much higher when addressed to a port that is not open. This may be due to the partial deactivation of network tasks when UDP buffers in the network stack are full.

The lateness results under bursts of traffic do not differ from those under continuous load, and no other effects on the systems were observed. Further burst results are therefore omitted from this work, as this was the case for all burst experiments.

\paragraph*{P0}
Figure~\ref{fig:plot6} shows the analog results of the experiments on the P0. The results of the simulated approach are very similar to the ESP32 equivalent, with the difference in slope explained by platform-specific pseudo driver implementations. The results of the experiments using real network packets show an impact slope of 2.2\%. The results also show a problem with the network driver being at the highest priority level. When the packet load is high enough that the network driver needs half of the processing resources for itself, the operating system scheduler distributes the processing resources equally between the driver and the observed (critical) task. This keeps the resources available to both tasks constant.
Unlike the ESP32 task, the monitored task does not starve if it has a lower priority. Before this can happen, the WiFi task crashes at 980 packets per second. No reconnection attempts were observed in this scenario.

\subsection{Utilization Experiments}

\begin{figure}[h]
	\centering
	\begin{subfigure}{.3\textwidth}
		\begin{tikzpicture}
			\begin{axis}[width=\textwidth,
				xmin=0,
				xmax=108000,
				x tick label style={/pgf/number format/1000 sep=},
				xlabel={packets / s},
				y tick label style={/pgf/number format/1000 sep=},
				ylabel={task cycles / s},
				legend pos=north east,
				grid=major,
				grid style={dashed},
				no markers
				]
	
				\addplot+ [dotted, color=teal] table[x=rcv_low, y=cc_low, col sep=comma]{data/plot8.csv};
				 \addlegendentry{p-}
				 \addplot+ [dashed, color=purple] table[x=rcv_high, y=cc_high, col sep=comma]{data/plot8.csv};
				 \addlegendentry{p+}
			\end{axis}
		\end{tikzpicture}
		\caption{P0: sim. network packets}
		\label{fig:plot8}
	\end{subfigure}
	\hfill
	\begin{subfigure}{.3\textwidth}
		\begin{tikzpicture}
			\begin{axis}[width=\textwidth,
				xmin=0,
				xmax=3500,
				x tick label style={/pgf/number format/1000 sep=},
				xlabel={packets / s},
				y tick label style={/pgf/number format/1000 sep=},
				ylabel={task cycles / s},
				legend pos=north east,
				grid=major,
				grid style={dashed},
				no markers
				]
	
				\addplot+ [dotted, color=teal] table[x=sent, y=cc_low, col sep=comma]{data/plot10.csv};
				 \addlegendentry{p-}
				 \addplot+ [densely dashed, color=violet] table[x=sent, y=cc_high, col sep=comma] {data/plot10.csv};
				 \addlegendentry{pd}
			\end{axis}
		\end{tikzpicture}
		\caption{ESP32: IP packets, unb. port}
		\label{fig:plot10}
	\end{subfigure}
	\hfill
	\begin{subfigure}{.3\textwidth}
		\begin{tikzpicture}
			\begin{axis}[width=\textwidth,
				ymin=70,
				xmin=0,
				xmax=2250,
				x tick label style={/pgf/number format/1000 sep=},
				xlabel={packets / s},
				y tick label style={/pgf/number format/1000 sep=},
				ylabel={task cycles / s},
				legend pos=south west,
				grid=major,
				grid style={dashed},
				no markers
				]
	
				\addplot+ [dotted, color=teal] table[x=sent, y=cc_low, col sep=comma]{data/plot11.csv};
				 \addlegendentry{p-}
				 \addplot+ [dashed, color=purple] table[x=sent, y=cc_high, col sep=comma] {data/plot11.csv};
				 \addlegendentry{p+}
			\end{axis}
		\end{tikzpicture}
		\caption{P0: real IP packets}
		\label{fig:plot11}
	\end{subfigure}
	\caption{\textbf{Performance results:} Observed task priority lower than networking task (p-), on critical (p+), and equal with WiFi driver (pd).}
	\label{fig:ccplots}
	\end{figure}
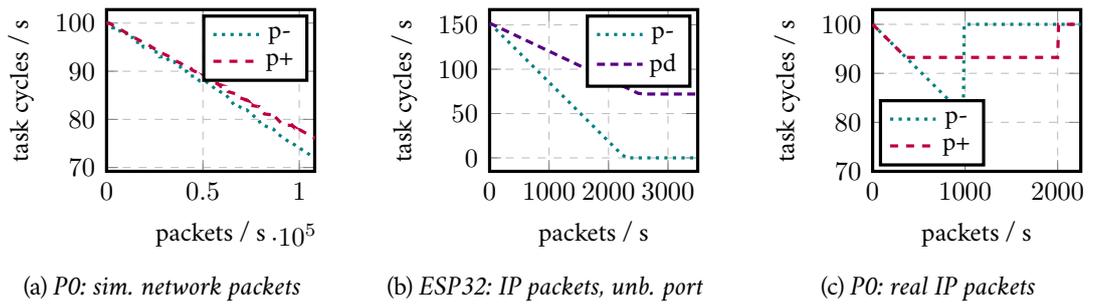

Figure~\ref{fig:ccplots} shows the parallel results for real network packets and software. The task priorities of the results shown were chosen to be equal to and lower than the WiFi driver. Task performance decreases by 2.15\% and 3.8\%, respectively. When packets are received on an unused port, the impact is even greater, with performance drops of 3.17\% and 6.67\%, respectively.

Figure~\ref{fig:plot8} shows the results for the network driver simulation on P0, which are comparable to the lateness results. With real network packets, the system crashes at 980 packets per second when the observed task is given a lower priority than the WiFi driver, and at 2,000 packets per second with the same priority. The performance of the task decreases by 2\% per packet per second until it reaches equal resource utilization with the driver (same priority) or is gradually displaced by it (lower priority).

\section{Discussion}
\label{sec:2_2_discussion}
This section discusses the findings from the experiments and the feasibility of IP networking on \acp{mcu}.

\subsection{Insights from the Experiment Results}

The evaluation results show that the performance and the realtime performance are directly dependent on the incoming IP network load. The overhead of receiving packets is high for both continuous floods and short bursts. The WiFi driver and the network stack use all the computational resources they need to handle packets unless they are preempted. However, the results also show that NIC-triggered ISRs have no impact on the observed systems. This observation is made when incoming packets are not handled because the driver is preempted by a task of critical priority. In the case of P0, it was not possible to assign a higher priority to a task than the priority of the WiFi driver. Therefore, it was not possible to measure NIC-generated ISR durations.

To mitigate the observed real-time guarantee violations on the tested devices, developers still have some options. The ESP IDF provides a priority level above the WiFi task's priority. Running a critical task here will preempt the driver. For critical code snippets, it is also possible to disable all or only the WiFi interrupt. This is the only option under Particle's DeviceOS.

\subsection{Feasibility of IP Networking}
The key takeaway is that all mission-critical operations must be incorporated into a critical task that is independent of any signals received over the network. Therefore, running command and control operations over the network introduces a high safety risk to hard real-time systems. In systems where the only way to disable network interrupts is to disable all interrupts, the critical task must be completely independent of any I/O signals. However, this invalidates most real-time embedded system use cases.

Network driver tasks, which are responsible for much of the impact on timing predictability and thus latency, have very high priority in the frameworks of the systems tested. There is no obvious way to change their priorities. This severely limits the leeway that critical tasks have. While short and independent tasks and code blocks can be executed in a safe manner, this is also where it ends.

This leads to the conclusion that network driver tasks are currently given a priority level in RTOSs that is not suitable for critical real-time systems. Using network connectivity only for monitoring largely timing independent tasks may still be an option if one is willing to provide the resources and prioritize the necessary tasks appropriately.

\subsection{Networking Software Robustness}
As discussed, current programming frameworks for IoT devices seem to be designed for certain connectivity guarantees rather than for the real-time of critical applications. However, this does not apply to the evaluated systems either, as can be seen from the results. Under the (rather general) conditions of the test setup, even low traffic loads lead to WiFi driver crashes and, depending on the framework configuration, even to total system failures. This could be reliably reproduced on P0 (see Figure~\ref{fig:plot11}), where the presence of a task with critical priority also prevented the driver from reconnecting after a connection loss. Similar driver crashes were observed in some iterations of the tests on the ESP32.

\section{Excursion: The PIERES Testbed}
\label{sec:2_2_testbed}

While implementing the system for our analysis, we recognized a need to make it easier to run experiments for specific real-time applications on connected embedded systems. For future research on the identified interrupt issues, we have designed a testbed. Its goal is to allow researchers to perform experiments and engineers to test their code with respect to network interrupts. The testbed should fulfill the following requirements. It has to
\begin{compactitem}
	\item run on real IoT hardware,
	\item be capable of simulating multiple NIC implementations,
	\item be able to simulate multiple network traffic scenarios,
	\item be easily configurable, and
	\item have minimal performance impact on the tested process.
\end{compactitem}

\subsection{Operation of the Testbed}
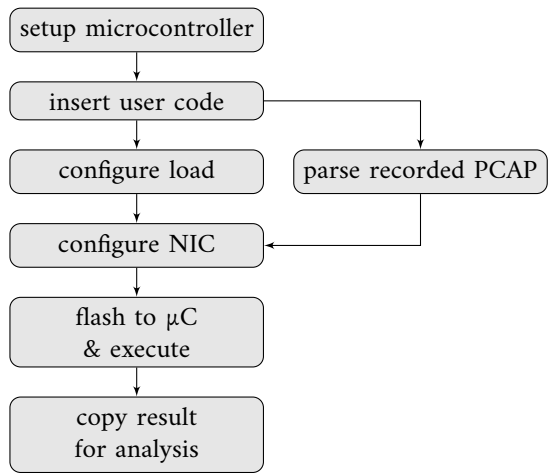
\begin{figure}[h]
	\begin{center}
	\tikzstyle{block} = [rectangle, draw, fill=black!10, 
    text width=8em, text centered, rounded corners, minimum height=1em]
	\tikzstyle{line} = [draw, -latex']
    
	\begin{tikzpicture}[node distance = 1cm, auto, every node/.style={font=\small}]
		
	    \node [block] (setup) {setup microcontroller};
    	\node [block, below = 1em of setup] (insert) {insert user code};
	    \node [block, below = 1em of insert] (randLoad) {configure load};
	    \node [block, right = 1em of randLoad, node distance=3cm] (pcap) {parse recorded PCAP};
    	\node [block, below = 1em of randLoad] (nic) {configure NIC};
    	\node [block, below = 1em of nic] (flash) {flash to \textmu{}C \& execute};
    	\node [block, below = 1em of flash] (analysis) {copy result for analysis};

    	\path [line] (setup) -- (insert);
	    \path [line] (insert) -- (randLoad);
    	\path [line] (randLoad) -- (nic);
	    \path [line] (pcap) |- (nic);
    	\path [line] (insert) -| (pcap);
	    \path [line] (nic) -- (flash);
	    \path [line] (flash) -- (analysis);
	\end{tikzpicture}
	\end{center}
	\caption{\textbf{Testbed operation:} Process of testbed usage.}
	\label{process}
\end{figure}

The operation of the testbed is shown in Figure~\ref{process}. The testbed is set up on a microcontroller. The user then inserts the code of the critical process under test and selects either a uniform load configuration, a random load configuration, or parses a recorded network trace. The user then configures the NIC and flashes the testbed to the device. Measured metrics can be configured by the user and range from execution time and number of interrupts to more complex metrics such as the ratio between interrupt sources (see Section~\ref{sec:2_2_experiments}).

\subsubsection*{NIC Implementations}
The testbed needs to model different \acp{nic} that can be configured by the user. The simplest NIC would notify the \ac{mcu} of the arrival of each packet by throwing an interrupt for each one. Alternatively, some modern NICs offer interrupt moderation. To incorporate this into the testbed, we offer a choice between a simple NIC model without interrupt moderation and several smarter NIC implementations.

For the simple NIC, the duration $d(l)$ of an interrupt depends on the packet length $l$ and is simply modeled as a length-dependent delay $d_l$ that is invoked $l$ times plus a constant length-independent delay $d_c$ that is an overhead that is invoked for each interrupt:

\begin{equation*}
	d(l) = d_l \cdot l + d_c
\end{equation*}

Different simple NIC implementations can be characterized by the user by setting the values for the length dependent delay $d_l$ and the length independent delay $d_c$.

For more sophisticated NIC simulations with interrupt moderation (cf. Section~\ref{sec:interrupt_moderation}), we support out-of-the-box definition of NICs with a counter mode, a timer mode, or a combination. In this case, each part of the interrupt duration model, the packet length and the corresponding dependent and independent delay, is modeled twice, once for the simulated ISR and once for a simulated receiver task.

A counter mode NIC does not generate an interrupt for every packet received, but instead counts the incoming packets, stores them in a buffer, and, after a certain number of packets, generates an interrupt for all of them. The counter and buffer are then reset. Another option is a NIC with timer mode. In this case, a delay timer of a specified duration is set for an incoming packet. When the timer expires, an interrupt is generated. If more packets arrive before the timer expires, the timer is reset without triggering an interrupt. However, problems may arise if the timer is constantly reset by incoming packets, never allowing an interrupt to be generated. The combination of both modes avoids this problem.

\subsubsection*{Network Traffic Scenarios}
The arrival of packets with corresponding timestamps over an observed period of time constitutes a load scenario. Since we want to simulate different scenarios, the testbed offers uniform loads, random loads, and user-defined/recorded loads.

The uniform loads have a constant receive frequency. Random loads use a Poisson distribution to model the arrival of new packets.

The Poisson distribution is obtained by inverse transform sampling with a uniform distribution. Assuming that the number of incoming packets per interval $p_i$ is Poisson distributed, the inter-arrival time $d_i$ is exponentially distributed:

\begin{equation*}
\begin{split}
	p_i &\sim Poisson(\lambda), \; d_i = p_{i+1} - p_i\\
	\Rightarrow d_i &\sim Exp(\lambda)\\
	\hat{d}_i &= F^{-1}(u_i) = - \frac{1}{\lambda}\ln(1-u_i) \hat{=} -\frac{1}{\lambda}\ln(u_i).
\end{split}
\end{equation*}

Using inverse transform sampling (as described in~\cite{Murphy2012} Section 23.2), we determine the empirical delays between packets $\hat{d}_i$ by sampling $u_i \sim U(0,1)$ and calculating $\hat{d}_i$. By setting the parameter $\lambda$, different randomized Poisson loads can be specified.

As a third option, the testbed allows you to replay recorded network scenarios.

\subsection{Test Execution}
\label{sec:2_2_experiments}

Two validation and two demonstration experiments were performed using a large summation (in a loop) with a conditional statement at each step of the iteration as the user code.

\subsubsection*{Prototype Implementation}
The testbed is implemented on an ESP32. The parsing script for the recorded network scenarios (PCAPs) is written in Python and generates C++ code. The last requirement of the testbed is met by using both cores to separate the computational load of the testbed code from the tested user code.

\subsubsection*{Validation}

All validation tests were performed with a Poisson distributed random load. The following results show only a selection of the experiments performed on the testbed.

\begin{figure}
	\centering
	\begin{subfigure}{.49\textwidth}
		\includegraphics[width=\textwidth]{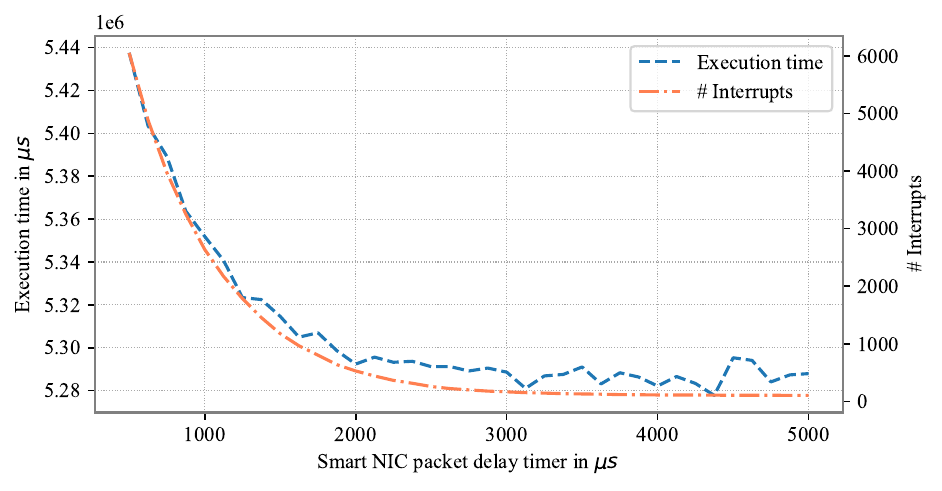}
		\caption{}
		\label{packetTimeoutVsExcTimeAndIntrs}
	\end{subfigure}
	\hfill
	\begin{subfigure}{.49\textwidth}
		\includegraphics[width=\textwidth]{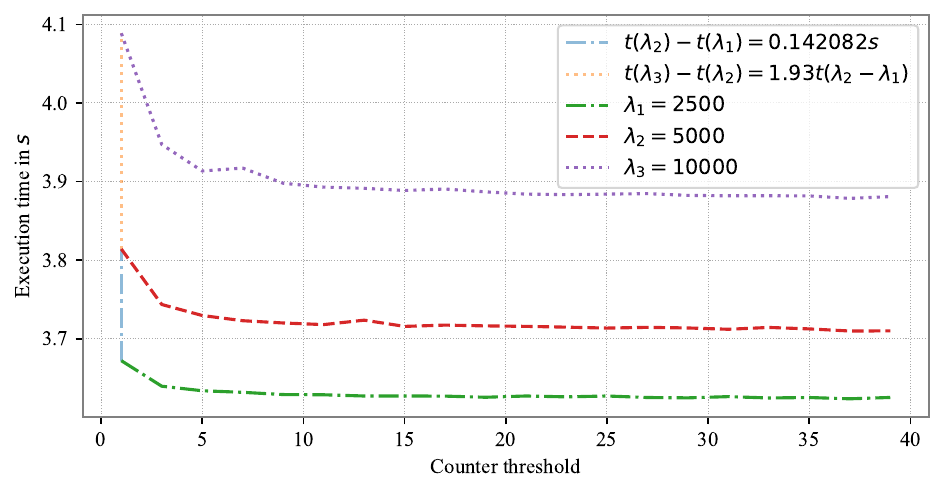}
		\caption{}
		\label{counterRatioPreserv}
	\end{subfigure}
	\caption{\textbf{Validation results:} Left: Execution time and number of interrupts decrease with the increase of the timer threshold when using the NIC packet delay timer for interrupt moderation. Right: Comparison of the execution time of three Poisson-distributed random network load scenarios for different $\lambda$ in dependence of the counter threshold.}
\end{figure}

First, the impact of the packet delay timer on execution time and interrupt count was measured, as shown in Figure~\ref{packetTimeoutVsExcTimeAndIntrs}. This test was performed using the combined interrupt moderation mode. %
Both the execution time and the number of interrupts decrease as the packet delay timer increases. Since the number of packets is constant, the execution time is reduced by using interrupt moderation. This gain from increasing the packet delay timer comes with the caveat of higher packet latency. %

In the second validation experiment, three different parameters $\lambda$ for the Poisson-distributed random load generation were compared, as shown in Figure~\ref{counterRatioPreserv}. This test was performed in counter mode. A higher parameter value corresponds to a higher load. The graph plots the counter ratio against the execution time. %
$\lambda_1$ is half the size of $\lambda_2$ and $\lambda_3$ is twice the size of $\lambda_2$. As shown in the legend, the difference between the execution times of $\lambda_3$ and $\lambda_2$ is twice as large as the difference between $\lambda_2$ and $\lambda_1$. This ratio stays roughly the same along the x-axis. This shows that the counter mode scales linearly with the load.

\subsubsection*{Practical Examples}

\begin{figure}
	\centering
	\includegraphics[width=0.6\textwidth]{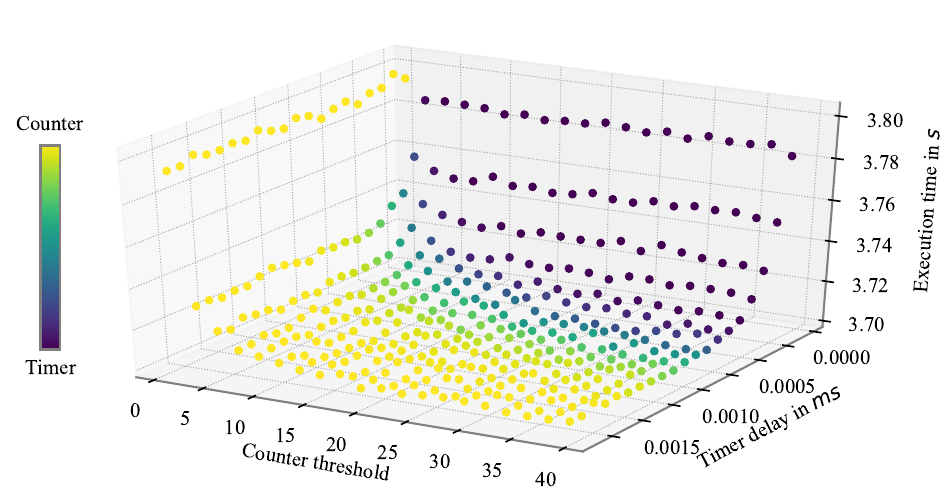}
	\caption{\textbf{Mixed interrupt moderation: }Ratio of the reasons for triggering interrupts, shown in relation to counter threshold, timer delay and execution time when using the NIC with mixed mode moderation.}
	\label{prac2}
\end{figure}

In the first practical example, the causes of interrupts were investigated using the combined interrupt moderation mode of the smart NIC model, as shown in Figure~\ref{prac2}. A Poisson distributed random load was used for this test. The coloring of the data points indicates the reason why an interrupt was generated. An area can be observed in the plane of the data points where the coloring indicates a balance between the two reasons. The area extends in both the counter threshold and timer delay directions, but drifts in the direction of the counter threshold axis, indicating that as the counter threshold increases, it plays less of a role in causing interrupts than does the timer delay.

In the second practical example, we take a look at recorded loads. We compare the execution time of user code using a mixed-mode NIC with a Spotify network load to a Zoom conference load that has been pre-recorded. The load is much less intensive compared to the previous experiments. We use these two loads because they have two different packet arrival patterns: the Spotify load is bursty, while the Zoom load is more continuous. Note that we use a longer running user code here (more iterations) to allow for a longer measurement.

\begin{figure}
	\centering
	\includegraphics[width=0.6\textwidth]{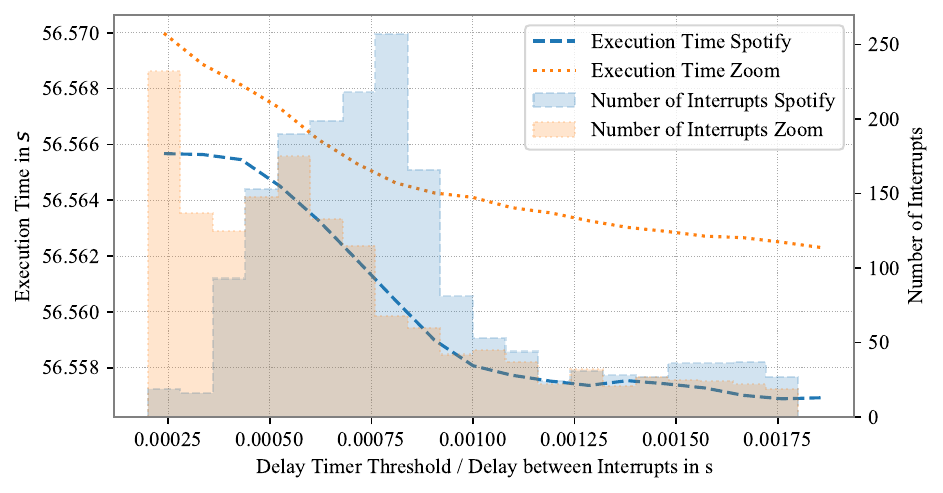}
	\caption{\textbf{Execution time results:} Run time of the test code in two replayed network scenarios from prerecorded PCAPs is shown for diverse delay timer thresholds.}
	\label{pcapDiag}
\end{figure}

Figure~\ref{pcapDiag} is a combination of two plots: the lines show the execution time for different delay timer thresholds, while the filled area is a histogram showing the intervals at which packets arrive in the two load scenarios. We can see that the execution time in the Spotify scenario benefits more from relaxing the timeout delay, while the behavior in the Zoom scenario more closely resembles that of a Poisson load.

Comparing the results of the two scenarios, it becomes clearer that the expected load behavior can be used to find appropriate interrupt mitigation parameters.

\section{Summary}
In this chapter, we presented an analysis of the impact of IP networking (or more specifically, packet reception) on computationally weak embedded real-time devices. We have also presented a testbed for experiments in this context.
The following points can be deduced:
\begin{compactitem}
\item Without explicit care, networking has a huge impact on the real-time performance of embedded systems, making real-time guarantees invalid.
\item Making mission critical real-time systems dependant on command and control signals sent via IP networks is not feasible without introducing high risk.
\item Sending command and control signals over an IP network is not feasible for tasks that require hard real-time guarantees.
\item The inflexibility of current network stack implementations and the high network overhead suggest reserving a separate core for network software.
\item The limited access to network-related tasks prevents holistic real-time system design and cooperative scheduling between them and application tasks.
\end{compactitem}

The use of the frameworks used to deploy the devices under study is not necessary for their programming. However, the development effort increases significantly when WiFi and IP networking functionality is manually added to a bare-metal design. Yet, cyber-physical systems that could pose a threat to their environment would justify this. At the same time, much of the problem lies in the general interdependency between real-time processes and the networking subsystem. Any real-time task that is, for example, waiting for IP-delivered control commands cannot be effectively isolated from the network, as isolation would render the function useless. 

In the next three chapters, we present several algorithms and system designs to mitigate these problems. However, simultaneous guarantees for both real-time deadlines and continuous operation are not possible when deploying devices in (wireless) IP networks.

\chapter{Real-Time Aware Packet Processing}
\label{cha:2_3_software}
\minitoc
Our analysis of networking impact on timeliness has shown that two factors of packet reception lead to reduced predictability in real-time systems: The hardware interrupts and the subsequent \ac{isr} execution generated by the \ac{nic} as well as the processing of incoming packets by the network stack task. It has been shown that an unmodified embedded \ac{ip} subsystem can have disastrous effects on the timing characteristics of real-time tasks.
Under high traffic, almost arbitrary delays can be induced, up to the point where the system is busy processing \ac{ip} packets.
Therefore, connecting embedded systems to \ac{ip} networks puts real-time properties at risk.
However, embedded devices are increasingly connected to \ac{ip} networks for remote control, measurement data reporting, software maintenance, and diagnostics~\cite{jazdi2014cyber}.

In order to secure critical device functions, typical \acp{rtos} provide the option of critical code sections where all interrupts are disabled. Depending on the remaining functions of the device, these sections can vary in size. The problem with turning off interrupt handling is that any function that depends on external input is disabled for the time being. Unfortunately, this is the default for embedded control systems, i.e., for remote control over IP, neither the network subsystem nor the relevant interrupt sources can be turned off, even though the associated physical motion is time-critical. 
The same issue exists for other traditional network security methods such as firewalls. As long as critical aspects of its functionality depend on receiving network packets, the \ac{rtos} must both provide connectivity and predictably process packets and forward them to the receiving critical real-time task. A firewall cannot do this in a prioritizing manner.    

As for mitigating the contention problem from the software side (the operating system), we cannot influence or moderate the hardware interrupts except by rejecting them altogether. Hardware solutions range from dedicating another processor core to network tasks~\cite{hermes} to sophisticated \acs{nic} offloading~\cite{nics, behnke2022priority} and \enquote{smart \acs{nic}s} equipped with their own fully featured processing system~\cite{smartnics, humphries2019mind}. In addition, advanced interrupt controllers also address priority space unification~\cite{task_aware}. Best-effort IP networks with wireless links that implement specialized real-time network technologies (such as \ac{tsn} as proposed in~\cite{8412458}) are still an active research topic~\cite{bruckner_introduction_2019} (cf. Chapter~\ref{cha:1_3_survey}). How interrupt moderation on the hardware side (the \ac{nic}) can help us with this is discussed in Chapter~\ref{cha:2_4_hardware}. 

In this chapter, we examine and discuss how we can use the tools provided by an \ac{rtos} to both identify congestion situations and mitigate their effects. Mitigating the effects of network-generated interrupts from software in a real-time operating system is challenging because the timing impact of the mitigation techniques themselves must be kept to a minimum. 
To address this issue, this chapter presents the following contributions that have been peer-reviewed for three publications~\cite{danicki2021detecting, blumschein2022differentiating, behnke2023towards}.

\begin{compactitem}
    \item Identification of three metrics for detecting amounts of interrupts that may jeopardize the local real-time guarantees and four algorithms to mitigate the impact of high interrupt loads while maintaining the network services on a best-effort basis. 
    \item A priority-differentiating driver extension for typical embedded \acp{rtos} and \acs{ip}-stacks that facilitates the deployment of IP networking in real-time scenarios combining the properties of: (1) Protection against network-induced system overloads, facilitating real-time systems,
    (2) Optimal processing latency for well-behaved \acl{hp} flows, and
    (3) Best-effort performance for \acl{lp} flows.
    \item Experimental evaluations of the four mitigation algorithms and the packet receive architecture.
\end{compactitem} 

Section~\ref{sec:2_3_considerations} discusses the preliminary considerations for our solutions. 
Section~\ref{sec:2_3_mitigation} presents the metric-based detection and mitigation algorithms, followed by their evaluation and discussion in Section~\ref{sec:2_3_mitigation_evaluation}.
Section~\ref{sec:2_3_differentiating} presents our priority-differentiating network driver extension followed by its evaluation in Section~\ref{sec:2_3_ipstack_eval}.
Section~\ref{sec:2_3_conclusion} concludes this chapter.

\section{Preliminary Considerations}
\label{sec:2_3_considerations}

For the purposes of the remainder of this thesis we consider \emph{devices} to be embedded systems running an \ac{rtos} and a small number of processes with fixed priorities managed by a preemptive scheduler. The embedded device is a constrained node in an IP network serving real-time applications as well as best-effort networking. 
In the network stack, a driver controls \ac{dma} transfers, establishes cache coherency and passes packet buffers to a networking task by means of a queue.
The device contains an embedded \ac{nic} used to connect to an IP network which itself might be connected to the Internet. 

The proposed approaches make use of IP flow information for real-time aware packet prioritization. Packets are received by real-time processes on the embedded system.

\begin{mydef}
\label{def:1}
Let $\mathcal{F}$ be the set of IP flows received by the regarded device. An IP flow $f  \in \mathcal{F}$ is a sequence of IP packets arriving at the regarded device and for the purposes of this thesis characterized by the tuple $(Src, Dst_{port}, P, t_P )$.
\begin{compactitem}
\item $Src$ is the source node, identified by its IP address.
\item $Dst_{port}$ is the destination port number.
\item $P$ is the priority of the flow as observed by the regarded device.
\item $t_P$ is the minimum expected period of the flow, meaning the interarrival time between two packets in the flow. 
\end{compactitem}
\end{mydef}

The priority of a flow is a parameter assigned by the receiving system, as the real-time implications to this system need to be considered. The period of a flow depends on the sender, application, and network infrastructure. 

\begin{figure}[h]
\centering
\includegraphics[width=0.7\columnwidth]{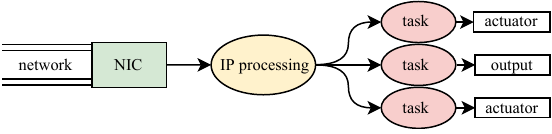} 
\caption{\textbf{Data flow:} Path of received messages through the regarded device.}
\label{fig:dataflow}
\end{figure}

\paragraph*{Data Flow}
Each task listens on a separate socket for messages. Each message is delivered using one IP packet. The tasks process the message and produce an output or control a physical actuator. Figure~\ref{fig:dataflow} shows the resulting data flow model. 

\subsection{Packet Processing Time Analysis}
\label{sec:2_3_proc_times}
To effectively save processing time, we need to determine the stages through which a received packet passes and their respective costs in terms of processing time. On both considered IP stack implementations (FreeRTOS+TCP and LwIP), we find an \ac{isr}, an Ethernet driver task (acting as \acs{ist}) and a protocol processing task. Relevant activities are

\begin{compactenum}
	\item Acknowledge \ac{irq}
	\item Remove received buffer from \ac{bd}-ring and invalidate its cache
	\item Refill \ac{bd} ring with newly allocated and invalidated buffer
	\item Parse packet contents, OSI layers 2 through 4, and act accordingly
\end{compactenum}

These activities are distributed over the stages as shown in Table~\ref{tab:work_in_stages}. Note that the division of \ac{isr}/\ac{ist} in the FreeRTOS+TCP driver is done in the recommended way, which reduces the schedule latencies caused by uninterruptible \acp{isr}. 

\begin{table}[h]
	\centering
    \caption{Distribution of necessary work over different stages in lwIP and FreeRTOS+TCP}
	\begin{tabular}{@{}l|ll@{}}\toprule
        Stage & \acs{lwip} & FreeRTOS+TCP \\\midrule
        \acs{isr} & (1) (2) (3) & (1) \\
        \acs{ist} & & (2) (3) \\
        \acs{ip}-task & (4) & (4) \\\bottomrule
    \end{tabular}
	\label{tab:work_in_stages}
\end{table}

When aiming to improve system performance by rationalizing away some parts of a program, it is imperative to follow a special case of \emph{Amdahl's Law}, where removing the need to execute a particular piece of code is like applying an infinite parallelization factor to it:
We need to eliminate parts that require a lot of processing time in order to successfully achieve a decent speedup. To establish reliable measurements, we performed a set of simple preliminary experiments evaluating the two network stack implementations. Timing tests were performed on a Xilinx Zynq-7000 FPGA-SoC combinig an ARMv7 Cortex-A9 dual-core processor and an \ac{fpga}.

As a first experiment, we measured how much \ac{udp} traffic had to be received to cause a \ac{cpu} load of 10\%.
In general, it was of little importance whether the packets were \ac{udp} packets sent to bound/unbound ports, \ac{tcp} packets sent to unbound ports, or even ARP requests, as long as there was no immediate response such as \ac{icmp} port unreachable, \ac{arp} response, or \ac{tcp} RST is provoked. 

To get an estimate of how much \ac{cpu} time each stage takes, we successively disabled more and more of the processing pipeline by making small changes to the code. The results are shown in Table~\ref{fig:packets_in_stages}.

\begin{table}[h]
	\centering
    \caption{Incoming \acs{udp} packets per second incurring a \acs{cpu}-load of 10\%}
	\label{fig:packets_in_stages}
	\begin{tabular}{@{}l|rr@{}}\toprule
	    Stage & \acs{lwip} & FreeRTOS+TCP \\\midrule
        \acs{isr} without cache invalidation & 40,000 & - \\
        \acs{isr} & 24,000 & 125,000 \\
        \acs{isr}, \acs{ist} & 12,000 & 14,500 \\
        \acs{isr}, \acs{ist}, \acs{ip}-task & 6,600  & 8,000 \\\bottomrule
    \end{tabular}
\end{table}

Notice the entry \emph{ISR without cache invalidation}.
Filled packet buffers have to be cache-invalidated after retrieval out of and freshly allocated ones prior to their insertion into the \ac{bd} ring in order for the \ac{dma} to work correctly. 

Since initial profiling showed a significant time overhead in the corresponding routines, we performed a test with the \acs{lwip} stack that omitted the cache invalidation of the freshly allocated buffer to be inserted into the \ac{bd} ring.
The cache invalidation of the retrieved packet is not as big an issue because its payload is zero length and the driver only invalidates the used portion of a packet buffer.

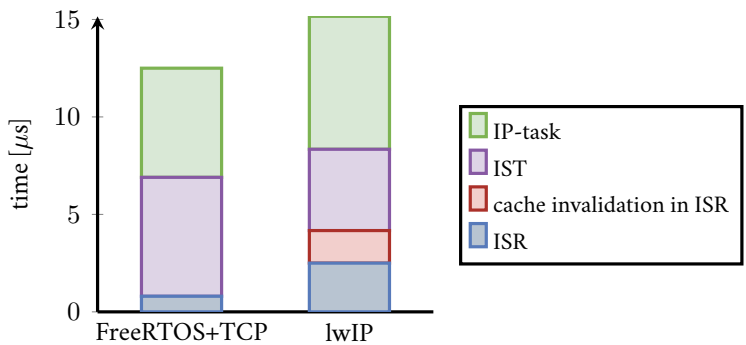
\begin{figure}[h]
	\centering
	\begin{tikzpicture}
    \begin{axis}
        [
            reverse legend,
            ybar stacked,
            width=6cm,
            height=5.5cm,
        	bar width=30pt,
        	ymin=0,
            enlargelimits=0.5,
            legend style={
                cells={anchor=west},
                at={(1.5,0.7)},
                anchor=north,
                font=\footnotesize},
            ylabel={time [$\mu$s]},
            symbolic x coords={FreeRTOS+TCP, lwIP},
            xtick=data,
            axis x line*=none,
            axis y line=left,
        ]
        \addplot+[ybar, draw = LineBlue, fill = FillBlue] plot coordinates {(FreeRTOS+TCP,0.8) (lwIP,2.5)};
        \addplot+[ybar, draw = LineRed, fill = FillRed] plot coordinates {(FreeRTOS+TCP,0) (lwIP,1.67)};
        \addplot+[ybar, draw = LinePurple, fill = FillPurple] plot coordinates {(FreeRTOS+TCP,6.1) (lwIP,4.17)};
        \addplot+[ybar, draw = LineGreen, fill = FillGreen] plot coordinates {(FreeRTOS+TCP,5.6) (lwIP,6.82)};
        \legend{\strut \acs{isr}, \strut cache invalidation in \acs{isr}, \strut \acs{ist}, \strut \acs{ip}-task}
    \end{axis}
    \end{tikzpicture}
	\caption{\textbf{Processing time differentiation:} Measured \acs{cpu} time per packet by stages in FreeRTOS+TCP and \acs{lwip}}
	\label{fig:cputime_in_stages}
\end{figure}

These numbers can be used to reconstruct the stage processing times (see Figure~\ref{fig:cputime_in_stages}). We can see that switching to and from the \ac{ist} in \ac{lwip} causes some measurable overhead.
Also, we see that a large portion of the processing time is actually spent in the ethernet driver compared to the \ac{ip} task. In particular, cache invalidation accounts for a significant portion of this time.

To summarize the key takeaway: To successfully save \ac{lp}-flow packet processing time, we need to start very early in the driver and develop a strategy to defer some per-packet driver handling, such as cache invalidation in our case. This is discussed in more detail in Section~\ref{sec:2_3_differentiating} after our metric-based detection and mitigation algorithms are introduced.

\section{Metric-based Detection and Mitigation Algorithms}
\label{sec:2_3_mitigation}
This section presents detection techniques and mitigation algorithms to handle network-generated interrupt floods in real-time systems. The metrics used to achieve this are defined before the algorithms are discussed and evaluated. 

\subsection{Detection}
To prevent a critical task from being drowned by network interrupts, we first need to detect such a situation. There are several different explicit and implicit metrics we can use to do that.

\paragraph*{Early- or Lateness} The most explicit metric is the real-time task's early- or lateness.
In many real-time systems, a process periodically performs a critical computation, targeting to finish it within a fixed duration.
When the critical task completes this computation in time (i.e. in less time than the target duration), we have positive earliness, defined as the target duration minus the actual time spent. When modeling real-time tasks, this is considered the task \emph{laxity}. 
When, however, the critical task exceeds its target duration, it incurs lateness, defined as how much longer it took than targeted. We will express lateness as a ratio of the critical task's target duration, e. g. $100\%$ lateness means it took twice as long as intended.

This metric directly corresponds to what we are trying to detect but also has drawbacks: 
Firstly, its measurement introduces latency as processes can only report earliness/lateness once per task cycle.
Thus, mitigation techniques might need to be overly cautious because otherwise, it will react too late.
Additionally, there is the more practical concern that the metric may be hard to come by in real systems since somehow, the metric must be reported. We thus call mitigation techniques relying on this \emph{cooperative}.

\paragraph*{Network Interrupt Count} A less direct metric is the number of incoming interrupts, discretized by dividing time into fixed time slices and identify the number of network interrupts in them.

Since many network interrupts occur per time slice, the metric's resolution is equally high. Timing precision is not crucial because misattributing the first few packets to the passed time slice does not introduce significant errors.

The drawback of this metric is that it only correlates with the situations we are trying to prevent if certain preconditions are true:
The network interrupt count shows approximately how much workload the interrupts are putting on a CPU.
We can thus estimate overall system resource usage if the resource requirements of the critical task stay approximately the same for each of its cycles.
The first precondition, therefore, is that the critical task has to require the same amount of resources at all times --- mitigation techniques relying on this can not react elastically to load change.
Furthermore, this metric assumes that the processing of each packet takes about the same time.
Mitigation techniques using this metric can only be effective if the ensemble of incoming packets is homogeneous enough such that the assumption that each packet takes approximately the same time to process is either true or practical because of the regular distribution of packet response time.

\paragraph*{Network Receive Queue Fill State} \label{sec:queue-fill-metric}
A third possible metric is detecting the queue fill state of received packets to be processed by the network driver.
The queue fill state shows whether the network driver can keep pace with the incoming packets.
If more packets arrive than the network driver can handle, the queue will fill up until it reaches maximum capacity; it would empty in the opposite case.

Of course, the queue's capacity impacts the quality of this metric.
With a queue capacity that is too low, the metric will show saturation even for short bursts of packets that are not representative of the overall traffic pace.
Low queue sizes will also lead to saturation if the scheduler did not wake up the network driver for some time.
If, conversely, the queue is too large it acts as a cushion and will delay alerting by filling up, allowing the network routines to stay activated for longer than is prudent.

The metric scales well in terms of packet response time deviation and elastic critical loads because it directly mirrors the ability of the network driver to process incoming traffic.
It has to be coupled with a mechanism like scheduler priority that moderates network driver execution such that an appropriate resource ceiling is found for the tasks.

\subsection{Mitigation Techniques}
\label{sec:mitigations}
In the following, we present four different mitigation techniques. The goal of these methods is to protect one or more time-critical tasks from interrupt overloads.
With the \emph{Burst Mitigation}, we suggest a technique that prevents unresponsiveness in case of short packet bursts.
The \emph{Hysteresis Mitigation} controls the networking tasks based on the early- or lateness of the critical task.
This idea taken further, the novel \emph{Budget Mitigation} attempts to balance out the networking part and the critical task by assigning a networking budget to the driver that is derived from the critical task's earliness.
Finally, the novel \emph{Queue Mitigation} takes a step back and explores a different way to detect high load, enabling a simple yet effective mitigation technique.

\begin{figure}[h]
    \centering
    \includegraphics[width=.9\columnwidth]{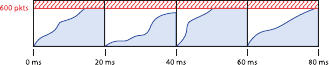}
    \caption{\textbf{Burst Mitigation:} Visualization of time slices.}
    \label{fig:burst}
\end{figure}

\paragraph*{Burst Mitigation}
Burst Mitigation is a simple approach to deal with very high packet loads in short periods.
Conceptually, time is split into fixed time slices, each having a fixed maximum packet capacity.
The network ISR tracks these time slices by querying the operating system for the tick count each time it is invoked, starting a new slice if enough time has passed since the start of the last one.
In each slice, the number of packets is counted.
If this packet count surpasses the fixed capacity, network interrupts are disabled until the start of the next time slice.
We visualize this in Figure~\ref{fig:burst}: The length of each time slice is $20ms$ and the curve plots the number of received packets in each slice.
The red zone on top is off-limits --- when $600$ packets are reached, interrupts are disabled.

\paragraph*{Hysteresis Mitigation}
When the network load exceeds the trigger capacity for the Burst Mitigation by a small amount, network interrupt activation migh oscillate.
We used the number of packets per time frame to detect whether we had to act to maintain the local real-time guarantees.
This, however, is just an imperfect proxy for the metric we are actually interested in: The time-critical task's real-time performance, i. e. the lateness of the time-critical task. Therefore, we based this mitigation on it.

\begin{figure}[h]
    \centering
    \includegraphics[width=.8\columnwidth]{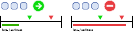}
    \caption{\textbf{Hysteresis Mitigation:} Late- / earliness on a scale with the block and unblock thresholds.}
    \label{fig:hysteresis-explain}
\end{figure}

Hysteresis is a popular technique for controlling processes~\cite{krasnoselskiiStaticHysteron1989}.
It defines two thresholds: A maximum allowable threshold for a metric above which whatever action contributing to the rise of that metric will be ceased and a minimum threshold below which the action will be re-started.
We use the lateness/earliness metric reported by the critical task, i.e. how much time is left in the time slice when the task has been accomplished as the hysteresis control metric.
Once the earliness falls below the minimum allowable value, we stop processing new packets in the network driver, deactivate interrupts from there and wait (in a loop that sleeps for some time in every iteration) for the critical task to report an earliness higher than the threshold.

\paragraph*{Budget Mitigation}
This mitigation technique ties the earliness metric more closely to the amount of work that is permissible within the network subsystems.
For example, there could be a local critical task load that uses up to $90\%$ of computation time per cycle.
Even a minuscule network load could then lead to the Hysteresis Mitigation permanently disabling the network.
Furthermore, we wanted to use the late-/earliness metric to react more responsively to elastic loads.

\begin{figure}[h]
    \centering
    \includegraphics[width=.9\columnwidth]{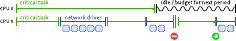}
    \caption{\textbf{Budget Mitigation: }Timelines of tasks on CPU. The network driver can preempt the critical task until its budget is depleted.}
    \label{fig:budget-explain}
\end{figure}

Budget Mitigation is another cooperative approach.
The critical task reports its earliness to the network driver after each completed cycle of computation.
This earliness is interpreted as the time budget of the network driver.
The driver will measure the time of its operations and subtract that from the latest budget.
Once the budget is depleted, the network subsystem (including the interrupts) is suspended until a new earliness notification is issued.
For this mitigation to work, we set both the critical and the network driver task to equal priorities such that the network driver has a chance to deplete its budget after which it will actively yield to the critical task.
This is an important tweak since the Budget Mitigation acts as a specialized scheduler by deciding when the network subsystem has to cease to operate.
It would otherwise be defeated by the operating system scheduler's time slice logic.

\paragraph*{Queue Mitigation}
We looked at three mitigation techniques so far ---~a very simple approach that requires a manual definition of a capacity limit and two cooperative ones that require communication with the critical task.
With Queue Mitigation, we propose a more universal, yet effective non-cooperative mitigation. 

\begin{figure}[h]
    \centering
    \includegraphics[width=.9\columnwidth]{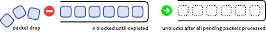}
    \caption{\textbf{Queue Mitigation:} After the packet queue fills up, all incoming packets are dropped until it is empty again.}
    \label{fig:queue-explain}
\end{figure}

This mitigation technique is based on a simple observation concerning the network queue we made in Section~\ref{sec:queue-fill-metric}:
The network driver can keep up with the traffic when the packet queue is not full.
Thus, with queue mitigation, we simply disable network interrupts if they fail to put a packet into the queue (because it was already full) and only re-enable the interrupts from the driver once it has processed all queued packets.

When using this mitigation technique, we assign a higher priority to the critical task.
As a result, the network driver is the first process running out of time when CPU resources get scarce or the interrupt frequency rises.
The direct effect is that the queue fills up.
With a reasonable queue size, the networking is then disabled before the critical task is too stressed, thus protecting the critical task from interrupt overload.

\section{Effectivity of the Mitigation Algorithms}
\label{sec:2_3_mitigation_evaluation}

To investigate the effectiveness of the presented metrics and techniques, we test on the ESP32 microcontroller. The controller is a common ARM-based development board with two CPU cores and the lightweight FreeRTOS operating system. FreeRTOS provides basic task scheduler and interrupt management as well as some data structures for our application.
The multi-core platform allows us to separate the traffic generator from the traffic consumer by placing each on one of the two cores.
A local task simulating time-critical computation is additionally placed on the consuming core such that both compete for CPU time.

In our test setup, the interrupt handler fills up a FreeRTOS queue while the network driver empties it, thus processing the incoming data.
This allows us to analyze the effect of both, the interrupts and the driver on the critical computation load.

\begin{figure}
    \centering
    \includegraphics[scale=0.9]{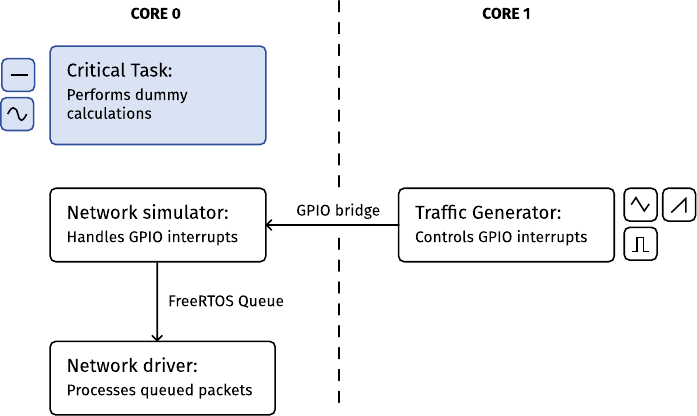}
    \vspace{0.5em}
    \caption{\textbf{Test setup: }Processes used to simulate different interrupt load scenarios on the ESP32 SoC.}
    \label{fig:overview}
\end{figure}

The traffic generator running on the second core generates network loads with a pyramid-shaped load pattern.
It sets a GPIO pin to high for each received packet, triggering a GPIO interrupt on the other core over a GPIO bridge.
There, the network simulator synthesizes a TCP SYN packet and places it into a queue for consumption by the network driver which acknowledges the packet's arrival.
Meanwhile, the observed critical task (also running on the first core) calculates an ascending series of binomial coefficients to generate an equal work load for each task cycle.
Its goal is to reach a target $(n, k)$ in a small time frame of $10ms$.
If the target is reached before the permissible time expired, the critical task will sleep the remaining time and start over in the next period.
If, however, the critical task fails to reach its target in time, it will accumulate lateness, i.e. continue until it reached the target values and then start its new period immediately. Figure~\ref{fig:overview} provides a graphical overview of this setup.

The amount of
\begin{compactitem}
    \item interrupts triggered,
    \item \acp{isr} executed,
    \item packets processed in the network driver,
    \item critical task cycles, and
    \item accumulated critical task latenesses
\end{compactitem}
are saved into atomic variables of a monitoring routine running on core 1.
Every second, the controller reports these variables via the serial interface and then clears the counters.

\begin{figure}[t]
    \centering
    \begin{subfigure}[t]{0.48\columnwidth}
    \includegraphics[width=\textwidth]{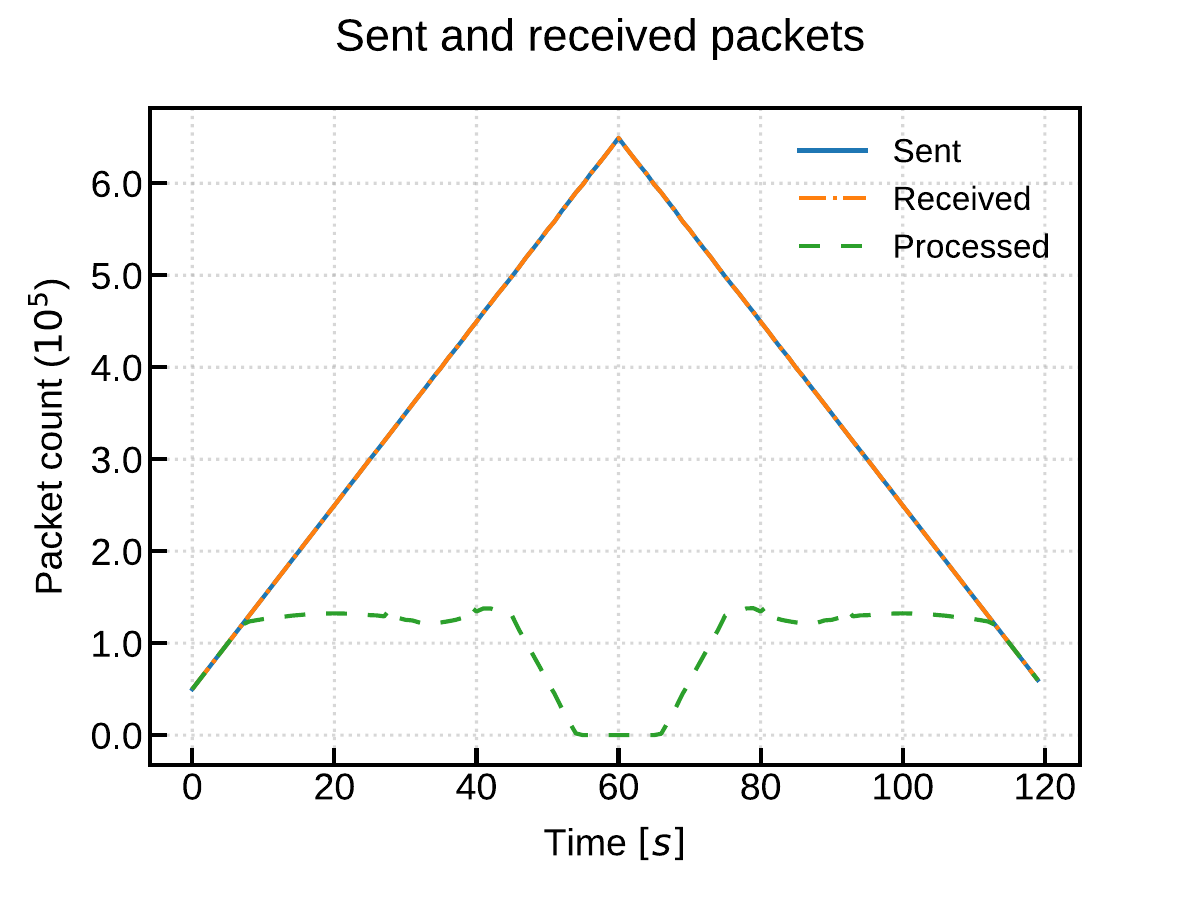}
    \caption{Scheduler: Number of sent, received, and processed packets (higher priority for critical task).}
    \label{fig:scheduler-packets}
    \end{subfigure}
    \hfill
    \begin{subfigure}[t]{0.48\columnwidth}
    \includegraphics[width=\textwidth]{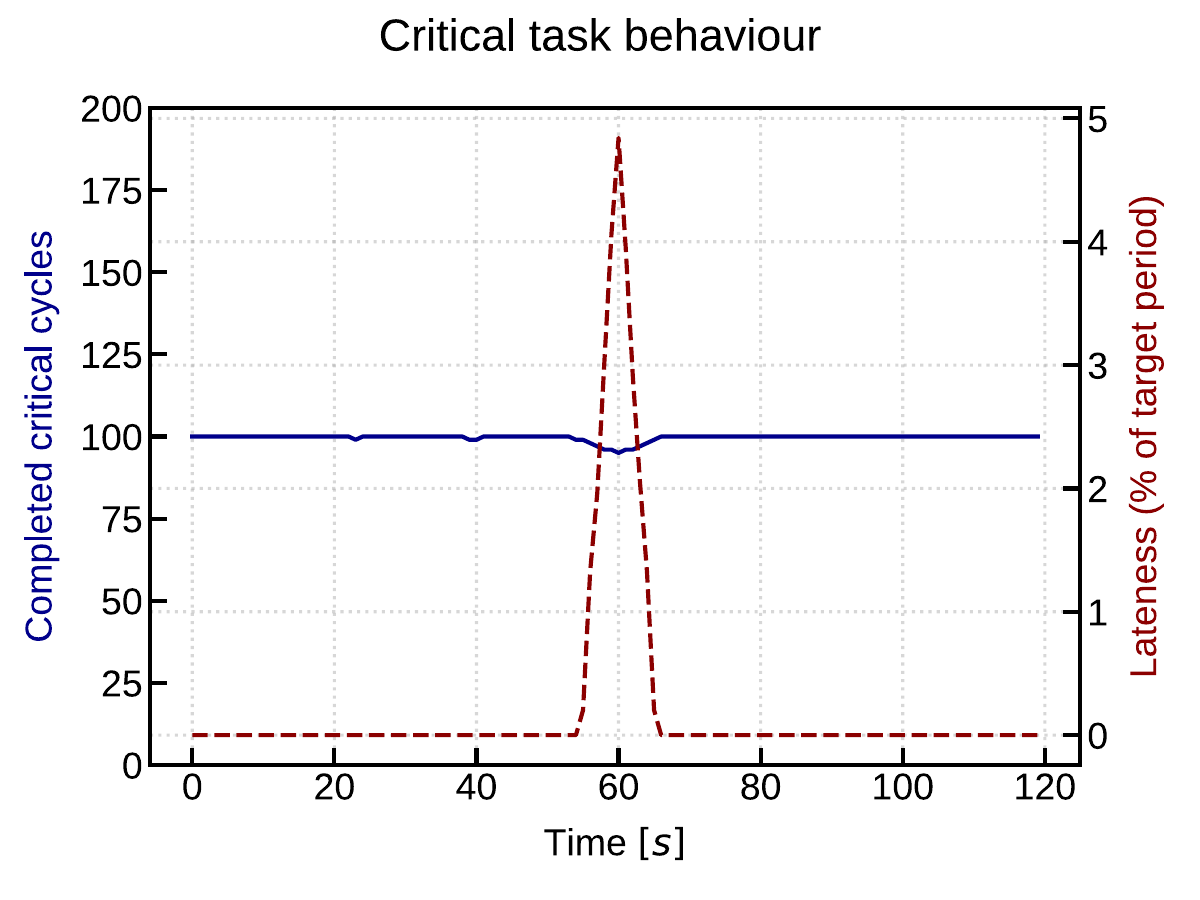}
    \caption{Scheduler: Number of completed critical cycles, accumulated lateness (same priorities as in Fig.~\ref{fig:scheduler-packets}).}
    \label{fig:scheduler-critical}
    \end{subfigure}
    \\
    \begin{subfigure}[t]{0.48\columnwidth}
    \includegraphics[width=\textwidth]{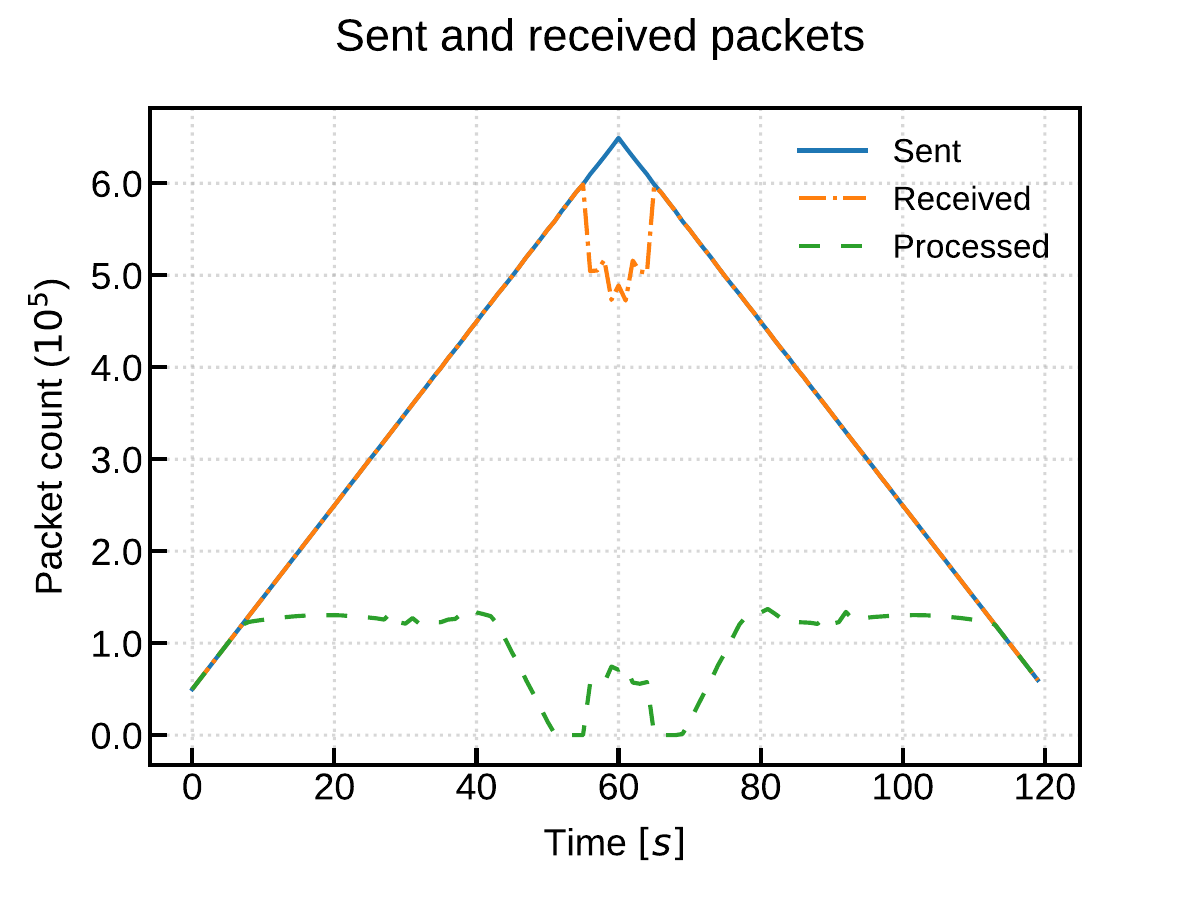}
    \caption{Burst Mitigation: Number of sent, received, and processed packets (capacity of $600$ packets per $20ms$, higher priority for critical task, queue size $100$).}
    \label{fig:burst-packets}
    \end{subfigure}
    \hfill
    \begin{subfigure}[t]{0.48\columnwidth}
    \includegraphics[width=\textwidth]{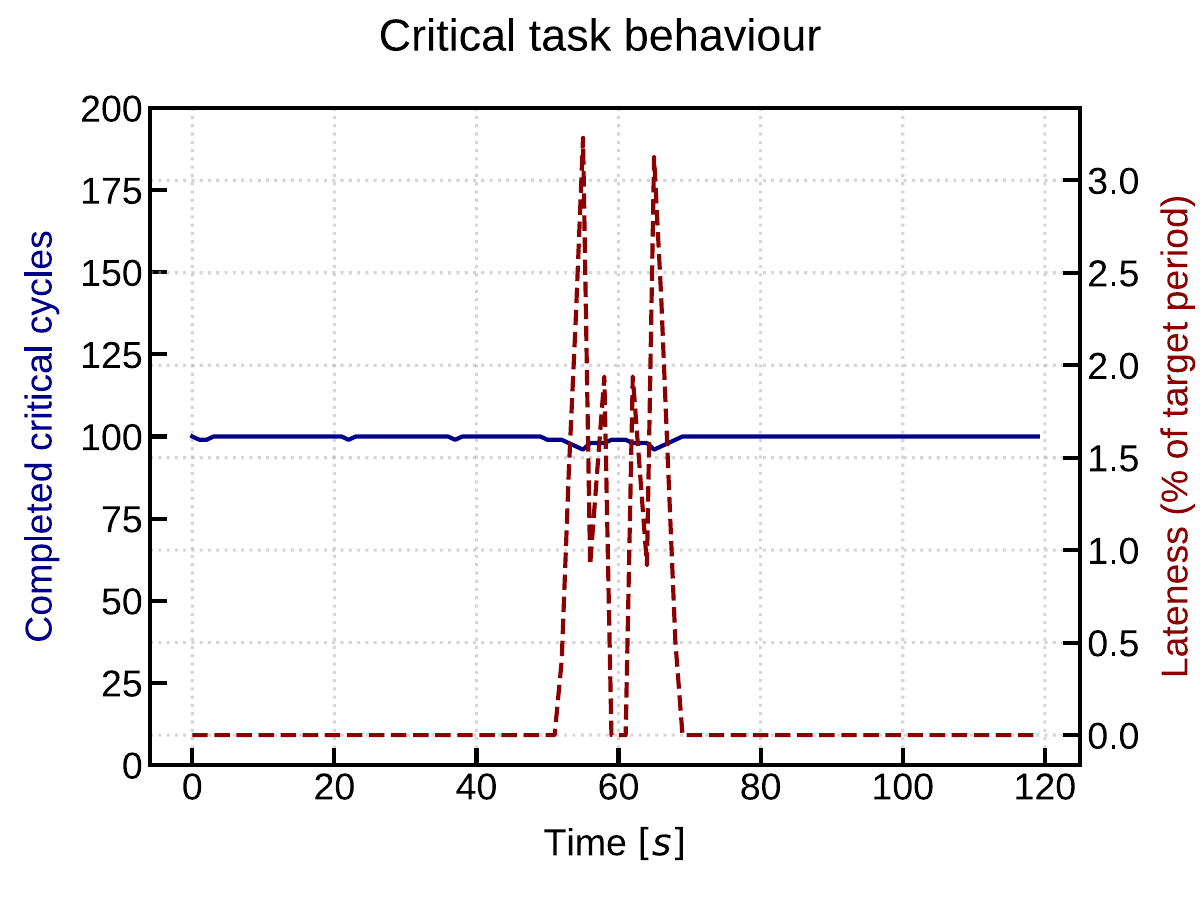}
    \caption{Burst Mitigation: Number of completed critical cycles, accumulated lateness (same parameters as in Fig.~\ref{fig:burst-packets}).}
    \label{fig:burst-critical}
    \end{subfigure}
    \caption{\textbf{Performance results:} OS scheduler and Burst Mitigation}
\end{figure}

\paragraph*{Baseline Scheduler}
We use the built-in FreeRTOS-Scheduler as a baseline. The option to assign (different) priorities to the critical task and network driver is a first, simple way to balance them out.

As a baseline, we tested how the system performs when both tasks are assigned equal priority.
As expected, the critical task started to incur significant lateness (up to $500\%$), while the network driver was able to process almost all packets.
Next, we tested a configuration where the critical task has higher priority than the network driver.
Interestingly, this alone brought the lateness down to almost zero.

A slight lateness of about $8\%$ is observed when the number of packets per second exceeds $50,000$, cf. Fig.~\ref{fig:scheduler-critical}).
At the same time, the number of packets processed by the driver is far lower than previously, dropping even more as the interrupt count rises (cf. Fig.~\ref{fig:scheduler-packets}).

\paragraph*{Burst Mitigation}
We evaluated Burst Mitigation with a capacity of $600$ packets per $20ms$. 
We assigned a higher scheduler priority to the critical task since even though we disable the network interrupts, the network driver still processes queued packets. 
By setting the priorities in favor of the critical task, we inhibit not only the ISR but also the driver from taking too much time off the critical task. 

Figure~\ref{fig:burst-packets} shows the number of sent, received, and processed packets we measured with the stated parameters.
The number of received packets per second does not exceed $30,000$, as is expected from $600$ packets per $20ms$.
In this experiment, the critical task did not introduce lateness, demonstrating the effectiveness of the mitigation when the parameter value is well chosen.
In further tests, we measured that raising the packet capacity significantly leads to lateness, showing that $30,000$ is the optimal condition for our setup.

\begin{figure}[t]
    \centering
    \begin{subfigure}{0.48\columnwidth}
    \includegraphics[width=\textwidth]{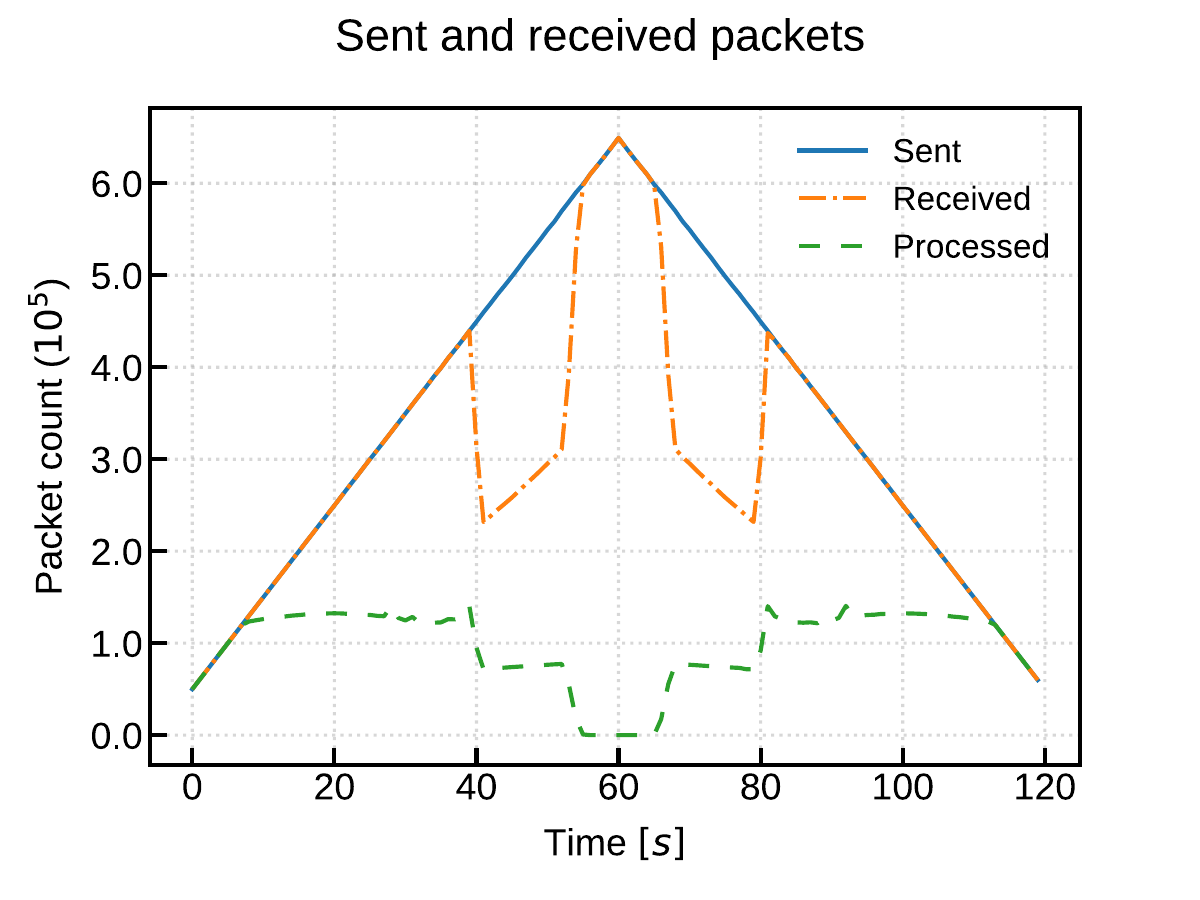}
    \caption{Hysteresis Mitigation: Number of sent, received, and processed packets (network driver deactivation at $90\%$ of allowable time expired at task completion)}%
    \label{fig:hyst-packets}
    \end{subfigure}
    \hfill
    \begin{subfigure}{0.48\columnwidth}
    \includegraphics[width=\textwidth]{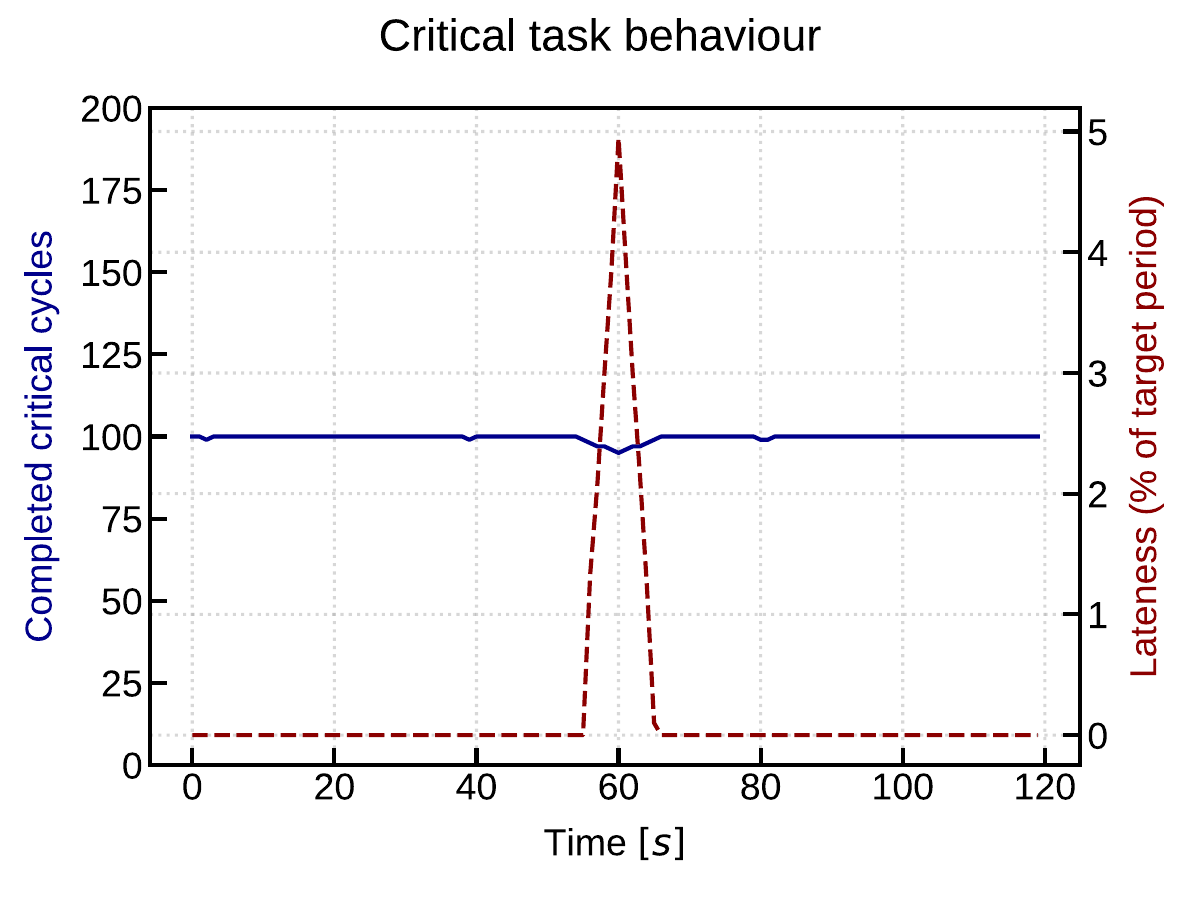}
    \caption{Hysteresis Mitigation: Number of completed cycles, accumulated lateness (same parameters as in Fig.~\ref{fig:hyst-packets}).}
    \label{fig:hyst-critical}
    \end{subfigure}
    \\
    \begin{subfigure}{0.48\columnwidth}
    \includegraphics[width=\textwidth]{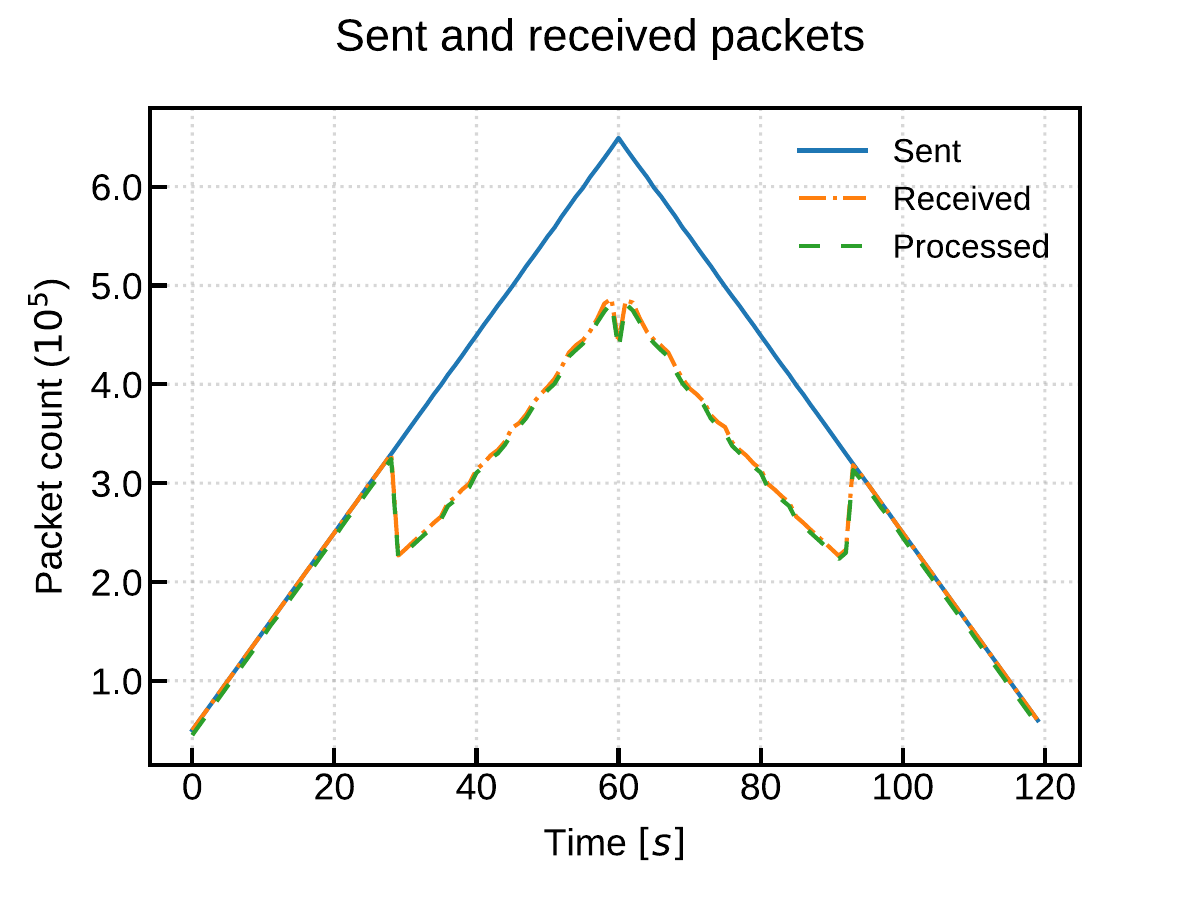}
    \caption{Budget Mitigation: Number of sent, received, and processed packets (both tasks have equal priority, queue size $100$).}
    \label{fig:budget-packets}
    \end{subfigure}
    \hfill
    \begin{subfigure}{0.48\columnwidth}
    \includegraphics[width=\textwidth]{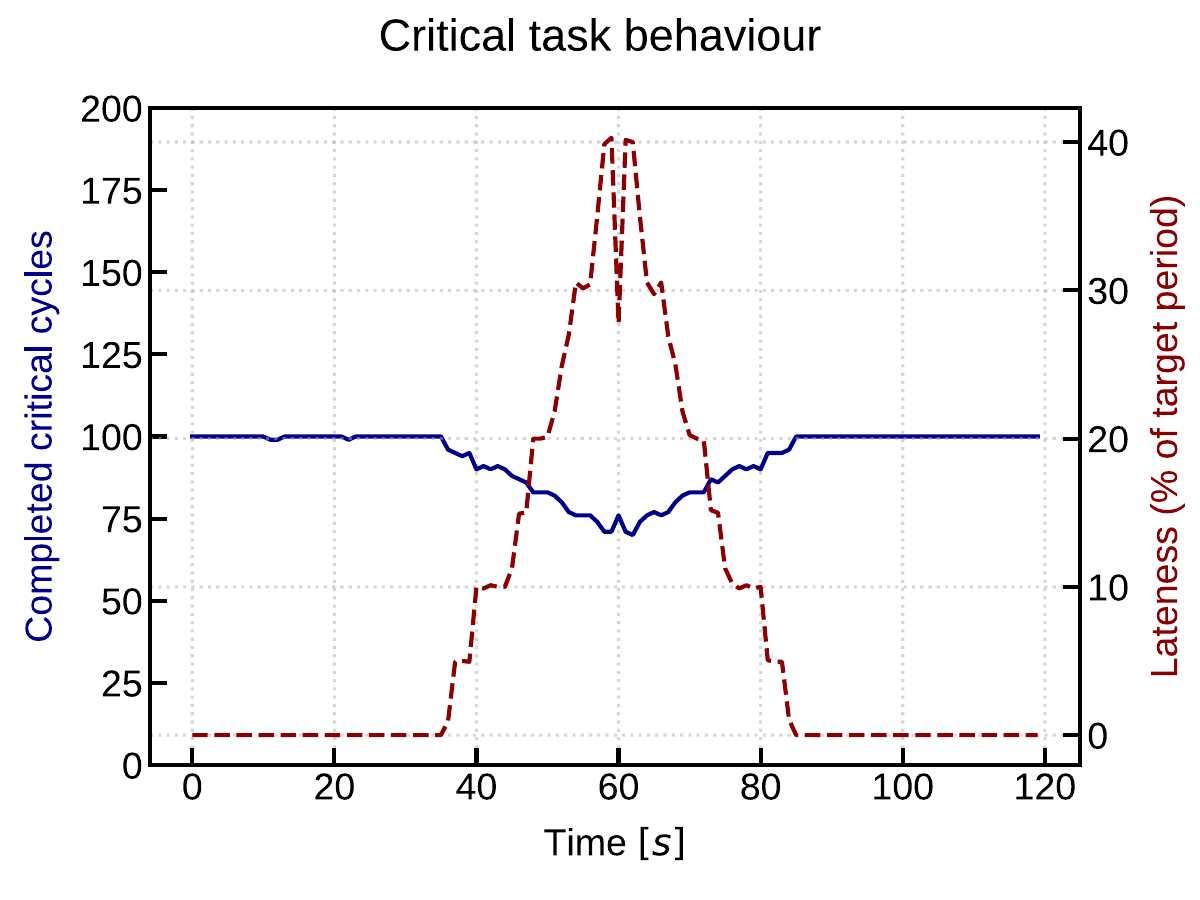}
    \caption{Budget Mitigation: Number of completed critical cycles, accumulated lateness (same parameters as in Fig.~\ref{fig:budget-packets}).}
    \label{fig:budget-critical}
    \end{subfigure}
    \caption{\textbf{Performance results:} Hysteresis and Budget Mitigation}
\end{figure}

\paragraph*{Hysteresis Mitigation}
The network driver's throughput does not stabilize as it does with the Burst Mitigation but instead plummets as the increased interrupt count puts sustained load on the CPU core Figure (cf.~\ref{fig:hyst-packets}).
This does not meet expectations as interrupts should be turned off together with the processing in the network driver by the mitigation algorithm.
It turns out that the interrupts, once reactivated, will drown out the network driver where the thresholds are checked.
Once the network driver is permanently preempted by ISR executions, the mitigation algorithm can no longer take into effect, explaining why the interrupt count curve starts to fit the sent packet curve again.
Nevertheless, Figure~\ref{fig:hyst-critical} shows that the Hysteresis Mitigation is quite effective in preventing lateness except when the load of interrupts itself throttles network driver execution.

\paragraph*{Budget Mitigation} \label{sec:results-budget}
Figure~\ref{fig:budget-packets} shows that once the budget is depleted for the first time, there is a drop in received and processed packets, beyond which both curves continue to track the trend of sent packets.
The slope of the received/processed curves is less steep than that of the sent packets, suggesting that the budget has a moderating impact but cannot lower the network subsystem activity enough for the real-time guarantees to be maintained.
For each additional packet received, $0.76$ additional packets are processed.

Figure~\ref{fig:budget-critical} shows that the overestimation of the budget per additional incoming packet leads to eventually breaking the real-time guarantees for high loads.

\begin{figure}[h]
    \centering
    \includegraphics[scale=0.45]{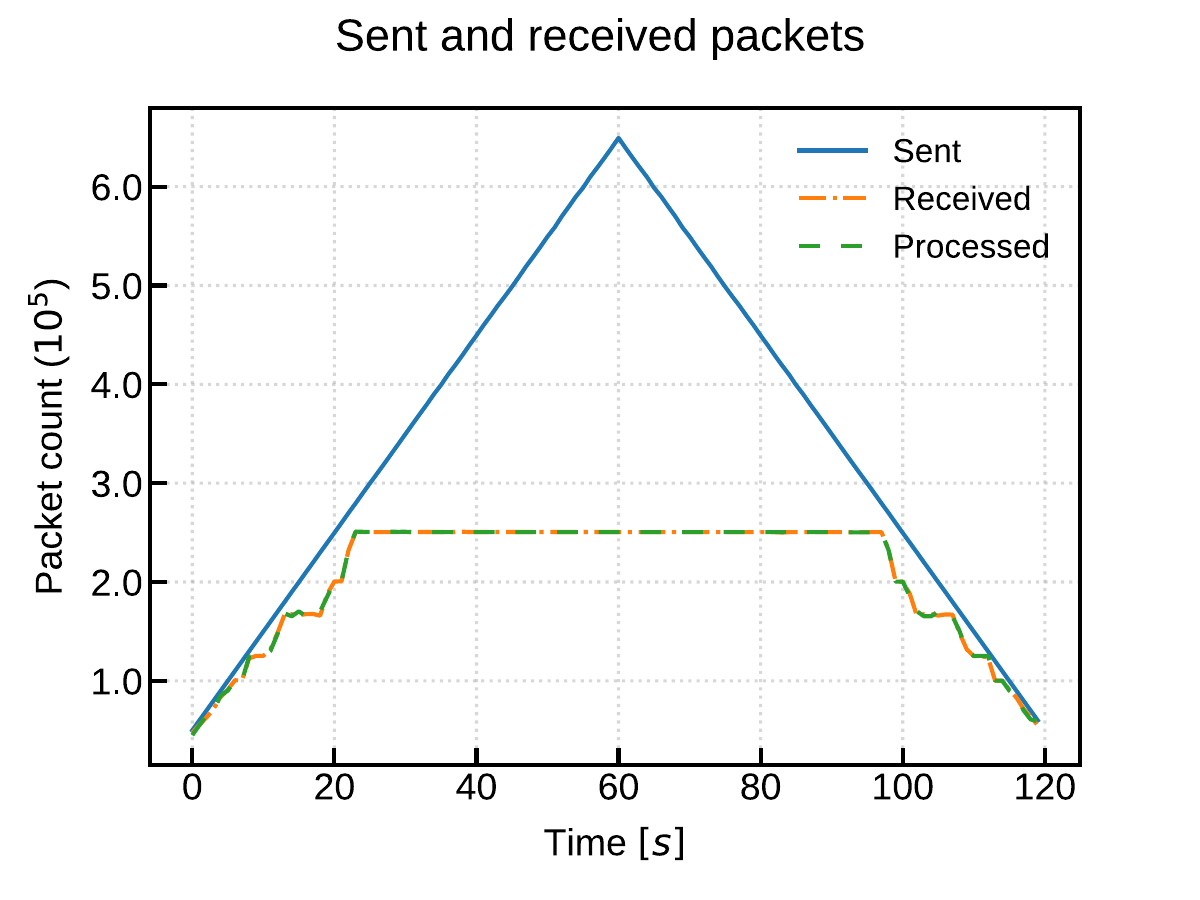}
    \caption{\textbf{Queue Mitigation results:} Number of sent, received, and processed packets (higher priority for critical task, queue size $500$). The lateness remained at zero.}
    \label{fig:queue-packets}
\end{figure}

\paragraph*{Queue Mitigation}
We evaluated the Queue Mitigation with different queue sizes.
Figure~\ref{fig:queue-packets} depicts the packet numbers with a queue size of $500$.
While the network stack is not able to cope with all incoming packets once they rise above $30,000$ per second, the processed packets nicely trace the received packets.

At the same time, the critical task incurred no lateness in this setup.
Queue Mitigation was thus able to process far more packets than Burst and Hysteresis Mitigation while protecting the critical task more effectively.

We also tested smaller and larger queue sizes. Reducing the queue size to $100$ diminished packet throughput, but added no lateness.
This is because a smaller queue makes the mitigation more cautious as the queue fills up faster.
We observed the opposite effect with a queue size of $750$.
Lateness started to emerge because the mitigation reacted belatedly.

\subsection{Discussion}
\label{sec:2_3_mitigation_discussion}
Using only the scheduler, no priority configuration led to particularly good results---with either lateness or very little packet throughput.
Additionally, its priorities cannot mitigate the problem of \emph{interrupts} drowning the process, since \acp{isr} always run above the process priority space.
However, the scheduler priorities can still aid actual mitigation techniques in balancing out the critical task and network driver, as we will show later.
Burst Mitigation allows to avoid drowning all other computation in interrupts, but depending on how the threshold was chosen, much time is spent in the \ac{isr} receiving packets that cannot be processed in time.

Hysteresis Mitigation can be effective in environments without network interrupt loads able to drown the network driver task.
In a high-load environment, it has to be coupled with an interrupt-reducing mechanism like Burst Mitigation in the ISR, which removes some of the advantages of stand-alone usage tested above.

Unlike the Burst Mitigation, the mechanism does not require any knowledge about system throughput -- the required threshold constants generalize better across platforms with differences in processing power and are more closely aligned with the goal of minimizing lateness.

The Budget Mitigation, not unlike the FreeRTOS scheduler, accumulates more lateness the higher the load, the link is approximately linear.
This is due to its underestimation of packet processing time.
The behavior can partially be explained by the fact that only the activity in the driver and not in the \ac{isr} is timed. 
The \ac{isr} activity thus does not impact budget consumption.

The characteristics of earliness as a reporting mechanism are also crucial: Our critical task reports the difference of the timestamps between cycle termination and cycle target, \emph{not} the time spent on the critical task alone.
If, therefore, the network driver has been allocated a large budget in one cycle, it will compete with the critical task for processing time, leading to it closely matching its deadline, and reporting little earliness.
The critical task can then terminate early the next cycle because the network task had a small budget, introducing oscillations in the budget ceiling.

Queue Mitigation always keeps the critical task on time while the packet throughput remains at a constant level.
The number of packets received by the \ac{isr} and processed by the driver is almost equal.
This indicates that interrupt handling and processing in the driver are well-balanced.

Alas, with Queue Mitigation there is still a hyper-parameter we need to tune, as with Burst Mitigation.
However, we argue that the queue size is a parameter that must be defined in any case, with any mitigation, and as argued in Section~\ref{sec:queue-fill-metric} defining a fixed queue length still leaves the system more flexible and elastic than defining a fixed burst capacity.

\paragraph*{Priority Differentiation}
The presented algorithms are able to detect packet floods and mitigate their effects to decrease the likelyhood of computation failures and deadline misses of the running processes. However, the general issue of a priority inversion between \ac{isr}/network stack and critical real-time processes cannot be solved by this.
In the following section, we investigate the lower half of the driver (the \ac{isr}) to map IP flows to processes and derive their actual real-time priority. Using this information, a driver and network stack extension adjusts priorities on a per-packet-basis while securing the system from unwanted packet floods.

\section{Differentiating Network Flows for Priority-Aware Packet Processing}
\label{sec:2_3_differentiating}
As shown in the previous section and in~\cite{niedermaier_you_2018}, a first approach to prevent congestion caused by received packets is to introduce budget enforcement on the entire \ac{rx} path. However, when limiting the reception of incoming packets, this affects every communication of the device, regardless of its importance. Yet, devices might be running multiple processes, each with different criticality levels for both their timely termination and their communication. One might include soft real-time traffic that, while not critical to the integrity of the controlled cyber-physical system, is relevant for basic functionality or monitoring. Note that we expect multiple IP flows with different levels of timeliness requirements and packet frequencies, allowing for a different view of the problem. 

\paragraph*{RX Path Priority Inversion}
We observe a problem of incorrect priority enforcement in the receive path of the network subsystem. In general, an incoming packet is used by (at most) one specific task. However, as long as its purpose is unknown, we need to schedule the processing of each packet equally. For the practical implementation of the network subsystem in an RTOS, this means assigning a fixed priority to the protocol processing server task, which inevitably creates a priority inversion situation:

\begin{compactitem}
    \item Using a high priority, as is common in current embedded frameworks, can starve high-priority tasks in favor of low-priority packets.
    \item Using a low priority, a receiving high priority task may wait for a medium priority task to finish executing. This is observed as network latency.
\end{compactitem}

The above priority inversion cannot be completely eliminated due to its inherent nature. On the one hand, we do not want to spend computing resources before we know if it is worth it given the current scheduling situation. On the other hand, incorporating the use of a particular packet requires prior protocol handling.

\paragraph*{IP Flow Classification}
The goal of this section is to present a network driver extension that distinguishes between different IP flows and their priorities in order to always prioritize the communication necessary for critical tasks and to apply the actual process priorities of real-time tasks to the processing of their packets. Thus, we aim to solve the problem of priority inversion in packet reception. In addition, the proposed architecture aims to protect against network congestion and improve overall network performance.

Past work has already proposed to do packet classification in the networking driver as early as possible~\cite{druschel1996lazy}, then apply a priority for each \acs{udp}-packet by receiving process and defer the subsequent packet processing~\cite{lee_interrupt_2010, lee_priority-based_2015} based on its assigned priority.
Others have analyzed and compared various network task architectures for enabling protocol processing prioritization with nicely integrating scheduling-properties that avoid priority inversion situations~\cite{mercer1991evaluation}.

In this section, we propose a packet reception architecture for typical embedded \acp{rtos} and \acs{ip} stacks that facilitates the deployment of IoT hardware in real-time scenarios without the need for specialized hardware. More formally, the architecture aims to combine the following properties:

\begin{compactitem}
    \item Protection against network-induced system overloads, facilitating real-time systems.
    \item Optimal processing latency for well-behaved \acl{hp} flows.
    \item Best-effort performance for \acl{lp} flows.
\end{compactitem}

This section also introduces a prototypical implementation modifying a common embedded network stack and presents a set of experiments that evaluate basic performance properties, showing the effectiveness of our architecture in various setups. Namely, we evaluate 

\begin{compactenum}
    \item the usefulness of a software-only approach in terms of CPU efficiency,
    \item the suitability for constrained scenarios, regarding possible overhead in CPU time and memory consumption, and
    \item improvements regarding real-time schedulability.
\end{compactenum}

\subsection{Requirements}
A modification of the packet reception subsystem for real-time IoT devices should satisfy the following requirements.

\begin{compactenum}
\item[1.1] \emph{Priority Inversion Mitigation.} Packets of high-priority flows must be processed and delivered before those of low-priority flows. 
\item[1.2] \emph{Overload Protection.} Packet floods must not lead to system overloads. In case of a packet flood, low-priority flows must be  throttled in favor of high-priority processes. A general rate limitation must protect the real-time properties of running tasks.
\item[1.3] \emph{Performance Retention.} The approach should not introduce a longer packet processing delay. However, predictability and overload protection must be prioritized.
\end{compactenum}

\subsection{Overview}
\begin{figure}[h]
    \centering
    \includegraphics[width=.9\columnwidth]{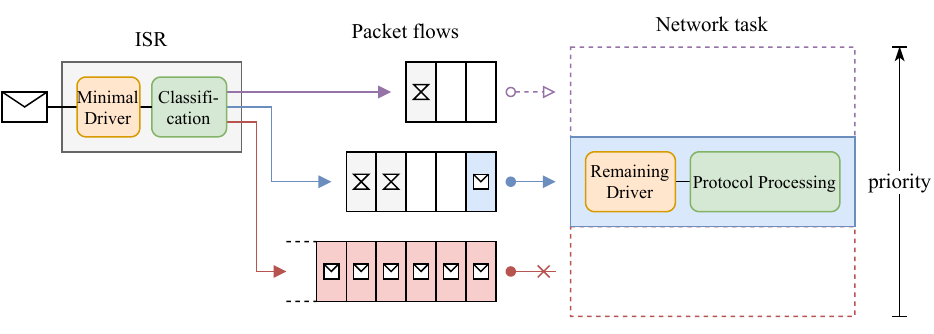}
    \caption{\textbf{Architecture overview:} 
   Packets are classified early and enqueued by their flow, with individual periodic capacity restrictions applied. Further processing is scheduled by inherited priority.}
    \label{fig:approach_overview}
\end{figure}

The proposed architecture is designed around a data structure of differentiated flow queues, which replaces the simple frame queue.
Each flow defines a priority and a period, to affect the further processing of its packets.
For the prioritization of processing, we add a priority manipulation mechanism to the protocol processing task.
In order to gain a maximal advantage from scheduling the subsequent processing stage, the driver is modified to do only the necessary work of classifying incoming packets to flows by their header entries.
The remaining activity is then executed on packet retrieval by the scheduled protocol processing task as presented in Figure~\ref{fig:approach_overview}.

The proposed architecture combines three concepts:

\begin{compactenum}
    \item \emph{Soft Early Demultiplexing} into receiver-centric flows is necessary for differentiating flows on an End-to-End basis, without reliance on network \ac{qos} and as a result satisfy \emph{Requirement 1.1}.
    \item \emph{Prioritized Protocol Handling} facilitates best-effort communication processes that utilize background resources on the same system and is necessary to satisfy \emph{Requirements 1.2 and 1.1}.
    \item \emph{Rate Limitation} applied to differentiated flows as well as on packet reception in general as a last resort protects the system from being vulnerable to unexpectedly high traffic in \ac{hp} flows satisfying \emph{Requirement 1.2}.
\end{compactenum}

Thus, being able to fully defend the considered system against flooding induced overload, while at the same time ensuring high connectivity for particular well-behaved \ac{hp}-flows even in scenarios with overall high incoming traffic, and handling \ac{lp}-flows with best-effort resources, this combination forms an IoT real-time aware overload protection.

In the following subsections we introduce the concept of each of the three basic building blocks of our architecture and discuss relevant implementation aspects.

\subsection{Soft Early Demultiplexing}
\label{sec:2_3_approach}

In order to minimize the effort spent until after classification, we employ Early Demultiplexing~\cite{druschel1996lazy}.
By peeking into key header entries, a packet is assigned to its eventual receiver process.

\begin{figure}[h]
    \centering
    \includegraphics[scale=.8]{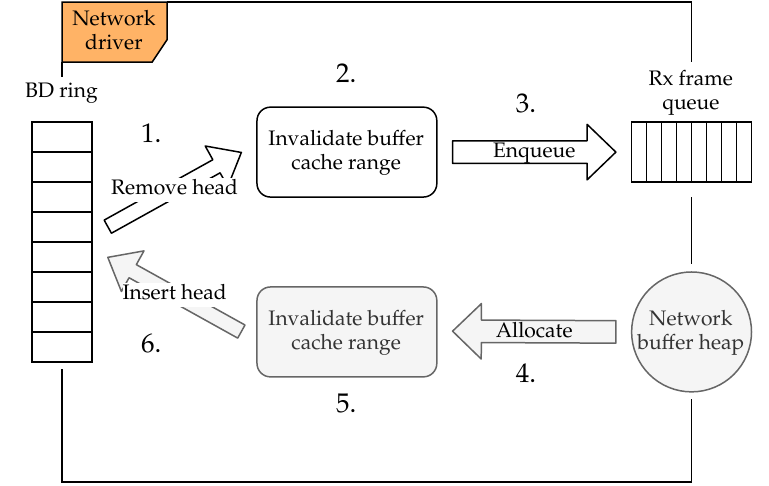}
    \caption[Receive activity in the original driver]{\textbf{Receive activity in the original driver:} 
    Once an interrupt occurs, the driver code moves a packet buffer from the \ac{bd} ring to a simple queue, and fills the vacant position with a newly allocated one. Due to cache coherency requirements of the memory system, the buffer caches have to be invalidated for both the retrieved and the replacement buffer.}
    \label{fig:driver_buffer_flow}
    
\end{figure}

The benefit of demultiplexing performed in software depends heavily on the amount of work that can be saved by mere demultiplexing compared to full protocol processing.
Since the packet scheduling in our architecture can only influence the processing that follows after Early Demultiplexing, the achievable degree of partial network liveness in overload scenarios depends on its quick execution.

Starting from the existing driver receive path, depicted in Figure~\ref{fig:driver_buffer_flow}, we introduce two changes: Packet classification and lazy cache invalidation.

\subsubsection*{Packet Classification}

The classification differentiates incoming packets into flows defined by the protocols ARP, ICMP, TCP and UDP.
While the former two form a single flow, the latter are further differentiated by local port numbers in order to implement the receiver task association.

Depending on the used network stack, the lookup from the port to a flow may either be performed using the existent network stack's list of bound socket control blocks, or else requires an additional data structure managed by the driver.
In our prototype based on FreeRTOS+TCP, the socket managing code in the original network stack can easily be locked in a critical section, leaving the \ac{isr} safe to access it.

If a scenario requires anticipating a large number of bound sockets, a sophisticated data structure with better complexity should be employed.
However, with only a few sockets bound at any particular point in time, a linear linked list lookup as found in typical embedded network stacks suffices.

\begin{figure}[h]
    \centering
    \includegraphics[scale=0.8]{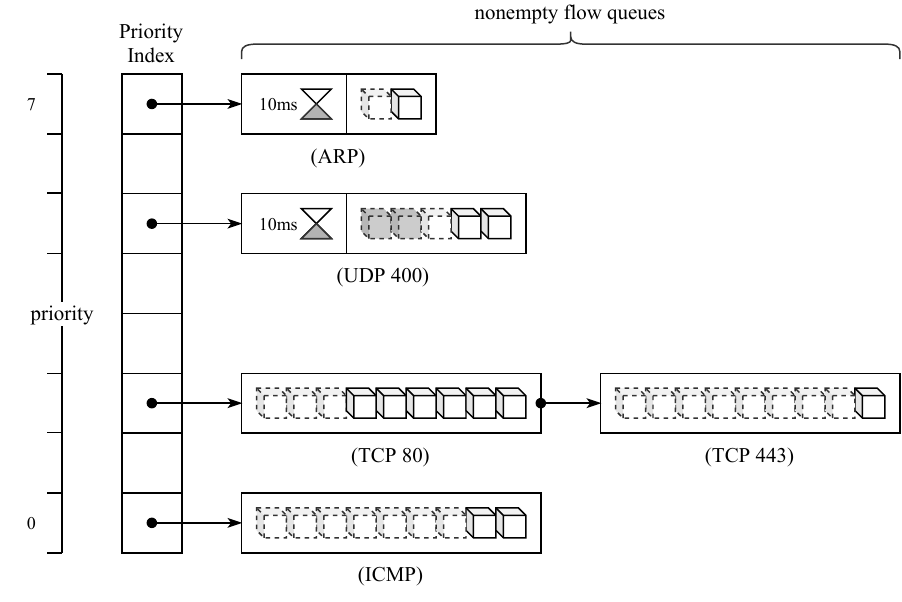}
	\caption[Differentiated flow queues]{\textbf{Differentiated flow queues:}
	Between reception by the driver and further driver and protocol processing, packets get stored in a queue according to their identified flow. These queues are organized by the flow priority, to facilitate fast retrieval of the highest/lowest prioritized packet.
	}
	\label{fig:pdest_depq}
\end{figure}

Instead of enqueueing every received packet to the same \ac{rx} frame queue, each packet is inserted into a specific queue according to the result of the classification.
Because the packets do not necessarily get processed in bounded time, the network subsystem might experience buffer starvation.
To avoid this, buffers of low priority packets are \emph{recycled} when the buffer memory reaches its limit. Buffer recycling here means that the allocated buffer element is freed and a new empty element is appended to the \ac{bd} ring. The packet is effectively dropped.
To this end, the differentiated flow queues are organized in a priority queue structure as depicted in Figure~\ref{fig:pdest_depq}.

The priority of a flow is defined by its respective receiver task and the overall priority space is equal to the one used by the \ac{rtos} task scheduler. This way, packet processing priorities are inherited according to the real-time considerations made for the running processes starting with the enqueueing of packets. Another feasible option is the utilization of network priorities such as the Differentiated Services Field~\footnote{RFC 2474}, flow priorities from real-time protocols, or traffic classes inside TSN-based networks. Yet, we assume environments that do not necessarily provide a network priority, thus relying on receiver priorities.

\subsubsection*{Lazy Cache Invalidation}

\begin{figure}[h]
    \centering
    \includegraphics[width=\columnwidth]{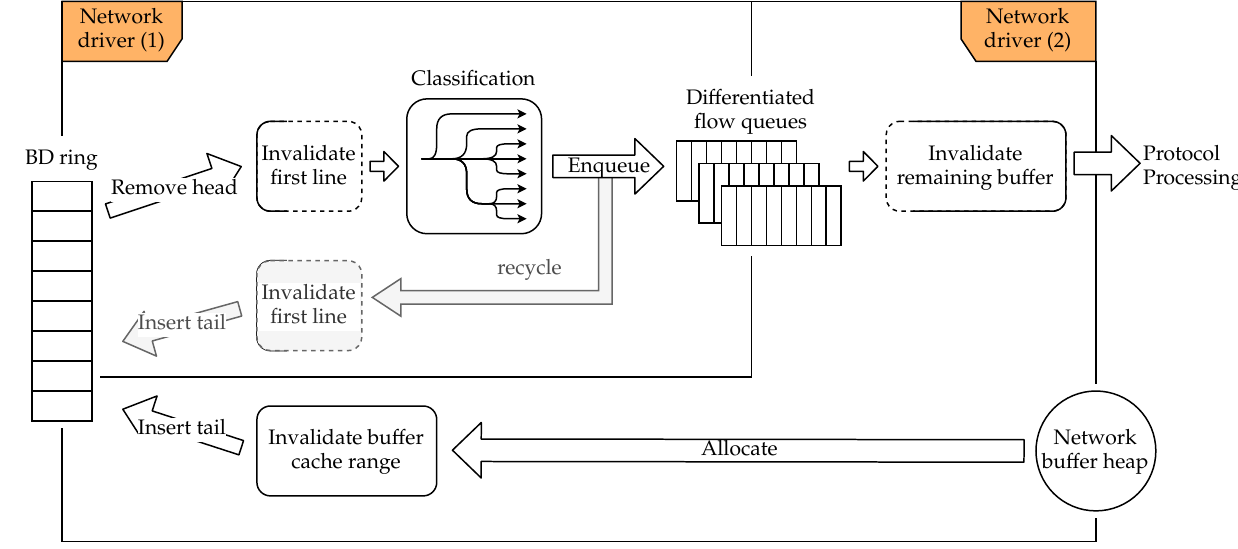}
    \caption{\textbf{Receive driver activity in our approach: }
    The driver is separated into two halves. In the \emph{eager driver} (1), a minimal effort is taken to classify each packet into a flow. As part of the scheduled subsequent protocol processing, the \emph{deferred driver} (2) establishes cache coherency and refills the \ac{bd} ring once the packet is needed.}
    \label{fig:my_driver_buffer_flow_with_recycling}
\end{figure}

On embedded systems that feature CPU-caches, the commonly cache incoherent \ac{dma} introduces a significant cost with the obligation to invalidate the transferred buffer cache lines (cf. \ref{sec:2_3_proc_times}).
In our case, the memory architecture requires the network driver to invalidate buffer cache lines prior to and after the processing by the \ac{nic} \ac{dma} engine, as shown in Figure~\ref{fig:driver_buffer_flow}.
As cache management noticeably prolongs the execution time of Early Demultiplexing, we incorporate a lazy cache coherency establishment scheme into the driver to retain the highest possible performance as per \emph{Requirement 1.3}.
The driver is therefore split into two halves, as depicted in Figure~\ref{fig:my_driver_buffer_flow_with_recycling}.

\begin{compactenum}[(1)]
    \item \textbf{Eager driver:} An immediately processed layer, executed as part of the \ac{isr}, classifies and enqueues packets.
    \item \textbf{Deferred driver:} A schedulable layer, executed in the network task according to the packet priority, establishes full cache coherency of received packets and prepares fresh replacement buffers.
\end{compactenum}

\begin{figure}[h]
    \centering
    \includegraphics[scale=0.8]{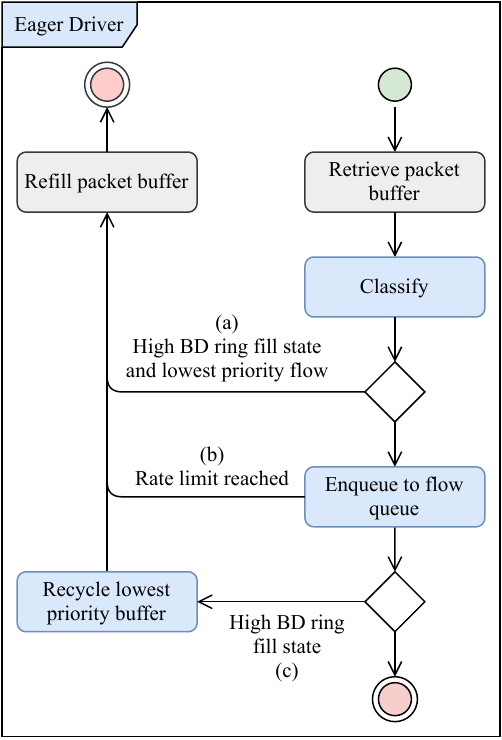}
    \caption{\textbf{Eager driver ISR:} Key execution paths that determine whether and when a packet buffer is recycled to save execution time in high-load scenarios.}
    \label{fig:top_half_activity}
\end{figure}

Prior to classification, only the first cache lines of the packet buffer containing the relevant header fields are invalidated.
Once the packet is chosen to be processed further, the remaining part is invalidated and a fresh packet buffer prepared and appended to the \ac{dma} \ac{bd} ring (cf. Section~\ref{sec:dma}).
This implies that as the differentiated flow queues fill up with packets, the \ac{bd} looses free packet buffers, forming a closed pool of packets shared by the \ac{bd} ring and the differentiated flow queues.
To prevent starvation of the \ac{bd}-ring caused by \ac{lp}-packets in the differentiated flow queues, the eager driver recycles lowest priority packet buffers once the \ac{bd}-ring hits a critical threshold (e.g. $\frac{1}{2}$).
This can be carried out with little cost, since only the accessed header cache lines have to be invalidated again.

The resulting activity in the eager half driver is depicted in Figure~\ref{fig:top_half_activity}.
Notable are the three different execution paths that might be taken:
If, due to a high \ac{bd} ring fill state a packet buffer has to be recycled, and the currently considered packet is of lowest priority, it gets recycled in a short-circuiting branch \textsuperscript{(a)}.
A flow queue may decline further packets to prevent overload by this particular flow \textsuperscript{(b)}.
Lastly, if the short-circuit branch was not taken but the \ac{bd} ring fill state is high, another packet buffer has to be recycled and inserted into the \ac{bd} ring \textsuperscript{(c)}.

\subsection{Prioritized Protocol Handling}
\label{sec:prioritedprotocolhandling}

Once the heterogeneous incoming packets are demultiplexed into differentiated flow queues, the protocol processing can be carried out with the receiver priority, as proposed in~\cite{druschel1996lazy,lee_interrupt_2010}.

We apply a priority inheritance scheme to the single protocol processing task~\cite{mercer1991evaluation}.
It allows the task scheduler to preempt the packet processing at any point in time. Additionally, it keeps a low footprint in terms of task resources and can be integrated into embedded network stacks that commonly use a single network task.

To implement this scheme, the priority of the network task has to be moderated depending on the currently processed packet and waiting packets, in order to avoid priority inversion.
Consider $\mathcal{F}$ as in Definition~\ref{def:1}, $P(f)$ the priority of flow $f$ and $p(f)$ a packet of $f$. Let further $\mathcal{W}$ be the set of currently waiting packets and $\mathcal{E}$ the set of packets in processing. The priority of the network task ($P_{IP-task}$) must be assigned as follows.

\[
P_\text{IP-task} = \max ( P(f): f \in \mathcal{F} \; \land \; \exists \:  p(f) \in \mathcal{W} \cup \mathcal{E} )
\]

This assignment implies the network task priority is recomputed every time a packet gets queued or a packet has been processed.
On packet reception, the priority needs to be elevated \textit{iff} the respective flow priority is higher than the current priority assigned to the network task.
On finished packet processing, the priority needs to be decreased \textit{iff} the priority of the highest priority packet waiting in the differentiated flow queues is lower than the current network task priority.
This operation is supported by the ability of the differentiated flow queue data structure to efficiently provide the highest enqueued priority (reconsider Figure~\ref{fig:pdest_depq}).

It may appear that by using priority inheritance carried out per packet, we put an overly high computational burden on the fixed-priority task scheduler.
Yet, among all possible designs that involve the task scheduler in packet scheduling decisions by correctly signaling the current priority demand at each time, this design has the lowest scheduler data structure manipulation overhead:
Another design could use multiple processing tasks with constant priority, to which packets are assigned according to their flow.
The unblocking operation triggered when the first packet of a particular priority is enqueued then adds at least the same overhead: The task has to be moved into the priority-respective ready task list, and moved out once blocked again.
Among all possible designs that properly communicate the current packet processing priority demand to the task scheduler, the priority-inheriting one has therefore the lowest possible scheduler data structure manipulation overhead.

In order to also have the deferrable parts of the driver processing scheduled according to packet flows, the networking task dequeues a highest priority packet buffer from the differentiated flow queues and executes the second half of the driver before continuing with the regular processing procedure.

\subsection{Rate Limitation}
To take advantage of Early Demultiplexing while at the same time keeping the system protected from overload conditions, deterministic mitigation techniques~\cite{danicki2021detecting} are applied to all but the low priority best-effort flows.
Additionally, the unconditionally executed \ac{isr} that demultiplexes incoming packets could incur a high load even if the subsequent scheduling cuts off further processing.
Hence, an additional, global rate limitation needs to be present.

To apply the rate limitation, each flow is scheduled by a conceptual aperiodic events server with each incoming packet being modelled as an aperiodic request.
In our prototype we use the deferrable server scheme (cf. Section~\ref{sec:aperiodic}).
Beyond the server capacity, packets are discarded.
For the individual flow queues, this happens as part of the inserting operation (reconsider Figure~\ref{fig:top_half_activity}), in order to avoid a situation with a paused \ac{hp} flow queue full of packets blocking all other processing.

To enforce a global rate limitation, once the capacity has been reached in one period, the driver processing switches from \ac{isr}-based execution to a polling driver task, staying in this mode until the capacity is not immediately reached at the begin of a period anymore.
When not processing packet receive \acp{irq} issued by the \ac{nic}, the \ac{bd} ring is filled until eventually packets are discarded by the \ac{nic}.

\subsection{Policy Integration}

In order to control the scheduling properties \emph{capacity}, \emph{period} and \emph{priority} for a flow, we expose these to the user for each socket via the \texttt{setsockopt}-\ac{api}.
Special flows such as for the protocols ARP, ICMP and those that are managed by the network stack, such as DNS and DHCP, can be configured using C macro definitions.
Similarly, the scheduling properties for the global rate limitation can be defined.

\subsection{Limitations}

The ability to proceed with deferred packet processing after a phase of higher system load depends on the number of available packet buffers.
As these buffers have to be prepared for immediate \ac{dma} operation and therefore a constant amount is dedicated to the lower levels of processing, additional memory might be necessary.

IP fragmentation cannot be dealt with properly in our architecture.
To demultiplex fragmented packets, their reassembly had to be done in the \ac{isr}, jeopardizing its \ac{wcet}.
This design treats all packet fragments as belonging to a background priority flow.
Yet, IP fragmentation is discouraged, as it introduces robustness, reliability and security issues~\cite{kent1987fragmentation, gilad2011fragmentation}.

\section{Evaluation of the Driver Extension}
\label{sec:2_3_ipstack_eval}

In this section we present empirical results collected from our prototypical IP stack implementation and subsequently discuss the effectiveness of the approach.

\subsection{Test Setup}

The test setup contains the FreeRTOS operating system with a modified FreeRTOS+TCP stack running on a Xilinx Zynq-7000 processing system containing a dual-core ARM Cortex A9.
Networking is done through a Gigabit-class Ethernet interface controlled by a Marvell 88E1518 \ac{phy} controller. Notable features are \ac{dma} and TX/RX-checksum offloading.
Measurements are taken on a single core.

Two methods for measuring the effect on system load under high packet loads were pursued:

\begin{compactenum}
    \item \emph{Passive:} A background worker carries out CPU intensive work and monitors its performance.
    \item \emph{Active:} The software is instrumented to indicate notable events, i.e. task switches, \acp{irq}, and packet processing.
\end{compactenum}

The former is suitable for precisely estimating the average load that a particular scenario puts on the CPU.
While the latter introduces some overhead in the range of 1-5\% to the processing and misses some of the \ac{irq} switching, it allows us to evaluate the distribution of processing-induced latency.

\subsection{Experiment 1: CPU-Time Saved with Early Demultiplexing}
In this scenario two UDP sockets are bound, one with a low and one with a high priority receiver process.
To not alter the results, the capacity of all flows as well as the overall \ac{irq} limitation is set to infinity.

Multiple system configurations were confronted with a zero-length UDP-packet load of a constant rate for 60 seconds. Through passive measurement performed by a medium-priority task, the average CPU processing time per packet was then calculated (Figure~\ref{fig:cpu_time_per_packet}).
In this experiment we observed that the CPU costs for processing a single packet are rather independent from the magnitude of incoming traffic, staying approximately constant in the range from $10^2$ to $10^6$ pkt/s.

\paragraph*{Results}
The results show the difference in processing time between the packet processing paths. 
When \ac{lp} packets get no chance to be scheduled, the executed activity is only that of the Early Demultiplexing \ac{isr} with an average processing duration of $1.62 \mu s$ per packet.
Compared to the original stack as a baseline, which needs $12.1 \mu s$ to fully process a packet, this results in a speedup of 7.5x.
However, due to the short-circuiting logic depicted in Figure~\ref{fig:top_half_activity} \textsuperscript{(a)}, in this constant \ac{lp}-flow measurement the packet buffers are discarded without being placed into a flow queue.
When disabling the short-circuiting code path, the per-packet processing time increases to $1.75 \mu s$, still yielding a seven-fold speedup compared to the full processing in the original stack.

\begin{figure}
    \centering
    \resizebox{0.6\columnwidth}{!}{
    \input{images/2_3_software/cpu_time_per_packet_udp-zero.pgf}
    }
    \caption{\textbf{Processing impact: }CPU time per zero-length UDP packet under loads between $10^2$ and $10^6$ pkt/s with different configurations.\\
    ¹Short-circuiting branch disabled. ²Cache invalidation deferral disabled.}
    \label{fig:cpu_time_per_packet}
\end{figure}

In this scenario, the \ac{hp} packets are processed the entire network stack and cause a processing time of $12.3 \mu s$ per packet, decreasing receive performance by $1.7$ \% compared to the baseline stack.
This already small relative difference would decrease further if the subsequent (obligatory) reception by the receiver task was taken into account.

By modifying the prototype to again eagerly establish cache coherency in the \ac{isr}, the time spent for \ac{lp} packets increases notably to $4.4 \mu s$.
Hence, we conclude that incorporating a driver deferral mechanism into the architecture is essential to the performance on cached systems.

\subsection{Experiment 2: Packet Processing Latency}

The second experiment deals with the predictability of packet processing latencies in the modified IP stack.
Using the active approach, the reconstruction of precise execution times of each packet is possible.
Additionally, this allows us to differentiate between the execution paths of the modified driver.
Since we instrumented the \ac{isr} entry, some constant \ac{irq} overhead due to context saving is not included in this analysis.
Compared to Eperimant 1, where the overall impact on CPU time is measured, this experiment measures the duration of the eager driver per packet.

\begin{figure}
    \centering
    \resizebox{0.7\columnwidth}{!}{
    \input{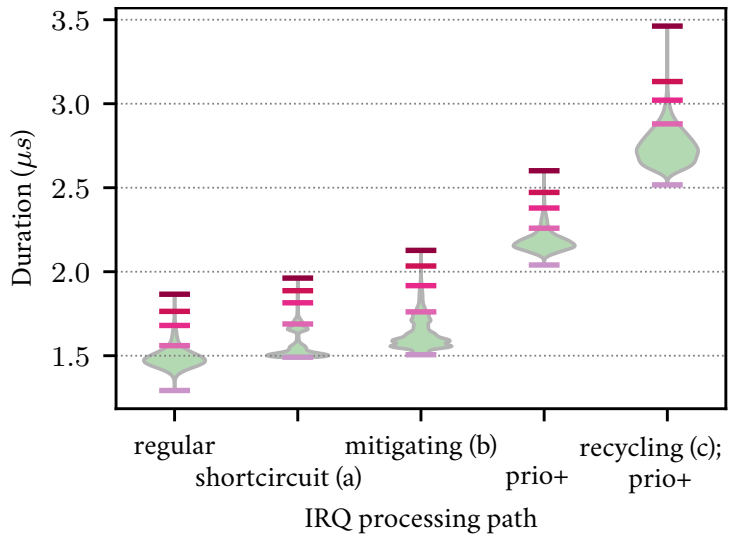}
    }
    \caption{\textbf{Eager driver: }Latency distributions for different execution paths in our modified stack:
    The horizontal bars indicate the percentiles $0\%$, $90\%$, $99\%$, $99.9\%$, $99.99\%$}
    \label{fig:packet_processing_latency}
\end{figure}

The system was flooded with $10^5$ zero-length UDP packets of two different priorities successively.
Figure~\ref{fig:packet_processing_latency} visualizes the distributions of \ac{isr} processing duration for specific processing paths. For each distribution, the quantiles $0\%, 90\%, 99\%, 99.9\%, 99.99\%$ are visualized as horizontal bars, in order to estimate a probabilistic \ac{wcet}.

\paragraph*{Results}
\ac{lp} packets initially take the fastest path ("regular"), where incoming packets are enqueued without any other processing.
Once the \ac{bd} ring has reached a high fill state, packet buffers have to be recycled.
Since the incoming packets are already at the lowest level present in the differentiated flow queues, the short-circuiting path ("shortcircuit", \textsuperscript{(a)} in Figure~\ref{fig:top_half_activity}) is taken.

\ac{hp} packets in contrast can cause a noticeable increase in \ac{isr} processing time.
At each occurrence of such a packet, the network task priority has to be increased in order to be scheduled subsequently ("prio+").
In case the \ac{bd} ring is already filled by previously received \ac{lp} packets now waiting inside their flow queue, a revocation is needed, adding further processing time ("recycling; prio+").
We also investigated on \ac{hp} packets that get rejected from their flow queue ("mitigating"), yet they behave similarly as shortcircuited packets.

The results show that the first three execution paths are similarly fast, while the ones that include an increase in priority or recycling operations are more costly.
As we discussed in section~\ref{sec:prioritedprotocolhandling}, a priority increase can only happen if the flow priority of a received packet is higher than the one of all the currently enqueued ones.
Without the network task being active to process packets and lower the highest enqueued priority again, this is only possible once for each flow in a cascade of increasingly prioritized flows.
Thus, when the system is flooded for some time and \ac{lp} packets start building up in their queues, eventually the faster paths of the \ac{isr} will be taken.

\subsection{Experiment 3: Mitigation and Prioritization}

\begin{figure}[h]
    \centering
    \resizebox{.8\columnwidth}{!}{
    \input{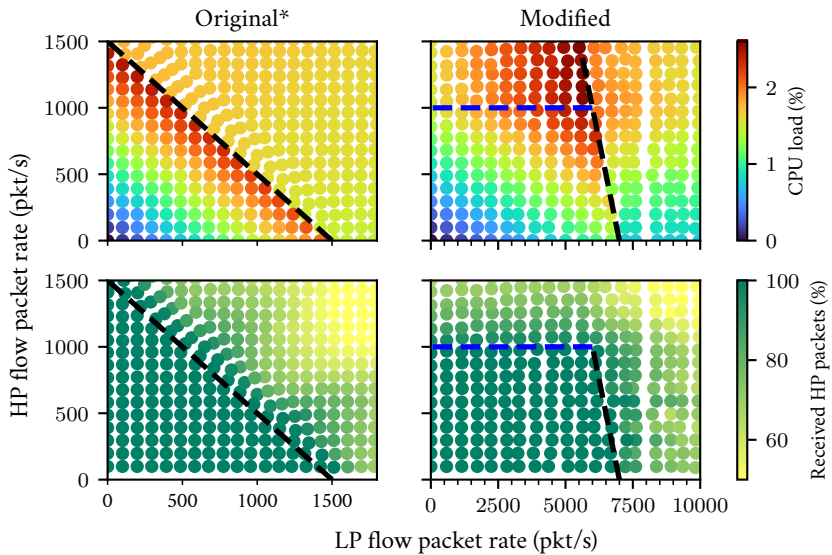}
    }
    \caption{\textbf{Mitigation Map: }CPU utilization and \ac{hp} flow liveness at various packet rates on our modified system versus the original system employing only an overall rate limitation. The blue and black lines mark flow-specific and overall rate limitations respectively.}
    \label{fig:mitigation_map}
\end{figure}

The final experiments show the effect of protocol processing prioritization and rate limitation, applied both for an individual flow and globally.
Experiments were conducted for multiple combinations of packet flood rates for a \ac{hp}- and \ac{lp}-flow, respectively, over a duration of 3 seconds each.
Again, a medium prioritized task measured the CPU load passively, preventing the scheduling of \ac{lp} packets.
Additionally, a receiver task was employed for the \ac{hp} flow in order to count the packets that arrived at their destination.
Figure~\ref{fig:mitigation_map} shows the CPU utilization and the ratio of successfully received \ac{hp} packets to sent ones, as a function of both packet rates.

The original stack was slightly modified to feature an overall \ac{isr} rate limitation, in order to allow a meaningful comparison to our approach.
It is implemented by switching to polling mode once the capacity is reached for a certain period, similar to the one employed in our prototype. In this experiment, the limitations is set at 3 packets per 2 milliseconds.

\paragraph*{Results}
The CPU utilization increases linearly along with both packet rates, until the global limit of 1500 pkt/s is reached.
Once the polling mode is active, the CPU load drops noticeably.
This can be accounted to the performance improvements gained by switching to a polling-based retrieving activity that handles multiple packets at once.
Further increasing the packet rate causes more \ac{hp} packets to be discarded by the \ac{nic}.

For the modified stack, parameter values anticipating a similar worst case CPU utilization were chosen.
We configured a high priority flow to allow one packet per millisecond and an unbounded low priority flow.
The \ac{isr} was limited to processing 7000 packets per second.

The CPU load also increases linearly with both packet rates.
As we would expect from the results of the first experiment, the load increases much slower with increasing \ac{lp} packet rates (notice the denser scale).
Above 1000 pkt/s (blue line) of \ac{hp} packets, the utilization stagnates as processing of further packets is cut off by the flow queue.
The additional triggered \ac{isr} executions are negligible at this scale.
When the sum of both rates exceeds 7000 pkt/s (black line), the CPU utilization also drops with polling activated.
Regarding the liveness of the \ac{hp} flow, we can see how it continuously decreases above the flow-specific rate of 1000 pkt/s.
Additionally, the global limitation impacts the \ac{hp} flow.
So, independent of the \ac{hp} flow rate itself, the communication liveness drops as the system is flooded with \ac{lp} packets.

When comparing the approach to a simple mitigating stack as a baseline, it becomes clear that it cannot help with processing higher rates of important packets.
Instead, supported by fast Early Demultiplexing and individual prioritization against the remaining tasks, it allows to postpone an overall limitation.
This way, a system can sustain a much higher load of less important packets before real-time disturbing effects start to occur.

\section{Conclusion}
\label{sec:2_3_conclusion}

This chapter presents several software-based approaches to detect and mitigate the real-time violating effects of incoming IP packets. The first half analyzes the network interrupt count and queue fill state, as well as the lateness of a critical task running on an embedded SoC while it is flooded with network traffic. Four mitigation strategies were developed to maintain smooth operation and timeliness of a critical process using signals derived from the data obtained during the analysis: Burst Mitigation, Hysteresis Mitigation, Budget Mitigation, and Queue Mitigation.

The quality of the different mitigation strategies was measured through the lateness of a critical task that runs concurrently to a network simulation triggering a high amount of interrupts. The evaluation indicates that Queue Mitigation is the most effective method as it processes the highest number of network packets while also safeguarding critical tasks from delays.

To further incorporate real-time task priority information on mixed-criticality devices, we presented an \ac{ip} stack design to individually schedule packet processing for differently prioritized \ac{ip} flows after early demultiplexing.
The problem of expensive processing in the network driver is addressed by incorporating the option of deferred buffer processing into our architecture.
Established embedded IP stacks, such as FreeRTOS+TCP and lwIP, can be adjusted to fit the proposed design with minimal modifications.

On our test system, the CPU load caused by low-priority packets in an already busy system is reduced by 86\% (a 7x speedup), even when dealing with packet buffers that travel through CPU caches.
Our approach enables system designers to anticipate packet rates of certain soft real-time flows, including those not belonging to any single receiver task, and derive an estimation for the respective \ac{wcet} through limitation parameters.
Compared to simpler overall mitigation algorithms, our design provides better isolation of processing time allocations for each flow, ensuring that important flows remain connected even when other flows exceed their rate limitation.
By allocating the same CPU resources to the processing of incoming packets, the networking subsystem can handle packets of a high-priority flow with up to 600\% higher overall traffic loads.

\chapter{A Real-Time Aware Network Interface Controller}
\label{cha:2_4_hardware}
\minitoc
In the past chapter we showed that differentiating and classifying IP packets as they arrive allows embedded real-time systems to prioritize their processing. 
By performing the classification as early as possible, limited resources can be used effectively for time-critical tasks. However, classifying packets in software also means that the \ac{irq} triggered by their arrival and inevitable \ac{isr} executions are unmoderated and might overwhelm the system by sheer volume (i.e. high interrupt frequencies as shown in Chapter~\ref{cha:2_2_problem}). Since the \acp{irq} are triggered outside the operating system's sphere of influence, effectively moderating these interrupts requires some modification to the generating hardware, in this case the \ac{nic}. Hence, in this chapter we present a modified \ac{nic} for embedded real-time devices that is interfaced with the \ac{rtos} to minimize interrupts in a real-time aware manner.

While there is extensive research on enabling technologies in the areas of communication and integration in IoT~\cite{bansal2020iot}, device architectures have received less attention~ (cf. Chapter~\ref{cha:1_3_survey} and~\cite{alcacer2019scanning, qiu2020edge}). Recent approaches focus on blocking or processing traffic before it arrives at the device~\cite{mandalari2021blocking, niedermaier2019secure, haar2019fane} while previous works failed to include network-specific factors and concentrated on interrupt management~\cite{task_aware, multisloth}. To the best of our knowledge, there is no published research on extending NICs and network stack implementations for real-time embedded systems.

This chapter proposes an extension to the receive functionality of \acp{nic} to minimize \acp{irq} under high load while maintaining short packet receive delays for critical tasks. The findings of this chapter have been published in~\cite{behnke2023towards, behnke2022priority}.
Specifically, we present the following contributions:

\begin{compactitem}
\item A priority-aware multiqueue \ac{nic} design for embedded real-time devices using interrupt moderation parameters.
\item A method for mapping real-time task priorities to IP flows which can be configured from the operating system.
\item A configurable \ac{nic} simulator emitting interrupt traces from network traces and a prototypical OS-side implementation to evaluate the effects of the proposed \ac{nic} extension on a real-time system.
\end{compactitem}

Section~\ref{sec:2_4_requirements} details the requirements of the multiqueue \ac{nic} and gives an overviwe over the proposed solution. 
Section~\ref{sec:2_4_multiqueue} presents the proposed \ac{nic} extension. 
Section~\ref{sec:2_4_evaluation} evaluates the approach in a isolated manner.

\section{Requirements \& Overview}
\label{sec:2_4_requirements}
We specify four requirements to the \ac{nic} adaptation appropriate for real-time devices with control over physical processes. 
    
\begin{compactenum}
    \item[2.1] \emph{Interrupt Moderation.} The danger of network-generated interrupt floods and unpredictable networking overheads should be mitigated by reducing the number of interrupts triggered by the \ac{nic}. Due to the limited processing capabilities of microcontrollers, their process schedules can be disrupted with little effort when each incoming packet triggers an interrupt. 
    \item[2.2] \emph{Packet Prioritization.} Solely limiting the number of interrupts leads to an analogous increase in receive delays as packets are accumulated before a notification occurs. Critical real-time tasks might not receive their packets in time. 
    Furthermore, the system can still be flooded with unrelated packets forcing the operating system to handle them. 
    Packets need to be classified and filtered before they reach the operating system.
    \item[2.3] \emph{NIC Parametrization.} To effectively prioritize and filter packets in different environments, the introduced \ac{nic} needs to be configurable. It needs to be possible to affect generated receive delays per process and tune them for the scenario's real-time requirements.
    \item[2.4] \emph{Continuous Configurability.} The necessary configuration of the \ac{nic} needs to be possible dynamically during runtime to facilitate changes in processes and the environment.
\end{compactenum}

The problem of network-generated interrupts affecting system performance can be solved by interrupt moderation (\emph{Requirement 2.1}). However, the techniques presented also have drawbacks. While they increase the overall efficiency of interrupt processing, they also increase the resulting packet delays and make them less predictable since packets are held back for a variable amount of time. In real-time systems, where process runtimes depend on incoming traffic, the occurrence of inaccuracies must be minimized. Therefore, a potential \ac{nic} design must attempt to reduce network overhead while guaranteeing low and constant latency for critical packets. As presented in Section~\ref{sec:2_6_related_nic} some specialized hardware exists running on FPGAs for specific real-time application types as well as multiqueue NICs for better multicore performance in data centers. However, to the best of our knowledge none exist for embedded hardware or real-time processing in general IP networks. 

To this end, the interrupt moderation parameters are designed to be reconfigurable. By minimizing the relative packet delay, the time a packet dwells in memory before it is processed is reduced. At the same time, however, this increases the total number of interrupts, which reduces throughput. This problem is not unique to real-time IoT devices, but is a natural consequence of interrupts. 
In bounded environments, however, we can take advantage of this fact. Since embedded systems typically perform a fixed set of specific tasks, we can use their metadata to filter and manage incoming packets before interruption at the hardware level. These are the priorities and timing requirements of the protocols or packet-receiving processes and their associated IP flows. Interrupt moderation can thus become a tool to enforce priority-compliant traffic scheduling before they enter the realm of the operating system.

\section{A Multiqueue NIC for Real-Time-aware Packet Reception}
\label{sec:2_4_multiqueue}

\begin{figure}
    \centering
    \includegraphics[width=0.8\textwidth]{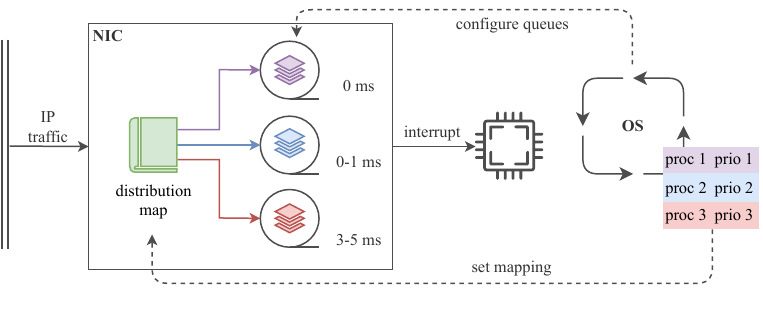}
    \caption{\textbf{Multiqueue NIC:} Traffic is organized into different queues with exemplary delay values attached; configurations of queues and mapping are performed by the operating system upon socket binding.} \label{fig:hwdesign}
\end{figure}

The modifications made to the NIC concern only the reception of packets and begin after frame validation at the MAC layer. An illustration of the design can be seen in Figure~\ref{fig:hwdesign}. To accommodate incoming packets belonging to different real-time processes, the receive buffer holding packet descriptors of the NIC is divided into multiple queues realized as ring buffers. This way, packet descriptors are assigned to different queues depending on their destination process and its priority, accounting for \emph{Requirement 2.2}. 

The metadata of validated packets is compared to a list of registered ports residing in a distribution map on the NIC. Here, packets are assigned to queues which hold packets of one IP flow each. According to the process priority and expected packet load, different interrupt moderation configurations (e.g. delay timers and counter threshold) are applied to them through the operating system. 

This way, packets for time-critical processes trigger interrupts immediately upon reception while less important packets (packets with low priority receiving tasks) are coalesced before one interrupt is triggered for all packets in the respective queue, indicated by the millisecond specifications in Figure~\ref{fig:hwdesign}. Packets with no associated process can be dropped before an interrupt is triggered since these packets would be dropped by the operating system at a later point in any case, but after generating unnecessary ISR and network stack work. This is especially important under high unanticipated traffic loads targeting the device and potentially leading to a denial of service. Finding appropriate queue configurations for different process priorities is part of the design process of the embedded real-time system as introduced delays need to be part of scenario modeling.

\subsection{Synthetically Added Bursts}
\label{sec:2_4_bursts}
Coalescing packets in the NIC reduces the number of interrupts triggered, ISRs executed and context switches performed. However, the amount of data to be processed by the network stack remains unchanged for packets registered for one of the running processes. Depending on the number of packets coalesced, interrupt moderation might lead to an accumulation of network stack workload into bursts. The necessary runtime to process an incoming packet is a lot smaller than the delay introduced by coalescing packets. A meaningful delay through these bursts can hence only happen under very high packet rates. We consider this when choosing the queue parameters as follows.

\subsection{Relevant Parameters}
\label{sec:2_4_parameters}
The multiqueue NIC introduces four parameters affecting packet delays and resource utilization as posed in \emph{Requirement 2.3}:

\begin{compactitem}
    \item \emph{Number of queues $m$.} The number of queues the receive buffer is divided into depends on the number of currently active processes accepting packets and supported protocols.
    \item \emph{Size of a queue $n_q$.} The number of elements of a queue $q$ corresponds to its expected packet load, available memory, and moderation parameters.
    \item \emph{Absolute queue timer values $t_{abs}(q)$.} Periodic duration until an interrupt is triggered by the queue $q$.
    \item \emph{Packet timer values $t_{pack}(q)$.} Amount of time after a packet is received by the queue $q$ that triggers an interrupt if not reset by another incoming packet.
\end{compactitem}

Additionally, the system introduces one implicite parameter:

\begin{compactitem}
    \item \emph{Maximum expected packet rate $R_{max}(q)$.} The maximum expected packet rate of the flow corresponding to a queue $q$. Equal to $\frac{1}{t_P(f)}$ as per Definition~\ref{def:1}.
\end{compactitem}

The timer values are used to span a time window of how long a packet remains in the queue. Depending on the packet rate, a variable number of packets is then coalesced to be announced by one interrupt. 
As these parameters have a high impact on the timeliness of incoming traffic and generated workload on the real-time device, the accuracy of their configuration is of high importance. While the number of queues is directly dependent on the current number of active (i.e. socket binding) processes, timer values and queue sizes have to be cautiously chosen. To be able to sensibly choose the parameters knowledge about expected packet rates and slack times of real-time processes is necessary.

The added packet processing time needs to be accounted for when developing an IP-connected real-time system. Process deadlines must allow for enough slack time for the system to handle concurrent packet reception. The higher the slack times are, the more packets can be processed without resulting in deadline misses. The worst-case scenario is subject to high interrupt rates affecting the process with the smallest slack time. For the calculation of appropriate queue parameters this value has to be factored in. The parameters must be chosen respecting the following considerations. 

\paragraph*{Queue size}
Choosing an appropriate queue size affects memory consumption as well as the maximum number of packets that can be coalesced in one interrupt. Applying interrupt moderation generally means holding more unprocessed packets in memory ultimately increasing the demand for the whole system. Memory implications for queue structures behave analogous but in a much smaller scale as only descriptors are held. 

The more packets can be held by one queue, the bigger the packet burst to be processed by the IP stack may become. Hence, this parameter also enforces an upper limit for the incoming packet rate per queue as elements are dropped when new packets arrive at a full queue. Additionally, this value has to be kept small enough for the IP stack processing time to be smaller than the minimum slack time when all queues generate a burst at the same time.   
The resulting delay corresponds to the Worst-Case Packet Processing Delay (WCPD) which depends on the per packet processing time $t_{netstack}$.

\[\text{WCPD} = t_{netstack} \sum_{q=0}^{m-1} n_q \]

\paragraph*{Absolute timer value}
The absolute queue timer realizes the upper latency bound of the interrupt rate window. To minimize the number of interrupts, this parameter needs to be maximized. At the same time, a higher absolute timer value also leads to a higher added latency an incoming packet might experience. Hence, the absolute timer value is limited by the maximum additional delay the underlying process can handle while still meeting its deadline. As a high value increases the burst of packets to be processed under high loads, the chance of the queue filling up increases, leading to packet loss. The maximum expected packet rate $R_{max}$ per queue $q$ needs to be factored in.
$t_d$ is the process-specific maximum allowed delay.

\[ t_{abs}(q) \le max(t_{d}(q), \; t_{\text{qf}}(q)) - \text{WCPD}  \]
\[ t_{\text{qf}}(q) = \frac{n_q}{R_{max}(q)} \]

\paragraph*{Packet timer value}
The packet timer realizes the lower latency bound of the interrupt rate window. Increasing this value generally decreases the number of interrupts as there is more time available for a new packet to arrive and reset the timer. This also means, that this value directly influences the minimum additional latency an incoming packet experiences. At the same time the amount of interrupts is highly dependent on the incoming traffic shape. The worst case in terms of interrupts generated is a packet rate corresponding to the packet timer value (one interrupt per packet). Hence, the available slack time needs to suffice to handle the number of interrupts generated by all packet timers combined when every timer iteration of each queue triggers an interrupt. $t_P(f)$ is the expected packet arrival period as per Definition 1.

\[
t_P(f) <= t_{pack}(q) <= t_{abs}(q)
 \]

\subsection{Configuration}
\label{sec:2_4_configuration}
As shown in Figure~\ref{fig:hwdesign}, two interfaces are used for the NIC configurations. One for setting the queuing parameters mentioned earlier and a second to write process-to-IP flow mappings to the distribution map. Both configurations are performed when a socket is bound using the network stack API (see Section~\ref{sec:rx_path}). For this purpose, the socket API is extended with driver calls that make the specific changes. Whenever a new process registers or releases a socket, the operating system transparently adjusts the number of queues and their parameters. The delay times and the size of the queues must be set for specific scenarios. The required information is passed to the driver as socket binding parameters.

As stated in Requirement 2.4, the system must be dynamically tunable at runtime to facilitate changes in processes or IP flows. Since the configuration process is linked to the socket API, the required tuning parameters can be passed at any time by the registering process. In the same way, NIC queues are released when a socket is no longer bound.

Since a network packet does not contain explicit information about the receiving process, a mapping must be made between the packet metadata and the processes. For this purpose, a mapping between IP flows and processes is created and placed on the NIC. In this design, the destination port is used to map a packet to a process. 
However, it could also be extended to match specific tags. This is be useful when all components of the distributed system are under the control of the developer to add security measures. The map must be on the NIC itself to cause minimal additional distribution delay, and still be configurable by the operating system to reflect the current set of existing processes.

\subsection{Memory Implications}
The presented approach has implications for memory usage on two levels: Firstly, the network buffer on the host system needs to be able to hold packet contents until they are processed, even when multiple packets are coalesced. Depending on the timer values and packet rate this might be a multiple of the usually necessary space. The network buffer resides on system RAM and is accessed by \ac{dma}. 
The second level is the required memory on the \ac{nic}. In an example implementation with 32~bit addresses, 32~bit timers, and a conservatively chosen maximum packet queue length of $65,536 KB$ the on-NIC memory necessary for one table entry is 30 Bytes as broken down in Table~\ref{tab:memory}.

\begin{table}
\centering
\caption{Required register memory per queue on the NIC for an example implementation.}
\begin{tabular}{l|c}
Component & Size \\ \hline
\emph{port\_id} & 16 bit \\ 
\emph{base\_address} & 32 bit \\
\emph{buffer\_size} & 16 bit \\
\emph{offset} & 16 bit  \\
\emph{next\_base\_address} & 32 bit \\
\emph{packet\_timer} & 32 bit \\
\emph{absolute\_timer} & 32 bit \\
\emph{packet\_timer\_expiration} & 32 bit \\
\emph{queue\_timer\_expiration} & 32 bit \\
\end{tabular}
\label{tab:memory}
\end{table}

The \emph{base\_address} field contains the RAM address of the beginning the the queue packet memory. The \emph{offset} is incremented for each incoming packet by its size. The \emph{next\_base\_address} is switched for the \emph{base\_address} when an interrupt for the queue occurs to be able to receive new packets while the buffered ones are processed.
When reserving one queue for non-transport layer protocols such as ARP (not requiring interrupt moderation fields and hence being 14 Bytes broad), the total memory requirement is

$$m \cdot 30B + 14B$$

for $m$ table entries on the NIC.

\section{Evaluation of the Multiqueue NIC}
\label{sec:2_4_evaluation}
The proposed NIC extension reduces the number and frequency of interrupts caused by incoming packets. Yet, as packets that belong to registered processes are not dropped, driver and network stack workloads remain in a time shifted manner. 
We evaluate the resulting timing implications by conducting three sets of experiments: The first explores the ability to reduce interrupts. The second compares the robustness of the real-time system against high traffic loads. The third analyzes the effects of different queue configurations under expected loads.

\subsection{Test Setup}
\begin{figure}[h]
    \centering
    \includegraphics[scale=1]{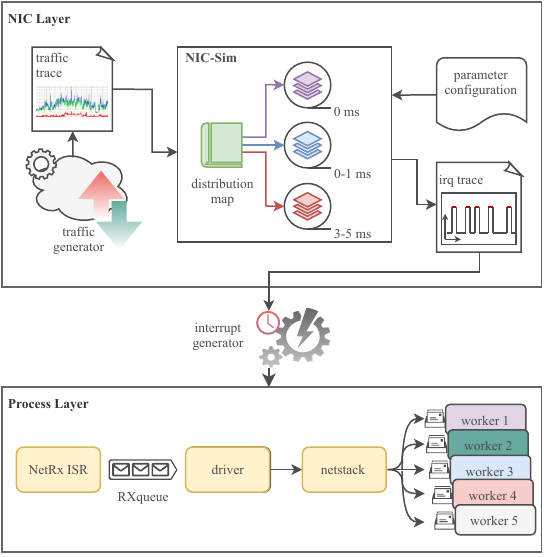}
    \caption{\textbf{Evaluation setup on two layers:} Interrupt/packet trace generation by a NIC simulator and process observation on an embedded real-time device.} \label{fig:setup}
\end{figure}

As the design proposes changes to hardware and an evaluation on a real embedded device is necessary for plausibility, the evaluation platform comprises of two layers. One layer assuring plausibility by providing a device running real-time processes (process layer) and one allowing hardware and configuration changes to the NIC (NIC layer). The evaluation setup is depicted in Figure~\ref{fig:setup}.

To evaluate the real-time behavior of a running system, we used an ESP32\footnote{\url{https://www.espressif.com/en/products/socs/esp32}} microcontroller for the process layer. It is equipped with a dual-core CPU and widely used for IoT tasks that involve communication via WiFi and Bluetooth. The two cores of the device permit a separation between observed processes and testing system. The observed processes run on the real-time operating system FreeRTOS where scheduling is performed preemptively on basis of process priorities. To generate realistic network loads, traffic traces from common industrial control systems are used. 

\subsection{NIC Layer: Simulator Implementation}
\label{sec:2_4_simulator}
The upper half of Figure~\ref{fig:setup} illustrates the NIC layer. A traffic generator pre-processes captured network traces and synthetic load patterns to generate a receive traffic trace for the NIC simulator. The simulator is written in Python using the event-based simulation library \emph{SimPy}. 
It is configured for each experiment run (as explained in Section~\ref{sec:2_4_multiqueue}) depending on relevant IP flows and processes. 
With the possibility to change NIC parameters, different interrupt traces can be created from the same network packet stream.
These traces additionally contain packet metadata for use by the process layer.

\subsection{Process Layer: Network Stack Implementation}
\label{sec:2_4_processlayer}
The interrupt traces generated by the simulator are applied to the processing layer by an interrupt generator implemented on the ESP32.
Due to the interrupt moderation, each interrupt notifies the network driver of a batch of one or more incoming packets. The NIC interrupt service routine (NetRx ISR in Figure~\ref{fig:setup}) receives this batch of packet descriptors and appends them to an operating system queue to be fetched by the network driver. From here, each packet is processed by the network stack task. If the packet destination port has a socket registered to it, the packet descriptor is forwarded to the socket mailbox and the associated task is notified. Otherwise, the packet is dropped.

The receiving real-time worker processes get access to sockets through the socket API. 
Using a receive function, the processes can then read data from the socket mailbox.
This approach implementation of Berkeley sockets corresponds to the mapping of IP flows to processes.
The receiving processes are workers of different priority. 
Each incoming packet is tracked from its time of arrival at the NIC until processing in its worker process where it triggers the task workload.

\subsection{Experiment 1: Interrupt Generation}

In all experiments, the IoT device runs four worker processes of different priority. The processes receive traffic using the widely used industrial communication protocol MODBUS/TCP\footnote{MODBUS/TCP traces provided by~\cite{frazao2018denial}.}. 
As each of the processes binds their own socket, four queues with different interrupt moderation configurations are set up in the NIC.

\begin{table}[h]
    \centering
    \caption{Queue configurations for baseline process IP flows.}
    \footnotesize
    \begin{tabular}{p{1cm}||p{1.5cm}|p{1.5cm}}
    	queue & absolute timer [ms] & packet timer [ms]\\
		\hline
        0 & \multicolumn{2}{c}{unmoderated} \\
		\cline{2-3}
		1 & 30 & 20 \\
        2 & 40 & 30 \\
        3 & 80 & 70 \\
    \end{tabular}
    \label{tab:configs}
\end{table}

Table~\ref{tab:configs} shows the queue configurations for the worker task flows. Queue 0 is configured to receive the IP flow of the critical task, hence no moderation is applied to this flow and packets are forwarded immediately upon arrival. Queues 1-3 are moderated with increasing delay values. Queue capacities are kept at a constant 128 packets per queue. Additionally, one baseline experiment is performed without any interrupt moderation. We observe the progression of interrupts generated in respect to packets received.

\paragraph*{Results}

\begin{figure}
    \centering
      \includegraphics[width=0.8\textwidth]{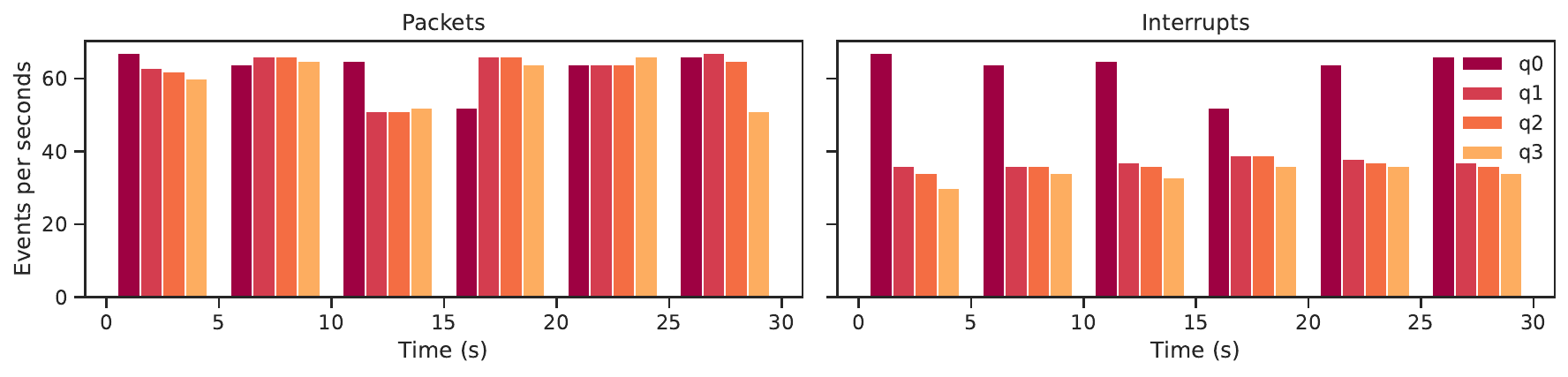}
      \caption{\textbf{Interrupt reduction:} Histogram of packets and caused interrupts over time with bin size of 5 seconds.}
      \label{fig:pckirq}
\end{figure}

The number of interrupts generated depends on the number of packets received in each queue and their configuration. 
Figure~\ref{fig:pckirq} shows a comparison of packet and interrupt numbers for the baseline experiment without additional load. Queues 1 - 3 moderate interrupts in different time windows, so they generate fewer interrupts than queue 0, which is receiving packets for a critical task. 

\subsection{Experiment 2: Unfiltered Packet Flood}

To observe the system under high traffic, it is subjected to packet floods ranging from 0 to 15000 packets per second. The worker setup on the device stays as defined in Experiment 1. To observe the effects of packet floods when they are targeted at unregistered sockets they are subjected to a separate moderated NIC queue. We evaluate the compute load of the flood on the system.  The experiments are repeated on four different absolute delay timer values for the added packet floods.
The absolute timer values range from $800\,\mu s$ to $3200\,\mu s$ resulting in the designations \emph{nomod} (for unmoderated flood traffic), \emph{d800}, \emph{d1600}, \emph{d2400}, and \emph{d3200}. As we are testing the system under higher than expected load, the packet timer can be disregarded for this experiment (cf. Section~\ref{sec:interrupt_moderation}). Each experiment runs for a duration of 30 seconds. 
We observe the additional runtime of the processes incurred by the network traffic.%

\paragraph*{Results}

The total rate of interrupts per packet ranged from 70\,\% in the undisturbed experiment (Experiment 1) to 2\,\% with high additional load of 15000 packets per second and $3200\,\mu s$ absolute timer value. The absolute moderation timer is an effective tool to moderate the high load as more packets are coalesced into interrupts while the critical task is unaffected.

\begin{figure}[h]
\centering
\includegraphics[width=0.5\columnwidth]{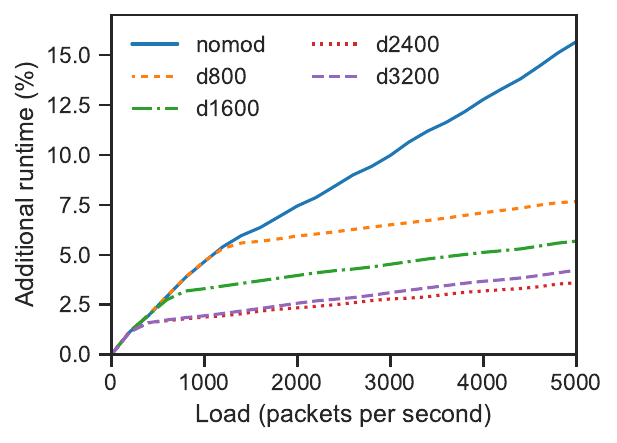}
\caption{\textbf{Timing impact:} Additional runtime of critical process (queue 0) induced by increased network load.}
\label{fig:runtimeq0}
\end{figure}

Next, we observe the interrupt-induced runtime increase of the critical process. A significant mitigation of the harmful effects of packet floods can be obtained in all moderation configurations for the critical process. Processes of lower priority also benefit from the approach, as the CPU is freed up from unnecessary ISRs. Figure~\ref{fig:runtimeq0} shows the mitigating effects for the critical task under variable additional load. The visible linear increase continues throughout all experiments. Further, it can be seen that there is a scenario-specific optimal configuration between \emph{d2400} and \emph{d3200} due to the effects of packet burst processing (cf. Section~\ref{sec:2_4_bursts}). By increasing the delay parameters, more packets are coalesced for each interrupt, meaning that the networking tasks are confronted with larger bursts of packets per notification. This has negative effects on CPU load starting at a critical packet count. For the maximum depicted packet load of 5000 packets per second the additional runtime could be decreased by 80\,\% (or 12 percentage points) resulting from the prevention of 93\,\% of interrupts.

\subsection{Experiment 3: Expected Load}
Since the NIC retains packet notifications for low priority tasks, it causes an additional interrupt delay depending on the delay timer configurations. To investigate this delay, the second set of experiments was performed on a stable system with no unexpected traffic floods. The combination of absolute and packet delay timers spans a window for the period length of interrupts. Using the approximated incoming packet rate of a flow $1/t_P(f)$, developers can adjust the values to minimize the introduced interrupt delay as described in Section~\ref{sec:2_4_configuration}. 

In this experiment, the four processes receive IP flows ranging from about 50 (for queue 0) to 200 (for queue 3) packets per second. We compare three different NIC configurations:

\begin{compactitem}
	\item \emph{No moderation}: All queues trigger interrupts as soon as a packet arrives.
	\item \emph{Medium moderation}: Queue 0 triggers an interrupt as soon as a packet arrives. Queues 1, 2, and 3 are configured to coalesce 2, 3, and 4 packets per interrupt, on average.
	\item \emph{Strict moderation}: Queue 0 triggers an interrupt as soon as a packet arrives. Queues 1, 2, and 3 are configured to coalesce 5, 6, and 7 packets per interrupt, on average.
\end{compactitem}

\paragraph*{Results}

\begin{figure}
    \centering
    \includegraphics[width=0.5\columnwidth]{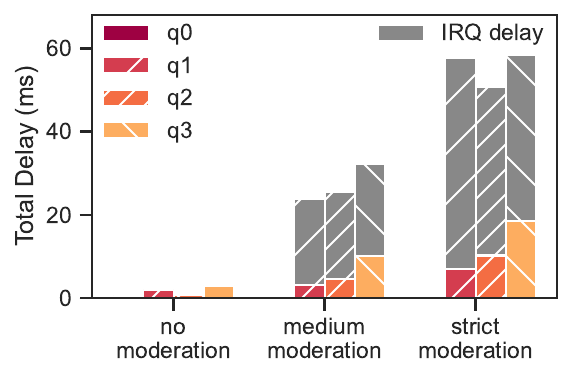}
    \caption{\textbf{Median delay times:} The IRQ delay specifies the average coalescing delays.}
    \label{fig:nodos}
\end{figure}

The results, depicted in Figure~\ref{fig:nodos}, show that the critical process (queue 0) does not suffer from any additional delay.
The delay added to processes of less priority is directly dependent on the chosen moderation parameters and process-specific load and can be chosen to fit the requirements of each process before or during runtime.
At the same time, the delay window prevents interrupt frequencies to climb to a critical level.
An increase in the OS-induced delay can be observed when too many packets are coalesced to one interrupt as the resulting bursts in packet processing increase the processing time for individual packets in the networking and worker processes.

While the queue moderation can effectively reduce the impact of high packet loads, the tuning of moderation parameters requires care.
Task deadlines, as well as packet loads and latencies for each process should be identified to reduce the impact of interrupt moderation under normal conditions while at the same time ensuring operability for critical tasks under unexpectedly high loads.

\section{Conclusion}
Unexpected floods of network traffic can delay process flows in real-time systems, putting critical real-time requirements at risk. Specialized networking hardware for real-time embedded devices can mitigate the risk.
This chapter presented a multiqueue \ac{nic} design shifting the early demultiplexing discussed in Chapter~\ref{cha:2_3_software} from the lower part of the driver to hardware. This way, the number of network-triggered IRQs can be moderated depending on packet priorities and load. By configuring the extended NIC via an adapted Berkley socket API, currently active processes waiting for packets can register their priorities with the hardware, effectively mapping IP flows to processes when a socket is bound and changing queue parameters accordingly. 
We evaluated the hardware design using a \ac{nic} simulation and an IoT device running a real-time operating system. The results of these experiments show that our approach significantly reduces the impact of traffic floods on critical process runtimes by saving 93\,\% of interrupts and 80\,\% of processing delay under packet rates of 5000 per second while the configuration of multiqueue parameters requires knowledge about expected network traffic and real-time requirements.

\chapter{A Hardware/Software Co-Design for Real-Time Aware Packet Processing}
\chaptermark{Co-Design for RT-Aware Packet Processing}
\label{cha:2_5_codesign}
\minitoc
The proposed designs discussed previously were exclusively software or hardware based, modifying either network driver or \ac{nic}. In an effort to mitigate the problems introduced into the real-time system by IP networking, they each developed distinct advantageous approaches to different parts of the issue. However, as the problem involves the \ac{nic} as receive hardware in conjunction with the network driver and \ac{rtos} task prioritization, a unified design combining the advantages of both approaches is warranted. Such a design is introducted and implemented in this chapter. Section~\ref{sec:2_5_considerations} presents the added considerations for the combined design. Section~\ref{sec:2_5_design} details the co-design. Section~\ref{sec:2_5_implementation} outlines our implementation specifics. Section~\ref{sec:2_5_evaluation} presents an evaluation of the implementation and Section~\ref{sec:2_5_conclusion} concludes this chapter.

\section{Preliminary Considerations}
\label{sec:2_5_considerations}
A unified design can combine the advantages of the two presented designs while also mitigate some of the individual disadvantages. These shortcomings can be summarized as follows. 

\paragraph*{NIC Design}
\begin{compactitem}
    \item Only the amount of \ac{isr} runs is reduced in a prioritized manner. Once packets are received by the OS queue, no reordering can happen and packet processing can still lead to an inversion of priorities.
    \item The design could create bursts of low priority packets that, once inside the operating system, might block the processing of high priority packets and lead to priority inversion.
    \item Interrupt moderation and multiqueue parameters have to be set by developers apriori and for each process individually.
\end{compactitem}

\paragraph*{Driver Design}
\begin{compactitem}
    \item The number of interrupts can not be reduced other than by switching to polling mode. 
    \item In the case of a flood of incoming unwanted packets the system can still be overwhelmed by \acp{irq} and \ac{isr} runs or be forced into polling mode with the remaining packet processing overhead.
    \item The classification of packets and packet-wise cache invalidation adds workload to the (preempting) \acp{isr}.
\end{compactitem}

To implement a unified design, the networking task, driver, and hardware components should be designed holistically and make configuration parameters available in a standardized manner. This design is subject to the same assumptions and requirements as the individual approaches. The preliminary considerations of the two designs remain relevant here and can be compiled as follows to act as a guideline for the design stage.

\begin{compactenum}
    \item[3.1] \emph{Early Demultiplexing of Packets into Flows.} Packets need to be sorted into recipient-associated flows early in hardware. Every flow should adapt the receiver task priority to extend the task priority space into the receive path. This early separation into flows is required to avoid resource contention in packet reception. Moreover, packet filtering at reception allows for fast rejection of irrelevant packages, thereby reducing the load on protocol processing in software.
    \item[3.2] \emph{Mitigation of Priority Inversion.} Whenever packets of multiple flows await processing, all packets of a lower-priority flow must wait for any packets of higher-priority flows to be processed and delivered. This avoids priority inversion, as high-priority recipient tasks will not be blocked by lower-priority recipients serviced earlier. To this end, the network task processing queued packets in the IP stack needs to always dequeue the high-priority packets available and change its task priority according to the flow.
    \item[3.3] \emph{Overload Protection through Interrupt Moderation.} Network-induced packet floods should never cause system overloads. Hence, the effects of unpredictable packet delivery events on the real-time system need to be mitigated. This entails decoupling the strict consecution of packet reception and \ac{dma} transfer from the subsequent processing of the packet.
    \item[3.4] \emph{Extensive Configurability.} In order to configure the modified network solution to integrate it effectively into the real-time application, the moderation parameters for all packet flows are required to be configurable from software. They need to be adjustable by the user in a similar way the real-time parameters for the \ac{rtos} are.

\end{compactenum}

\section{Co-Design}
\label{sec:2_5_design}
\begin{figure}[h]
    \centering
    \includegraphics[width=\textwidth]{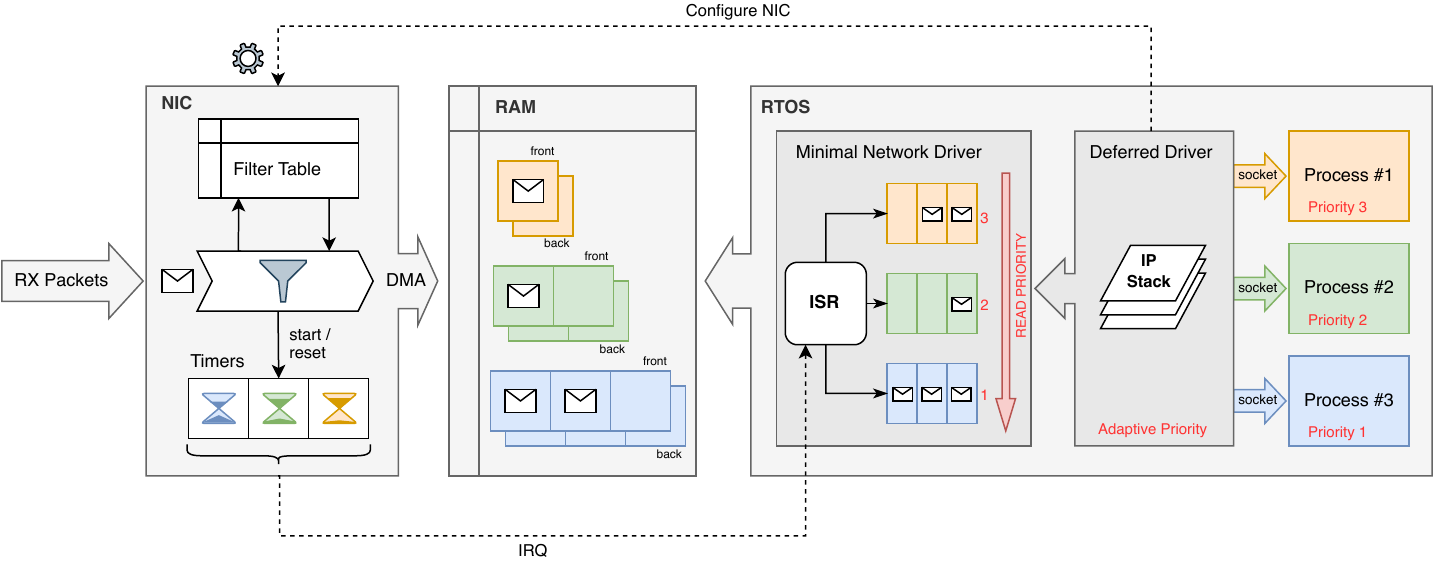}
    \caption{\textbf{Co-Design:} Schematic of the unified design displays how hardware and software components are co-integrated to form a holistic solution for the receive path. Three flows (orange, green and blue) associated with recipient processes (nos. 1 -- 3) are displayed, assuming their priorities for packet reception and processing. Each flow is configured with a different timer and queue size configuration.}
    \label{fig:arch}
  \end{figure}

The combined design unifies the priority spaces of real-time tasks, packet processing, and network-generated interrupts. An abstract representation of the design can be seen in Figure~\ref{fig:arch}. 

The design extends from hardware to software, demultiplexing the packets received from the \ac{mac} into application-centric flows early in hardware and processing them in their flows until they are delivered to the destination application. 
Depending on developer-defined tasks and their priorities, incoming packets can be dropped and coalesced to fewer \acp{irq} before any workload emerges in the \ac{rtos}. Every packet received by the \ac{nic} is classified based on its metadata and filtered according to the configuration stored inside the filter table. Matched packets will be transferred into the corresponding packet queue buffers in \ac{ram} via \ac{dma}, with every queue buffer holding the packets for a specific flow (designated \emph{front buffer}). A queue buffer will accumulate packets until expiration of the queue is indicated by associated timers managed in the \ac{nic}. At this point, the front buffer is replaced by a new queue buffer (\emph{back buffer}) as target for \ac{dma} transactions and the software is informed of a packet queue ready to be processed.

The software side of the design implements the network driver and is split into two segments. In its design, the same structural approach was used as is proposed in Chapter~\ref{cha:2_3_software}, since partitioning the driver into an eager and a deferred part allows the workload in the \ac{isr} to be kept small by offloading the IP stack processing into a separate network task. Similarly, using multiple differentiated flow queues was adopted to implement packet processing by flow priority inside the deferred driver task. When a packet queue is ready to be processed, an IRQ is raised. The subsequently executed \ac{isr} inside the minimal network driver extracts all packets stored in the buffer and appends them to the associated flow queue. Once packets are available, the network task in the deferred driver is responsible for IP processing, ultimately submitting them to the sockets of the receiver processes.

Flow configuration and back buffer resupply for the \ac{nic} are performed through the socket \ac{api} and network driver, respectively. When a socket is bound, a flow inheriting the priority of the responsible process is established, with the driver supplying initial configuration to the \ac{nic}. Configuration of interrupt moderation parameters such as packet queue buffer size or interrupt timers may be performed via socket options in the user processes. This allows the user to adjust these parameters to the requirements of their real-time software, as per \emph{Design Requirement 3.4}.

\subsection{Changes to Individual Implementations}
To accommodate the combined design the individual implementations have to be adapted. This section discusses the challenges of a new design and presents necessary changes.

\subsubsection*{NIC Design}
\paragraph*{Demultiplexing Packets using the Filter Table}
Every entry in the filter table is associated with a packet flow, mapping the packet metadata employed in classification to the flow distribution parameters. This metadata will typically include the protocol and transport layer destination port number and is used for matching incoming packets to a flow. The distribution parameters constitute a memory address identifying the next available slot inside the front packet queue buffer. In every filter table entry two packet queue buffers are referenced (designated front and back). Using a back buffer facilitates fast buffer swapping on timer expiration, since a replacement for the expired front buffer is available immediately. Thus, the technique prevents packets from being dropped for the duration of the queue swap.

Although the proposed solution focuses on transport layer packets, lower-layer packets like ARP, ICMP or IGMP need to be addressed as well. Since these are not associated with any sockets, a default flow is proposed to accommodate them. It should be handled like any other flows inside the hardware, with the associated task priority being that of the network driver process.
Demultiplexing packets early in hardware offers performance advantages: since non-matching packets would otherwise be discarded later on in the IP stack anyway, dropping them in hardware reduces the packet processing overhead in software. Furthermore, it mitigates the danger of packet flooding with unanticipated packets, therefore meeting \emph{Design Requirement 3.1}.

\paragraph*{Timers for Interrupt Moderation}
As previously established, the generation of individual receive interrupts on a per-packet basis is detrimental to the real-time capability of a system. The solution proposed in Chapter~\ref{cha:2_4_hardware} stipulates generating interrupts using timers. Since this approach has been proven effective at interrupt moderation, the idea will be adopted in this design. Using timers for IRQ generation effectively coalesces all interrupts generated for received packets into a single interrupt triggered when the timer is due. This approach couples the receive interrupt execution only loosely to packet reception, with any danger of network induced overloads mitigated, thus realizing \emph{Design Requirement 3.3}.

As in the original \ac*{nic} desgin, two timers are proposed for each flow. The packet timer and the queue timer. Both timers dictate the minimum and maximum possible duration of the time frame in which interrupts may be raised. The packet timer is started when the first packet is filtered into a flow and reset on every subsequently filtered packet. Thus, it dictates the lower bound of the time frame, which is the minimum time a packet will remain queued. The queue timer on the other hand dictates the maximum duration packets may stay queued before being passed to software. Through configuration of these timers, the additional delay incurred for the received packets may be adjusted. Furthermore, they may also be tuned to the packet reception rates expected by the corresponding processes running on the \ac{rtos}. For example, a high-priority task awaiting steady rates of real-time data over the network may configure both timers to a very short duration or even zero to force immediate packet reception. On the other hand, a low-priority process awaiting reception of periodic telemetry information may configure the timer much more relaxed.

\subsubsection*{Driver Design}
The driver only has to be changed minimaly compared to the design from Chapter~\ref{cha:2_3_software}. The following two paragraphs descibe the elements of the driver necessary to fulfil the requirements stated above.

\paragraph*{ISR in the Network Driver}
The minimal network driver is responsible for extracting all packets from the available packet queue buffers and moving them to the flow queues. Processing steps like cache invalidation or validation of offloaded checksums are also performed in the \ac{isr}. However, cache invalidation is not required for the complete buffer at this point, since only the metadata such as packet size (and optionally the checksum) need to be extracted from the packet queues buffers. As explained in Section~\ref{sec:2_3_approach}, it is sufficient in this stage to only invalidate the parts of the cache where the data resides. The invalidation of the remaining cache lines is delayed for the deferred driver to handle, therefore reducing the workload to be performed in the \ac{isr} (i.e. \emph{lazy cache invalidation}).
The beneficial effects of lazy cache invalidation on the run time of the \ac{isr} are estimated to be even greater in this design. Since the packet queue buffers usually contain multiple packets, cache invalidation for the complete buffer range will demand much more run time than merely invalidating small ranges for each of the packets in the buffer.

\paragraph*{Deferred Driver}
The deferred driver will fetch the packets received from the flow queues, processing them in the IP stack. Packets are always read from the populated queue of highest priority, adhering to the strict flow prioritization to accommodate \emph{Design Requirement 3.2}. As concomitant measure, the network process changes its own task priority to the priority of the flow queue being processed, hence mitigating the possibility of priority inversion as has been described in Section~\ref{sec:prioritedprotocolhandling}.

\section{Implementation Details}
\label{sec:2_5_implementation}
This section describes implementation details for the previously introduced design. The referenced implementation is a working prototype developed for this thesis. The prototype was implemented on a Xilinx Zynq \ac{soc}, which features a \ac{fpga} integrated with an ARM Cortex-A9 processor as \ac{ps}.

\subsection{The NIC}
An out-of-the-box Ethernet \ac{nic} design on the \ac{fpga} was used as the basis for implementing the modifications to the network interface controller. The standard design features an Ethernet subsystem and an \ac{axi} \ac{dma} \ac{ip-core} which connects the Ethernet Subsystem to a peripheral high-performance \ac{axi} bus interface of the processing system.

\begin{figure}
    \centering
    \includegraphics[width=\textwidth]{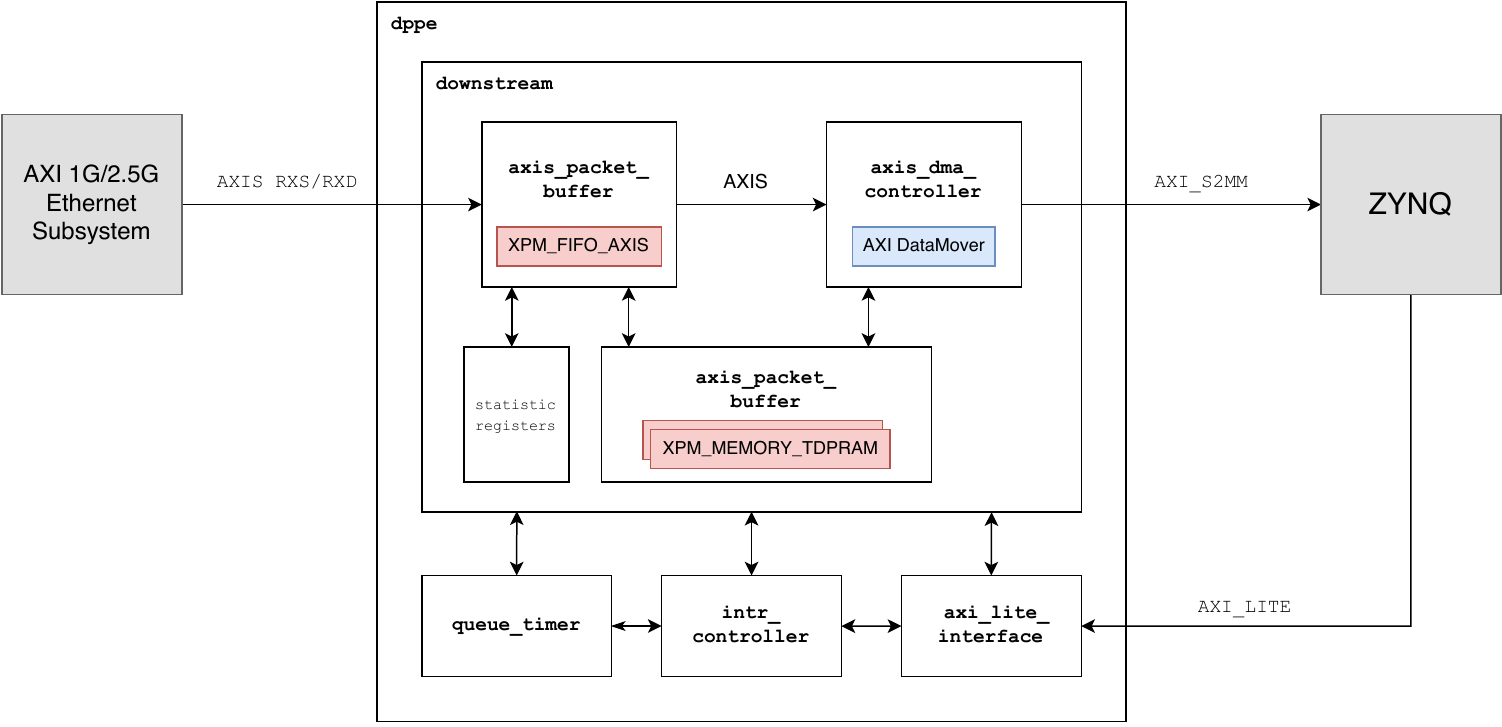}
    \caption{\textbf{Module hierarchy:} Hardware implementation of the RX Path depicting the hierarchy and interconnection of all modules. Modules shown in blue are parameterized macros or instantiated IP-Cores distributed by Xilinx.}
    \label{fig:hw-arch}
  \end{figure}

The module hierarchy of the hardware design for the custom \ac{nic} is displayed in Figure~\ref{fig:hw-arch}. All modifications to the existing hardware are encapsulated in an additional \ac{ip-core} designated the \ac{mq-dsppe}. This \ac{ip-core} is connected directly to the output of the Ethernet subsystem, receiving status information on the the \ac{rxs} and the actual frame payload on the \ac{rxd}. The buses are connected directly to the \ac{axi} Stream Packet Buffer module, described in more detail in the following subsection. Since the proposed design requires unique \ac{dma} functionality, a custom-built \ac{dma} controller is included in the \ac{ip-core}. Details are described in Section~\ref{sec:2_5_dma_module}. Therefore, the receive functionality of the \ac{axi} \ac{dma} \ac{ip-core} is not required in the custom \ac{nic} implementation and the custom \ac{dma} controller module is instead connected to the peripheral bus interface of the \ac{ps}.

All modules directly involved in processing and forwarding the received Ethernet frames are encapsulated in the Downstream module. The \ac{axi} Stream Packet Buffer module receives the frame data as well as additional status information directly from the Ethernet Subsystem. It scans the frames and extracts all relevant information required for filtering the packets. Once it has latched all relevant data from the Ethernet frame stream, it sends a lookup query to the Filter Table module, receiving information on whether the packet may be forwarded to the \ac{dma} controller or dropped. Additionally, parts of the data looked up in the filter table are passed to the \ac{dma} controller. This module transfers the Ethernet frame received to the selected frame queue buffer resident in the \ac{ps} \ac{ram}. The frames are stored in series inside the buffer, with one word of metadata such as frame length and raw checksum placed before each frame.

During operations, the Downstream module also increments statistics timers, collecting information on the number of packets forwarded or dropped and the number of \ac{dma} errors encountered in transmission.

\subsubsection*{The AXI Stream Packet Buffer Module}
This module is responsible for scanning and classifying the frames received from the Ethernet subsystem. The parameters extracted from the received streams are the frame length as well as the port number in transport layer packets. Scanning the Ethernet frames while they are being transmitted word for word on the \ac{axi} stream bus is realized using a \ac{fsm} and stream word counters. The counters are incremented for every word received on the bus, enabling the \ac{fsm} to identify relevant fields by their byte offsets inside the nested packet headers. Since all packet headers need to be scanned, a portion of the frame needs to be received before deciding whether to forward or drop the frame. Thus, it cannot be forwarded to the \ac{dma} controller as of yet and needs to be buffered inside this module. An \ac{axi} Stream FIFO\footnote{\url{https://docs.xilinx.com/r/2022.1-English/ug953-vivado-7series-libraries/XPM_FIFO_AXIS}} is utilized for this purpose, allowing the frameto be forwarded or dropped when sequentially shifting out the frame on the bus interface connected to the \ac{dma} controller.

\begin{figure}
    \centering
    \includegraphics[width=0.8\textwidth]{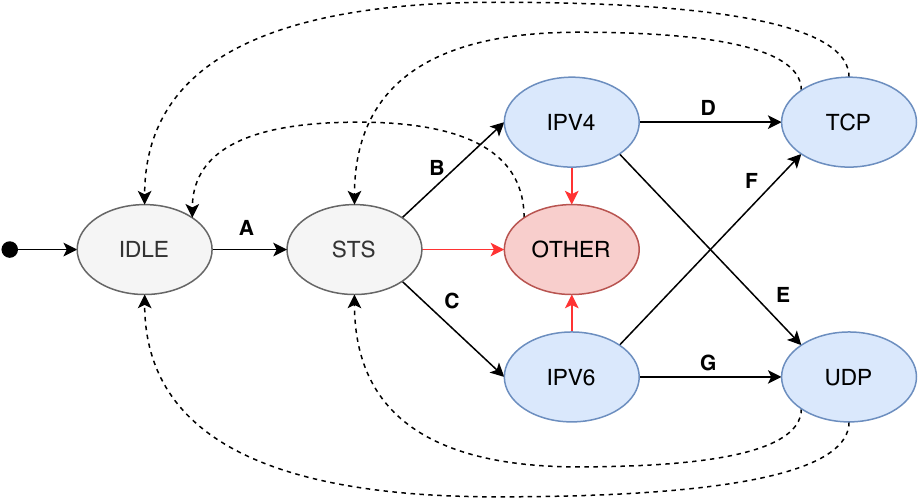}
    \caption{\textbf{Frame classification:} Simplified state diagram of the FSM controlling the scanning process of the frame.}
    \label{fig:eth_fsm}
\end{figure}

A state diagram for the \ac{fsm} is provided in Figure~\ref{fig:eth_fsm}. The IDLE state is the default state, where the module awaits receiving frame data. Status information about a received frame is the first thing the Ethernet subsystem transfers, causing the \ac{fsm} to switch to state STS on the status stream bus, providing information such as the total frame length and the EtherType from the frame header. Based on the value stored, the \ac{fsm} switches into states IPv4 or IPv6, respectively. In case of any other EtherType, the packet is not relevant and the \ac{fsm} switches into state OTHER, ignoring the remainder of the frame transmission. By the time the \ac{fsm} has switched into the network layer state, the actual frame data is being transmitted on the \ac{rxd}. Dependent on the network layer protocol, the IP header fields Protocol (Ipv4) or Next Header (IPv6) are inspected to determine the transport layer protocol (\ac{tcp} or \ac{udp}), again switching into state OTHER when any other values than the expected ones are encountered. Once the headers are transmitted, the port numbers are latched and the module is able to post its query to the filter table, providing the port number and frame length. In case the \ac{fsm} is in state OTHER, a query is still sent but containing the value zero instead of a port number. The \ac{fsm} then either returns to state IDLE in case no further frame transmission was initiated or it switches into state STS in case the status stream information on the next Ethernet frame is already pending.

\subsubsection*{The Filter Table Module}
This module stores the information determining where frames with a certain id are transferred to in memory via \ac{dma}. Every entry stored inside the table has the following fields:

\begin{compactitem}
    \item \emph{id} stores the port number (16 bit)
    \item \emph{buffer length} holds the total length of the packet queue buffer (configurable, default 16 bit)
    \item \emph{front buffer address} stores the memory address of the front buffer (configurable, default 32 bit)
    \item \emph{back buffer address} stores the memory address of the back buffer (configurable, default 32 bit)
    \item \emph{offset pointer} holds the number of bytes already stored inside the table (configurable, default 16 bit)
\end{compactitem}

The offset pointer points indicates the next free byte in the buffer and is always incremented by the length of the frame transferred. Thus, the actual address to which a frame will be transferred into memory is calculated as the sum of the front buffer address and the offset pointer. If the buffer length minus the sum is less than the frame length, the frame will dropped since there is not sufficient memory left in the buffer to store the complete frame.

\subsubsection*{The DMA Controller Module}
\label{sec:2_5_dma_module}
This module implements the \ac{dma} functionality required for transferring the received packets to memory. For the actual data movement, the module uses an \ac{axi} DataMover\footnote{\url{https://docs.xilinx.com/r/en-US/pg022_axi_datamover}}. This module is capable of transferring data received on an \ac{axi} Stream bus into \ac{ps} Memory. Since the \ac{dma} engine itself buffers transfer requests, care was taken to track all pending \ac{dma} operations to prevent the hardware from releasing a filled front packet queue buffer to software while \ac{dma} transfers into the buffer are still in progress.

\subsection{The Driver}
The software solution was implemented into the Xilinx port of LwIP 2.1.1. Primarily modifying the vendor-specific lower-level part of the network stack as well as the socket module. The minimal driver is implemented in an \ac{isr}, which is registered with the processor's interrupt controller to be executed when the Ethernet \ac*{mq-dsppe} \ac{ip-core} raises an \ac{irq}. The deferred driver functionality was implemented into the vendor-specific network interface code of the network stack, superimposed over a task responsible for moving packets from the frame queue to the LwIP stack. Furthermore, the single frame queue was replace with a data structure, holding the frame queues for all flows as well as additional information on the associated task priorities.

Much care was taken to avoid concurrency issues such as race conditions or deadlocks between tasks. Mutexes were used for locking access to the central data structure holding the queue information. This was necessary, as this data structure is accessed from the deferred driver task, the \ac{isr} and any user task via the Berkeley socket \ac{api}. To avoid race conditions with the \ac{isr}, interrupts are deactivated in critical sections where a task accesses driver-related memory which the \ac{isr} shares access to.

Inside LwIP, the socket \ac{api} was modified to allow interrupt moderation configuration to be performed via the socktop() function, which is part of the Berkeley Socket \ac{api}. Moreover, core locking was enabled for LwIP. This configuration option ensures all accesses to the IP stack are thread safe by using a mutex to lock the central data structure. The positive corollary of using core locking is the effective elimination of a user-space worker task (designated \emph{tcpip}), which ordinarily is employed in handling packet input. Since the proposed design featured an adaptive-priority task in the deferred driver, which already handles packet input to the IP stack, eliminating the packet reception capability of the \emph{tcpip} thread was necessary.

\section{Evaluation}
\label{sec:2_5_evaluation}
In this section the implementation of the custom hardware-software solution is evaluated. For this purpose, the design was implemented on an \ac{fpga} evaluation board and tested in a scenario using real network traffic. The empirical evidence provided in this section will substantiate the design choices made and prove the effectiveness of the developed approach.

\subsection{Experiment Setup}
The experiment was performed on an Alinx AX7Z035 evaluation board featuring the Xilinx XC7Z7035 Zynq \ac{soc}, which merges a dual-core ARM Cortex-A9 with a Kintex-7 \ac{fpga}. The evaluation board features a Micrel KSZ9013 Gigabit Ethernet PHY connected to the \ac{fpga} fabric on which the custom hardware design is implemented. To simulate a low-power embedded device as typically employed in IoT environments, the ARM Cortex was configured to operate in single-core mode with a clock rate of 300 MHz.

To collect precise information on the amount of processing time utilized by each task, FreeRTOS was compiled with run time statistics enabled. Additionally, the statistics collection tick rate in FreeRTOS was increased 100-fold to enhance the accuracy of the measurements.

In the experiment, the impact of high rates of low-priority packets on a high-priority network-dependent task were evaluated. For this scenario, two worker tasks were defined, a low-priority task (LP) and a high-priority task (HP). Each task held an open socket, anticipating \ac{udp} network traffic. To simulate the packet processing delay for the worker tasks, the task would engage in busy-waiting for 5 ms after every packet received on the socket.
The experiment was performed for 3 setups, once for the custom implementation and twice for the baseline implementation. Two configurations were tested on the baseline implementation, one with a default frame queue size of 32 packets and another with an infinite frame queue size. This was done to examine the effect a limited frame queue size has on packet reception. Measurements were taken for a time period of 70 seconds per run, with packets sent for 60 of those seconds.

During the experiment, a constant rates of zero-length \ac{udp} packets was sent to the high-priority and low-priority tasks. Throughout the 7 runs performed, the rate of packets sent to the high-priority task was held constant at 100 packets per second. The rate packets sent to the low-priority task was doubled on every run. Thus, the packet rates sent to the task were 0, 100, 200, 400, 800, 1600 and 3200 packets per second (pps). The processing time every task was scheduled as running was measured as well as the number packets received be the high-priory task.

\subsection{Results}

\pgfplotstableread[col sep=comma,trim cells=true]{data/data_hptask.csv}\hptaskTable
\pgfplotstableread[col sep=comma,trim cells=true]{data/data_lptask.csv}\lptaskTable
\pgfplotstableread[col sep=comma,trim cells=true]{data/data_nwtask.csv}\nwtaskTable
\pgfplotstableread[col sep=comma,trim cells=true]{data/data_idletask.csv}\idletaskTable

\begin{figure}[t]
\centering
\begin{tikzpicture}[
  /pgfplots/every axis/.style={ %
    ybar stacked,
    ymin=0, ymax=119,
    ymajorgrids=true,
    major grid style={dotted,black},
    xmin=-0.6, xmax=6.6,
    xtick=data,
    xticklabels={
        0 pkt/s, 100 pkt/s, 200 pkt/s, 400 pkt/s,
        800 pkt/s, 1600 pkt/s, 3200 pkt/s
    },
    x tick label style={rotate=45,anchor=east,yshift=-2mm},
    bar width=10pt,
    legend style={
        cells={anchor=west},
        legend pos=north west,
        column sep=1pt,
        /tikz/every even column/.append style={column sep=4pt}
    },
    legend columns=-1,
    ylabel={Total processing time (\%)},
  },
]
\pgfplotsset{width=\textwidth, height=0.45\textwidth}
\pgfplotsset{x tick style={white}}

\definecolor{color_hp}{rgb}{0.722, 0.329, 0.314}
\definecolor{color_lp}{rgb}{0.424, 0.557, 0.749}
\definecolor{color_nw}{rgb}{0, 1, 0.502}
\definecolor{color_idle}{rgb}{0.77, 0.77, 0.77}

\begin{axis}[bar shift=-13pt,hide axis,]
    \addplot[draw=LineRed,fill=FillRed] table[x=X,y=s0] {\hptaskTable};
    \addplot[draw=LineBlue,fill=FillBlue] table[x=X,y=s0] {\lptaskTable};
    \addplot[draw=LineGreen,fill=FillGreen] table[x=X,y=s0] {\nwtaskTable};
    \addplot[draw=color_idle,fill=color_idle] table[x=X,y=s0] {\idletaskTable};
\end{axis}

\begin{axis}[]
    \addplot+[draw=LineRed,fill=FillRed] table[x=X,y=s1] {\hptaskTable};
    \addplot+[draw=LineBlue,fill=FillBlue] table[x=X,y=s1] {\lptaskTable};
    \addplot+[draw=LineGreen,fill=FillGreen] table[x=X,y=s1] {\nwtaskTable};
    \addplot+[draw=color_idle,fill=color_idle] table[x=X,y=s1] {\idletaskTable};
\end{axis}

\begin{axis}[bar shift=13pt, hide axis]
    \addplot+[draw=LineRed,fill=FillRed] table[x=X,y=s2] {\hptaskTable};
    \addlegendentry{HP task}
    \addplot+[draw=LineBlue,fill=FillBlue] table[x=X,y=s2] {\lptaskTable};
    \addlegendentry{LP task}
    \addplot+[draw=LineGreen,fill=FillGreen] table[x=X,y=s2] {\nwtaskTable};
    \addlegendentry{network task}    
    \addplot+[draw=color_idle,fill=color_idle] table[x=X,y=s2] {\idletaskTable};
    \addlegendentry{idle task}
\end{axis}
\end{tikzpicture}
\caption{\textbf{Processing time distribution:} For each of the seven clusters of bars, the order of the 3 setups displayed is the following: left is the baseline implementation, center is the baseline implementation with unrestricted queue size and right is the custom implementation.}
\label{fig:diag1}
\end{figure}
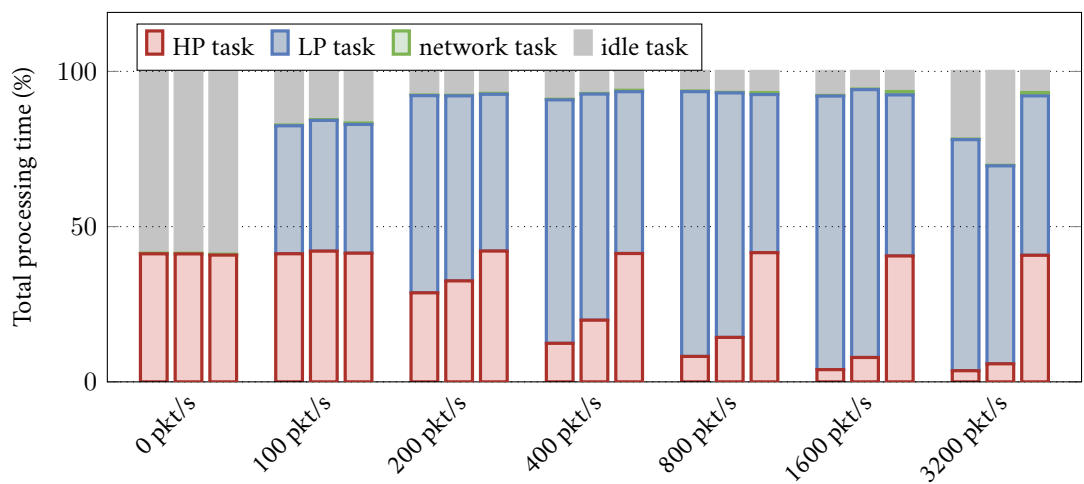

\begin{figure}
\centering
\begin{tikzpicture}
\begin{axis}[
    xlabel=Number of LP packets sent,
    ylabel=Number of HP packets received,
    xmin=-100, xmax=3500,
    ymin=0, ymax=7990,
    legend columns=-1,
    legend style={
        column sep=1pt,
        /tikz/every even column/.append style={column sep=4pt}
    },
]
\pgfplotsset{width=0.6\textwidth, height=0.45\textwidth}

\addplot[smooth,mark=*,LinePetrol] plot coordinates {
    (0, 5772)
    (100, 5775)
    (200, 4019)
    (400, 1742)
    (800, 1151)
    (1600, 555)
    (3200, 505)
};
\addlegendentry{baseline}

\addplot[smooth,mark=*,LineGreen] plot coordinates {
    (0, 5769)
    (100, 5897)
    (200, 4558)
    (400, 2785)
    (800, 2009)
    (1600, 1103)
    (3200, 817)
};
\addlegendentry{baseline-uq}

\addplot[smooth,mark=*,LineRed] plot coordinates {
    (0, 5714)
    (100, 5802)
    (200, 5903)
    (400, 5794)
    (800, 5835)
    (1600, 5692)
    (3200, 5723)
};
\addlegendentry{custom}

\end{axis}
\end{tikzpicture}
\caption{\textbf{Responsiveness:} Number of packets received by the high-priority task in relation to the number low-priority packets sent to the system.}
\label{fig:diag2}
\end{figure}
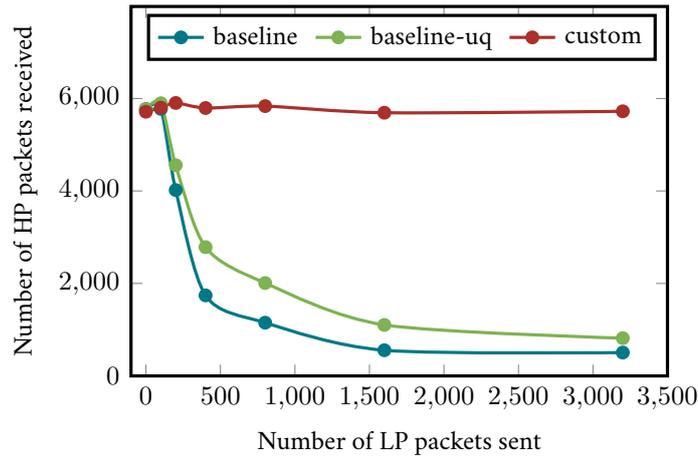

The results of the measurements are shown in Figures~\ref{fig:diag1} and~\ref{fig:diag2}. They clearly establish the custom implementation in the \ac{nic} to be the only system where the high-priority task is not affected by the increasingly higher rates of low-priority packets sent. In comparison, the high priority task in both baseline setups is allotted increasingly less running time by the scheduler. This is likely due to priority inversion blocking the high-priority task, while the low-priority task is able to keep running due to large amount of low-priority packets received. Furthermore, it may be ascertained that a limited queue size degrades the performance of the high-priority task even more. Thus, the empirical findings establish the effectiveness of the design proposed in this thesis and support the implementation choices.

\section{Conclusion}
\label{sec:2_5_conclusion}
This chapter presented a hardware/software co-design mitigating the negative effects of IP packet reception and ensuring the real-time performance of time-critical tasks running on an embedded device.
Devised as a unified hardware-software solution, the design combines aspects of the exclusively software-based approach presented in Chapter~\ref{cha:2_3_software} with those of and the exclusively hardware-based approach presented in Chapter~\ref{cha:2_4_hardware}. To this end, the two established proposed approaches were reviewed and their shortcomings ascertained. The combined design was developed building on individual ideas from the approaches and adapting them. 

The design features a network interface driver which employs early packet demultiplexing into prioritized packet flows. Packets are scanned and their metadata is compared with the configurations of all flows to realize sorting them into prioritized packet queues in memory. Interrupt generation for processing the queue buffers is timer-based. Two timers are used, defining the time frame for how long packet will stay queued before the interrupt service routine fetches the buffer and passes the packets along to the network driver.

The design has been implemented in a working prototype. The implementation required the development of a custom hardware design on an \ac{fpga}-\ac{soc} and adaptation of an existing IP Stack. The hardware design is based on a standard Ethernet configuration. The novel functionality in the network interface controller was implemented inside a custom \ac{ip-core} integrated into the existing base design. The implementation of the software involved the creation of a driver for interaction with the custom \ac{ip-core}. Furthermore the IP stack was modified, with the vendor-specific network interface portion substantially changed and the socket \ac{api} of LwIP adjusted.

A test scenario was chosen to contrast the custom implementation with a standard Ethernet configuration as baseline in a realistic test scenario. The results show that the combined design is able to protect time critical tasks from the real-time violating effects of high lower priority packet loads. Moreover, the reception of high priority packets is not hindered by low priority packet floods, showing that the previously existing priority inversion (cf. Chapter~\ref{sec:2_2_intro}) could be successfully mitigated.

\chapter{Offloading Real-Time Tasks in Distributed Systems}
\label{cha:3_3_offloading}
\minitoc
In this chapter, the application scope of this thesis is extended to include  distributed computing in local area networks. In the previous approaches, we aimed to introduce IP networking to embedded real-time systems in a safe(r) manner. In the following, we put these devices and their connectivity to use, by sharing their real-time tasks with more powerful resources in the vicinity. We present an architecture for scalable real-time task offloading as well as a distributed real-time scheduler for this purpose. 

The integration of real-time tasks in distributed systems introduces challenges that impact system performance, reliability, and predictability~\cite{yen1998performance, zhou2020dependable, akesson2020empirical}. Achieving precise synchronization across distributed nodes, accounting for network latency and communication overhead, ensuring fault tolerance and reliability, and optimizing resource allocation are among the key challenges associated with real-time task execution. Furthermore, scalability issues may arise as the number of nodes increases.
Ongoing research efforts in these areas, along with the exploration of novel algorithms and architectures, hold the potential to advance the integration of real-time tasks in distributed systems, making them more robust, scalable, and capable of supporting mission-critical applications across diverse domains.

To make a contribution to these efforts, we identify the tradeoffs and implications of offloading real-time tasks to local edge resources connected by wireless networks.
For this purpose, a system architecture is designed and a simulator implemented to test the scenario with various parameters related to the properties of the network of actors and tasks. The basic architecture includes mobile embedded devices acting as clients that offload real-time tasks, a cluster of worker nodes computing these tasks, and a scheduler distributing the offloaded tasks incorporating task deadlines and network latencies into its decision making. Further, the architecture is extended with a distributed scheduler design eliminating the single point of failure a single scheduler node constitutes. 
With this system, different scenarios will be run in order to test the possibilities and limits and finally to evaluate the circumstances under which the real-time conditions can still be guaranteed. 
The focus of this work lies on the incorporation of network uncertainties.

The first section of this chapter introduces the motivation and the specific challenges concerning the offloading of real-time tasks in IP networks. Section~\ref{sec:3_3_sysarch} explains the system architecture with which we aim to meet these challenges. Section~\ref{sec:3_3_scheduler} proposes a scheduler to perform latency-aware offloading. Section~\ref{sec:3_3_evaluation} evaluates different system configurations to quantify trade-offs between performance and reliability. In Section~\ref{sec:3_3_tolerance} we discuss fault-tolerance in such systems and introduce a distributed scheduler. Section~\ref{sec:3_3_conclusion} concludes the chapter.
Parts of this chapter have been peer-reviewed and published in~\cite{behnke2023offloading}.

\section{Offloading Real-Time Tasks}
\label{sec:3_3_introduction}
The \ac{iiot} aims to connect embedded devices in industrial environments to enable remote control, mobile computing, monitoring, and collaborative problem solving.
Examples can be found in the cyber-physical systems used in smart factories, logistics, and autonomous vehicles~\cite{bock2021performance,barzegaran_fogification_2020}, where devices often operate under real-time constraints.
While real-time computing in automation and embedded systems is well researched and deployed, its combination with packet-based wireless IP networking presents new challenges and opportunities. 

As already motivated in Chapter~\ref{cha:1_1_introduction}, embedded devices typically have resource and power constraints that limit the type and amount of local computation.
For example, resource-intensive tasks such as object recognition in video may require specialized hardware and more power than an embedded device can reasonably provide. Another example for this can be found in close-proximity model training, where sensor data is accumulated and processed on edge servers~\cite{becker2021local}.
A common way to deal with these limitations in practice is to offload individual tasks to more powerful computers in the local area~\cite{qiu2020edge,iiot_offloading_energy,hossain_edge_2020}.

However, for tasks with real-time requirements, this can be very challenging due to unpredictable latencies in wireless networks, especially when considering mobile devices. Some recent works on the issue of real-time task offloading have been identified in the survey in Chapter~\ref{cha:1_3_survey}.

Scheduling real-time tasks in distributed systems has received increasing attention~\cite{roy2020contention,yi2020task,behnke2020interrupting}, but related work generally assumes very reliable and predictable communication.
However, especially with wireless connections, the exact prediction of the expected latency is difficult.
Various efforts have been made to develop methods to guarantee end-to-end delay in wireless networks, but some form of uncertainty always remains~\cite{e2e_delay_1,e2e_delay_2,e2e_delay_3,behnke2023towards}.
In addition, efficiently deploying workloads in a distributed system is difficult, especially when these workloads occur sporadically~\cite{scheduling_difficult}.

\section{System Architecture}
\label{sec:3_3_sysarch}

\begin{figure}
    \centering
    \includegraphics[width=.7\columnwidth]{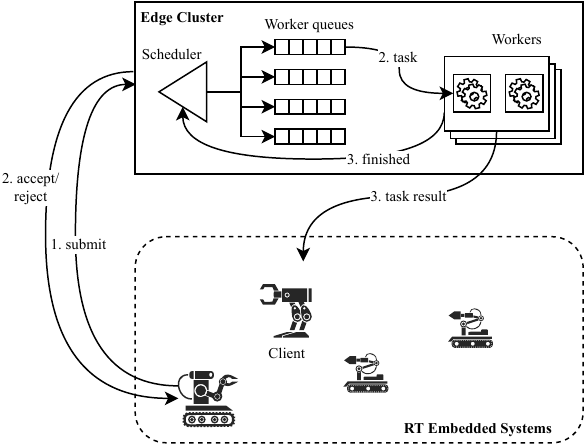}
    \caption{\textbf{System architecture:} Overview of actors and communication channels.}
    \label{fig:arch_overview}
\end{figure}

The goal of this chapter is a system design for real-time task offloading for mobile embedded systems and the development and evaluation of a scheduler for this purpose. Figure~\ref{fig:arch_overview} shows an overview of the design. Potentially mobile machines (clients) submit tasks with deadlines to the scheduler, which checks time feasibility and -- if accepted -- schedules the tasks to worker queues. The workers run on the same compute cluster as the scheduler and can be preempted by it. Each worker computes one task at a time and returns the result directly to the source client while signaling task completion to the scheduler.

The considered environments contain a scalable number of (semi-) autonomous devices that reside and potentially move in a local area spanned by an IP network. Based on these settings, the following task and scheduler properties can be derived.

\begin{compactitem}
	\item Tasks arrive \textit{sporadic}: The devices that submit the tasks are fully or semi autonomous and therefore the arrival of tasks happens during runtime, irregular and not predictable for the scheduler.
	\item The scheduling must therefore happen \textit{online} during system run-time, with \textit{dynamic priorities}.
	\item The clients might not be fully independent in the overall setting, however, their interaction is assumed to happen outside of the task context. Therefore the tasks are independent and \textit{no precedence constraints} need to be considered.
	\item Tasks have \textit{hard deadlines}, that are known at the time the task is submitted. 
	\item Tasks are \textit{fully preemptable}
	\item Finally, the computation model is that of a \textit{homogeneous multiprocessor}. However, with the major difference of distributed processing entities (workers) and task dispatch via local network.
\end{compactitem}

The following subsections describe the design of the involved entities. Design decisions and the operation of the scheduler are explained in detail in Section~\ref{sec:3_3_scheduler}.

\subsection{Tasks}
\label{sec:3_3_tasks}

The \textit{task} is the central data structure that holds all the necessary information about the workload that a client wants to offload.
It is modeled after real-time operating system processes and is shared and sent between the different entities in the system.
In an IIoT environment such as a smart factory, we assume that all entities and workloads are known before the system is set up.
Therefore, clients can assume that the programs or binaries necessary to compute all the workloads that a client wants to offload are present on the workers.
This makes a task submission much like a \textit{Remote Procedure Call (RPC)}.
A task is generated initially by the client, who sends it to the scheduler, which in turn eventually sends it to a worker.
It is serialized and deserialized for transportation over the network, but all entities work on the same shared Task data structure.
\newpage

\begin{mydef}
\label{def:flow}
Let $\mathcal{T}$ be a task offloaded by a client, scheduled, and computed by a worker. For the purpose of this work it is characterized by the tuple $$(C, T_d, t_r, t_\text{cs}, t_w, t_e, \mathcal{P})$$
\begin{compactitem}
\item $C$ the source client. 
\item $T_d$ the absolute deadline.
\item $t_r$ the initial relative deadline.
\item $t_\text{cs}$ the connection setup time.
\item $t_w$ the worst-case execution time.
\item $t_e$ the elapsed execution time.
\item $\mathcal{P}$ the set of parameters.  
\end{compactitem}
\end{mydef}

The client $C$ specifies its own id and its IP address and port on which it will listen for the connection that the worker will establish.
$t_r$ is the relative deadline at the time the client creates the task and sends it to the scheduler.
This value, along with the connection setup time $t_\text{cs}$ measured by the client, is used by the scheduler to calculate network latency.
The duration $t_e$ is set and used by the scheduler to keep track of the execution time when a task is computed by a worker and possibly preempted.
The set of parameters $\mathcal{P}$ contains the command line arguments and payload data that the worker passes to the actual executable.

\paragraph*{Task Dispatch}
There are two sensible design choices for when to dispatch consecutively scheduled tasks to the assigned workers:

\begin{compactenum}
    \item Dispatch a task to the worker as soon as it is assigned to it, and let the worker maintain an internal task queue.
    \item Dispatch the next task only after the previous one is finished.
\end{compactenum}

The first option has the advantage that consecutively scheduled tasks are present on the worker and can be computed with minimal interruption in between.
On the other hand, the scheduler sends lower priority tasks to the worker to add to its queue while it is working on the highest priority task.
The second option interrupts the worker only when a task with a higher priority than the one the worker is currently processing arrives.
For the remainder of this work, the second approach is used to ensure minimal disruption and maximum predictability of the execution time of the running task. 
We assume that the connection between the workers and the scheduler is wired and very fast (compared to the wireless connection of the clients), so that task dispatching causes little idle time for the workers.
In a scenario where tasks come with large input payloads, such as multimedia, the first approach may be more appropriate.

\paragraph*{Task Payloads}
Tied to the task dispatch approaches is the question of what path the task payloads take.
Task payloads refer to the input parameters that the client sends with the task and the task result that is sent back to the client.
There are three possible paths for the input parameters:

\begin{compactenum}
    \item The client passes only the task metadata to the scheduler, which then communicates the assigned worker back to the client, so it can send the input parameters directly to the worker.
    \item The client sends the task together with its input parameters to the scheduler, which eventually passes them on to the assigned worker.
    \item The client uploads the input parameters to a shared storage outside the scheduler that all workers have access to.
\end{compactenum}

Based on the task dispatch handling decision, the second option is chosen for task input parameter handling.
In a scenario where the task input parameters consist of large amounts of data, such as multimedia, the scheduler might become a bottleneck if it has to handle the distribution of this data.
In this case, one of the other approaches might be more appropriate.
The task result payload is only relevant to the client. Therefore, the worker sends it directly to the client.

\subsection{Clients}
\textit{Clients} are the entities that generate \textit{tasks} they want to offload.
They are designed as a software library that provides an interface for submitting tasks.
Client machines are mobile embedded devices that perform some kind of physical action and are connected to the rest of the network via a wireless link. They implement fallback behavior in case a task is rejected by the scheduler.

\paragraph*{Task configuration}

The clients are responsible to set the following properties of offloaded tasks as defined in Section~\ref{sec:3_3_tasks}. 

\begin{compactitem}
    \item Execution time
    \item Deadline
    \item Result payload size
\end{compactitem}

\paragraph*{Task Rejections} \label{sec:3_3_design_client_task_rejections}
Submitted tasks are either accepted or rejected by the scheduler.
In case a task is accepted, the client will assume that the task result is available to the client before the deadline.
If it arrives after the deadline, this means a failure of the system and is to be generally avoided.
So, if the scheduler is not able to schedule the task on one of the workers without nearing the deadline by a certain uncertainty window, it rather rejects the task completely.
The client has some fallback behavior for this case, which causes less disruption than a missed deadline.

\subsection{Workers}
Workers compute the offloaded tasks distributed by the scheduler.
They can be thought of as distributed processing cores: A worker can process exactly one task at a time and is fully utilized with that execution.
Worker machines are edge resources that are physically close to the scheduler host and on the same local area network.
However, it listens for incoming tasks even while processing one, since tasks need to be preemptable in a setting with deadlines.
They ensure predictable task execution times.%

Workers connect to the scheduler at startup and listen for tasks.
Upon receiving a task, the worker begins executing the task with the parameters provided by the client, but continues to listen for incoming tasks to allow for preemption by the scheduler.
When the worker receives a task while still processing an earlier task, the earlier task gets preempted until the new task is completed.

\paragraph*{Worker Operation} \label{sec:3_3_impl_worker_functionality}

After the worker successfully connects to the scheduler, it enters the main loop, where it waits for tasks to be sent by the scheduler through the open connection.
When a task is received, a new thread is started to handle that task while the main loop continues to listen for messages from the scheduler.

\paragraph*{Worker Preparation}

The task handler running in the new thread then starts to prepare the task computation. 
The TCP 3-way handshake takes a non-negligible amount of time when performed on a connection with a latency in the double-digit millisecond range~\cite{pei2016wifi}.
For this reason, the worker starts establishing the connection to the client immediately after receiving a task, in parallel with the actual task computation.
The worker sends the computation results of a task directly to the client.
This ensures that the task result can be sent with the shortest possible delay once the task has finished computing.

\paragraph*{Worker Computation}
The worker computes the tasks by starting the task executable as a separate operating system process.%
Once the process is started and a process id (PID) is returned by the operating system, the running task and its process ID are pushed into the internal task queue.
The thread then sleeps until the process has exited.
If there is already a task running, the worker lets the operating system suspend the process execution and marks the process as preempted in the task queue.
Only then the new process to compute the new task is started and the thread goes to sleep until the new process has exited.

Once the process has exited, the worker reads the computation result and packages it for transport to the client. 
The worker also records the actual execution time (excluding the time spent being preempted).
The worker notifies the scheduler of having finished the task and sends the task result to the client.
However, this needs an established connection to the client, which was started in parallel with the task computation.
If the connection setup takes longer than the task computation, the worker must wait until the open connection is available.
Finally, if some task was preempted, the halted process is resumed.

\section{The Scheduler}
\label{sec:3_3_scheduler}
This section explains the design choices and operation of the scheduler, which dispatches tasks submitted by clients to workers.
After parsing a task message into the corresponding task structure, the \textit{time to deadline} is calculated from the absolute deadline $T_d$ of the task and the current system time $T_s$.
For each incoming task, the scheduler then derives the laxity~$= T_d - T_s - (t_w - t_e)$ and density~$= (t_w - t_e) / (T_d - T_s)$ to make a scheduling decision.

\subsection{Scheduling Algorithm: Partitioned EDF}
Based on the task model and the assumed environment, only dynamic priority, preempting scheduling algorithms such as \textit{Earliest Deadline First (EDF)} and \textit{Least Laxity First (LLF)} are suitable. LLF has the major drawback of thrashing~\cite{llt}, which is exacerbated in this scenario. Since context switches are expensive, the high frequency of context switches caused by the thrashing effect is already problematic in a general-purpose operating system.
In our scenario, where each preemption not only causes a context switch on the worker, but also requires communication, the effect has an even greater impact on overall system performance.
For this reason, EDF was considered a more reasonable algorithm for our scheduler.
Yet, for the sake of completeness, there also exist enhanced versions of \textit{LLT} which aim to avoid thrashing.

In a multiprocessor environment, EDF works either with a global task queue (global EDF) or with partitioned queues for each worker (partitioned EDF).
In addition to the non-optimality of global EDF due to Dhall's effect~\cite{dhall}, a global queue also makes task acceptance testing more difficult:
When a task is added to the global task queue and given its priority, it is not immediately clear which worker will eventually pick up the task, at what time, and whether this is early enough to meet the deadline.
It would be necessary to simulate separate run queues for each worker to find the next worker to become free when the new task eventually becomes the highest priority task in the global queue.
Since the scheduler must decide whether it can guarantee the timely execution of tasks as they arrive, partitioned scheduling is more appropriate.
Partitioned EDF requires an algorithm or heuristic to assign a task to a worker when multiple workers can meet the task deadline.
First-fit, worst-fit, and best-fit are the heuristics considered. Which one to use will depend on the circumstances and will be further explored in the evaluation.

\subsection{Deadline Adjustment \& Acceptance Checks}
\label{sec:3_3_adjustment}
When a new task arrives at time $t$, the scheduler adjusts the deadline set by the client to account for network latency.
In our setting, the communication delay between a client and the scheduler is roughly the same as between a worker and the client.
Therefore, the scheduler uses the delay that occurs in the client-scheduler connection to calculate the expected delay $d_\text{exp}$ that will occur when the worker sends the task result to the client:
The client provides the \textit{connection setup time} $t_\text{cs}$ and the initial time to deadline $t_r$ at the time the client generated the task in the task message.

\begin{equation*}
d_\text{exp}=t_r-T_d-t 
\end{equation*}

However, in a high latency environment with short task execution times, it may take longer to establish the connection than to complete the task computation.
In this case, the difference between $t_\text{cs}$ and $t_w$ must also be considered for the delay adjustment $d_\text{adj}$:
\begin{equation*}
d_\text{adj} = \begin{cases}
d_\text{exp}+t_\text{cs}-t_w & \text{, }t_\text{cs} > t_w\\
d_\text{exp} &\text{, else}
\end{cases}
\end{equation*}
The \textit{adjusted delay} is then multiplied by the scheduler's uncertainty factor $\mathcal{U}$ to account for latency jitter.
$\mathcal{U}$ is a configuration option of the scheduler and the effect of different values is one of the subjects of the evaluation.
Finally, the resulting value is subtracted from the deadline. 

\begin{equation*}
T_{d\text{-new}}=T_{d\text{-old}} - \mathcal{U} \cdot d_\text{adj}
\end{equation*}

This adjusted deadline can now be used for scheduling and assessed to see if it leaves enough time for the task result to be sent to the client and not miss the original deadline set by the client.
A first acceptance check is then performed to ensure that the time to the adjusted deadline is greater than the task execution time.
The scheduler maintains a queue for each worker according to the partitioned EDF scheme. To make a scheduling decision, a copy of each queue is scheduled using EDF and evaluated for potential deadline misses. Queues with predicted misses are eliminated. In addition, the overall density of the queue is calculated from the individual task \textit{densities}.
If the scheduler is configured to use the \textit{first-fit} strategy to select the worker, it will select the first worker without a missed deadline.
For the other strategies, each worker is tested.
In the case of the \textit{best-fit} or \textit{worst-fit} strategies, the scheduler selects the worker with the highest or lowest total density, respectively.
If the new task has the highest priority, the currently running task is preempted. 
The scheduler records the elapsed execution time $t_e$ of the task, and the worker saves its context.
If no worker is found for which its queue can be scheduled without predicted missed deadlines, the task is rejected.

\section{System Simulation}
\label{sec:3_3_evaluation}

We simulate three scenarios to investigate the impact of our architecture on offloaded real-time tasks in wireless IIoT environments. 
We explore the limits at which tasks or network environments may or may not meet real-time requirements.

\subsection{Experimental Setup}
We implemented the proposed architecture and partitioned EDF scheduler in Rust and made the code publicly available\footnote{\url{https://github.com/dos-group/Real-Time-Offloading-Simulator-IIoT}}.
We implemented a reference scheduler for comparison. It runs EDF on a global task queue, rejecting tasks only if the time to deadline is less than the WCET and without deadline adjustment.

\paragraph*{Network}
We use \textit{Mininet}\footnote{\url{https://mininet.org}} to emulate a network of clients, workers, and the scheduler.
The virtual network links between hosts can be configured with arbitrary latency, jitter, and bandwidth limits.
The basic network layout is the same for all test setups:
Client hosts are connected wirelessly to an access point, which is connected to a switch via Ethernet.
Edge cluster hosts (running workers and the scheduler) are also connected to this switch via Ethernet.

\paragraph*{Test Scenarios}

All tasks in a test run have the same predefined execution time of 30 seconds and have been run 5 times each.
The relative deadlines, on the other hand, are only configured in terms of mean and variance, and are dynamically sampled from a normal distribution by the clients at runtime.
This results in more or less preemptions, depending on the configuration, because the deadlines of the tasks differ accordingly for the same execution time.

The task submission frequency is sampled from a Poisson distribution for which $\lambda = 1$ is configured for all clients.
However, clients only submit one task at a time, so the minimum time between task submissions is limited by the task execution time.
The worst case task execution times are set to 100ms for all experiments. 

\begin{table}[h]
    \centering
    \caption{\label{tab:parameters} Test Scenario Parameters}
    \begin{tabular}{l || c|c|c|}
    \textbf{Scenario} &\textbf{1}&\textbf{2}&\textbf{3}\\
    \hline
    number of clients & 10 -- 50 & 30 & 30 \\
    submission freq. mean [1/s] & 1 &1 &1\\
    latency mean [ms] & 30&30&30\\
    latency variance [ms] &10&10&10 -- 50 \\
    laxity mean [ms] & 100 &180 -- 60&100\\
    \end{tabular}
\end{table}

As described in Section~\ref{sec:3_3_adjustment}, the scheduler is configured with an uncertainty factor $\mathcal{U}$ and the task assignment heuristic.
The scheduler approximates the worker-to-client latency to be considered when creating the schedule based on the client-to-scheduler latency measured during task submission.
The measured client to worker latency is multiplied by $\mathcal{U}$ to account for variations in latency.
The following test scenarios explore how much variance needs to be accounted for with which uncertainty factor, and the impact this has on overall throughput.
Each scenario is tested with $\mathcal{U}$ ranging from 0.25 to 5.0. 
Table~\ref{tab:parameters} lists the chosen parameters for the three scenarios.

\subsection{Scenario 1: Number of Clients}

The first scenario examines the behavior of the system under varying loads.
Ten to fifty clients each create an average of one task per second.

The different task assignment heuristics were compared to see if the tasks were distributed appropriately.
All reported results use the worst-fit allocation strategy, as first-fit and best-fit simply concentrate the workload on a single worker.

\paragraph*{Results}

\begin{figure}[h]
    \centering
    \includegraphics[width=1\linewidth]{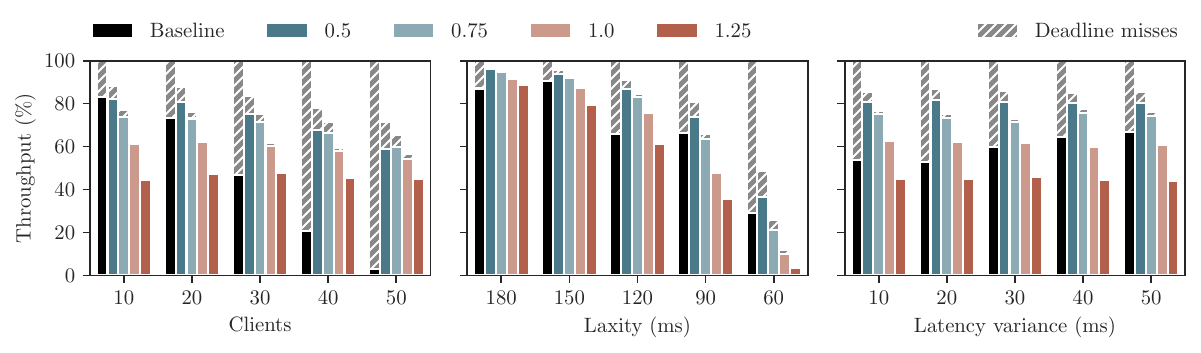}
    
    \vspace{-6mm}\hspace{.05\linewidth}\subfloat[\label{fig:scalability-a} Scenario 1]{\hspace{.28\linewidth}}
    \hspace{.03\linewidth}
    \subfloat[\label{fig:scalability-b} Scenario 2]{\hspace{.28\linewidth}}
    \hspace{.03\linewidth}
    \subfloat[\label{fig:scalability-c} Scenario 3]{\hspace{.28\linewidth}}
    \vspace{-2mm}
    \caption{For each scenario, we report the throughput (successfully finished tasks) and the fraction of tasks that have been accepted but missed their deadline. We compare the reference baseline vs. our latency-aware approach at $\mathcal{U}=\{0.5, 0.75, 1.0, 1.25\}$.}
    \label{fig:throughput}
\end{figure}

All experimental results can be seen in Figure~\ref{fig:throughput}. We present only the most relevant results for uncertainty factors between 0.5 and 1.25. For higher values, the acceptance rate drops without any further gain in reliability. We present the ratio of successfully offloaded tasks, where a submission is considered \emph{successful} if the scheduler accepts it and the computation result is returned before its deadline has passed. 

The results show that as the load increases, the reference miss rate increases, reducing the actual throughput of successful tasks. Depending on the uncertainty factor used, our approach minimizes the number of missed deadlines for any load. However, for higher loads and higher uncertainty factors, the number of rejected tasks increases, leading to an expected decrease in relative throughput. There are few to no missed deadlines in any of the experiments using our design and a success rate of up to 60\% for the highest tested submission load. Rejected tasks are returned to clients as early as possible for fallback processing. 

\subsection{Scenario 2: Task Laxity}
Next, we test the effect of task laxity on the acceptance rate and the miss rate. The lower the laxity, the less communication and scheduling time is available. Experiments were conducted with laxities ranging from 180ms down to 60ms. 

\paragraph*{Results}
It is evident that task laxity plays an important role in real-time processing. With sufficient laxity, the proposed design is able to meet all deadlines while rejecting only a small fraction of tasks, resulting in a relative throughput of 95\%. As the laxity becomes smaller, fewer tasks can be safely accepted. The proposed design outperforms the reference implementation for all laxities and for uncertainty factors below 1.0, meaning that some risk of missed deadlines must be accepted. The best results could be achieved with $\mathcal{U} = 0.5$, with a successful throughput difference of 5 - 20 percentage points compared to the reference scheduler. By choosing $\mathcal{U} > 1.0$, missed deadlines can be avoided for all tested laxities. However, in the most extreme case of 60ms submission laxity, only 5\% of submissions are accepted to accommodate this predictability. With $\mathcal{U} = 1.0$ and submission laxities above 90ms, more tasks are successful than with the reference scheduler.

\subsection{Scenario 3: Latency variance}
The final scenario tests the behavior of the system under increasing network latencies. We tested latency variances ranging from 10ms to 50ms.

\paragraph*{Results}
The results for this scenario show the least impact of increasing the parameter. Again, we see that higher uncertainty factors lead to higher deadline hit rates and lower task acceptance rates. The relative throughput of our design outperforms the reference implementation with somewhat risky uncertainty factor values below 1.0.

\subsection{Summary}
\label{sec:3_3_eval_summary}

For very tight deadlines, where the laxity is in the same range as the latency, the real-time scheduler relies heavily on correct latency estimates to generate efficient schedules.
For low-variance latencies, it has been shown that high throughput can still be achieved while preventing missed deadlines and hence gaining overall system predictability.

For pessimistic latency estimates, stronger guarantees of meeting deadlines can be provided.
However, this comes at the expense of throughput, since a correspondingly large number of tasks must be discarded.
Still, high throughput can be achieved if isolated misses are acceptable. 
The choice of an acceptable uncertainty factor must be made taking into account client fallback mechanisms. 

The highest advantages of the proposed design can be seen under high task submission loads, where the reference design is only able to successfully compute 5\% of submitted tasks while missing the deadlines of the remaining 95\% of tasks. Here, our design has still a relative throughput of 45\% with no deadline misses.

\section{Removing the Single Point of Failure}
\label{sec:3_3_tolerance}
In the realm of computing systems, one critical aspect that deserves close attention is the potential occurrence of a \ac{spof}. An \ac{spof} represents a part of a system, which, if it fails, would cause the whole system to stop functioning. These failures can occur in any part of the system and can be due to various reasons such as hardware failure, software bugs, network congestion, or even human error. Addressing this vulnerability is essential, particularly in systems where continuous operation is integral to their purpose, such as real-time systems. In real-time systems, especially those involving safety-critical operations, the implications of an \ac{spof} can be severe. Real-time systems are often characterized by stringent timing constraints and, therefore, any component failure can lead to missed deadlines, resulting in potential system failure. These implications can be even more pronounced in networks used for distributed real-time systems. 

In our case, the centralized access and scheduling node presents an \ac{spof}. When it becomes unavailable, task offloading requests cannot be forwarded and no answers can be returned. In this section, we present a fault-tolerant  architecture for real-time task offloading. A self-managing network of edge nodes schedules and processes tasks in a distributed manner. Each node hereby acts as a scheduler and worker and is connected to a wireless access point. 

\subsection{Risk Distribution}
To provide a fault-tolerant system design, some basic architecture paradigms have to be adjusted. In the following, we introduce the advantages of self-organized wireless networks and distributed scheduling approaches.

\subsubsection*{Self-Organized Distributed Environments}
The goal of the distributed approach is to create an environment where each participating node is able to autonomously provide computation capabilities while cooperating with any other nodes in the same network. This resmbles the paradigm of \acp{son}.

\acp{son} represent a paradigm shift in communication networks, moving away from traditional centralized network architecture towards more dynamic and decentralized configurations. In centralized networks, a single central node oversees the communication between all other nodes. This design can create potential issues such as single points of failure and bottlenecks given the central node's crucial role.
\acp{son} are composed of devices, known as nodes, which can both transmit and receive data. A fundamental aspect of \acp{son} is the absence of fixed infrastructure; each node operates independently and can connect freely with other nodes. This allows the network to adapt and reorganize itself as nodes join or leave, ensuring constant connectivity and operational continuity~\cite{peltomaki2011algorithms}.

In \acp{son}, the functions of the network are distributed across various nodes. Each node can operate as a host for local data and services and act as a router, forwarding data from other nodes. As a result, if a node fails or leaves the network, the remaining nodes can reconfigure themselves and route data through different paths, thereby ensuring the network continues to function seamlessly. This enhances network reliability. Distributing the load over multiple nodes also helps to avoid performance bottlenecks.

However, despite the numerous advantages, \acp{son} also bring about their unique challenges. The dynamic and decentralised nature of \acp{son} results in frequent changes in network topology. This necessitates an efficient routing protocol that can quickly adapt to changes and ensure data is efficiently transmitted between nodes. Additionally, because each node has its own resources and capabilities, coordinating and synchronizing activities amongst nodes can be complex.
Another challenge lies in the fairness of resource allocation. Without a central authority to oversee and enforce resource distribution, ensuring that all nodes contribute and consume resources fairly can be a complex task. Consequently, a suitable consensus and coordination mechanism is required to manage resource allocation.

\subsubsection*{Distributed Scheduling}
Unlike centralized scheduling algorithms, where a single node or scheduler takes on the responsibility of managing resources and tasks, \acp{dsa} decentralize this responsibility. Each node in the distributed system has the ability to allocate its resources and manage its tasks with local scheduling. Decisions are made based on localized information, although nodes may communicate to share load or information when necessary. This decentralized approach offers several advantages: 

\begin{compactenum}
    \item \acp{dsa} improve system scalability by sharing the responsibility of managing tasks and resources as the network expands and more nodes are added.
    \item \acp{dsa} improve system reliability and resiliency. Even if a node fails or leaves the network, the other nodes continue with their local scheduling, thereby keeping the system operational. 
    \item \acp{dsa} improve communication delays. Each node makes its scheduling decisions, eliminating the need for constant communication with a central scheduler.
\end{compactenum}

Despite the advantages of \acp{dsa}, they are not without challenges. One major challenge is balancing local and global optimization. While each node can make decisions based on its own information, these decisions may not result in the best overall solution or performance for the entire system. This is connected to maintaining consistency among nodes. In real-time systems, where tasks have strict timing constraints, ensuring all nodes have a synchronized and accurate perception of time is crucial. This becomes more complex in a distributed setting and requires appropriate synchronization protocols. Constant communication between nodes to maintain synchronized states consumes network resources and might result in increased latency variance. Therefore, finding efficient synchronization mechanisms that minimize this overhead becomes imperative in the design of \acp{dsa}, particularly in real-time systems operating within \acp{son}.

In the following sections, we present a self-managing distributed architecture for real-time task offloading based on the centralized architecture in Section~\ref{sec:3_3_sysarch}. 

\subsection{System Design: A Self-Managing Network}
In the context of the proposed system for managing offloaded real-time tasks in self-organized wireless networks, the architectural blueprint plays a critical role in tracing the underlying operational mechanisms and their interactions. The system's architecture consists of three significant modules: Client devices, \acp{ap}, and edge nodes containing a worker and a scheduler each.
The subsequent sections will probe deeper into the specifics of the architectural elements.

\subsubsection*{Device Communication and Node Interaction}

The communication and interaction between the various components of the system form the architecture backbone.

The edge nodes maintain a dedicated wired network (LAN) for internal communication while being part of a broader network that comprises devices and Access Points.

\begin{figure}
    \centering
    \includegraphics[width=0.9\textwidth]{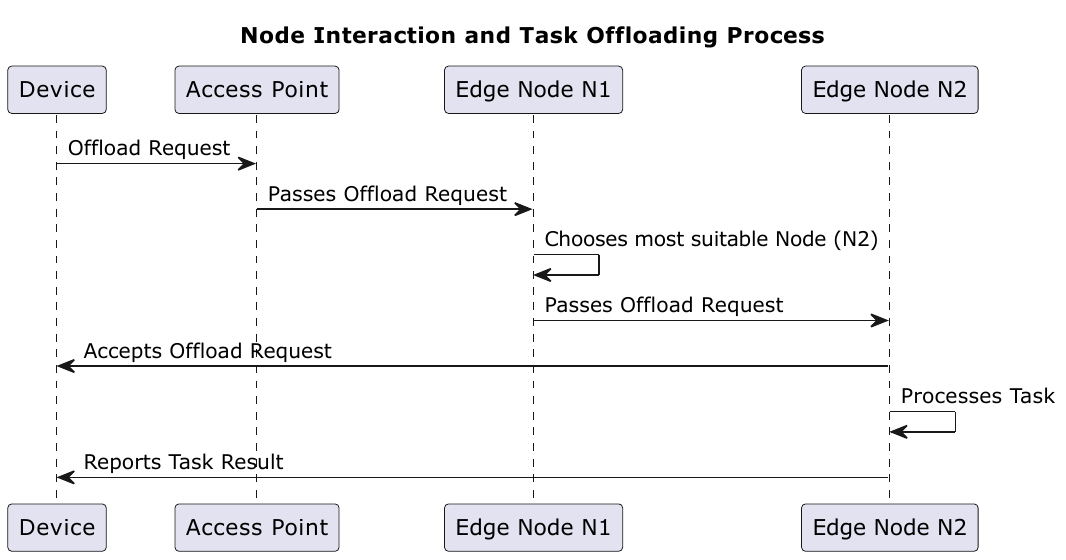}
    \caption{\textbf{Sequence diagram: } Task offloading process.}
    \label{fig:communication_sequence}
\end{figure}

The client devices in the system send offloading requests to the 'closest' node they are aware of. This 'closeness' is determined not by physical proximity, but is based on the latency of communication. This mechanism ensures that tasks are offloaded efficiently, taking into account the uncertainty associated with the nondeterministic communication channels between devices and nodes.

Upon receipt of an offload request from a device, the designated node, as shown in \figurename\ref{fig:communication_sequence}, takes on the responsibility of distributing the task to the most suitable node for execution. This selection process is driven by the scheduling algorithm in use within the system. This task distribution strategy is instrumental in ensuring that the real-time requirements of the tasks are adequately fulfilled.

When a node accepts an offload request and commences its execution, it sends a notification to the device that originated the request, indicating the initiation of the processing phase. This interaction forms a feedback loop, keeping the devices informed about the status of their offloaded tasks.

\subsection{Clients}
The clients stay mostly unchanged from the original architecture. Running on mobile devices such as plant floor robots, they change thei rlocation over time.
This mobility directly impacts the association between the devices and \acp{ap}, influencing the decisions of task offloading. 

\paragraph*{Closest Nodes}
\label{sub:device_closest_nodes} 
Devices maintain a record of their closest edge node via the latency of the last successful communication between the device and the respective edge node. Each device keeps a simple array of the three closest nodes based on their communication latency. When a new node enters the network and reports its status, or an existing node communicates with the device, the respective latency is calculated and the list of closest nodes is updated.

\paragraph*{Offloading Requests}
The device identifies the closest edge node to offload the task, based on the last known communication latency with each node. 

In case there has been no prior communication with the chosen node, a new TCP socket is opened to the respective edge node. If communication has occurred before, the existing open socket is reused, preserving network resources and time.

The device then sends the task offloading request to the selected edge node using this TCP connection. Upon receiving the request, the edge node identifies the most appropriate node (according to the worst-fit strategy) to execute the task and forwards the offload request to it over a dedicated TCP socket. 

This receiving node assesses whether it can accept the request. If it can, it sends an ACCEPT result back to the device using a dedicated TCP socket. If it is unable to accept the request, it forwards the offload request to the second most appropriate node, and this process continues. If the last node in the chain is unable to accept the request, it returns a REJECT message to the device using a dedicated TCP socket.

Moreover, in case a device receives no response (either ACCEPT or REJECT) within a defined threshold period, it picks the second-closest edge node and sends the same Offload Request to it accordingly. 

\subsection{Access Points} 
The \acp{ap} play a vital role in the distributed scheduler model, serving as the connection hubs between the devices and the edge node network.
In this system, APs establish wireless connections with devices. Unlike the devices, APs are stationary entities, ensuring constant coverage and efficient utilization of the network space. They are organized in a grid pattern that spans the physical area, facilitating wide network coverage.

Notably, APs also function as an abstraction layer for the edge node infrastructure, simplifying the offloading process for devices. By connecting to the APs, devices can offload tasks without needing direct interaction with the complex edge node network. It is the responsibility of the APs to handle the communication with the edge node network. 

\subsection{Edge Nodes} 
In the context of the distributed architecture, the edge nodes function as an interconnected network of computation resources. They form a self-organized distributed environment. Computing nodes can join and leave the network based on factors such as availability, resource allocation, and system requirements.
By leveraging socket-based communication, the node is able to receive information regarding the state of the entire network, including node densities and task overloading scenarios. Once the connection is successfully set up, the node completes its initialization, announces itself to the network and starts to accept task offloading requests.

\paragraph*{Joining the Network} 
Every node starts its lifecycle by sending a join request to the network. The request is then processed by all other nodes, who in turn broadcast their own densities and update their node lists to include the new member. 
After a new node joins the network successfully, it sends a message to all the other nodes to signal it is ready to accept offloading requests. This information is also sent to all active clients and they trigger their process for deciding whether the new node should be considered when creating offloading requests based on the latency of the communication.

\paragraph*{Node Density Transmission}
A node's density is the queue density of the worker as defined in Section~\ref{sec:3_3_adjustment}.
Node density broadcasts are used to manage and maintain the up-to-date state of the edge worker network in the distributed scheduler. Each worker node regularly transmits its current task density to the other nodes in the network. 
Node density broadcasts are used to manage and maintain the up-to-date state of the edge worker network in the distributed scheduler. Each worker node regularly transmits its current task density to the other nodes in the network. 
Additionally to the transmission when a new node joins the network, the density information is updated and transmitted to all other nodes when a node accepts a new task. Furthermore, when a node completes a task, it broadcasts an update, allowing the network to adjust to the node's reduced load. These transmissions are transmitted over the LAN using a TCP socket. Hence, they are no broadcasts in the technical sense but rather individual connections to all known nodes.

\paragraph*{Node Advertisement}
Worker nodes periodically (in the magnitude of seconds) advertise themselves to the network.
Once received by an \ac{ap}, the node advertisement is forwarded to the WiFi subnet where the client machines are connected. This propagation ensures that all devices are kept informed about the active edge nodes in the network.
Each device, upon receiving a node advertisement, measures the communication latency to the broadcasting node. They might consider offloading tasks to a node that has the lowest communication latency, thus reducing task completion time.
It is important to note that this measured latency is end-to-end, from the worker to the client, rather than just from the AP to the device. 

\paragraph*{Handling Offloading Requests}
When an offload request is received, the worker node determines whether it can accept the job based on its current task schedule. If the node cannot accept the job due to time constraints, it finds the worker with the least density and forwards the request to it. If no available node can accept the job, the original node will reject the job and inform the device. 
After the completion of a task, the executing worker node reports the result back to the device that originally offloaded the task. This communication completes the cycle of task offloading and ensures that the computation results are relayed back to the originating device.

\subsection{Evaluation}
The comprehension of the functioning and efficiency of the implemented distributed scheduler for offloaded real-time tasks in self-organized wireless networks is achieved via the conduction of five key simulation scenarios. Each scenario varies a specific parameter and is designed for enhanced understanding of the system's performance, behaviour, and robustness under different conditions.

\begin{table}[h]
    \centering
    \caption{\label{tab:dist_parameters} Test Scenario Parameters for Distributed Scheduler}
    \begin{tabular}{l || c|c|c|c|}
    \textbf{Scenario} &\textbf{1}&\textbf{2}&\textbf{3}&\textbf{4}\\
    \hline
    number of clients & 10 -- 50 & 30 & 30 & 30\\
    number of nodes & 4 & 4 & 4 -- 28 & 3-4\\
    laxity mean [ms] & 100 & 180 -- 60 & 200 & 200\\
    laxity var [ms] & 33 & 60 -- 20 & 67 & 67\\
    \end{tabular}
\end{table}

In each of the described scenarios, a single simulation parameter is altered while keeping others constant, to allow for the precise analysis and evaluation of the impact of different simulation parameters on the system's performance. Each simulation is programmed to run for a length of 30 seconds, incorporates an uncertainty factor and task submission frequency of 1, and a \ac{wcet} of 100ms, equal to the evaluation parameters of the centralized scheduler. The first two scenarios are designed with a resemblance to the scenarios evaluated in Section~\ref{sec:3_3_evaluation} for a comparative study between the functionality and performance of the centralized scheduler against the distributed scheduler. The scenario parameters are defined in Table~\ref{tab:dist_parameters}.%

The third scenario is designed to assess the communication overhead resulting from the coordination of the distributed network. It provides valuable insights into the impacts of increasing the \acp{en} from 4 to 28 on overall task latency and communication latency. This evaluation will shed light on the scalability of the system, and how well it can handle increased communication overhead.

The fourth scenario aims to evaluate the robustness of the network. It probes the ability of the system to adapt and manage the abrupt loss of \acp{en} due to software or hardware errors or connectivity issues. This scenario operates under conditions identical to those of Scenario 3, but with a critical alteration; one of the \acp{en} is programmatically terminated at the seventh second of the simulation, simulating the unexpected loss of an \ac{en}. The aftermath of this event and its impact on the network performance are thoroughly inspected.

Following this, the same scenario is rerun but this time with only three \acp{en} from the onset. The objective of this experiment is to compare the system's behaviour under both conditions. The premise here is that the results from both cases should be relatively similar, given that the \ac{en} was lost early in the simulation. However, due to the seven seconds of operation with four \acp{en} in the first case, the performance is expected to be slightly better. The overall comparison aims to provide insights into the workload distribution and overall system performance in the face of spontaneous alterations in node availability.

\subsubsection*{Results}
\begin{figure}
\centering
\includegraphics[width=0.5\textwidth]{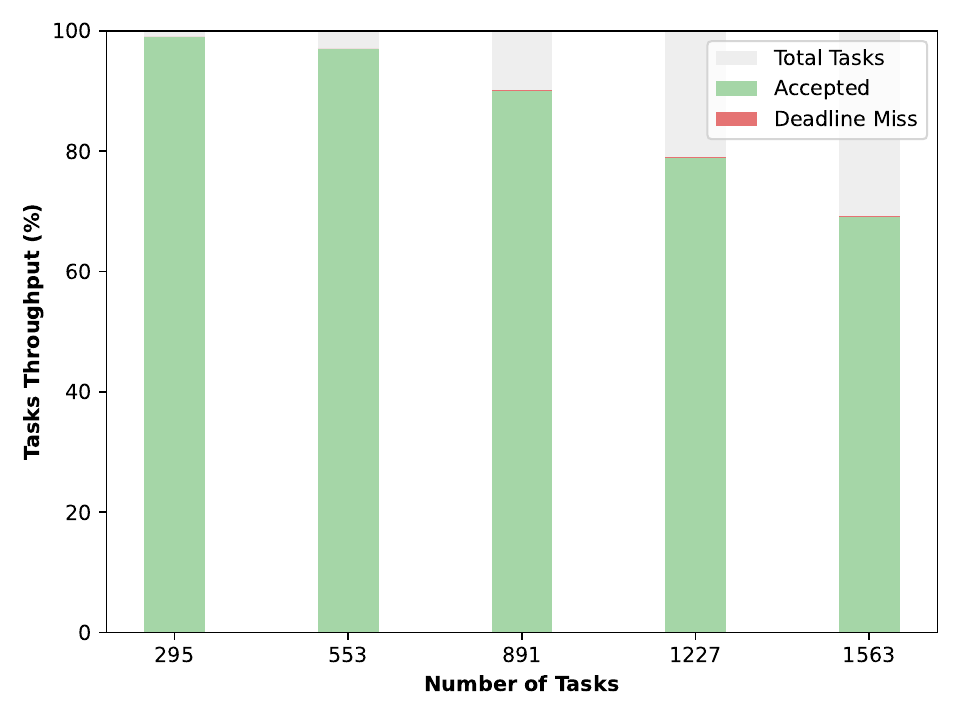}
\caption{\textbf{Scenario 1:} Task throughput chart.}
\label{fig:scenario1_accept}
\end{figure}

\paragraph*{Scenario 1}\label{subsub:scenario_1} Figure~\ref{fig:scenario1_accept} demonstrates the relationship between three variables: the total number of tasks generated, the proportion of accepted tasks, and the percentage of deadline misses from the accepted tasks.

It is observable that the rate of acceptance is significantly higher when fewer tasks are present. However, as the number of devices increases, a consequential decrease is perceived in the acceptance rate of tasks. Specifically, when the device count peaks at 40 - generating 100\% theoretical workload - the acceptance rate records a decline, dropping to 80\%. Moreover, an observation of 0.2\% missed deadlines can also be detected in this scenario.

The trend continues as the device count further increases to an extreme of 50. Though the acceptance rate diminishes further, there is no noticeable increase in the deadline misses. This observation suggests that the recorded misses could be attributed more to the consequence of \ac{en} communication overhead under higher loads (notably between 80\% and 90\% for 40 and 50 devices respectively) rather than any perceived inability of the distributed scheduler to effectively schedule the tasks.

\paragraph*{Scenario 2}\label{subsub:scenario_2}  

\begin{figure}
    \centering
    \includegraphics[width=0.5\textwidth]{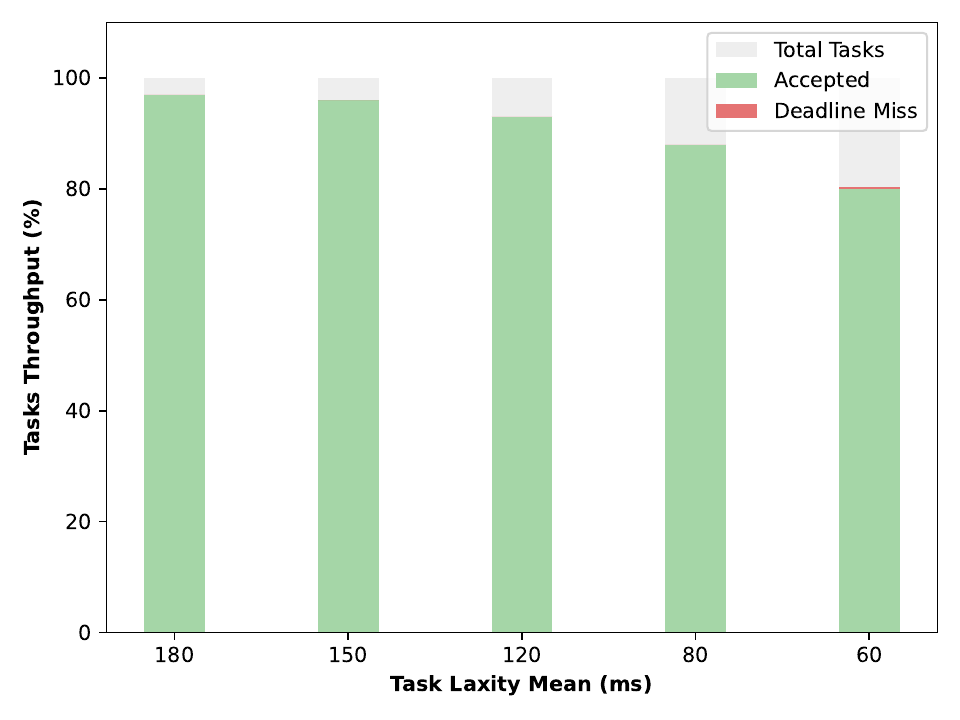}
    \caption{\textbf{Scenario 2:} Task throughput chart.}
    \label{fig:scenario2_accept}
\end{figure}

The initial expectation is that the increase in task laxity should correspond with an increase in the acceptance rate of tasks. The empirical observations visualized in \figurename\ref{fig:scenario2_accept} align with the initial presumption. For a task laxity expressed in terms of its mean and variance up to 120ms and 40ms respectively, we have documented an acceptance rate exceeding 90\%. 
However, the situation alters as we go lower than 120ms mean task laxity. In those instances, there is a marked reduction in the acceptance rate, which starts to diminish noticeably. In the most extreme case, a mean task laxity of 60 culminated in a non-negligible deadline miss rate amounting to 0.3\%.

Moreover, sequential observations disclosed that the \ac{en} utilisation rate mirrors the fluctuations in the task acceptance rate. Specifically, utilisation proportionately decreases as the acceptance rate drops, underscoring the direct correlation between these two key performance indicators. In conclusion, it's evident how task laxity influences the performance of the distributed scheduler under scrutiny.

\paragraph*{Scenario 3} In this scenario, our attention is directed towards the network overhead resultant from an expanded number of \acp{en}. This scenario maintains a consistent load across different executions by using 30 client devices. To ensure tasks are theoretically schedulable, the task laxity is set at a mean of 200ms and a variance of 67ms. 

\begin{figure}
    \centering
    \includegraphics[width=0.8\textwidth]{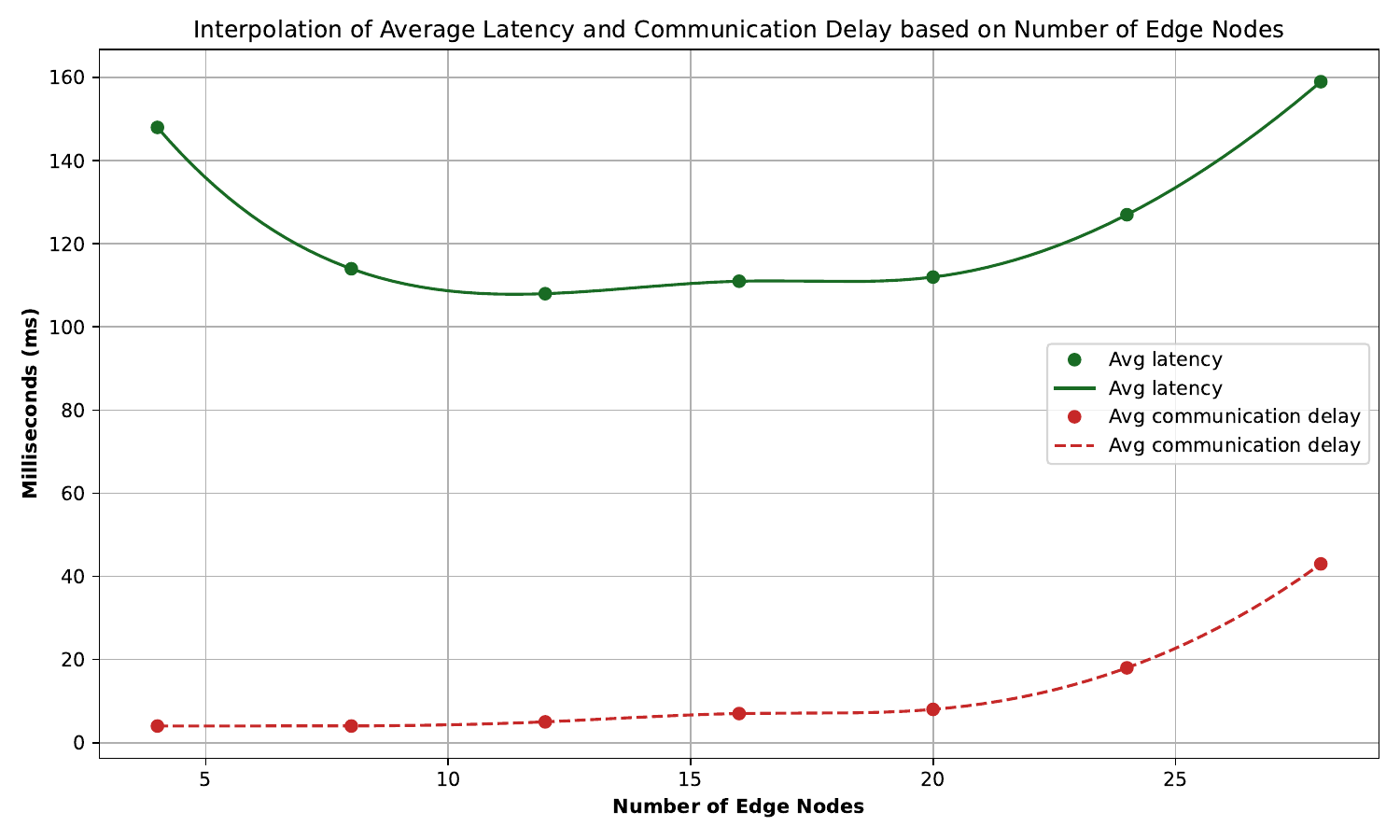}
    \caption{\textbf{Scenario 3:} Average task latency (end-to-end) and average communication delay.}
    \label{fig:scenario4_latency}
\end{figure}

Figure~\ref{fig:scenario4_latency} presents the interpolation of average end-to-end task latency and the average communication delay. This delay encompasses both devices - \ac{en} communication and the internal communication delays amongst the \acp{en}. The end-to-end task latency seems to decrease with the expansion of \acp{en} up to 12, whilst the communication delay remains steady at around 4-6ms. However, post this point, communication delays start to increase, signalling heightened network overhead. At 28 \acp{en}, a substantial communication overhead of 43ms is observed, significantly impairing system performance.

Two additional observations arise in this scenario. Firstly, the average task latency begins to increase parallel to the communication delay beyond 20 \acp{en}. This can be rationalized, given that the task number and laxity remain fixed, implying that the time tasks spend in queues should remain relatively consistent, and, hence, the observed increase is attributed to the communication delay.

In summary, the system performance appears to peak at around 20 \acp{en}, beyond which the overhead begins to significantly influence network functionality. Past this point, the detrimental effects of overhead become prominently visible, thereby affecting system performance and task scheduling in the network simulation. In practice, a network of offloading nodes would have to be segregated into multiple partitions at this point, as the number of nodes increases the number of information sharing transmissions exponentially.

\paragraph*{Scenario 4} The evaluation of this scenario is focused particularly on the performance under varying network conditions, specifically, the difference in task execution when one of the four \acp{en} was forcibly removed during the simulation as opposed to running the simulation with only three \acp{en} from the beginning.

Remarkably, the outcomes of the two executions were similar, with task acceptance rates registering 91\% (with node crash) and 89\% (without node crash) respectively, and no instances of deadline misses in either execution. 

It is important to note that the \ac{en} termination during the first execution led to the loss of only one task. Predictably, this was the single task in the process of execution on the \ac{en} under termination. Despite a few more offloading requests being directed to the terminated \ac{en} post hoc, the network detected the absence of a response and successfully managed to execute all the forwarded tasks on the still active \acp{en}.

\begin{figure}
\centering
\includegraphics[width=\textwidth]{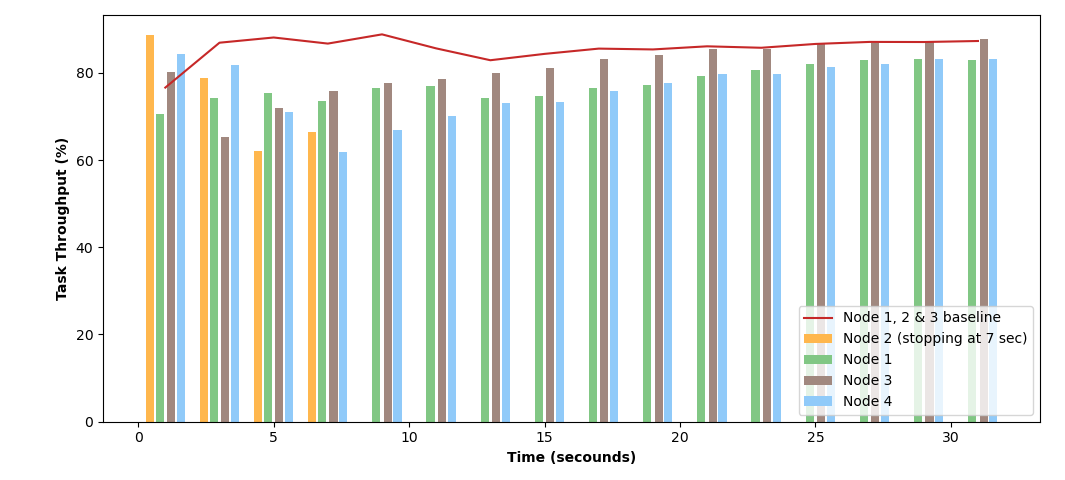}
\caption{\textbf{Scenario 4:} Node utilization chart.}
\label{fig:scenario5_util}
\end{figure}

The latency of task execution was observed to be comparable in both experiment runs. In addition, there was a noticeable increase in the average \ac{en} utilization, rising from 67\% to 87\%. Nonetheless, upon close inspection of \figurename\ref{fig:scenario5_util}, it can be observed that up until the seventh second the average utilization of the four \acp{en} (indicated with bars) is significantly lower compared to the baseline measurements with three \acp{en} from the start (indicated with the red line). After \ac{en} 2 (orange bar) is terminated, the average utilization of the other \acp{en} starts to rise until it reaches the baseline utilization.

\subsection{Comparison with Centralized Approach}

The implemented distributed scheduler eliminates the single point of failure. This attribute significantly improves network resilience by ensuring that the failure of a single entity does not compromise the entire system. Despite this, the centralized scheduler offers advantages in terms of zero communication overhead for the upkeep of the workers and for distributing tasks across multiple workers.

\paragraph*{Comparison - Scenario 1:} 
The acceptance rate in the centralized approach is roughly 0.82, 0.78, 0.74, 0.68 and 0.6 for 10-50 client devices respectively. Although numerically different from the ones observed in the distributed scheduler (cf.~\ref{fig:scenario1_accept}), the results are comparable due to the similar pattern of acceptance rate variation with the increased number of clients. The most noticeable difference is that the distributed scheduler commences at a much higher rate for 10 devices (namely 99\%), directly related to the fact that under such a low system load (average 24\%), the communication network is not overly loaded, rendering the simulated communication delays significantly lower than the ones assumed in the basic implementation.
A similar observation could be made for the deadline misses. Although we record a higher percentage of deadline misses (roughly 0.05, 0.07, 0.1, 0.12 and 0.15) with the centralized scheduler, it is indeed apparent that the relative percentage of deadline misses increases proportionally with the increased number of clients.

\paragraph*{Comparison - Scenario 2:} Similarly, the second scenario is designed to reflect the second scenario of the centralized approach. Here, we observed ratios of (0.95, 0.91, 0.85, 0.70, 0.35) for accepted tasks and (0.00, 0.01, 0.04, 0.06, 0.1) for deadline misses, for 10-50 clients respectively. When compared to the distributed scheduler results, although the range of absolute values in this scenario differs, the trend is similarly reflected – the percentage of accepted tasks starts very high for larger task laxities (also compared to scenario 1) and reduces significantly for stricter deadlines.
One significant difference, however, is the relative percentage of deadline misses. In the results observed, no deadline misses were observed in Scenario 2, except from a very moderate 0.3\% of the accepted tasks that missed their deadline in the execution with 50 clients. The results acquired with the central scheduler have a significantly larger portion of tasks that miss their deadline. This discrepancy could be explained by the considerably higher communication delay assumed – with a 30ms mean delay (one way), a 60ms mean task laxity, and a delay factor of 0.5 (compensating for the one-way communication), it is anticipated that some tasks will miss their deadline due to the divergence in the communication delay between the worker and the clients.

\section{Conclusion}
\label{sec:3_3_conclusion}
In this chapter, we investigated the trade-offs and implications of offloading real-time tasks over wireless networks to local edge resources in the context of IIoT environments.
On this basis, a system architecture was designed that integrates networking uncertainties.
Three interacting actors have been conceived:
(1) The client machines, which are mobile devices offloading tasks with deadlines.
(2) The scheduler, which creates a schedule based on the partitioned earliest deadline first algorithm, on which the tasks are finally executed.
(3) And the worker nodes that reside at the edge of the local network and execute the offloaded tasks as well as return their results to the clients. 
The focus here was the inclusion of an unreliable network in the creation of the schedule. 
This system was implemented and extensively tested.
Given the broadness of the proposed architecture optimizations and adaptations towards specialized use-cases can be applied to the scheduler and real-time networking concepts can be incorporated in future work. Scheduling algorithms could incorporate network bandwidth, memory consumption, and payload sizes. 

Additionally, the system design was extended to be fully distributed. Edge nodes were used to combine worker nodes and a scheduler each, where all nodes share the same system state information.  The resulting self-managing offloading network successfully demonstrates that such a distributed set-up can effectively mitigate the single point of failure problem of centralized schedulers, ensuring no tasks are lost in the event of sudden \ac{en} failure. Furthermore, such an event does not affect the overall performance of the system or its real-time guarantees as long as the number of participating edge nodes does not introduce too much additional communication. When scaling such a system further, the design needs to be segregated in multiple subnetworks.

\chapter{Related Work}
\minitoc
This chapter contains the related work concerning the contributions of this thesis. 
In Section~\ref{sec:2_6_measuring} works related to Chapter~\ref{cha:2_2_problem} concerning the measurement and detection of interrupt overloads are discussed. Section~\ref{sec:2_6_interrupts} we present work related to the same chapter but with focus on the management of interrupts in operating systems. Section~\ref{sec:2_6_processing} applies this topic to software solutions specific to network packet reception and is related to Chapter~\ref{cha:2_3_software}. Section~\ref{sec:2_6_hardware}, related to Chapter~\ref{cha:2_4_hardware} presents hardware solutions for interrupt and packet processing. 
Related work considering the real-time task offloading in Chapter~\ref{cha:2_5_codesign} is presented in Section~\ref{sec:2_6_offloading}.
Finally, Section~\ref{sec:2_6_surveys} presents surveys comparable to our Chapter~\ref{cha:1_3_survey}.

\section{Measuring and Detecting the Real-Time Impact of Interrupts}
\label{sec:2_6_measuring}
In Chapter~\ref{cha:2_2_problem} we present a quantitative analysis of the impact of interrupts on real-time performance followed by detection algorithms for this in Chapter~\ref{cha:2_3_software}. 
Although sophisticated mitigation approaches are being developed, only a few other works quantify the impact of \ac{irq} loads on real-time tasks.

In~\cite{regnier08evaluation}, two Linux-based \acp{rtos} are compared for real-time performance under high interrupt loads. The authors explicitly specify the \ac{irq} load caused with network packets (20 Hz) while measuring \ac{isr} latencies. Temporal variations were identiefied on a system with a multiple of processing capabilities compared to resource constrained embedded systems.
Furthermore, Regehr et al. present a collection of approximate \ac{irq} frequencies which might be caused by faulty components or network devices~\cite{regehr05preventing}.
However the authors focus on mitigating \ac{irq}-caused CPU overload through interrupt rate limitation strategies.
This is achieved by limiting the \ac{irq} frequencies with three strategies: either via a software-implement hard limit, a software-implemented limit which allows bursts, or a hardware-implemented hard limit.
The authors present the CPU load measured for an unloaded and loaded system each.
Josef Strnadel enhanced the concept from~\cite{regehr05preventing} with an improved hardware-implemented algorithm~\cite{strnadel12monitoring}.
In his work, the \ac{irq} rate limits are determined dynamically through an attached \ac{fpga} which monitors the overall system load.

In~\cite{niedermaier_you_2018} Niedermaier et. al. investigate the negative impact of packet flooding on the timing behavior in \acp{plc}.
Similar to our work in Chapter~\ref{cha:2_2_problem}, they find severe task delays caused by the load of unimportant packet flows.
They also considered port-scanning scenarios, which also measurably degraded task performance.
Ironically, the mere introduction of an \ac{rtos} is suggested as a possible solution.
However, \acp{rtos} apparently only enforces strict policies that have to take different flows into account in the first place.
Therefore, the only remedy would be to assign a background priority to the entire \acs{rx}-path.
Their second approach limits the total bandwidth of incoming traffic.
Both of these approaches are simple mitigation techniques that exhibit unsatisfactory connectivity behavior for given \ac{hp}-packet flows and their receiving processes.

Overload detection can be refined by using LSTMs and CNNs for traffic classification~\cite{jia2020flowguard}. Here, packets are passed through the closest edge server for execution of the models.

\paragraph*{All in all:} These works quantify or detect either the \ac{irq} load, the resulting system load, or both.
However, the employed software systems and the underlying hardware are complex and heterogeneous.
It would hence be implausible to use the available numbers to deduce the effects of a high \ac{irq} load on real-time tasks running on \ac{cots} embedded systems.

\section{Interrupt Management in Operating Systems}
\label{sec:2_6_interrupts}
Addressing the unwanted impact of interrupts is being researched since decades.
A popular example of unwanted effects is priority inversion, where the execution of a high-priority task is preempted by an \ac{isr} which belongs to a lower-priority task.
To understand today's common interrupt handling mechanisms, it is helpful to understand the evolution of those mechanisms.
The first part of this section will briefly outline implementations that pioneered concepts which are relevant up until today.
Thereafter, we present scientific works which address the impact of \acp{irq} on embedded real-time systems and applications.

\subsubsection*{Basic Interrupt Handling Mechanisms}

The \textit{LynxOS}\footnote{\url{https://www.lynx.com/products/lynxos-178-do-178c-certified-posix-rtos}} \ac{rtos} has introduced the concept of running re-entrant network protocol stacks in the priority spectrum of a receiver's process.
By de-multiplexing network traffic at the \ac{nic} level, low latency for processing real-time network traffic can be guaranteed. Simultaneously, priority inversions caused by packets directed at non-real-time processes is avoided.
LynxOS claims to be the first system built on an architecture that separates \acp{isr} and \acp{ist} (i.e., light weight kernel tasks)~\cite{bunnell95operating}.

The \textit{SUN} (now \textit{Oracle}) \textit{Solaris} \ac{os}~\cite{oracle19oracle} is another prominent example for implementing interrupt handling within the priority spectrum of regular process and thread scheduling.
Solaris distinguishes real-time, system, and time-sharing scheduling classes.
\acp{irq} which cannot be controlled by software, have the highest priority in the system.
Processes in the real-time class have a the second-highest priority spectrum, a time quantum, and they are scheduled strictly on the basis of these parameters.
As long as an \texttt{RT} process is ready to run, no \texttt{SYS} or \texttt{TS} process can run.
The system class exists to schedule the execution of special system processes, such as paging or \textit{streams} (i.e., the networking code).
Processes of the \texttt{SYS} class of processes have fixed priorities below \texttt{RT}.
Regular time-sharing processes are mapped into the lowest priority spectrum.
With these priority spectrums, Solaris was able to effectively isolate real-time processing from networking.

\textit{Windows Embedded Compact} (\textit{Windows CE})~\cite{microsoft13history}, is an \ac{os} family developed for handheld consumer electronics (CE) devices.
Windows CE up to version 5 was based on a microkernel design and has used a slot model for memory management.
Interrupt handling is divided among the kernel and a process running all device drivers.
Starting with version 3, Windows CE has been providing 256 priority levels for thread scheduling.
From the beginning, Windows CE has separated \acp{isr} and \acp{ist} for I/O and network processing.
By running network handlers with a priority lower than the one used for real-time tasks, Windows CE effectively is able to implement isolation and call admission for incoming network activities.

\paragraph*{All in all:} These \acp{os} served millions of industrial applications in practice and the influences of their concept can still be found today.
Nevertheless, applications, their surrounding environments and their underlying hardware have changed.
Nowadays, even entry-level hardware has complex performance features, such as multi-level caching, pipelining, pre-fetching and out-of-order execution.
Manufacturing is heavily cost-optimized and the transistor sizes decrease constantly.
All together, these developments complicate the timing predictability and existing knowledge needs to be confirmed or refined with up-to-date numbers.

\subsubsection*{Advanced Interrupt Management}

Prominent work regarding the unification of priority spaces is the approach implemented in the \textit{Sloth} OS~\cite{sloth}~\cite{sleepy_sloth}. Sloth implements a general abstraction for software threads and ISR, abolishing their distinction. Instead of using a software scheduler for threads, every control flow is designed as a thread-related system call using the hardware interrupt system. By letting the hardware manage all control flows, context switches have less overhead and --- which is more interesting here --- ISR and (other) threads preempt each other in accordance with their priorities. While this abolishes the problem of priority inversion, high packet loads still lead to high interrupt frequencies, impacting real-time tasks. This overall approach has later been further extended for performance~\cite{sloth_time} and security~\cite{safer_sloth}.

The priority inversion impact of interrupts in real-time systems has been identified and tackled by Amiri et. al. by employing priority inheritance protocols for interrupt service threads~\cite{amiri2015predictable}. This approach however only works for the schedulable part of interrupt handling of device drivers. 

Using interrupt moderation to relieve the CPU in high traffic scenarios is a method studied mainly for high throughput devices, as systems connected to Gigabit Ethernet networks are subject to potentially millions of packets per second~\cite{gebert2016performance}. However, some work also exists studying embedded devices running the Linux kernel. Spanos et al. evaluate the performance implications of advanced interrupt handling techniques in the "New API" Linux device driver extension~\cite{spanos2008internals}. 

\paragraph*{All in all:} The research described shows that there are promising solutions for an improved latency or efficiency of interrupt handling.
However, none of the examined papers focus on the impact of network \acp{irq} on real-time tasks and do not quantify the effects directly.
Considering that large monolithic \acp{os} such as Linux are used for realizing networked hard real-time systems~\cite{penaflor01real} and that there are real-time kernels specifically developed for wireless sensor networks~\cite{will09real}, this is a surprising observation and in our opinion a gap in research.

\section{Real-Time Aware Packet Processing}
\label{sec:2_6_processing}
This section presents past works on the reception and processing of network packets in real-time systems. It focuses on solutions implemented in software, e.g. the network driver and IP stack implementation.

\subsubsection*{Software Networking Subsystem}
The network stack architecture \emph{Lazy Receiver Processing} (LRP) introduced an important and widely used approach that is still of interest today~\cite{druschel1996lazy}. It improves performance, stability, and fairness on server systems with high incoming network throughput. Processing incoming packets can cause application processes to become persistently congested and unable to receive data. The packets must then be discarded while the applications continue to starve. 
This situation is prevented by early multiplexing of incoming packets, prioritized execution of the rest of the protocol stack in the context of the receiving process, and thus early discarding of packets on congested paths. This approach would benefit from hardware packet presorting and the overall interrupt reduction achieved by our multi-queue NIC (cf. Chapter~\ref*{cha:2_4_hardware}). This can prevent or mitigate throughput degradation as the packet rate increases. By consistently differentiating among network flows, they take an elegant approach that can also efficiently improve the real-time behavior of IoT devices.

Building on the idea of LRP, Lee et al. investigated reducing the impact of \ac{lp} packets on the real-time behavior of a network-independent task by introducing port-based prioritization of protocol processing~\cite{lee_interrupt_2010, lee_priority-based_2015}.
To accomplish this, they classify a UDP packet by its port in a \enquote{top half} interrupt handler.
By looking at a special port-priority table, whose data is obtained from all bound UDP sockets, its top half infers a priority for each packet.
In Linux's \texttt{softirq} scheduling unit, which belongs to the kernel, they introduce a gate functionality:
Packets are processed only as long as their priority is higher than the system's currently active priority.
Otherwise, they are delayed until at least the next regular \texttt{softirq} call.
Thus, they show how their modification is able to measurably reduce the latency of a long-running critical task.

The time-predictable IP stack tpIP~\cite{schoeberl_tpip_2018} addresses the challenge of real-time communication in cyber-physical systems. To enable timing predictability and \ac{wcet} analysis the proposed stack uses polling functions in the socket API with non-blocking read and write operations. While focusing on timing analysis and predictability, no measures are taken towards processing performance, interrupt scheduling or the issue of traffic overloads.

The Linux kernel provides advanced networking capabilities for routing and traffic control (tc)~\cite{hubert2002linux}. The latter provides mechanisms to control IP traffic such as traffic shaping and forwarding. Using queuing disciplines (qdiscs), incoming and outgoing packets can be queued for each network device. Traffic rates can be moderated and packets can be assigned priorities on the basis of meta information.  Similar to the approaches presented in this thesis, packets can be matched to certain descriptions and inserted to a qdisc depending on the packet classification. Recent work also shows that the novel Linux kernel features eBPF and EDT can be used to improve the scalability of filters in Linux-based networking~\cite{becker2022network}. However, due to the packet-proportional overhead, \acp{rtos} do not commonly provide similar tools.   

The approach of~\cite{closebutnotcloseenough} simulates control and background traffic to a hypothetical critical power plant control system that is exposed to the internet and as such a target of DoS attacks. 
The paper presents an interrupt-overload detection. However, its mitigation techniques have the same limitations as our Burst and Hysteresis mitigation presented in Section~\ref{sec:mitigations}. The high processing power requirements are addressed by putting a more powerful router between the network and the embedded devices. 

\subsubsection*{User-Level Stacks}

To facilitate the evolution of transport protocols, Honda et. al. advocate user-level stacks~\cite{honda2014userlevelstacks}.
By reducing the \ac{os} responsibilities to managing \ac{nic} sharing and packet multiplexing, these can be referenced as a library by each application independently.
While the code size overhead can be avoided by using shared libraries, the main challenge is to enable elegant and efficient multiplexing and to reduce the overhead caused by user/kernelspace transitions.
They implemented their \enquote{MultiStack} in Linux and then benchmarked it for evaluation.
Good performance was achieved through efficient early demuxing and the novel \texttt{netmap} framework~\cite{rizzo2012netmap}.
The latter allows user-space processes to directly access packet buffer memory by mapping through the respective memory regions.
In addition, similar to the newly introduced \texttt{io\_uring}-\acs{api}, it helps to amortize system call costs over many packets by mapping management data structures to userspace as well.
While their development considers more complex \acp{os}, which traditionally do packet handling in privileged kernel mode, and also does not consider protecting devices from overload, the proposed network stack architecture would work well with our overload shedding architecture.

\paragraph*{All in all:} 
With the works by Druschel et al.~\cite{druschel1996lazy} and Lee et al.~\cite{lee_interrupt_2010} elegant solutions towards real-time aware packet processing have been proposed.  
However, their implementation is limited by the inappropriate scheduling behavior of the \texttt{softirq} handler in Linux, which can't be pre-empted by even the most critical processes and is rescheduled in a similar way to polling, adding unnecessarily high network latency once packets are no longer being eagerly processed.
Also, their work only considers UDP packets.
Finally, when considering congestion scenarios, mere flow differentiation and prioritization is not sufficient to protect execution guarantees, since packets may also arrive at a highly prioritized task port in large quantities.

\section{Specialized Hardware}
\label{sec:2_6_hardware}
A class of approaches tries to mitigate the unwanted effects with additional hardware.
This hardware usually intercepts the \acp{irq} between their source and the CPU.

\subsubsection*{Interrupt Controllers}
The Advanced Interrupt Controller~\cite{task_aware} monitors the priority of the currently running process to determine if an interrupt should be triggered or held back by comparing it to the interrupt priority. A simple extension of the interrupt controller unifies the priority spaces of attached interrupts and operating system processes. However, this does not facilitate for the circumstances around network packets since interrupt priorities of all packets are the same and different packets cannot trigger interrupts of different priorities. 
In a proposal by Leyva-del-Foyo et al. an \ac{fpga} is used as a custom interrupt controller~\cite{leyvadelfoyo04custom} to realize dynamic priorities for the \ac{irq} lines.
Additionally, all \acp{isr} become \acp{ist}, which enables the \ac{fpga} to unify their synchronization and scheduling.
Scheler et al. present an enhancement to~\cite{leyvadelfoyo04custom} in concepts and hardware where a co-processor takes care of the \ac{irq} handling~\cite{scheler09parallel}.
Together with an extended conceptual approach, this solution allows for multiple task activation, multiple tasks per priority and stack sharing.

\subsubsection*{SmartNICs}
\label{sec:2_6_related_nic}
Katsikas et. al. and Döring et. al. have recently published papers that survey current trends in (smart) network interfaces~\cite{smartnics, doring2021smartnics}. According to these surveys, standard \acp{nic} cannot keep up with increasing network speeds and more complex network structures. As a result, \acp{nic} must adapt to increasing line rates, offload an increasing number of networking tasks, and provide greater flexibility to adapt to different applications. As a result, researchers and commercial players are increasingly interested in Smart\acp{nic} that are partially or fully implemented on FPGAs. Commercial examples and research reflect the above requirements as they are mainly focused on high performance applications in network servers, data centers and others.
Commercial examples can be found from many companies, including Intel, Nvidia, Cisco, and others. They generally target cloud, server, and data center solutions, offering data rates of 100Gb/s and higher, as well as high flexibility for virtualized network functions. Examples of commercial Smart\acp{nic} relevant to the problem of this thesis, i.e. for small embedded systems with real-time requirements, were not found during the research on related work.
The Intel PAC N3000, for example, fully implements a SmartNIC on FPGAs and is mainly promoted to telecommunication service providers. Similarly, Nvidia promotes its BlueField computing unit to data centers, but it is implemented as a combination of ASIC and FPGA components combined in one SoC.

Similar trends can be seen in research. Corundum~\cite{forencich2020corundum} is an open source SmartNIC prototyping platform for high-speed networking research, such as packet scheduling, congestion control, or new parallel network architectures. It can support over 10,000 queues for transmission scheduling. With Corundum, Forencich et al. focus on high line rates and offloading a wide range of tasks from the CPU, rather than targeting embedded real-time systems. 

In contrast, simpler platforms are available for research and education. RiceNIC implements an extensible Gigabit Ethernet NIC that can be used as a baseline for NIC prototyping~\cite{shafer2006reconfigurable}. Similarly, NetFPGA, first developed for educational purposes at Stanford University, is a platform for developing FPGA-based \acp{nic}, among other networking devices, and is increasingly being used for networking research~\cite{watson2006netfpga}.

Network Interface Controllers with multiple transmit and receive queues have been introduced by Intel as early as 2007. The goal is to make use of multicore systems by parallelizing network load on the different queues. The trend is to increase the number of queues to facilitate cloud computing as Zhu et. al. showed in 2020~\cite{zhu2020data}. Multiqueueing in general exists for different high-throughput I/O devices but to our best knowledge is not a common sight among real-time or embedded architectures.

Some modern \acp{nic} support serving received packets into multiple \ac{bd} rings~\cite{nics}. For once, this is useful to distribute packet reception work over multiple CPU cores in high-performance scenarios. The assignment of packets to these queues can also result from a hardware classification based on packet header fields, which may be configurable by the driver~\cite{shinde2013unicorn, tripathi2009crossbow}. However, this is a feature only found in advanced NIC hardware~\cite{honda2014userlevelstacks}.

Loom~\cite{stephens2019loom} is a multiqueue NIC design that moves per-flow scheduling decisions from the software network stack into the NIC. This way, high throughput and homogeneous policy enforcement can be guaranteed while also providing isolation in multi-tenant cloud data centers.

In~\cite{lonardo2015fpga} Lonardo et al. present an application specific NIC design run on FPGAs for high energy physics experiments. The design allows a remote \ac{dma} to CPU and GPU memories, relieving the OS from data transfer management, allowing real-time processing on data received over the network. 

Many approaches to solving the problem of high interrupt counts breaking real-time priorities include extending the system with additional hardware. The Peripheral Control Processor is a proposed co-processor that executes interrupts and remaps priorities to unify the priority space between tasks and interrupts~\cite{scheler09parallel}.

\paragraph*{All in all:} To mitigate unpredictable processing delays by reducing or moderaing the amount of interrupts, the hardware responsible for the interrupts has to be modified. The presented approaches show that there are solutions both from industry and academia. However, they either do not consider real-time systems with mixed priority processes and/or are high performance devices for datacenters designed for maximum (best-effort) networking.

\section{Distributed Real-Time Scheduling}
\label{sec:2_6_offloading}
In this section, we present works in the area of distributed real-time scheduling and task offloading.~\cite{lin2019computation} presents an extensive overview of the trend of shifting process offloading toward Edge Computing environments, including a review of different aspects of application partitioning, task allocation as well as resource management, and distributed execution. 
Directly addressing the actual problem of process offloading, the work in~\cite{wang2019edge} presents an article about key issues and methods of the problem and adopts a model for studying and categorizing different aspects of it. Existing research is summarized to draw a big picture.

\subsubsection*{Scheduling Architectures}
Shukla and Munir~\cite{shukla_efficient_2017} present an offloading architecture for IoT devices. Computers on the same local network as the IoT device process data-intensive tasks while meeting their real-time deadlines.

Liu et al. address the problem of offloading data- and compute-intensive computations for vehicular networks~\cite{liu_real-time_2020}. Using a fog/cloud architecture, tasks are classified and distributed according to latency requirements. Roadside units as well as other vehicles can act as compute nodes. They formulate a real-time task offloading model that maximizes the task service ratio and propose a real-time task offloading algorithm that works cooperatively.

In~\cite{dist_sched_1985}, the authors present a system they developed for distributed dynamic scheduling of real-time tasks with hard deadlines.
The system consists of an acceptance checking algorithm that determines whether a task can be guaranteed to meet its deadline.
In addition, a distributed scheduling algorithm was developed that can preempt tasks while respecting deadlines and taking into account various overheads such as communication overhead.
The authors assume a rather loosely connected network of nodes, where tasks can arrive at any node.
The actual scheduling is done in a distributed manner, i.e., if the node at which a task arrives cannot meet the deadline, the nodes communicate directly with each other to find a suitable node to execute the task.
Similarities to the architecture presented in this thesis lie in the chosen distributed real-time scenario and that different forms of overhead are explicitly considered in the scheduling algorithm.

In~\cite{relwork_1} a procedure is also presented in which real-time tasks are scheduled on a distributed system connected by a CAN bus.
Here, however, not only the CPU time is scheduled, but also the network bandwidth with the help of a constant bandwidth server.
In this way, an end-to-end delay for task execution can be derived, which in turn provides very reliable guarantees that deadlines will be met.

The authors of~\cite{dist_sched_1994} also address dynamic scheduling of tasks with end-to-end time constraints on distributed hardware.
Tasks may have precedence constraints in their scenario, which are resolved offline in the form of pseudo deadlines using a specially developed algorithm.
Tasks can be generated on any node in the network and are distributed throughout the system by a local scheduler.
The local schedulers interact with each other to reserve the necessary resources for task execution. 
The entire system, consisting of network, nodes, and software, is assumed to be fault-free and predictable.

\subsubsection*{Offloading Decisions}

Chen et al.~\cite{iiot_offloading_energy} use fog computing for energy-optimal dynamic computation offloading in IIoT environments.
Computation time, energy, and resource utilization are considered and optimized. Real-time guarantees are not taken into account. Similarly, Elashri and Azim~\cite{elashri_energy-efficient_2020} use offloading of real-time tasks to cloud and fog resources for energy efficiency. They propose two algorithms for making an efficient offloading decision for soft and weakly hard (firm) real-time applications while guaranteeing task schedulability.

Ma et al.~\cite{ma_reliability_2022} identify a trade-off problem between reliability and latency in IIoT applications.
In their work, they propose an offloading framework for visual applications that considers both performance metrics.
A scheme is presented that uses and maximizes a utility function that balances reliability and latency.

Chen, Wang, and Lee deal with offloading tasks in the area of mobile edge cloud computing~\cite{mobile_edge_offloading}.
With respect to energy consumption, an approach is developed and evaluated to make the most energy-efficient offloading decisions.
Centralized and distributed greedy maximum scheduling algorithms are introduced and evaluated.
However, the offloaded tasks are not real-time critical.

A concrete use-case of process offloading in real-time computing environments is shown in~\cite{nimmagadda2010real}, which presents real-time tracking and recognition of moving objects, utilizing computation offloading. The work includes an estimation of involved computation as well as communication and incorporates the development of an offloading decision frame work for dividing computations between the server and robots.

\paragraph*{All in all:} The amount of the related work presented shows, that even though the offloading of tasks from computationally weak systems to distributed systems is an old topic the issue has been revived in recent years. Both, the real-timeness of tasks as well as the mobility of devices have given the field new challenges to master. Our contribution to this field in Chapter~\ref{cha:3_3_offloading} proposes a simple offloading architecture and scheduler. It considers networking uncertainties while assuming relatively simple network designs in delimited locations such as smart factories.

\section{Industrial Networking Surveys}
\label{sec:2_6_surveys}
In Chapter~\ref{cha:1_3_survey} we present a survey of real-time networking technologies. Several other works have looked at network requirements for industrial use cases. In the following, we present a selection of surveys from the past years that compare and analyze industrial networking solutions.

Xu et al.~\cite{xu2018survey} surveyed IIoT research from a systems perspective in 2018. In their work, they divide the wide research area into the three key aspects of control, networking, and computing before categorizing and investigating related works. While incorporating networking they do not cater to any real-time requirements. A more real-time focused survey was presented by Kim et al.~\cite{kim2017survey} in 2017.
In their work, the authors analyzed real-time communication in wireless sensor networks, which are in large part relevant to IIoT scenarios. Another survey on IIoT research was presented by Sisinni et al.~\cite{sisinni2018industrial} in 2018 with a focus on open challenges. Next to the need for energy efficiency, coexistence, interoperability, and security they also identify real-time performance as a future research challenge. 

Park et al.~\cite{park_wireless_2018} presented requirements and challenges with using wireless networks in control applications. In particular, the focus lies on network design and relevant protocols.
Key benefits such as ease of installation and reduced operating costs are discussed. Several use cases, both relevant to current technologies (e.g. building automation, tire pressure sensors, wireless sensor coverage in industrial applications) and more futuristic examples where the avionics are brought out as an example where reducing cabling can provide 2-5\% weight reduction alone.
As wireless networks cannot provide the same isolation from interference as corresponding wired technologies, control loops must be adapted to tolerate packet loss and variable delays.
A joint design approach where control loops and the wireless network are jointly optimized to arrive at an acceptable performance is presented as one promising approach to improve the performance of wireless control systems.

In~\cite{zunino_factory_2020}, Zunino et al. investigate the status of network technologies relevant to the Industrial Internet, or Industry 4.0.
They find that although most technologies needed to realize Industry 4.0 are present, they fall short on flexibility, scalability, reliability, real-time behavior, and security, to name a few.
A key point is to realize that "Industry 4.0" is a reaction to the change in market demands especially seen in developed countries.
A shorter time to market, a higher degree of customization, and small production batches are brought forward as key enablers to stay competitive.
To realize this, production lines must be highly configurable, have a high degree of situational awareness, and be extremely energy efficient.
This manifests itself as a need for the total system to be highly flexible, i.e. reorganization of tool placements, production cadence, and overall component flow. In turn, this requires the supporting communications network to be highly configurable and flexible to meet all the different demands.
With the pervasive use of sensors that tie directly into control systems, logistics, and overall planning as a central tenet of what is known as "Industry 4.0", the network must be scalable and accommodate a large number of devices.
Where current OT systems are tailored specifically for a particular task and hard real-time requirements are met by careful analysis and provisioning, future networks cannot be designed with such static scenarios in mind.
The communications networks must be able to provide real-time guarantees to a subset of nodes whilst also providing other QoS levels for various nodes, sensors, and systems.

Chen et al.~\cite{chen_wireless_2021} argue how smart manufacturing must use wireless technologies to orchestrate multi-robot systems to obtain the needed flexibility.
The third industrial revolution improved the flexibility of the \emph{quantity} produced, with the fourth, the flexibility of type, quantity, and quality being at the center of attention.
To change the production setup in such fundamental ways means that multi-robot systems must be highly configurable, the network must adapt to varying loads, and automated delivery systems like AGVs must be just as adaptable.
URLLC networks are a must, and to support such low-latency demands, SDN and NFV are essential in the network architecture to handle roaming, error correction, edge computing, etc.

\chapter{Conclusion}
Over the last decade, new technologies in communications, artificial intelligence, and big data engineering have enabled a variety of new operational scenarios where embedded devices with real-time requirements need to be connected to packet-switched networks. Examples include the convergence of IT and OT in smart factories and inter- and intra-communication in autonomous vehicles. This thesis presents work on the challenges of connecting computationally weak real-time devices to these networks.

In a review of real-time communication technologies, we survey Industry 4.0 use case scenarios of the foreseeable future and classify them according to their networking and real-time requirements. In addition, we review existing standards and technologies used for real-time communication on the industrial plant floor and evaluate the requirements of the scenario classes against them. By reviewing recent research from academia and industry, we identify current trends and gaps. The review shows that while there is much interest in IIoT architectures and communication technologies, there is no real-time hardened communication standard for wireless mobile machines. In addition, there is almost no research on the real-time embedded devices used in highly connected scenarios. Devices and operating systems seem to be extended with IP and wireless communication capabilities without proper integration into real-time scheduling.

Since network packet reception happens arbitrarily, its required processing workload, namely \ac{irq}, \ac{isr}, and driver executions, is unpredictable and depends on the incoming packet load. We devise a method to analyze the timing behavior of real-time tasks with different priorities under multiple loads that vary in volume and receiver task priorities. The systems under test, two computationally weak \ac{cots} \ac{iot} microcontrollers running FreeRTOS, were shown to prioritize not only unscheduled \acp{irq} and \acp{isr}, but also network protocol processing over real-time tasks. The workload induced by incoming packets increases proportionally with the packet load, consuming up to 6.67\% CPU capacity per packet per second under the worst configuration. A preempted real-time task loses this processing time to meet its deadline.

The problem can be addressed either by detecting this implicit priority inversion and responding with load limiting, or by designing a real-time priority aware network subsystem on the embedded device. Using metrics such as the network queue fill state, we detect risky loads. We propose four techniques to mitigate the problem through rate limiting.

In a second software-based approach, we propose a novel network driver for embedded real-time systems that demultiplexes packets as they arrive and prioritizes them based on their receiving process. We combine this with subsequent per-flow aperiodic scheduling. By instrumenting existing embedded IP stacks, strict prioritization with minimal latency is implemented without the need for additional task resources. On a per-flow basis, simple mitigation techniques are applied that cause barely measurable overhead while protecting the system from congestion. Our IP stack adaptation can reduce the processing time of low priority packets by over 86\% compared to an unmodified stack. This allows the network subsystem to remain active at 7 times the general traffic load before disabling the receive \acp{irq} as a last resort to ensure deadlines.

The proposed software mitigation techniques prove useful in protecting real-time tasks from network congestion, but must eventually shut down the network connection when the \acp{irq} load alone becomes too high. The only method to further mitigate the load is to moderate the amount of interrupts generated by the hardware network interface controller. A real-time aware multi-queue \ac{nic} is designed and evaluated. It pre-orders incoming packets into different queues and moderates interrupts based on the requirements of the running real-time task. In this way, the amount of interrupts can be reduced while ensuring the timely arrival of critical packets to high-priority tasks. In addition, packets that do not belong to a real-time task can be filtered before any work is generated.

In an attempt to combine the software- and hardware-based approaches, a hardware-software co-design for a custom network interface controller and corresponding driver is developed. The central goal of the design is to address and mitigate the problems real-time systems face when connected to IP networks. Advantageous aspects of both solutions are combined with new techniques developed in this thesis into a holistic design.
A working implementation developed in parallel with this thesis is presented and outlined, providing an example of how such a combined software/hardware approach can be realized in practice. Furthermore, the implementation is used in an experiment to provide empirical evidence of its effectiveness. The results show that the implementation prevents priority inversion and succeeds in providing predictable resources for high-priority tasks even under high load.

Offloading is a common method to address the resource and performance limitations of networked embedded devices, which are becoming more prevalent in industrial environments. It involves transferring computationally intensive tasks to a more powerful device on the network, often located nearby and using wireless communications. As mentioned earlier, devices used in most industrial scenarios operate under real-time constraints. Offloading real-time tasks over a wireless network with latency uncertainties presents challenges.
We seek to better understand these challenges by proposing a system architecture and scheduler for real-time task offloading in wireless IIoT environments. Our design was able to prevent missed deadlines under heavy load and network uncertainties, and outperformed a reference scheduler in terms of successful task throughput. Specifically, under high task load, where the reference scheduler had a success rate of 5\%, our design achieved a success rate of 60\%. In addition, the design is extended to be fully distributed, eliminating any single points of failure.

\section{Implications, Limitations \& Future Work}
Connecting devices to large networks naturally involves the risk a gateway to the outside entails.
The severity of this risk depends on the environment being considered. For example, IoT devices connected directly to the internet are much more vulnerable to packet flooding than embedded systems used in closed factory networks.
The second axis of risk is real-time criticality. Even if a device is located in a secured factory network, a failure of the braking routine of a moving robot is highly critical and must be prevented by decoupled safety systems.
With the contributions of this thesis, we have been able to reduce the risks to real-time properties at the embedded device level. We have shown that real-time aware reduction of network generated interrupts and dynamic prioritization of network processing can protect time-critical processes. 

At the same time, we are cautious about adding IP networking to critical real-time devices and emphasize that, in practice, mission-critical real-time applications should be designed holistically and secured with redundancies to prevent catastrophic damage. Therefore, the survey in Chapter \ref{cha:1_3_survey} is specific to manufacturing and smart factory scenarios, comparing current research (prototypes) with mature standards and practices from critical industrial applications. The surveyed works as well as the methods and designs presented in this thesis can only be seen as proofs of concept before intensive testing, standardization and integration.

We propose interrupt and rate limiting methods that can be subject to misconfiguration. They perform best when configured with the most realistic network load metrics, which requires knowledge of the running processes and their communication behavior. Future research in this area therefore includes methods for automatic and adaptive parameter configuration: On the one hand, parameter inference from application- and environment-specific expected network loads needs to be supported by an appropriate tool, and on the other hand, autonomous reconfiguration methods benefit our approaches in more dynamic environments.

In Chapter \ref{cha:3_3_offloading} we present a system architecture for offloading real-time tasks to local edge nodes. Future work should extend this system to enable more computational layers, such as eBPF-based low-latency responses and cloud computing resources for best-effort tasks. The management of mixed criticalities and deadline sizes is a challenge that remains to be solved. In addition, enabling embedded real-time devices to make run-time offload decisions based on available computational resources needs more attention for better integration into distributed offloading systems.

\backmatter
\markboth{Bibliography}{bibliography}
\addcontentsline{toc}{chapter}{Bibliography}
\label{cha:bibliography}
\begin{singlespace}
  \printbibliography
  \listoffigures

  \renewcommand{\chaptermark}[1]{\markboth{#1}{#1}}
\chapter*{Acronyms}
\addcontentsline{toc}{chapter}{Acronyms}
\chaptermark{Acronyms}
\begin{acronym}
    \acro{ap}[AP]{Access Point}
    \acro{api}[API]{Application Programming Interface}
    \acro{arp}[ARP]{Address Resolution Protocol}
    \acro{axi}[AXI]{Advanced Extensible Interface}
    \acro{bd}[BD]{{Buffer Descriptor}}
    \acro{cots}[COTS]{{Commercial Off-The-Shelf}}
    \acro{cpu}[CPU]{Central Processing Unit}
    \acro{dhcp}[DHCP]{Dynamic Host Configuration Protocol}
    \acro{dma}[DMA]{{Direct Memory Access}}
    \acro{dns}[DNS]{Domain Name System}
    \acro{dsa}[DSA]{Distributed Scheduling Algorithm}
    \acro{edf}[EDF]{Earliest Deadline First}
    \acro{en}[EN]{Edge Node}
    \acro{fpga}[FPGA]{{Field-Programmable Gate Array}}
    \acro{fsm}[FSM]{Finite State Machine}
    \acro{hp}[HP]{{High Priority}}
    \acro{icmp}[ICMP]{Internet Control Message Protocol}
    \acro{iiot}[IIoT]{Industrial Internet of Things}
    \acro{iot}[IoT]{{Internet of Things}}
    \acro{ip-core}[IP Core]{Interlectual Property Core}
    \acro{ip}[IP]{Internet Protocol}
    \acro{irq}[IRQ]{{Interrupt Request}}
    \acro{isr}[ISR]{{Interrupt Service Routine}}
    \acro{ist}[IST]{{Interrupt Service Task}}
    \acro{it}[IT]{Information Technology}
    \acro{lan}[LAN]{Local Area Network}
    \acro{lp}[LP]{{Low Priority}}
    \acro{lwip}[LwIP]{Lightweight IP}
    \acro{mac}[MAC]{{Medium Access Control}}
    \acro{mcu}[MCU]{{Microcontroller Unit}}    
    \acro{mmio}[MMIO]{Memory-mapped I/O}
    \acro{mmu}[MMU]{Memory Management Unit}
    \acro{mq-dsppe}[MQ-DSPPE]{Ethernet Multi-Queue Downstream Packet Processing Engine}
    \acro{nic}[NIC]{{Network Interface Controller}}
    \acro{os}[OS]{{Operating System}}
    \acro{ot}[OT]{Operational Technology}
    \acro{phy}[PHY]{{Physical Layer}}
    \acro{plc}[PLC]{Programmable Logic Controller}
    \acro{posix}[POSIX]{Portable Operating System Interface}
    \acro{ps}[PS]{Processing Subsystem}
    \acro{qos}[QOS]{{Quality of Service}}
    \acro{ram}[RAM]{Random Access Memory}
    \acro{rms}[RMS]{Rate-Monotonic Scheduling}
    \acro{rtos}[RTOS]{{Real-Time Operating System}}
    \acro{rxd}[RXD]{Receive Data Stream Bus}
    \acro{rxs}[RXS]{Receive Status Stream Bus}
    \acro{rx}[RX]{{Receive}}
    \acro{soc}[SoC]{System-on-Chip}
    \acro{son}[SON]{Self-Organized Network}
    \acro{spof}[SPOF]{Single Point of Failure}
    \acro{tcp}[TCP]{Transmission Control Protocol}
    \acro{tsn}[TSN]{Time-Sensitive Networking}
    \acro{udp}[UDP]{User Datagram Protocl}
    \acro{wcet}[WCET]{{Worst-Case Execution Time}}
\end{acronym}

\end{singlespace}
\end{document}